\documentclass[sigconf,balance=false]{acmart}

\usepackage{booktabs}
\usepackage{float}
\usepackage{multirow}
\usepackage{array}
\usepackage{colortbl}   %
\usepackage{graphicx}
\usepackage{xcolor}
\usepackage{algorithm}
\usepackage{algorithmic}
\usepackage{subcaption}
\usepackage{enumitem}
\usepackage{xspace}
\usepackage{pifont}
\usepackage{tikz}
\usetikzlibrary{positioning,arrows.meta,fit,calc,backgrounds,shapes.misc,decorations.pathreplacing,matrix}
\colorlet{semC}{blue!55!black}      \colorlet{semF}{blue!6}
\colorlet{latC}{violet!60!black}    \colorlet{latF}{violet!6}
\colorlet{strC}{orange!70!black}    \colorlet{strF}{orange!9}
\colorlet{goldC}{yellow!75!orange}  \colorlet{goldD}{yellow!35!black}
\colorlet{winC}{teal!65!black}      %
\colorlet{docF}{black!7}            \colorlet{docD}{black!30}
\colorlet{okC}{green!45!black}      \colorlet{badC}{red!70!black}
\colorlet{inkL}{black!60}           \colorlet{tradC}{black!55}
\tikzset{
  paperfig/.style={>={Stealth[length=1.7mm,width=1.3mm]}, line width=0.5pt, font=\footnotesize},
  box/.style={draw=black!45, rounded corners=1.5pt, align=center, inner sep=3pt, fill=white},
  keybox/.style={box, draw=black!80, line width=0.75pt, fill=black!2},
  semB/.style={box, draw=semC, fill=semF}, latB/.style={box, draw=latC, fill=latF}, strB/.style={box, draw=strC, fill=strF},
  lbl/.style={font=\scriptsize, text=inkL, inner sep=1pt, align=center},
  flow/.style={->, black!55}, main/.style={->, black!80, line width=0.7pt},
  grp/.style={draw=black!25, rounded corners=3pt, dashed, dash pattern=on 1.5pt off 1.2pt, inner sep=3.5pt},
  panel/.style={draw=black!18, rounded corners=4pt, fill=black!1.5, inner sep=5pt},
  doc/.style={draw=docD, fill=docF, minimum width=2.6mm, minimum height=3.4mm, inner sep=0pt, line width=0.35pt},
  gold/.style={draw=goldD, fill=goldC, minimum width=2.6mm, minimum height=3.4mm, inner sep=0pt, line width=0.45pt},
  badge/.style={circle, draw=black!75, fill=white, font=\scriptsize\bfseries, inner sep=0.6pt, minimum size=3.6mm},
  good/.style={text=okC, font=\scriptsize\bfseries}, bad/.style={text=badC, font=\scriptsize\bfseries},
  win/.style={draw=black!85, line width=0.9pt, rounded corners=1.5pt},
}

\colorlet{lrC}{blue!55!black}      %
\colorlet{csC}{violet!60!black}    %
\definecolor{lrC}{HTML}{0F8ABA}   %
\definecolor{csC}{HTML}{7C6CF2}   %
\newcommand{\reroute}{{Link Rerouting}\xspace}
\newcommand{\chainctx}{{Chain-aware Context Selection}\xspace}

\DeclareRobustCommand{\secref}[1]{\hyperref[#1]{\S\ref*{#1}}}
\DeclareRobustCommand{\tabref}[1]{\hyperref[#1]{Table~\ref*{#1}}}
\DeclareRobustCommand{\figref}[1]{\hyperref[#1]{Figure~\ref*{#1}}}
\DeclareRobustCommand{\appref}[1]{\hyperref[#1]{Appendix~\ref*{#1}}}
\DeclareRobustCommand{\eqnref}[1]{\hyperref[#1]{Equation~(\ref*{#1})}}
\definecolor{refC}{HTML}{E8590C}   %
\definecolor{citeC}{HTML}{E64980}  %
\definecolor{gainC}{HTML}{6B7280}  %
\definecolor{detourC}{HTML}{7C6CF2}%
\AtEndPreamble{\hypersetup{colorlinks=true,linkcolor=refC,citecolor=citeC,urlcolor=refC,filecolor=refC}}
\newcommand{\method}{\textsc{DetouR}\xspace}
\theoremstyle{acmplain}

\newcommand{\tbd}[1]{\textcolor{red}{#1}}   %
\DeclareRobustCommand{\pnum}[1]{\ifcsname pn@#1\endcsname\csname pn@#1\endcsname\else\tbd{[pending]}\fi}
\expandafter\def\csname pn@all/hub/ctrl/range\endcsname{26--39}
\expandafter\def\csname pn@all/hub/detour/range\endcsname{16--21}
\expandafter\def\csname pn@all/rerankcalls/range\endcsname{24--36}
\expandafter\def\csname pn@assoc/e0/0/anchor@50\endcsname{5.7}
\expandafter\def\csname pn@assoc/e0/0/cf@100\endcsname{0.0}
\expandafter\def\csname pn@assoc/e0/0/cf@50\endcsname{0.0}
\expandafter\def\csname pn@assoc/e0/0/llm@50\endcsname{2.6}
\expandafter\def\csname pn@assoc/e0/0/n\endcsname{453}
\expandafter\def\csname pn@assoc/e0/0/pop@100\endcsname{0.0}
\expandafter\def\csname pn@assoc/e0/0/pop@50\endcsname{0.0}
\expandafter\def\csname pn@assoc/e0/0/sem_dense@50\endcsname{2.0}
\expandafter\def\csname pn@assoc/e0/1-2/anchor@50\endcsname{5.9}
\expandafter\def\csname pn@assoc/e0/1-2/cf@100\endcsname{3.2}
\expandafter\def\csname pn@assoc/e0/1-2/cf@50\endcsname{1.3}
\expandafter\def\csname pn@assoc/e0/1-2/llm@50\endcsname{3.0}
\expandafter\def\csname pn@assoc/e0/1-2/n\endcsname{831}
\expandafter\def\csname pn@assoc/e0/1-2/pop@100\endcsname{0.0}
\expandafter\def\csname pn@assoc/e0/1-2/pop@50\endcsname{0.0}
\expandafter\def\csname pn@assoc/e0/1-2/sem_dense@50\endcsname{2.6}
\expandafter\def\csname pn@assoc/e0/10-29/anchor@50\endcsname{6.8}
\expandafter\def\csname pn@assoc/e0/10-29/cf@100\endcsname{39.1}
\expandafter\def\csname pn@assoc/e0/10-29/cf@50\endcsname{24.5}
\expandafter\def\csname pn@assoc/e0/10-29/llm@50\endcsname{7.6}
\expandafter\def\csname pn@assoc/e0/10-29/n\endcsname{2{,}022}
\expandafter\def\csname pn@assoc/e0/10-29/pop@100\endcsname{0.0}
\expandafter\def\csname pn@assoc/e0/10-29/pop@50\endcsname{0.0}
\expandafter\def\csname pn@assoc/e0/10-29/sem_dense@50\endcsname{2.8}
\expandafter\def\csname pn@assoc/e0/3-9/anchor@50\endcsname{5.6}
\expandafter\def\csname pn@assoc/e0/3-9/cf@100\endcsname{16.5}
\expandafter\def\csname pn@assoc/e0/3-9/cf@50\endcsname{8.6}
\expandafter\def\csname pn@assoc/e0/3-9/llm@50\endcsname{6.8}
\expandafter\def\csname pn@assoc/e0/3-9/n\endcsname{2{,}054}
\expandafter\def\csname pn@assoc/e0/3-9/pop@100\endcsname{0.0}
\expandafter\def\csname pn@assoc/e0/3-9/pop@50\endcsname{0.0}
\expandafter\def\csname pn@assoc/e0/3-9/sem_dense@50\endcsname{2.7}
\expandafter\def\csname pn@assoc/e0/30+/anchor@50\endcsname{9.9}
\expandafter\def\csname pn@assoc/e0/30+/cf@100\endcsname{75.1}
\expandafter\def\csname pn@assoc/e0/30+/cf@50\endcsname{61.1}
\expandafter\def\csname pn@assoc/e0/30+/llm@50\endcsname{12.3}
\expandafter\def\csname pn@assoc/e0/30+/n\endcsname{1{,}431}
\expandafter\def\csname pn@assoc/e0/30+/pop@100\endcsname{86.2}
\expandafter\def\csname pn@assoc/e0/30+/pop@50\endcsname{52.3}
\expandafter\def\csname pn@assoc/e0/30+/sem_dense@50\endcsname{5.3}
\expandafter\def\csname pn@assoc/e0/all/anchor@50\endcsname{6.9}
\expandafter\def\csname pn@assoc/e0/all/cf@100\endcsname{32.8}
\expandafter\def\csname pn@assoc/e0/all/cf@50\endcsname{22.9}
\expandafter\def\csname pn@assoc/e0/all/llm@50\endcsname{7.5}
\expandafter\def\csname pn@assoc/e0/all/n\endcsname{6{,}791}
\expandafter\def\csname pn@assoc/e0/all/pop@100\endcsname{18.2}
\expandafter\def\csname pn@assoc/e0/all/pop@50\endcsname{11.0}
\expandafter\def\csname pn@assoc/e0/all/sem_dense@50\endcsname{3.2}
\expandafter\def\csname pn@assoc/e0/midshare\endcsname{49}
\expandafter\def\csname pn@assoc/e0/tailshare\endcsname{19}
\expandafter\def\csname pn@assoc/onehop/2wiki/IRCoT\endcsname{39}
\expandafter\def\csname pn@assoc/onehop/2wiki/IRCoT/pool\endcsname{22}
\expandafter\def\csname pn@assoc/onehop/2wiki/ReAct\endcsname{44}
\expandafter\def\csname pn@assoc/onehop/2wiki/ReAct/pool\endcsname{20}
\expandafter\def\csname pn@assoc/onehop/2wiki/SF\endcsname{79}
\expandafter\def\csname pn@assoc/onehop/2wiki/SF/pool\endcsname{30}
\expandafter\def\csname pn@assoc/onehop/2wiki/SR1\endcsname{60}
\expandafter\def\csname pn@assoc/onehop/2wiki/SR1/pool\endcsname{21}
\expandafter\def\csname pn@assoc/onehop/2wiki/poolrange\endcsname{20--30}
\expandafter\def\csname pn@assoc/onehop/2wiki/range\endcsname{39--78}
\expandafter\def\csname pn@assoc/onehop/hotpotqa/IRCoT\endcsname{53}
\expandafter\def\csname pn@assoc/onehop/hotpotqa/IRCoT/pool\endcsname{27}
\expandafter\def\csname pn@assoc/onehop/hotpotqa/ReAct\endcsname{55}
\expandafter\def\csname pn@assoc/onehop/hotpotqa/ReAct/pool\endcsname{28}
\expandafter\def\csname pn@assoc/onehop/hotpotqa/SF\endcsname{74}
\expandafter\def\csname pn@assoc/onehop/hotpotqa/SF/pool\endcsname{42}
\expandafter\def\csname pn@assoc/onehop/hotpotqa/SR1\endcsname{63}
\expandafter\def\csname pn@assoc/onehop/hotpotqa/SR1/pool\endcsname{25}
\expandafter\def\csname pn@assoc/onehop/hotpotqa/poolrange\endcsname{25--42}
\expandafter\def\csname pn@assoc/onehop/hotpotqa/range\endcsname{53--73}
\expandafter\def\csname pn@assoc/reach/2wiki/assoc\endcsname{83.9}
\expandafter\def\csname pn@assoc/reach/2wiki/req\endcsname{61.2}
\expandafter\def\csname pn@assoc/reach/2wiki/sem\endcsname{69.4}
\expandafter\def\csname pn@assoc/reach/2wiki/union\endcsname{88.1}
\expandafter\def\csname pn@assoc/reach/hotpotqa/assoc\endcsname{85.8}
\expandafter\def\csname pn@assoc/reach/hotpotqa/req\endcsname{66.6}
\expandafter\def\csname pn@assoc/reach/hotpotqa/sem\endcsname{76.7}
\expandafter\def\csname pn@assoc/reach/hotpotqa/union\endcsname{91.1}
\expandafter\def\csname pn@assoc/reach/ndr/assoc\endcsname{19.0}
\expandafter\def\csname pn@assoc/reach/ndr/req\endcsname{0.1}
\expandafter\def\csname pn@assoc/reach/ndr/sem\endcsname{1.7}
\expandafter\def\csname pn@assoc/reach/ndr/union\endcsname{20.5}
\expandafter\def\csname pn@assoc/reach/trec_tot/assoc\endcsname{11.3}
\expandafter\def\csname pn@assoc/reach/trec_tot/req\endcsname{10.0}
\expandafter\def\csname pn@assoc/reach/trec_tot/sem\endcsname{19.3}
\expandafter\def\csname pn@assoc/reach/trec_tot/union\endcsname{24.7}
\expandafter\def\csname pn@assoc/rec/cfvotes\endcsname{3.7}
\expandafter\def\csname pn@assoc/rec/cfvotes/hit\endcsname{46}
\expandafter\def\csname pn@assoc/rec/expand\endcsname{18.0}
\expandafter\def\csname pn@assoc/rec/llmstatic\endcsname{1.8}
\expandafter\def\csname pn@assoc/rec/llmstep\endcsname{2.2}
\expandafter\def\csname pn@assoc/rec/oracle/assoc\endcsname{19.0}
\expandafter\def\csname pn@assoc/rec/oracle/sem\endcsname{1.7}
\expandafter\def\csname pn@assoc/rec/random\endcsname{1.5}
\expandafter\def\csname pn@assoc/rec/ratio\endcsname{2.8}
\expandafter\def\csname pn@assoc/rec/react\endcsname{1.0}
\expandafter\def\csname pn@assoc/rec/req\endcsname{0.1}
\expandafter\def\csname pn@assoc/rec/semfirst\endcsname{1.3}
\expandafter\def\csname pn@assoc/rec/semfirst/hit\endcsname{19}
\expandafter\def\csname pn@assoc/spearman/absmax\endcsname{0.08}
\expandafter\def\csname pn@assoc/spearman/range\endcsname{$-0.08$ to $-0.01$}
\expandafter\def\csname pn@hot/EM/Decomp/detour\endcsname{37.3}
\expandafter\def\csname pn@hot/EM/Decomp/droprepeats\endcsname{36.0}
\expandafter\def\csname pn@hot/EM/Decomp/interleave\endcsname{35.5}
\expandafter\def\csname pn@hot/EM/Decomp/named\endcsname{34.9}
\expandafter\def\csname pn@hot/EM/Decomp/novelquery\endcsname{36.1}
\expandafter\def\csname pn@hot/EM/Decomp/randomroute\endcsname{36.1}
\expandafter\def\csname pn@hot/EM/Decomp/reranker\endcsname{36.6}
\expandafter\def\csname pn@hot/EM/Decomp/rerankertops\endcsname{35.9}
\expandafter\def\csname pn@hot/EM/Decomp/rerankerwin\endcsname{36.6}
\expandafter\def\csname pn@hot/EM/Decomp/rewired\endcsname{36.2}
\expandafter\def\csname pn@hot/EM/Decomp/routeonly\endcsname{36.0}
\expandafter\def\csname pn@hot/EM/Decomp/rrf\endcsname{34.0}
\expandafter\def\csname pn@hot/EM/Decomp/simgraph\endcsname{34.3}
\expandafter\def\csname pn@hot/EM/Decomp/titleroute\endcsname{36.7}
\expandafter\def\csname pn@hot/EM/Decomp/window\endcsname{36.1}
\expandafter\def\csname pn@hot/EM/IRCoT/detour\endcsname{44.5}
\expandafter\def\csname pn@hot/EM/IRCoT/droprepeats\endcsname{43.3}
\expandafter\def\csname pn@hot/EM/IRCoT/interleave\endcsname{42.5}
\expandafter\def\csname pn@hot/EM/IRCoT/named\endcsname{42.2}
\expandafter\def\csname pn@hot/EM/IRCoT/novelquery\endcsname{43.7}
\expandafter\def\csname pn@hot/EM/IRCoT/randomroute\endcsname{43.6}
\expandafter\def\csname pn@hot/EM/IRCoT/reranker\endcsname{40.9}
\expandafter\def\csname pn@hot/EM/IRCoT/rerankertops\endcsname{42.6}
\expandafter\def\csname pn@hot/EM/IRCoT/rerankerwin\endcsname{43.3}
\expandafter\def\csname pn@hot/EM/IRCoT/rewired\endcsname{43.9}
\expandafter\def\csname pn@hot/EM/IRCoT/routeonly\endcsname{42.9}
\expandafter\def\csname pn@hot/EM/IRCoT/rrf\endcsname{40.2}
\expandafter\def\csname pn@hot/EM/IRCoT/simgraph\endcsname{41.7}
\expandafter\def\csname pn@hot/EM/IRCoT/titleroute\endcsname{44.7}
\expandafter\def\csname pn@hot/EM/IRCoT/window\endcsname{43.3}
\expandafter\def\csname pn@hot/EM/ReAct/detour\endcsname{40.9}
\expandafter\def\csname pn@hot/EM/ReAct/droprepeats\endcsname{38.1}
\expandafter\def\csname pn@hot/EM/ReAct/interleave\endcsname{37.5}
\expandafter\def\csname pn@hot/EM/ReAct/named\endcsname{37.8}
\expandafter\def\csname pn@hot/EM/ReAct/novelquery\endcsname{38.6}
\expandafter\def\csname pn@hot/EM/ReAct/randomroute\endcsname{39.0}
\expandafter\def\csname pn@hot/EM/ReAct/reranker\endcsname{38.7}
\expandafter\def\csname pn@hot/EM/ReAct/rerankertops\endcsname{39.3}
\expandafter\def\csname pn@hot/EM/ReAct/rerankerwin\endcsname{39.3}
\expandafter\def\csname pn@hot/EM/ReAct/rewired\endcsname{38.7}
\expandafter\def\csname pn@hot/EM/ReAct/routeonly\endcsname{38.2}
\expandafter\def\csname pn@hot/EM/ReAct/rrf\endcsname{35.7}
\expandafter\def\csname pn@hot/EM/ReAct/simgraph\endcsname{37.8}
\expandafter\def\csname pn@hot/EM/ReAct/titleroute\endcsname{40.7}
\expandafter\def\csname pn@hot/EM/ReAct/window\endcsname{37.9}
\expandafter\def\csname pn@hot/EM/SR1/detour\endcsname{42.5}
\expandafter\def\csname pn@hot/EM/SR1/droprepeats\endcsname{42.5}
\expandafter\def\csname pn@hot/EM/SR1/interleave\endcsname{41.9}
\expandafter\def\csname pn@hot/EM/SR1/named\endcsname{41.5}
\expandafter\def\csname pn@hot/EM/SR1/novelquery\endcsname{42.4}
\expandafter\def\csname pn@hot/EM/SR1/randomroute\endcsname{42.2}
\expandafter\def\csname pn@hot/EM/SR1/reranker\endcsname{41.9}
\expandafter\def\csname pn@hot/EM/SR1/rerankertops\endcsname{42.5}
\expandafter\def\csname pn@hot/EM/SR1/rerankerwin\endcsname{42.4}
\expandafter\def\csname pn@hot/EM/SR1/rewired\endcsname{42.5}
\expandafter\def\csname pn@hot/EM/SR1/routeonly\endcsname{41.5}
\expandafter\def\csname pn@hot/EM/SR1/rrf\endcsname{41.3}
\expandafter\def\csname pn@hot/EM/SR1/simgraph\endcsname{41.6}
\expandafter\def\csname pn@hot/EM/SR1/titleroute\endcsname{42.3}
\expandafter\def\csname pn@hot/EM/SR1/window\endcsname{42.4}
\expandafter\def\csname pn@hot/EM/pool/detour\endcsname{41.3}
\expandafter\def\csname pn@hot/EM/pool/droprepeats\endcsname{40.0}
\expandafter\def\csname pn@hot/EM/pool/interleave\endcsname{39.4}
\expandafter\def\csname pn@hot/EM/pool/named\endcsname{39.1}
\expandafter\def\csname pn@hot/EM/pool/novelquery\endcsname{40.2}
\expandafter\def\csname pn@hot/EM/pool/randomroute\endcsname{40.2}
\expandafter\def\csname pn@hot/EM/pool/reranker\endcsname{39.5}
\expandafter\def\csname pn@hot/EM/pool/rerankertops\endcsname{40.1}
\expandafter\def\csname pn@hot/EM/pool/rerankerwin\endcsname{40.4}
\expandafter\def\csname pn@hot/EM/pool/rewired\endcsname{40.3}
\expandafter\def\csname pn@hot/EM/pool/routeonly\endcsname{39.6}
\expandafter\def\csname pn@hot/EM/pool/rrf\endcsname{37.8}
\expandafter\def\csname pn@hot/EM/pool/simgraph\endcsname{38.9}
\expandafter\def\csname pn@hot/EM/pool/titleroute\endcsname{41.1}
\expandafter\def\csname pn@hot/EM/pool/window\endcsname{39.9}
\expandafter\def\csname pn@hot/calls/Decomp\endcsname{6.00}
\expandafter\def\csname pn@hot/calls/IRCoT\endcsname{6.00}
\expandafter\def\csname pn@hot/calls/ReAct\endcsname{5.75}
\expandafter\def\csname pn@hot/calls/SR1\endcsname{3.39}
\expandafter\def\csname pn@hot/complete/Decomp/detour\endcsname{56.7}
\expandafter\def\csname pn@hot/complete/Decomp/droprepeats\endcsname{52.5}
\expandafter\def\csname pn@hot/complete/Decomp/interleave\endcsname{42.7}
\expandafter\def\csname pn@hot/complete/Decomp/named\endcsname{45.1}
\expandafter\def\csname pn@hot/complete/Decomp/novelquery\endcsname{52.9}
\expandafter\def\csname pn@hot/complete/Decomp/randomroute\endcsname{53.4}
\expandafter\def\csname pn@hot/complete/Decomp/reranker\endcsname{48.4}
\expandafter\def\csname pn@hot/complete/Decomp/rerankertops\endcsname{50.5}
\expandafter\def\csname pn@hot/complete/Decomp/rerankerwin\endcsname{53.9}
\expandafter\def\csname pn@hot/complete/Decomp/rewired\endcsname{48.8}
\expandafter\def\csname pn@hot/complete/Decomp/routeonly\endcsname{45.1}
\expandafter\def\csname pn@hot/complete/Decomp/rrf\endcsname{41.2}
\expandafter\def\csname pn@hot/complete/Decomp/rrfnorepeat\endcsname{41.9}
\expandafter\def\csname pn@hot/complete/Decomp/rrfskip\endcsname{41.2}
\expandafter\def\csname pn@hot/complete/Decomp/simgraph\endcsname{46.5}
\expandafter\def\csname pn@hot/complete/Decomp/titleroute\endcsname{55.4}
\expandafter\def\csname pn@hot/complete/Decomp/window\endcsname{52.5}
\expandafter\def\csname pn@hot/complete/IRCoT/detour\endcsname{76.7}
\expandafter\def\csname pn@hot/complete/IRCoT/droprepeats\endcsname{73.5}
\expandafter\def\csname pn@hot/complete/IRCoT/interleave\endcsname{68.3}
\expandafter\def\csname pn@hot/complete/IRCoT/named\endcsname{70.5}
\expandafter\def\csname pn@hot/complete/IRCoT/novelquery\endcsname{74.9}
\expandafter\def\csname pn@hot/complete/IRCoT/randomroute\endcsname{75.0}
\expandafter\def\csname pn@hot/complete/IRCoT/reranker\endcsname{66.3}
\expandafter\def\csname pn@hot/complete/IRCoT/rerankertops\endcsname{70.0}
\expandafter\def\csname pn@hot/complete/IRCoT/rerankerwin\endcsname{73.5}
\expandafter\def\csname pn@hot/complete/IRCoT/rewired\endcsname{72.5}
\expandafter\def\csname pn@hot/complete/IRCoT/routeonly\endcsname{68.7}
\expandafter\def\csname pn@hot/complete/IRCoT/rrf\endcsname{50.9}
\expandafter\def\csname pn@hot/complete/IRCoT/rrfnorepeat\endcsname{63.6}
\expandafter\def\csname pn@hot/complete/IRCoT/rrfskip\endcsname{50.9}
\expandafter\def\csname pn@hot/complete/IRCoT/simgraph\endcsname{71.4}
\expandafter\def\csname pn@hot/complete/IRCoT/titleroute\endcsname{76.3}
\expandafter\def\csname pn@hot/complete/IRCoT/window\endcsname{73.7}
\expandafter\def\csname pn@hot/complete/ReAct/detour\endcsname{67.5}
\expandafter\def\csname pn@hot/complete/ReAct/droprepeats\endcsname{58.4}
\expandafter\def\csname pn@hot/complete/ReAct/interleave\endcsname{51.3}
\expandafter\def\csname pn@hot/complete/ReAct/named\endcsname{53.9}
\expandafter\def\csname pn@hot/complete/ReAct/novelquery\endcsname{60.2}
\expandafter\def\csname pn@hot/complete/ReAct/randomroute\endcsname{61.5}
\expandafter\def\csname pn@hot/complete/ReAct/reranker\endcsname{53.3}
\expandafter\def\csname pn@hot/complete/ReAct/rerankertops\endcsname{55.6}
\expandafter\def\csname pn@hot/complete/ReAct/rerankerwin\endcsname{59.3}
\expandafter\def\csname pn@hot/complete/ReAct/rewired\endcsname{57.2}
\expandafter\def\csname pn@hot/complete/ReAct/routeonly\endcsname{56.0}
\expandafter\def\csname pn@hot/complete/ReAct/rrf\endcsname{39.6}
\expandafter\def\csname pn@hot/complete/ReAct/rrfnorepeat\endcsname{44.9}
\expandafter\def\csname pn@hot/complete/ReAct/rrfskip\endcsname{39.6}
\expandafter\def\csname pn@hot/complete/ReAct/simgraph\endcsname{55.1}
\expandafter\def\csname pn@hot/complete/ReAct/titleroute\endcsname{65.7}
\expandafter\def\csname pn@hot/complete/ReAct/window\endcsname{58.9}
\expandafter\def\csname pn@hot/complete/SR1/detour\endcsname{69.9}
\expandafter\def\csname pn@hot/complete/SR1/droprepeats\endcsname{68.6}
\expandafter\def\csname pn@hot/complete/SR1/interleave\endcsname{61.3}
\expandafter\def\csname pn@hot/complete/SR1/named\endcsname{63.1}
\expandafter\def\csname pn@hot/complete/SR1/novelquery\endcsname{68.7}
\expandafter\def\csname pn@hot/complete/SR1/randomroute\endcsname{68.9}
\expandafter\def\csname pn@hot/complete/SR1/reranker\endcsname{64.1}
\expandafter\def\csname pn@hot/complete/SR1/rerankertops\endcsname{66.9}
\expandafter\def\csname pn@hot/complete/SR1/rerankerwin\endcsname{69.7}
\expandafter\def\csname pn@hot/complete/SR1/rewired\endcsname{67.1}
\expandafter\def\csname pn@hot/complete/SR1/routeonly\endcsname{61.5}
\expandafter\def\csname pn@hot/complete/SR1/rrf\endcsname{58.4}
\expandafter\def\csname pn@hot/complete/SR1/rrfnorepeat\endcsname{59.6}
\expandafter\def\csname pn@hot/complete/SR1/rrfskip\endcsname{58.4}
\expandafter\def\csname pn@hot/complete/SR1/simgraph\endcsname{64.7}
\expandafter\def\csname pn@hot/complete/SR1/titleroute\endcsname{69.3}
\expandafter\def\csname pn@hot/complete/SR1/window\endcsname{68.7}
\expandafter\def\csname pn@hot/complete/pool/detour\endcsname{67.7}
\expandafter\def\csname pn@hot/complete/pool/droprepeats\endcsname{63.3}
\expandafter\def\csname pn@hot/complete/pool/interleave\endcsname{55.9}
\expandafter\def\csname pn@hot/complete/pool/named\endcsname{58.2}
\expandafter\def\csname pn@hot/complete/pool/novelquery\endcsname{64.2}
\expandafter\def\csname pn@hot/complete/pool/randomroute\endcsname{64.7}
\expandafter\def\csname pn@hot/complete/pool/reranker\endcsname{58.0}
\expandafter\def\csname pn@hot/complete/pool/rerankertops\endcsname{60.8}
\expandafter\def\csname pn@hot/complete/pool/rerankerwin\endcsname{64.1}
\expandafter\def\csname pn@hot/complete/pool/rewired\endcsname{61.4}
\expandafter\def\csname pn@hot/complete/pool/routeonly\endcsname{57.8}
\expandafter\def\csname pn@hot/complete/pool/rrf\endcsname{47.5}
\expandafter\def\csname pn@hot/complete/pool/rrfnorepeat\endcsname{52.5}
\expandafter\def\csname pn@hot/complete/pool/rrfskip\endcsname{47.5}
\expandafter\def\csname pn@hot/complete/pool/simgraph\endcsname{59.4}
\expandafter\def\csname pn@hot/complete/pool/titleroute\endcsname{66.7}
\expandafter\def\csname pn@hot/complete/pool/window\endcsname{63.4}
\expandafter\def\csname pn@hot/compose/EM/Decomp\endcsname{$+1.3$}
\expandafter\def\csname pn@hot/compose/EM/Decomp/abs\endcsname{1.3}
\expandafter\def\csname pn@hot/compose/EM/Decomp/ci\endcsname{[$-0.2$, $2.7$]}
\expandafter\def\csname pn@hot/compose/EM/Decomp/full\endcsname{$+1.3$ [$-0.2$, $2.7$]}
\expandafter\def\csname pn@hot/compose/EM/IRCoT\endcsname{$+1.6$}
\expandafter\def\csname pn@hot/compose/EM/IRCoT/abs\endcsname{1.6}
\expandafter\def\csname pn@hot/compose/EM/IRCoT/ci\endcsname{[$0.2$, $3.0$]}
\expandafter\def\csname pn@hot/compose/EM/IRCoT/full\endcsname{$+1.6$ [$0.2$, $3.0$]}
\expandafter\def\csname pn@hot/compose/EM/ReAct\endcsname{$+2.7$}
\expandafter\def\csname pn@hot/compose/EM/ReAct/abs\endcsname{2.7}
\expandafter\def\csname pn@hot/compose/EM/ReAct/ci\endcsname{[$1.2$, $4.3$]}
\expandafter\def\csname pn@hot/compose/EM/ReAct/full\endcsname{$+2.7$ [$1.2$, $4.3$]}
\expandafter\def\csname pn@hot/compose/EM/SR1\endcsname{$+1.1$}
\expandafter\def\csname pn@hot/compose/EM/SR1/abs\endcsname{1.1}
\expandafter\def\csname pn@hot/compose/EM/SR1/ci\endcsname{[$-0.4$, $2.5$]}
\expandafter\def\csname pn@hot/compose/EM/SR1/full\endcsname{$+1.1$ [$-0.4$, $2.5$]}
\expandafter\def\csname pn@hot/compose/EM/ext\endcsname{$+1.8$}
\expandafter\def\csname pn@hot/compose/EM/ext/abs\endcsname{1.8}
\expandafter\def\csname pn@hot/compose/EM/ext/ci\endcsname{[$0.8$, $2.8$]}
\expandafter\def\csname pn@hot/compose/EM/ext/full\endcsname{$+1.8$ [$0.8$, $2.8$]}
\expandafter\def\csname pn@hot/compose/EM/nsig\endcsname{1}
\expandafter\def\csname pn@hot/compose/EM/pool\endcsname{$+1.7$}
\expandafter\def\csname pn@hot/compose/EM/pool/abs\endcsname{1.1}
\expandafter\def\csname pn@hot/compose/EM/pool/ci\endcsname{[$0.8$, $2.6$]}
\expandafter\def\csname pn@hot/compose/EM/pool/full\endcsname{$+1.7$ [$0.8$, $2.6$]}
\expandafter\def\csname pn@hot/compose/EM/range\endcsname{$+1.1$ to $+2.7$}
\expandafter\def\csname pn@hot/compose/ans/Decomp\endcsname{$+5.5$}
\expandafter\def\csname pn@hot/compose/ans/Decomp/abs\endcsname{5.5}
\expandafter\def\csname pn@hot/compose/ans/Decomp/ci\endcsname{[$4.2$, $6.9$]}
\expandafter\def\csname pn@hot/compose/ans/Decomp/full\endcsname{$+5.5$ [$4.2$, $6.9$]}
\expandafter\def\csname pn@hot/compose/ans/IRCoT\endcsname{$+1.6$}
\expandafter\def\csname pn@hot/compose/ans/IRCoT/abs\endcsname{1.6}
\expandafter\def\csname pn@hot/compose/ans/IRCoT/ci\endcsname{[$0.8$, $2.5$]}
\expandafter\def\csname pn@hot/compose/ans/IRCoT/full\endcsname{$+1.6$ [$0.8$, $2.5$]}
\expandafter\def\csname pn@hot/compose/ans/ReAct\endcsname{$+4.9$}
\expandafter\def\csname pn@hot/compose/ans/ReAct/abs\endcsname{4.9}
\expandafter\def\csname pn@hot/compose/ans/ReAct/ci\endcsname{[$3.6$, $6.2$]}
\expandafter\def\csname pn@hot/compose/ans/ReAct/full\endcsname{$+4.9$ [$3.6$, $6.2$]}
\expandafter\def\csname pn@hot/compose/ans/SR1\endcsname{$+3.5$}
\expandafter\def\csname pn@hot/compose/ans/SR1/abs\endcsname{3.5}
\expandafter\def\csname pn@hot/compose/ans/SR1/ci\endcsname{[$2.5$, $4.6$]}
\expandafter\def\csname pn@hot/compose/ans/SR1/full\endcsname{$+3.5$ [$2.5$, $4.6$]}
\expandafter\def\csname pn@hot/compose/ans/ext\endcsname{$+3.3$}
\expandafter\def\csname pn@hot/compose/ans/ext/abs\endcsname{3.3}
\expandafter\def\csname pn@hot/compose/ans/ext/ci\endcsname{[$2.6$, $4.1$]}
\expandafter\def\csname pn@hot/compose/ans/ext/full\endcsname{$+3.3$ [$2.6$, $4.1$]}
\expandafter\def\csname pn@hot/compose/ans/nsig\endcsname{4}
\expandafter\def\csname pn@hot/compose/ans/pool\endcsname{$+3.9$}
\expandafter\def\csname pn@hot/compose/ans/pool/abs\endcsname{3.9}
\expandafter\def\csname pn@hot/compose/ans/pool/ci\endcsname{[$3.2$, $4.6$]}
\expandafter\def\csname pn@hot/compose/ans/pool/full\endcsname{$+3.9$ [$3.2$, $4.6$]}
\expandafter\def\csname pn@hot/compose/ans/range\endcsname{$+1.6$ to $+5.5$}
\expandafter\def\csname pn@hot/compose/complete/Decomp\endcsname{$+11.6$}
\expandafter\def\csname pn@hot/compose/complete/Decomp/abs\endcsname{11.6}
\expandafter\def\csname pn@hot/compose/complete/Decomp/ci\endcsname{[$9.9$, $13.3$]}
\expandafter\def\csname pn@hot/compose/complete/Decomp/full\endcsname{$+11.6$ [$9.9$, $13.3$]}
\expandafter\def\csname pn@hot/compose/complete/IRCoT\endcsname{$+8.0$}
\expandafter\def\csname pn@hot/compose/complete/IRCoT/abs\endcsname{8.0}
\expandafter\def\csname pn@hot/compose/complete/IRCoT/ci\endcsname{[$6.5$, $9.5$]}
\expandafter\def\csname pn@hot/compose/complete/IRCoT/full\endcsname{$+8.0$ [$6.5$, $9.5$]}
\expandafter\def\csname pn@hot/compose/complete/ReAct\endcsname{$+11.5$}
\expandafter\def\csname pn@hot/compose/complete/ReAct/abs\endcsname{11.5}
\expandafter\def\csname pn@hot/compose/complete/ReAct/ci\endcsname{[$9.8$, $13.2$]}
\expandafter\def\csname pn@hot/compose/complete/ReAct/full\endcsname{$+11.5$ [$9.8$, $13.2$]}
\expandafter\def\csname pn@hot/compose/complete/SR1\endcsname{$+8.3$}
\expandafter\def\csname pn@hot/compose/complete/SR1/abs\endcsname{8.3}
\expandafter\def\csname pn@hot/compose/complete/SR1/ci\endcsname{[$6.9$, $9.8$]}
\expandafter\def\csname pn@hot/compose/complete/SR1/full\endcsname{$+8.3$ [$6.9$, $9.8$]}
\expandafter\def\csname pn@hot/compose/complete/ext\endcsname{$+9.3$}
\expandafter\def\csname pn@hot/compose/complete/ext/abs\endcsname{9.3}
\expandafter\def\csname pn@hot/compose/complete/ext/ci\endcsname{[$8.2$, $10.4$]}
\expandafter\def\csname pn@hot/compose/complete/ext/full\endcsname{$+9.3$ [$8.2$, $10.4$]}
\expandafter\def\csname pn@hot/compose/complete/nsig\endcsname{4}
\expandafter\def\csname pn@hot/compose/complete/pool\endcsname{$+9.9$}
\expandafter\def\csname pn@hot/compose/complete/pool/abs\endcsname{9.9}
\expandafter\def\csname pn@hot/compose/complete/pool/ci\endcsname{[$8.8$, $10.9$]}
\expandafter\def\csname pn@hot/compose/complete/pool/full\endcsname{$+9.9$ [$8.8$, $10.9$]}
\expandafter\def\csname pn@hot/compose/complete/range\endcsname{$+8.0$ to $+11.6$}
\expandafter\def\csname pn@hot/cpu/range\endcsname{2.8--23.4}
\expandafter\def\csname pn@hot/depth/EM/IRCoT\endcsname{$+0.5$}
\expandafter\def\csname pn@hot/depth/EM/IRCoT/abs\endcsname{0.5}
\expandafter\def\csname pn@hot/depth/EM/IRCoT/ci\endcsname{[$-1.1$, $2.2$]}
\expandafter\def\csname pn@hot/depth/EM/IRCoT/full\endcsname{$+0.5$ [$-1.1$, $2.2$]}
\expandafter\def\csname pn@hot/depth/EM/nsig\endcsname{0}
\expandafter\def\csname pn@hot/depth/ans/Decomp\endcsname{$+6.1$}
\expandafter\def\csname pn@hot/depth/ans/Decomp/abs\endcsname{6.1}
\expandafter\def\csname pn@hot/depth/ans/Decomp/ci\endcsname{[$4.9$, $7.4$]}
\expandafter\def\csname pn@hot/depth/ans/Decomp/full\endcsname{$+6.1$ [$4.9$, $7.4$]}
\expandafter\def\csname pn@hot/depth/ans/IRCoT\endcsname{$+6.3$}
\expandafter\def\csname pn@hot/depth/ans/IRCoT/abs\endcsname{6.3}
\expandafter\def\csname pn@hot/depth/ans/IRCoT/ci\endcsname{[$5.0$, $7.5$]}
\expandafter\def\csname pn@hot/depth/ans/IRCoT/full\endcsname{$+6.3$ [$5.0$, $7.5$]}
\expandafter\def\csname pn@hot/depth/ans/ReAct\endcsname{$+8.3$}
\expandafter\def\csname pn@hot/depth/ans/ReAct/abs\endcsname{8.3}
\expandafter\def\csname pn@hot/depth/ans/ReAct/ci\endcsname{[$6.9$, $9.8$]}
\expandafter\def\csname pn@hot/depth/ans/ReAct/full\endcsname{$+8.3$ [$6.9$, $9.8$]}
\expandafter\def\csname pn@hot/depth/ans/SR1\endcsname{$+6.3$}
\expandafter\def\csname pn@hot/depth/ans/SR1/abs\endcsname{6.3}
\expandafter\def\csname pn@hot/depth/ans/SR1/ci\endcsname{[$5.1$, $7.5$]}
\expandafter\def\csname pn@hot/depth/ans/SR1/full\endcsname{$+6.3$ [$5.1$, $7.5$]}
\expandafter\def\csname pn@hot/depth/ans/ext\endcsname{$+7.0$}
\expandafter\def\csname pn@hot/depth/ans/ext/abs\endcsname{7.0}
\expandafter\def\csname pn@hot/depth/ans/ext/ci\endcsname{[$6.1$, $7.8$]}
\expandafter\def\csname pn@hot/depth/ans/ext/full\endcsname{$+7.0$ [$6.1$, $7.8$]}
\expandafter\def\csname pn@hot/depth/ans/nsig\endcsname{4}
\expandafter\def\csname pn@hot/depth/ans/pool\endcsname{$+6.7$}
\expandafter\def\csname pn@hot/depth/ans/pool/abs\endcsname{6.7}
\expandafter\def\csname pn@hot/depth/ans/pool/ci\endcsname{[$6.0$, $7.5$]}
\expandafter\def\csname pn@hot/depth/ans/pool/full\endcsname{$+6.7$ [$6.0$, $7.5$]}
\expandafter\def\csname pn@hot/depth/ans/range\endcsname{$+6.1$ to $+8.3$}
\expandafter\def\csname pn@hot/depth/complete/Decomp\endcsname{$+8.5$}
\expandafter\def\csname pn@hot/depth/complete/Decomp/abs\endcsname{8.5}
\expandafter\def\csname pn@hot/depth/complete/Decomp/ci\endcsname{[$7.1$, $10.0$]}
\expandafter\def\csname pn@hot/depth/complete/Decomp/full\endcsname{$+8.5$ [$7.1$, $10.0$]}
\expandafter\def\csname pn@hot/depth/complete/IRCoT\endcsname{$+21.8$}
\expandafter\def\csname pn@hot/depth/complete/IRCoT/abs\endcsname{21.8}
\expandafter\def\csname pn@hot/depth/complete/IRCoT/ci\endcsname{[$19.7$, $23.9$]}
\expandafter\def\csname pn@hot/depth/complete/IRCoT/full\endcsname{$+21.8$ [$19.7$, $23.9$]}
\expandafter\def\csname pn@hot/depth/complete/ReAct\endcsname{$+17.0$}
\expandafter\def\csname pn@hot/depth/complete/ReAct/abs\endcsname{17.0}
\expandafter\def\csname pn@hot/depth/complete/ReAct/ci\endcsname{[$15.1$, $18.9$]}
\expandafter\def\csname pn@hot/depth/complete/ReAct/full\endcsname{$+17.0$ [$15.1$, $18.9$]}
\expandafter\def\csname pn@hot/depth/complete/SR1\endcsname{$+10.6$}
\expandafter\def\csname pn@hot/depth/complete/SR1/abs\endcsname{10.6}
\expandafter\def\csname pn@hot/depth/complete/SR1/ci\endcsname{[$9.1$, $12.2$]}
\expandafter\def\csname pn@hot/depth/complete/SR1/full\endcsname{$+10.6$ [$9.1$, $12.2$]}
\expandafter\def\csname pn@hot/depth/complete/ext\endcsname{$+16.5$}
\expandafter\def\csname pn@hot/depth/complete/ext/abs\endcsname{16.5}
\expandafter\def\csname pn@hot/depth/complete/ext/ci\endcsname{[$15.3$, $17.7$]}
\expandafter\def\csname pn@hot/depth/complete/ext/full\endcsname{$+16.5$ [$15.3$, $17.7$]}
\expandafter\def\csname pn@hot/depth/complete/nsig\endcsname{4}
\expandafter\def\csname pn@hot/depth/complete/pool\endcsname{$+14.5$}
\expandafter\def\csname pn@hot/depth/complete/pool/abs\endcsname{14.5}
\expandafter\def\csname pn@hot/depth/complete/pool/ci\endcsname{[$13.5$, $15.5$]}
\expandafter\def\csname pn@hot/depth/complete/pool/full\endcsname{$+14.5$ [$13.5$, $15.5$]}
\expandafter\def\csname pn@hot/depth/complete/range\endcsname{$+8.5$ to $+21.8$}
\expandafter\def\csname pn@hot/detour/EM/Decomp\endcsname{$+1.8$}
\expandafter\def\csname pn@hot/detour/EM/Decomp/abs\endcsname{1.8}
\expandafter\def\csname pn@hot/detour/EM/Decomp/ci\endcsname{[$0.3$, $3.4$]}
\expandafter\def\csname pn@hot/detour/EM/Decomp/full\endcsname{$+1.8$ [$0.3$, $3.4$]}
\expandafter\def\csname pn@hot/detour/EM/IRCoT\endcsname{$+1.9$}
\expandafter\def\csname pn@hot/detour/EM/IRCoT/abs\endcsname{1.9}
\expandafter\def\csname pn@hot/detour/EM/IRCoT/ci\endcsname{[$0.5$, $3.4$]}
\expandafter\def\csname pn@hot/detour/EM/IRCoT/full\endcsname{$+1.9$ [$0.5$, $3.4$]}
\expandafter\def\csname pn@hot/detour/EM/ReAct\endcsname{$+3.4$}
\expandafter\def\csname pn@hot/detour/EM/ReAct/abs\endcsname{3.4}
\expandafter\def\csname pn@hot/detour/EM/ReAct/ci\endcsname{[$1.8$, $5.1$]}
\expandafter\def\csname pn@hot/detour/EM/ReAct/full\endcsname{$+3.4$ [$1.8$, $5.1$]}
\expandafter\def\csname pn@hot/detour/EM/SR1\endcsname{$+0.6$}
\expandafter\def\csname pn@hot/detour/EM/SR1/abs\endcsname{0.6}
\expandafter\def\csname pn@hot/detour/EM/SR1/ci\endcsname{[$-0.9$, $2.1$]}
\expandafter\def\csname pn@hot/detour/EM/SR1/full\endcsname{$+0.6$ [$-0.9$, $2.1$]}
\expandafter\def\csname pn@hot/detour/EM/ext\endcsname{$+2.0$}
\expandafter\def\csname pn@hot/detour/EM/ext/abs\endcsname{2.0}
\expandafter\def\csname pn@hot/detour/EM/ext/ci\endcsname{[$1.0$, $3.0$]}
\expandafter\def\csname pn@hot/detour/EM/ext/full\endcsname{$+2.0$ [$1.0$, $3.0$]}
\expandafter\def\csname pn@hot/detour/EM/nsig\endcsname{3}
\expandafter\def\csname pn@hot/detour/EM/pool\endcsname{$+1.9$}
\expandafter\def\csname pn@hot/detour/EM/pool/abs\endcsname{1.4}
\expandafter\def\csname pn@hot/detour/EM/pool/ci\endcsname{[$1.0$, $2.9$]}
\expandafter\def\csname pn@hot/detour/EM/pool/full\endcsname{$+1.4$ [$0.5$, $2.3$]}
\expandafter\def\csname pn@hot/detour/EM/range\endcsname{$+0.6$ to $+3.4$}
\expandafter\def\csname pn@hot/detour/ans/Decomp\endcsname{$+7.3$}
\expandafter\def\csname pn@hot/detour/ans/Decomp/abs\endcsname{7.3}
\expandafter\def\csname pn@hot/detour/ans/Decomp/ci\endcsname{[$5.8$, $8.8$]}
\expandafter\def\csname pn@hot/detour/ans/Decomp/full\endcsname{$+7.3$ [$5.8$, $8.8$]}
\expandafter\def\csname pn@hot/detour/ans/IRCoT\endcsname{$+2.3$}
\expandafter\def\csname pn@hot/detour/ans/IRCoT/abs\endcsname{2.3}
\expandafter\def\csname pn@hot/detour/ans/IRCoT/ci\endcsname{[$1.3$, $3.3$]}
\expandafter\def\csname pn@hot/detour/ans/IRCoT/full\endcsname{$+2.3$ [$1.3$, $3.3$]}
\expandafter\def\csname pn@hot/detour/ans/ReAct\endcsname{$+7.9$}
\expandafter\def\csname pn@hot/detour/ans/ReAct/abs\endcsname{7.9}
\expandafter\def\csname pn@hot/detour/ans/ReAct/ci\endcsname{[$6.3$, $9.6$]}
\expandafter\def\csname pn@hot/detour/ans/ReAct/full\endcsname{$+7.9$ [$6.3$, $9.6$]}
\expandafter\def\csname pn@hot/detour/ans/SR1\endcsname{$+3.2$}
\expandafter\def\csname pn@hot/detour/ans/SR1/abs\endcsname{3.2}
\expandafter\def\csname pn@hot/detour/ans/SR1/ci\endcsname{[$2.1$, $4.3$]}
\expandafter\def\csname pn@hot/detour/ans/SR1/full\endcsname{$+3.2$ [$2.1$, $4.3$]}
\expandafter\def\csname pn@hot/detour/ans/ext\endcsname{$+4.4$}
\expandafter\def\csname pn@hot/detour/ans/ext/abs\endcsname{4.4}
\expandafter\def\csname pn@hot/detour/ans/ext/ci\endcsname{[$3.6$, $5.3$]}
\expandafter\def\csname pn@hot/detour/ans/ext/full\endcsname{$+4.4$ [$3.6$, $5.3$]}
\expandafter\def\csname pn@hot/detour/ans/nsig\endcsname{4}
\expandafter\def\csname pn@hot/detour/ans/pool\endcsname{$+5.2$}
\expandafter\def\csname pn@hot/detour/ans/pool/abs\endcsname{5.2}
\expandafter\def\csname pn@hot/detour/ans/pool/ci\endcsname{[$4.3$, $6.0$]}
\expandafter\def\csname pn@hot/detour/ans/pool/full\endcsname{$+5.2$ [$4.3$, $6.0$]}
\expandafter\def\csname pn@hot/detour/ans/range\endcsname{$+2.3$ to $+7.9$}
\expandafter\def\csname pn@hot/detour/complete/Decomp\endcsname{$+13.9$}
\expandafter\def\csname pn@hot/detour/complete/Decomp/abs\endcsname{13.9}
\expandafter\def\csname pn@hot/detour/complete/Decomp/ci\endcsname{[$12.1$, $15.8$]}
\expandafter\def\csname pn@hot/detour/complete/Decomp/full\endcsname{$+13.9$ [$12.1$, $15.8$]}
\expandafter\def\csname pn@hot/detour/complete/IRCoT\endcsname{$+8.4$}
\expandafter\def\csname pn@hot/detour/complete/IRCoT/abs\endcsname{8.4}
\expandafter\def\csname pn@hot/detour/complete/IRCoT/ci\endcsname{[$6.9$, $9.9$]}
\expandafter\def\csname pn@hot/detour/complete/IRCoT/full\endcsname{$+8.4$ [$6.9$, $9.9$]}
\expandafter\def\csname pn@hot/detour/complete/ReAct\endcsname{$+16.1$}
\expandafter\def\csname pn@hot/detour/complete/ReAct/abs\endcsname{16.1}
\expandafter\def\csname pn@hot/detour/complete/ReAct/ci\endcsname{[$14.1$, $18.1$]}
\expandafter\def\csname pn@hot/detour/complete/ReAct/full\endcsname{$+16.1$ [$14.1$, $18.1$]}
\expandafter\def\csname pn@hot/detour/complete/SR1\endcsname{$+8.6$}
\expandafter\def\csname pn@hot/detour/complete/SR1/abs\endcsname{8.6}
\expandafter\def\csname pn@hot/detour/complete/SR1/ci\endcsname{[$7.1$, $10.1$]}
\expandafter\def\csname pn@hot/detour/complete/SR1/full\endcsname{$+8.6$ [$7.1$, $10.1$]}
\expandafter\def\csname pn@hot/detour/complete/ext\endcsname{$+11.0$}
\expandafter\def\csname pn@hot/detour/complete/ext/abs\endcsname{11.0}
\expandafter\def\csname pn@hot/detour/complete/ext/ci\endcsname{[$9.9$, $12.2$]}
\expandafter\def\csname pn@hot/detour/complete/ext/full\endcsname{$+11.0$ [$9.9$, $12.2$]}
\expandafter\def\csname pn@hot/detour/complete/nsig\endcsname{4}
\expandafter\def\csname pn@hot/detour/complete/pool\endcsname{$+11.8$}
\expandafter\def\csname pn@hot/detour/complete/pool/abs\endcsname{11.8}
\expandafter\def\csname pn@hot/detour/complete/pool/ci\endcsname{[$10.7$, $12.9$]}
\expandafter\def\csname pn@hot/detour/complete/pool/full\endcsname{$+11.8$ [$10.7$, $12.9$]}
\expandafter\def\csname pn@hot/detour/complete/range\endcsname{$+8.4$ to $+16.1$}
\expandafter\def\csname pn@hot/detourvsreranker/EM/pool\endcsname{$+1.8$}
\expandafter\def\csname pn@hot/detourvsreranker/EM/pool/abs\endcsname{1.8}
\expandafter\def\csname pn@hot/detourvsreranker/EM/pool/ci\endcsname{[$0.7$, $2.9$]}
\expandafter\def\csname pn@hot/detourvsreranker/EM/pool/full\endcsname{$+1.5$ [$0.4$, $2.6$]}
\expandafter\def\csname pn@hot/detourvsrerankerwin/EM/pool\endcsname{$+0.9$}
\expandafter\def\csname pn@hot/detourvsrerankerwin/EM/pool/abs\endcsname{0.9}
\expandafter\def\csname pn@hot/detourvsrerankerwin/EM/pool/ci\endcsname{[$0.0$, $1.9$]}
\expandafter\def\csname pn@hot/detourvsrerankerwin/EM/pool/full\endcsname{$+0.9$ [$0.0$, $1.9$]}
\expandafter\def\csname pn@hot/droprepeats/EM/Decomp\endcsname{$-0.1$}
\expandafter\def\csname pn@hot/droprepeats/EM/Decomp/abs\endcsname{0.1}
\expandafter\def\csname pn@hot/droprepeats/EM/Decomp/ci\endcsname{[$-0.3$, $0.2$]}
\expandafter\def\csname pn@hot/droprepeats/EM/Decomp/full\endcsname{$-0.1$ [$-0.3$, $0.2$]}
\expandafter\def\csname pn@hot/droprepeats/EM/IRCoT\endcsname{$0.0$}
\expandafter\def\csname pn@hot/droprepeats/EM/IRCoT/abs\endcsname{0.0}
\expandafter\def\csname pn@hot/droprepeats/EM/IRCoT/ci\endcsname{[$-0.3$, $0.3$]}
\expandafter\def\csname pn@hot/droprepeats/EM/IRCoT/full\endcsname{$0.0$ [$-0.3$, $0.3$]}
\expandafter\def\csname pn@hot/droprepeats/EM/ReAct\endcsname{$+0.1$}
\expandafter\def\csname pn@hot/droprepeats/EM/ReAct/abs\endcsname{0.1}
\expandafter\def\csname pn@hot/droprepeats/EM/ReAct/ci\endcsname{[$-0.4$, $0.7$]}
\expandafter\def\csname pn@hot/droprepeats/EM/ReAct/full\endcsname{$+0.1$ [$-0.4$, $0.7$]}
\expandafter\def\csname pn@hot/droprepeats/EM/SR1\endcsname{$+0.1$}
\expandafter\def\csname pn@hot/droprepeats/EM/SR1/abs\endcsname{0.1}
\expandafter\def\csname pn@hot/droprepeats/EM/SR1/ci\endcsname{[$-0.3$, $0.5$]}
\expandafter\def\csname pn@hot/droprepeats/EM/SR1/full\endcsname{$+0.1$ [$-0.3$, $0.5$]}
\expandafter\def\csname pn@hot/droprepeats/EM/ext\endcsname{$+0.1$}
\expandafter\def\csname pn@hot/droprepeats/EM/ext/abs\endcsname{0.1}
\expandafter\def\csname pn@hot/droprepeats/EM/ext/ci\endcsname{[$-0.2$, $0.3$]}
\expandafter\def\csname pn@hot/droprepeats/EM/ext/full\endcsname{$+0.1$ [$-0.2$, $0.3$]}
\expandafter\def\csname pn@hot/droprepeats/EM/nsig\endcsname{0}
\expandafter\def\csname pn@hot/droprepeats/EM/pool\endcsname{$+0.1$}
\expandafter\def\csname pn@hot/droprepeats/EM/pool/abs\endcsname{0.1}
\expandafter\def\csname pn@hot/droprepeats/EM/pool/ci\endcsname{[$-0.2$, $0.3$]}
\expandafter\def\csname pn@hot/droprepeats/EM/pool/full\endcsname{$+0.1$ [$-0.2$, $0.3$]}
\expandafter\def\csname pn@hot/droprepeats/EM/range\endcsname{$-0.1$ to $+0.1$}
\expandafter\def\csname pn@hot/droprepeats/ans/Decomp\endcsname{$0.0$}
\expandafter\def\csname pn@hot/droprepeats/ans/Decomp/abs\endcsname{0.0}
\expandafter\def\csname pn@hot/droprepeats/ans/Decomp/ci\endcsname{[$-0.3$, $0.3$]}
\expandafter\def\csname pn@hot/droprepeats/ans/Decomp/full\endcsname{$0.0$ [$-0.3$, $0.3$]}
\expandafter\def\csname pn@hot/droprepeats/ans/IRCoT\endcsname{$0.0$}
\expandafter\def\csname pn@hot/droprepeats/ans/IRCoT/abs\endcsname{0.0}
\expandafter\def\csname pn@hot/droprepeats/ans/IRCoT/ci\endcsname{[$0.0$, $0.0$]}
\expandafter\def\csname pn@hot/droprepeats/ans/IRCoT/full\endcsname{$0.0$ [$0.0$, $0.0$]}
\expandafter\def\csname pn@hot/droprepeats/ans/ReAct\endcsname{$-0.3$}
\expandafter\def\csname pn@hot/droprepeats/ans/ReAct/abs\endcsname{0.3}
\expandafter\def\csname pn@hot/droprepeats/ans/ReAct/ci\endcsname{[$-0.6$, $0.0$]}
\expandafter\def\csname pn@hot/droprepeats/ans/ReAct/full\endcsname{$-0.3$ [$-0.6$, $0.0$]}
\expandafter\def\csname pn@hot/droprepeats/ans/SR1\endcsname{$0.0$}
\expandafter\def\csname pn@hot/droprepeats/ans/SR1/abs\endcsname{0.0}
\expandafter\def\csname pn@hot/droprepeats/ans/SR1/ci\endcsname{[$-0.3$, $0.3$]}
\expandafter\def\csname pn@hot/droprepeats/ans/SR1/full\endcsname{$0.0$ [$-0.3$, $0.3$]}
\expandafter\def\csname pn@hot/droprepeats/ans/ext\endcsname{$-0.1$}
\expandafter\def\csname pn@hot/droprepeats/ans/ext/abs\endcsname{0.1}
\expandafter\def\csname pn@hot/droprepeats/ans/ext/ci\endcsname{[$-0.2$, $0.0$]}
\expandafter\def\csname pn@hot/droprepeats/ans/ext/full\endcsname{$-0.1$ [$-0.2$, $0.0$]}
\expandafter\def\csname pn@hot/droprepeats/ans/nsig\endcsname{0}
\expandafter\def\csname pn@hot/droprepeats/ans/pool\endcsname{$-0.1$}
\expandafter\def\csname pn@hot/droprepeats/ans/pool/abs\endcsname{0.1}
\expandafter\def\csname pn@hot/droprepeats/ans/pool/ci\endcsname{[$-0.2$, $0.1$]}
\expandafter\def\csname pn@hot/droprepeats/ans/pool/full\endcsname{$-0.1$ [$-0.2$, $0.1$]}
\expandafter\def\csname pn@hot/droprepeats/ans/range\endcsname{$-0.3$ to $0.0$}
\expandafter\def\csname pn@hot/droprepeats/complete/Decomp\endcsname{$+0.1$}
\expandafter\def\csname pn@hot/droprepeats/complete/Decomp/abs\endcsname{0.1}
\expandafter\def\csname pn@hot/droprepeats/complete/Decomp/ci\endcsname{[$-0.1$, $0.3$]}
\expandafter\def\csname pn@hot/droprepeats/complete/Decomp/full\endcsname{$+0.1$ [$-0.1$, $0.3$]}
\expandafter\def\csname pn@hot/droprepeats/complete/IRCoT\endcsname{$-0.2$}
\expandafter\def\csname pn@hot/droprepeats/complete/IRCoT/abs\endcsname{0.2}
\expandafter\def\csname pn@hot/droprepeats/complete/IRCoT/ci\endcsname{[$-0.5$, $0.0$]}
\expandafter\def\csname pn@hot/droprepeats/complete/IRCoT/full\endcsname{$-0.2$ [$-0.5$, $0.0$]}
\expandafter\def\csname pn@hot/droprepeats/complete/ReAct\endcsname{$-0.5$}
\expandafter\def\csname pn@hot/droprepeats/complete/ReAct/abs\endcsname{0.5}
\expandafter\def\csname pn@hot/droprepeats/complete/ReAct/ci\endcsname{[$-0.8$, $-0.1$]}
\expandafter\def\csname pn@hot/droprepeats/complete/ReAct/full\endcsname{$-0.5$ [$-0.8$, $-0.1$]}
\expandafter\def\csname pn@hot/droprepeats/complete/SR1\endcsname{$-0.1$}
\expandafter\def\csname pn@hot/droprepeats/complete/SR1/abs\endcsname{0.1}
\expandafter\def\csname pn@hot/droprepeats/complete/SR1/ci\endcsname{[$-0.4$, $0.3$]}
\expandafter\def\csname pn@hot/droprepeats/complete/SR1/full\endcsname{$-0.1$ [$-0.4$, $0.3$]}
\expandafter\def\csname pn@hot/droprepeats/complete/ext\endcsname{$-0.2$}
\expandafter\def\csname pn@hot/droprepeats/complete/ext/abs\endcsname{0.2}
\expandafter\def\csname pn@hot/droprepeats/complete/ext/ci\endcsname{[$-0.4$, $-0.1$]}
\expandafter\def\csname pn@hot/droprepeats/complete/ext/full\endcsname{$-0.2$ [$-0.4$, $-0.1$]}
\expandafter\def\csname pn@hot/droprepeats/complete/nsig\endcsname{1}
\expandafter\def\csname pn@hot/droprepeats/complete/pool\endcsname{$-0.2$}
\expandafter\def\csname pn@hot/droprepeats/complete/pool/abs\endcsname{0.2}
\expandafter\def\csname pn@hot/droprepeats/complete/pool/ci\endcsname{[$-0.3$, $0.0$]}
\expandafter\def\csname pn@hot/droprepeats/complete/pool/full\endcsname{$-0.2$ [$-0.3$, $0.0$]}
\expandafter\def\csname pn@hot/droprepeats/complete/range\endcsname{$-0.5$ to $+0.1$}
\expandafter\def\csname pn@hot/flag/Decomp\endcsname{8.2}
\expandafter\def\csname pn@hot/flag/IRCoT\endcsname{34.2}
\expandafter\def\csname pn@hot/flag/ReAct\endcsname{32.1}
\expandafter\def\csname pn@hot/flag/SR1\endcsname{13.5}
\expandafter\def\csname pn@hot/goldlink\endcsname{80}
\expandafter\def\csname pn@hot/interVSrrf/EM/Decomp\endcsname{$+1.5$}
\expandafter\def\csname pn@hot/interVSrrf/EM/Decomp/abs\endcsname{1.5}
\expandafter\def\csname pn@hot/interVSrrf/EM/Decomp/ci\endcsname{[$0.0$, $2.9$]}
\expandafter\def\csname pn@hot/interVSrrf/EM/Decomp/full\endcsname{$+1.5$ [$0.0$, $2.9$]}
\expandafter\def\csname pn@hot/interVSrrf/EM/IRCoT\endcsname{$+2.3$}
\expandafter\def\csname pn@hot/interVSrrf/EM/IRCoT/abs\endcsname{2.3}
\expandafter\def\csname pn@hot/interVSrrf/EM/IRCoT/ci\endcsname{[$0.6$, $4.0$]}
\expandafter\def\csname pn@hot/interVSrrf/EM/IRCoT/full\endcsname{$+2.3$ [$0.6$, $4.0$]}
\expandafter\def\csname pn@hot/interVSrrf/EM/ReAct\endcsname{$+1.8$}
\expandafter\def\csname pn@hot/interVSrrf/EM/ReAct/abs\endcsname{1.8}
\expandafter\def\csname pn@hot/interVSrrf/EM/ReAct/ci\endcsname{[$0.3$, $3.3$]}
\expandafter\def\csname pn@hot/interVSrrf/EM/ReAct/full\endcsname{$+1.8$ [$0.3$, $3.3$]}
\expandafter\def\csname pn@hot/interVSrrf/EM/SR1\endcsname{$+0.7$}
\expandafter\def\csname pn@hot/interVSrrf/EM/SR1/abs\endcsname{0.7}
\expandafter\def\csname pn@hot/interVSrrf/EM/SR1/ci\endcsname{[$-0.7$, $2.1$]}
\expandafter\def\csname pn@hot/interVSrrf/EM/SR1/full\endcsname{$+0.7$ [$-0.7$, $2.1$]}
\expandafter\def\csname pn@hot/interVSrrf/EM/ext\endcsname{$+1.6$}
\expandafter\def\csname pn@hot/interVSrrf/EM/ext/abs\endcsname{1.6}
\expandafter\def\csname pn@hot/interVSrrf/EM/ext/ci\endcsname{[$0.7$, $2.5$]}
\expandafter\def\csname pn@hot/interVSrrf/EM/ext/full\endcsname{$+1.6$ [$0.7$, $2.5$]}
\expandafter\def\csname pn@hot/interVSrrf/EM/nsig\endcsname{1}
\expandafter\def\csname pn@hot/interVSrrf/EM/pool\endcsname{$+1.6$}
\expandafter\def\csname pn@hot/interVSrrf/EM/pool/abs\endcsname{1.6}
\expandafter\def\csname pn@hot/interVSrrf/EM/pool/ci\endcsname{[$0.8$, $2.4$]}
\expandafter\def\csname pn@hot/interVSrrf/EM/pool/full\endcsname{$+1.6$ [$0.8$, $2.4$]}
\expandafter\def\csname pn@hot/interVSrrf/EM/range\endcsname{$+0.7$ to $+2.3$}
\expandafter\def\csname pn@hot/interVSrrf/ans/Decomp\endcsname{$+0.7$}
\expandafter\def\csname pn@hot/interVSrrf/ans/Decomp/abs\endcsname{0.7}
\expandafter\def\csname pn@hot/interVSrrf/ans/Decomp/ci\endcsname{[$-0.6$, $2.0$]}
\expandafter\def\csname pn@hot/interVSrrf/ans/Decomp/full\endcsname{$+0.7$ [$-0.6$, $2.0$]}
\expandafter\def\csname pn@hot/interVSrrf/ans/IRCoT\endcsname{$+4.9$}
\expandafter\def\csname pn@hot/interVSrrf/ans/IRCoT/abs\endcsname{4.9}
\expandafter\def\csname pn@hot/interVSrrf/ans/IRCoT/ci\endcsname{[$3.6$, $6.3$]}
\expandafter\def\csname pn@hot/interVSrrf/ans/IRCoT/full\endcsname{$+4.9$ [$3.6$, $6.3$]}
\expandafter\def\csname pn@hot/interVSrrf/ans/ReAct\endcsname{$+4.4$}
\expandafter\def\csname pn@hot/interVSrrf/ans/ReAct/abs\endcsname{4.4}
\expandafter\def\csname pn@hot/interVSrrf/ans/ReAct/ci\endcsname{[$3.1$, $5.7$]}
\expandafter\def\csname pn@hot/interVSrrf/ans/ReAct/full\endcsname{$+4.4$ [$3.1$, $5.7$]}
\expandafter\def\csname pn@hot/interVSrrf/ans/SR1\endcsname{$+2.3$}
\expandafter\def\csname pn@hot/interVSrrf/ans/SR1/abs\endcsname{2.3}
\expandafter\def\csname pn@hot/interVSrrf/ans/SR1/ci\endcsname{[$1.1$, $3.4$]}
\expandafter\def\csname pn@hot/interVSrrf/ans/SR1/full\endcsname{$+2.3$ [$1.1$, $3.4$]}
\expandafter\def\csname pn@hot/interVSrrf/ans/ext\endcsname{$+3.9$}
\expandafter\def\csname pn@hot/interVSrrf/ans/ext/abs\endcsname{3.9}
\expandafter\def\csname pn@hot/interVSrrf/ans/ext/ci\endcsname{[$3.1$, $4.6$]}
\expandafter\def\csname pn@hot/interVSrrf/ans/ext/full\endcsname{$+3.9$ [$3.1$, $4.6$]}
\expandafter\def\csname pn@hot/interVSrrf/ans/nsig\endcsname{3}
\expandafter\def\csname pn@hot/interVSrrf/ans/pool\endcsname{$+3.1$}
\expandafter\def\csname pn@hot/interVSrrf/ans/pool/abs\endcsname{3.1}
\expandafter\def\csname pn@hot/interVSrrf/ans/pool/ci\endcsname{[$2.4$, $3.8$]}
\expandafter\def\csname pn@hot/interVSrrf/ans/pool/full\endcsname{$+3.1$ [$2.4$, $3.8$]}
\expandafter\def\csname pn@hot/interVSrrf/ans/range\endcsname{$+0.7$ to $+4.9$}
\expandafter\def\csname pn@hot/interVSrrf/complete/Decomp\endcsname{$+1.5$}
\expandafter\def\csname pn@hot/interVSrrf/complete/Decomp/abs\endcsname{1.5}
\expandafter\def\csname pn@hot/interVSrrf/complete/Decomp/ci\endcsname{[$0.1$, $2.9$]}
\expandafter\def\csname pn@hot/interVSrrf/complete/Decomp/full\endcsname{$+1.5$ [$0.1$, $2.9$]}
\expandafter\def\csname pn@hot/interVSrrf/complete/IRCoT\endcsname{$+17.3$}
\expandafter\def\csname pn@hot/interVSrrf/complete/IRCoT/abs\endcsname{17.3}
\expandafter\def\csname pn@hot/interVSrrf/complete/IRCoT/ci\endcsname{[$15.2$, $19.5$]}
\expandafter\def\csname pn@hot/interVSrrf/complete/IRCoT/full\endcsname{$+17.3$ [$15.2$, $19.5$]}
\expandafter\def\csname pn@hot/interVSrrf/complete/ReAct\endcsname{$+11.7$}
\expandafter\def\csname pn@hot/interVSrrf/complete/ReAct/abs\endcsname{11.7}
\expandafter\def\csname pn@hot/interVSrrf/complete/ReAct/ci\endcsname{[$9.9$, $13.6$]}
\expandafter\def\csname pn@hot/interVSrrf/complete/ReAct/full\endcsname{$+11.7$ [$9.9$, $13.6$]}
\expandafter\def\csname pn@hot/interVSrrf/complete/SR1\endcsname{$+2.9$}
\expandafter\def\csname pn@hot/interVSrrf/complete/SR1/abs\endcsname{2.9}
\expandafter\def\csname pn@hot/interVSrrf/complete/SR1/ci\endcsname{[$1.7$, $4.1$]}
\expandafter\def\csname pn@hot/interVSrrf/complete/SR1/full\endcsname{$+2.9$ [$1.7$, $4.1$]}
\expandafter\def\csname pn@hot/interVSrrf/complete/ext\endcsname{$+10.6$}
\expandafter\def\csname pn@hot/interVSrrf/complete/ext/abs\endcsname{10.6}
\expandafter\def\csname pn@hot/interVSrrf/complete/ext/ci\endcsname{[$9.6$, $11.8$]}
\expandafter\def\csname pn@hot/interVSrrf/complete/ext/full\endcsname{$+10.6$ [$9.6$, $11.8$]}
\expandafter\def\csname pn@hot/interVSrrf/complete/nsig\endcsname{4}
\expandafter\def\csname pn@hot/interVSrrf/complete/pool\endcsname{$+8.4$}
\expandafter\def\csname pn@hot/interVSrrf/complete/pool/abs\endcsname{8.4}
\expandafter\def\csname pn@hot/interVSrrf/complete/pool/ci\endcsname{[$7.4$, $9.3$]}
\expandafter\def\csname pn@hot/interVSrrf/complete/pool/full\endcsname{$+8.4$ [$7.4$, $9.3$]}
\expandafter\def\csname pn@hot/interVSrrf/complete/range\endcsname{$+1.5$ to $+17.3$}
\expandafter\def\csname pn@hot/n\endcsname{1{,}500}
\expandafter\def\csname pn@hot/named/EM/Decomp\endcsname{$+1.2$}
\expandafter\def\csname pn@hot/named/EM/Decomp/abs\endcsname{1.2}
\expandafter\def\csname pn@hot/named/EM/Decomp/ci\endcsname{[$-0.1$, $2.5$]}
\expandafter\def\csname pn@hot/named/EM/Decomp/full\endcsname{$+1.2$ [$-0.1$, $2.5$]}
\expandafter\def\csname pn@hot/named/EM/IRCoT\endcsname{$+1.1$}
\expandafter\def\csname pn@hot/named/EM/IRCoT/abs\endcsname{1.1}
\expandafter\def\csname pn@hot/named/EM/IRCoT/ci\endcsname{[$-0.1$, $2.3$]}
\expandafter\def\csname pn@hot/named/EM/IRCoT/full\endcsname{$+1.1$ [$-0.1$, $2.3$]}
\expandafter\def\csname pn@hot/named/EM/ReAct\endcsname{$+0.1$}
\expandafter\def\csname pn@hot/named/EM/ReAct/abs\endcsname{0.1}
\expandafter\def\csname pn@hot/named/EM/ReAct/ci\endcsname{[$-1.1$, $1.3$]}
\expandafter\def\csname pn@hot/named/EM/ReAct/full\endcsname{$+0.1$ [$-1.1$, $1.3$]}
\expandafter\def\csname pn@hot/named/EM/SR1\endcsname{$+0.9$}
\expandafter\def\csname pn@hot/named/EM/SR1/abs\endcsname{0.9}
\expandafter\def\csname pn@hot/named/EM/SR1/ci\endcsname{[$-0.5$, $2.2$]}
\expandafter\def\csname pn@hot/named/EM/SR1/full\endcsname{$+0.9$ [$-0.5$, $2.2$]}
\expandafter\def\csname pn@hot/named/EM/ext\endcsname{$+0.7$}
\expandafter\def\csname pn@hot/named/EM/ext/abs\endcsname{0.7}
\expandafter\def\csname pn@hot/named/EM/ext/ci\endcsname{[$-0.2$, $1.6$]}
\expandafter\def\csname pn@hot/named/EM/ext/full\endcsname{$+0.7$ [$-0.2$, $1.6$]}
\expandafter\def\csname pn@hot/named/EM/nsig\endcsname{0}
\expandafter\def\csname pn@hot/named/EM/pool\endcsname{$+0.8$}
\expandafter\def\csname pn@hot/named/EM/pool/abs\endcsname{0.8}
\expandafter\def\csname pn@hot/named/EM/pool/ci\endcsname{[$0.1$, $1.6$]}
\expandafter\def\csname pn@hot/named/EM/pool/full\endcsname{$+0.8$ [$0.1$, $1.6$]}
\expandafter\def\csname pn@hot/named/EM/range\endcsname{$+0.1$ to $+1.2$}
\expandafter\def\csname pn@hot/named/ans/Decomp\endcsname{$+4.0$}
\expandafter\def\csname pn@hot/named/ans/Decomp/abs\endcsname{4.0}
\expandafter\def\csname pn@hot/named/ans/Decomp/ci\endcsname{[$2.9$, $5.2$]}
\expandafter\def\csname pn@hot/named/ans/Decomp/full\endcsname{$+4.0$ [$2.9$, $5.2$]}
\expandafter\def\csname pn@hot/named/ans/IRCoT\endcsname{$+0.7$}
\expandafter\def\csname pn@hot/named/ans/IRCoT/abs\endcsname{0.7}
\expandafter\def\csname pn@hot/named/ans/IRCoT/ci\endcsname{[$0.2$, $1.3$]}
\expandafter\def\csname pn@hot/named/ans/IRCoT/full\endcsname{$+0.7$ [$0.2$, $1.3$]}
\expandafter\def\csname pn@hot/named/ans/ReAct\endcsname{$+2.2$}
\expandafter\def\csname pn@hot/named/ans/ReAct/abs\endcsname{2.2}
\expandafter\def\csname pn@hot/named/ans/ReAct/ci\endcsname{[$1.3$, $3.1$]}
\expandafter\def\csname pn@hot/named/ans/ReAct/full\endcsname{$+2.2$ [$1.3$, $3.1$]}
\expandafter\def\csname pn@hot/named/ans/SR1\endcsname{$+2.2$}
\expandafter\def\csname pn@hot/named/ans/SR1/abs\endcsname{2.2}
\expandafter\def\csname pn@hot/named/ans/SR1/ci\endcsname{[$1.4$, $3.0$]}
\expandafter\def\csname pn@hot/named/ans/SR1/full\endcsname{$+2.2$ [$1.4$, $3.0$]}
\expandafter\def\csname pn@hot/named/ans/ext\endcsname{$+1.7$}
\expandafter\def\csname pn@hot/named/ans/ext/abs\endcsname{1.7}
\expandafter\def\csname pn@hot/named/ans/ext/ci\endcsname{[$1.2$, $2.2$]}
\expandafter\def\csname pn@hot/named/ans/ext/full\endcsname{$+1.7$ [$1.2$, $2.2$]}
\expandafter\def\csname pn@hot/named/ans/nsig\endcsname{4}
\expandafter\def\csname pn@hot/named/ans/pool\endcsname{$+2.3$}
\expandafter\def\csname pn@hot/named/ans/pool/abs\endcsname{2.3}
\expandafter\def\csname pn@hot/named/ans/pool/ci\endcsname{[$1.7$, $2.8$]}
\expandafter\def\csname pn@hot/named/ans/pool/full\endcsname{$+2.3$ [$1.7$, $2.8$]}
\expandafter\def\csname pn@hot/named/ans/range\endcsname{$+0.7$ to $+4.0$}
\expandafter\def\csname pn@hot/named/complete/Decomp\endcsname{$+7.3$}
\expandafter\def\csname pn@hot/named/complete/Decomp/abs\endcsname{7.3}
\expandafter\def\csname pn@hot/named/complete/Decomp/ci\endcsname{[$5.9$, $8.8$]}
\expandafter\def\csname pn@hot/named/complete/Decomp/full\endcsname{$+7.3$ [$5.9$, $8.8$]}
\expandafter\def\csname pn@hot/named/complete/IRCoT\endcsname{$+3.1$}
\expandafter\def\csname pn@hot/named/complete/IRCoT/abs\endcsname{3.1}
\expandafter\def\csname pn@hot/named/complete/IRCoT/ci\endcsname{[$2.2$, $4.1$]}
\expandafter\def\csname pn@hot/named/complete/IRCoT/full\endcsname{$+3.1$ [$2.2$, $4.1$]}
\expandafter\def\csname pn@hot/named/complete/ReAct\endcsname{$+4.9$}
\expandafter\def\csname pn@hot/named/complete/ReAct/abs\endcsname{4.9}
\expandafter\def\csname pn@hot/named/complete/ReAct/ci\endcsname{[$3.8$, $6.1$]}
\expandafter\def\csname pn@hot/named/complete/ReAct/full\endcsname{$+4.9$ [$3.8$, $6.1$]}
\expandafter\def\csname pn@hot/named/complete/SR1\endcsname{$+5.5$}
\expandafter\def\csname pn@hot/named/complete/SR1/abs\endcsname{5.5}
\expandafter\def\csname pn@hot/named/complete/SR1/ci\endcsname{[$4.4$, $6.7$]}
\expandafter\def\csname pn@hot/named/complete/SR1/full\endcsname{$+5.5$ [$4.4$, $6.7$]}
\expandafter\def\csname pn@hot/named/complete/ext\endcsname{$+4.5$}
\expandafter\def\csname pn@hot/named/complete/ext/abs\endcsname{4.5}
\expandafter\def\csname pn@hot/named/complete/ext/ci\endcsname{[$3.8$, $5.3$]}
\expandafter\def\csname pn@hot/named/complete/ext/full\endcsname{$+4.5$ [$3.8$, $5.3$]}
\expandafter\def\csname pn@hot/named/complete/nsig\endcsname{4}
\expandafter\def\csname pn@hot/named/complete/pool\endcsname{$+5.2$}
\expandafter\def\csname pn@hot/named/complete/pool/abs\endcsname{5.2}
\expandafter\def\csname pn@hot/named/complete/pool/ci\endcsname{[$4.5$, $6.0$]}
\expandafter\def\csname pn@hot/named/complete/pool/full\endcsname{$+5.2$ [$4.5$, $6.0$]}
\expandafter\def\csname pn@hot/named/complete/range\endcsname{$+3.1$ to $+7.3$}
\expandafter\def\csname pn@hot/novelquery/EM/Decomp\endcsname{$+1.2$}
\expandafter\def\csname pn@hot/novelquery/EM/Decomp/abs\endcsname{1.2}
\expandafter\def\csname pn@hot/novelquery/EM/Decomp/ci\endcsname{[$0.4$, $2.1$]}
\expandafter\def\csname pn@hot/novelquery/EM/Decomp/full\endcsname{$+1.2$ [$0.4$, $2.1$]}
\expandafter\def\csname pn@hot/novelquery/EM/IRCoT\endcsname{$+0.8$}
\expandafter\def\csname pn@hot/novelquery/EM/IRCoT/abs\endcsname{0.8}
\expandafter\def\csname pn@hot/novelquery/EM/IRCoT/ci\endcsname{[$-0.5$, $2.0$]}
\expandafter\def\csname pn@hot/novelquery/EM/IRCoT/full\endcsname{$+0.8$ [$-0.5$, $2.0$]}
\expandafter\def\csname pn@hot/novelquery/EM/ReAct\endcsname{$+2.3$}
\expandafter\def\csname pn@hot/novelquery/EM/ReAct/abs\endcsname{2.3}
\expandafter\def\csname pn@hot/novelquery/EM/ReAct/ci\endcsname{[$0.9$, $3.8$]}
\expandafter\def\csname pn@hot/novelquery/EM/ReAct/full\endcsname{$+2.3$ [$0.9$, $3.8$]}
\expandafter\def\csname pn@hot/novelquery/EM/SR1\endcsname{$+0.1$}
\expandafter\def\csname pn@hot/novelquery/EM/SR1/abs\endcsname{0.1}
\expandafter\def\csname pn@hot/novelquery/EM/SR1/ci\endcsname{[$-0.6$, $0.9$]}
\expandafter\def\csname pn@hot/novelquery/EM/SR1/full\endcsname{$+0.1$ [$-0.6$, $0.9$]}
\expandafter\def\csname pn@hot/novelquery/EM/ext\endcsname{$+1.1$}
\expandafter\def\csname pn@hot/novelquery/EM/ext/abs\endcsname{1.1}
\expandafter\def\csname pn@hot/novelquery/EM/ext/ci\endcsname{[$0.4$, $1.8$]}
\expandafter\def\csname pn@hot/novelquery/EM/ext/full\endcsname{$+1.1$ [$0.4$, $1.8$]}
\expandafter\def\csname pn@hot/novelquery/EM/nsig\endcsname{2}
\expandafter\def\csname pn@hot/novelquery/EM/pool\endcsname{$+1.1$}
\expandafter\def\csname pn@hot/novelquery/EM/pool/abs\endcsname{1.1}
\expandafter\def\csname pn@hot/novelquery/EM/pool/ci\endcsname{[$0.5$, $1.7$]}
\expandafter\def\csname pn@hot/novelquery/EM/pool/full\endcsname{$+1.1$ [$0.5$, $1.7$]}
\expandafter\def\csname pn@hot/novelquery/EM/range\endcsname{$+0.1$ to $+2.3$}
\expandafter\def\csname pn@hot/novelquery/ans/Decomp\endcsname{$+2.6$}
\expandafter\def\csname pn@hot/novelquery/ans/Decomp/abs\endcsname{2.6}
\expandafter\def\csname pn@hot/novelquery/ans/Decomp/ci\endcsname{[$1.7$, $3.6$]}
\expandafter\def\csname pn@hot/novelquery/ans/Decomp/full\endcsname{$+2.6$ [$1.7$, $3.6$]}
\expandafter\def\csname pn@hot/novelquery/ans/IRCoT\endcsname{$+0.1$}
\expandafter\def\csname pn@hot/novelquery/ans/IRCoT/abs\endcsname{0.1}
\expandafter\def\csname pn@hot/novelquery/ans/IRCoT/ci\endcsname{[$-1.0$, $1.1$]}
\expandafter\def\csname pn@hot/novelquery/ans/IRCoT/full\endcsname{$+0.1$ [$-1.0$, $1.1$]}
\expandafter\def\csname pn@hot/novelquery/ans/ReAct\endcsname{$+4.7$}
\expandafter\def\csname pn@hot/novelquery/ans/ReAct/abs\endcsname{4.7}
\expandafter\def\csname pn@hot/novelquery/ans/ReAct/ci\endcsname{[$3.2$, $6.1$]}
\expandafter\def\csname pn@hot/novelquery/ans/ReAct/full\endcsname{$+4.7$ [$3.2$, $6.1$]}
\expandafter\def\csname pn@hot/novelquery/ans/SR1\endcsname{$+0.8$}
\expandafter\def\csname pn@hot/novelquery/ans/SR1/abs\endcsname{0.8}
\expandafter\def\csname pn@hot/novelquery/ans/SR1/ci\endcsname{[$0.1$, $1.4$]}
\expandafter\def\csname pn@hot/novelquery/ans/SR1/full\endcsname{$+0.8$ [$0.1$, $1.4$]}
\expandafter\def\csname pn@hot/novelquery/ans/ext\endcsname{$+1.8$}
\expandafter\def\csname pn@hot/novelquery/ans/ext/abs\endcsname{1.8}
\expandafter\def\csname pn@hot/novelquery/ans/ext/ci\endcsname{[$1.1$, $2.6$]}
\expandafter\def\csname pn@hot/novelquery/ans/ext/full\endcsname{$+1.8$ [$1.1$, $2.6$]}
\expandafter\def\csname pn@hot/novelquery/ans/nsig\endcsname{3}
\expandafter\def\csname pn@hot/novelquery/ans/pool\endcsname{$+2.0$}
\expandafter\def\csname pn@hot/novelquery/ans/pool/abs\endcsname{2.0}
\expandafter\def\csname pn@hot/novelquery/ans/pool/ci\endcsname{[$1.4$, $2.7$]}
\expandafter\def\csname pn@hot/novelquery/ans/pool/full\endcsname{$+2.0$ [$1.4$, $2.7$]}
\expandafter\def\csname pn@hot/novelquery/ans/range\endcsname{$+0.1$ to $+4.7$}
\expandafter\def\csname pn@hot/novelquery/complete/Decomp\endcsname{$+3.8$}
\expandafter\def\csname pn@hot/novelquery/complete/Decomp/abs\endcsname{3.8}
\expandafter\def\csname pn@hot/novelquery/complete/Decomp/ci\endcsname{[$2.7$, $4.9$]}
\expandafter\def\csname pn@hot/novelquery/complete/Decomp/full\endcsname{$+3.8$ [$2.7$, $4.9$]}
\expandafter\def\csname pn@hot/novelquery/complete/IRCoT\endcsname{$+1.7$}
\expandafter\def\csname pn@hot/novelquery/complete/IRCoT/abs\endcsname{1.7}
\expandafter\def\csname pn@hot/novelquery/complete/IRCoT/ci\endcsname{[$0.4$, $3.1$]}
\expandafter\def\csname pn@hot/novelquery/complete/IRCoT/full\endcsname{$+1.7$ [$0.4$, $3.1$]}
\expandafter\def\csname pn@hot/novelquery/complete/ReAct\endcsname{$+7.3$}
\expandafter\def\csname pn@hot/novelquery/complete/ReAct/abs\endcsname{7.3}
\expandafter\def\csname pn@hot/novelquery/complete/ReAct/ci\endcsname{[$5.6$, $9.0$]}
\expandafter\def\csname pn@hot/novelquery/complete/ReAct/full\endcsname{$+7.3$ [$5.6$, $9.0$]}
\expandafter\def\csname pn@hot/novelquery/complete/SR1\endcsname{$+1.2$}
\expandafter\def\csname pn@hot/novelquery/complete/SR1/abs\endcsname{1.2}
\expandafter\def\csname pn@hot/novelquery/complete/SR1/ci\endcsname{[$0.4$, $2.0$]}
\expandafter\def\csname pn@hot/novelquery/complete/SR1/full\endcsname{$+1.2$ [$0.4$, $2.0$]}
\expandafter\def\csname pn@hot/novelquery/complete/ext\endcsname{$+3.4$}
\expandafter\def\csname pn@hot/novelquery/complete/ext/abs\endcsname{3.4}
\expandafter\def\csname pn@hot/novelquery/complete/ext/ci\endcsname{[$2.6$, $4.3$]}
\expandafter\def\csname pn@hot/novelquery/complete/ext/full\endcsname{$+3.4$ [$2.6$, $4.3$]}
\expandafter\def\csname pn@hot/novelquery/complete/nsig\endcsname{4}
\expandafter\def\csname pn@hot/novelquery/complete/pool\endcsname{$+3.5$}
\expandafter\def\csname pn@hot/novelquery/complete/pool/abs\endcsname{3.5}
\expandafter\def\csname pn@hot/novelquery/complete/pool/ci\endcsname{[$2.8$, $4.3$]}
\expandafter\def\csname pn@hot/novelquery/complete/pool/full\endcsname{$+3.5$ [$2.8$, $4.3$]}
\expandafter\def\csname pn@hot/novelquery/complete/range\endcsname{$+1.2$ to $+7.3$}
\expandafter\def\csname pn@hot/order/EM/Decomp\endcsname{$-1.3$}
\expandafter\def\csname pn@hot/order/EM/Decomp/abs\endcsname{1.3}
\expandafter\def\csname pn@hot/order/EM/Decomp/ci\endcsname{[$-2.5$, $-0.1$]}
\expandafter\def\csname pn@hot/order/EM/Decomp/full\endcsname{$-1.3$ [$-2.5$, $-0.1$]}
\expandafter\def\csname pn@hot/order/EM/IRCoT\endcsname{$-0.3$}
\expandafter\def\csname pn@hot/order/EM/IRCoT/abs\endcsname{0.3}
\expandafter\def\csname pn@hot/order/EM/IRCoT/ci\endcsname{[$-1.5$, $0.9$]}
\expandafter\def\csname pn@hot/order/EM/IRCoT/full\endcsname{$-0.3$ [$-1.5$, $0.9$]}
\expandafter\def\csname pn@hot/order/EM/ReAct\endcsname{$-0.9$}
\expandafter\def\csname pn@hot/order/EM/ReAct/abs\endcsname{0.9}
\expandafter\def\csname pn@hot/order/EM/ReAct/ci\endcsname{[$-2.2$, $0.3$]}
\expandafter\def\csname pn@hot/order/EM/ReAct/full\endcsname{$-0.9$ [$-2.2$, $0.3$]}
\expandafter\def\csname pn@hot/order/EM/SR1\endcsname{$-0.7$}
\expandafter\def\csname pn@hot/order/EM/SR1/abs\endcsname{0.7}
\expandafter\def\csname pn@hot/order/EM/SR1/ci\endcsname{[$-2.0$, $0.5$]}
\expandafter\def\csname pn@hot/order/EM/SR1/full\endcsname{$-0.7$ [$-2.0$, $0.5$]}
\expandafter\def\csname pn@hot/order/EM/ext\endcsname{$-0.7$}
\expandafter\def\csname pn@hot/order/EM/ext/abs\endcsname{0.7}
\expandafter\def\csname pn@hot/order/EM/ext/ci\endcsname{[$-1.5$, $0.2$]}
\expandafter\def\csname pn@hot/order/EM/ext/full\endcsname{$-0.7$ [$-1.5$, $0.2$]}
\expandafter\def\csname pn@hot/order/EM/nsig\endcsname{0}
\expandafter\def\csname pn@hot/order/EM/pool\endcsname{$-0.8$}
\expandafter\def\csname pn@hot/order/EM/pool/abs\endcsname{0.8}
\expandafter\def\csname pn@hot/order/EM/pool/ci\endcsname{[$-1.6$, $-0.1$]}
\expandafter\def\csname pn@hot/order/EM/pool/full\endcsname{$-0.8$ [$-1.6$, $-0.1$]}
\expandafter\def\csname pn@hot/order/EM/range\endcsname{$-1.3$ to $-0.3$}
\expandafter\def\csname pn@hot/order/ans/Decomp\endcsname{$0.0$}
\expandafter\def\csname pn@hot/order/ans/Decomp/abs\endcsname{0.0}
\expandafter\def\csname pn@hot/order/ans/Decomp/ci\endcsname{[$0.0$, $0.0$]}
\expandafter\def\csname pn@hot/order/ans/Decomp/full\endcsname{$0.0$ [$0.0$, $0.0$]}
\expandafter\def\csname pn@hot/order/ans/IRCoT\endcsname{$0.0$}
\expandafter\def\csname pn@hot/order/ans/IRCoT/abs\endcsname{0.0}
\expandafter\def\csname pn@hot/order/ans/IRCoT/ci\endcsname{[$0.0$, $0.0$]}
\expandafter\def\csname pn@hot/order/ans/IRCoT/full\endcsname{$0.0$ [$0.0$, $0.0$]}
\expandafter\def\csname pn@hot/order/ans/ReAct\endcsname{$0.0$}
\expandafter\def\csname pn@hot/order/ans/ReAct/abs\endcsname{0.0}
\expandafter\def\csname pn@hot/order/ans/ReAct/ci\endcsname{[$0.0$, $0.0$]}
\expandafter\def\csname pn@hot/order/ans/ReAct/full\endcsname{$0.0$ [$0.0$, $0.0$]}
\expandafter\def\csname pn@hot/order/ans/SR1\endcsname{$0.0$}
\expandafter\def\csname pn@hot/order/ans/SR1/abs\endcsname{0.0}
\expandafter\def\csname pn@hot/order/ans/SR1/ci\endcsname{[$0.0$, $0.0$]}
\expandafter\def\csname pn@hot/order/ans/SR1/full\endcsname{$0.0$ [$0.0$, $0.0$]}
\expandafter\def\csname pn@hot/order/ans/ext\endcsname{$0.0$}
\expandafter\def\csname pn@hot/order/ans/ext/abs\endcsname{0.0}
\expandafter\def\csname pn@hot/order/ans/ext/ci\endcsname{[$0.0$, $0.0$]}
\expandafter\def\csname pn@hot/order/ans/ext/full\endcsname{$0.0$ [$0.0$, $0.0$]}
\expandafter\def\csname pn@hot/order/ans/nsig\endcsname{0}
\expandafter\def\csname pn@hot/order/ans/pool\endcsname{$0.0$}
\expandafter\def\csname pn@hot/order/ans/pool/abs\endcsname{0.0}
\expandafter\def\csname pn@hot/order/ans/pool/ci\endcsname{[$0.0$, $0.0$]}
\expandafter\def\csname pn@hot/order/ans/pool/full\endcsname{$0.0$ [$0.0$, $0.0$]}
\expandafter\def\csname pn@hot/order/ans/range\endcsname{$0.0$ to $0.0$}
\expandafter\def\csname pn@hot/order/complete/Decomp\endcsname{$0.0$}
\expandafter\def\csname pn@hot/order/complete/Decomp/abs\endcsname{0.0}
\expandafter\def\csname pn@hot/order/complete/Decomp/ci\endcsname{[$0.0$, $0.0$]}
\expandafter\def\csname pn@hot/order/complete/Decomp/full\endcsname{$0.0$ [$0.0$, $0.0$]}
\expandafter\def\csname pn@hot/order/complete/IRCoT\endcsname{$0.0$}
\expandafter\def\csname pn@hot/order/complete/IRCoT/abs\endcsname{0.0}
\expandafter\def\csname pn@hot/order/complete/IRCoT/ci\endcsname{[$0.0$, $0.0$]}
\expandafter\def\csname pn@hot/order/complete/IRCoT/full\endcsname{$0.0$ [$0.0$, $0.0$]}
\expandafter\def\csname pn@hot/order/complete/ReAct\endcsname{$0.0$}
\expandafter\def\csname pn@hot/order/complete/ReAct/abs\endcsname{0.0}
\expandafter\def\csname pn@hot/order/complete/ReAct/ci\endcsname{[$0.0$, $0.0$]}
\expandafter\def\csname pn@hot/order/complete/ReAct/full\endcsname{$0.0$ [$0.0$, $0.0$]}
\expandafter\def\csname pn@hot/order/complete/SR1\endcsname{$0.0$}
\expandafter\def\csname pn@hot/order/complete/SR1/abs\endcsname{0.0}
\expandafter\def\csname pn@hot/order/complete/SR1/ci\endcsname{[$0.0$, $0.0$]}
\expandafter\def\csname pn@hot/order/complete/SR1/full\endcsname{$0.0$ [$0.0$, $0.0$]}
\expandafter\def\csname pn@hot/order/complete/ext\endcsname{$0.0$}
\expandafter\def\csname pn@hot/order/complete/ext/abs\endcsname{0.0}
\expandafter\def\csname pn@hot/order/complete/ext/ci\endcsname{[$0.0$, $0.0$]}
\expandafter\def\csname pn@hot/order/complete/ext/full\endcsname{$0.0$ [$0.0$, $0.0$]}
\expandafter\def\csname pn@hot/order/complete/nsig\endcsname{0}
\expandafter\def\csname pn@hot/order/complete/pool\endcsname{$0.0$}
\expandafter\def\csname pn@hot/order/complete/pool/abs\endcsname{0.0}
\expandafter\def\csname pn@hot/order/complete/pool/ci\endcsname{[$0.0$, $0.0$]}
\expandafter\def\csname pn@hot/order/complete/pool/full\endcsname{$0.0$ [$0.0$, $0.0$]}
\expandafter\def\csname pn@hot/order/complete/range\endcsname{$0.0$ to $0.0$}
\expandafter\def\csname pn@hot/qt/bridge/detour/EM\endcsname{$+2.6$}
\expandafter\def\csname pn@hot/qt/bridge/detour/EM/abs\endcsname{2.6}
\expandafter\def\csname pn@hot/qt/bridge/detour/EM/ci\endcsname{[$1.5$, $3.7$]}
\expandafter\def\csname pn@hot/qt/bridge/detour/EM/full\endcsname{$+1.8$ [$0.8$, $2.9$]}
\expandafter\def\csname pn@hot/qt/bridge/detour/complete\endcsname{$+12.3$}
\expandafter\def\csname pn@hot/qt/bridge/detour/complete/abs\endcsname{12.3}
\expandafter\def\csname pn@hot/qt/bridge/detour/complete/ci\endcsname{[$11.0$, $13.5$]}
\expandafter\def\csname pn@hot/qt/bridge/detour/complete/full\endcsname{$+12.3$ [$11.0$, $13.5$]}
\expandafter\def\csname pn@hot/qt/bridge/n\endcsname{1{,}197}
\expandafter\def\csname pn@hot/qt/bridge/route/EM\endcsname{$+1.7$}
\expandafter\def\csname pn@hot/qt/bridge/route/EM/abs\endcsname{1.7}
\expandafter\def\csname pn@hot/qt/bridge/route/EM/ci\endcsname{[$1.1$, $2.3$]}
\expandafter\def\csname pn@hot/qt/bridge/route/EM/full\endcsname{$+1.7$ [$1.1$, $2.3$]}
\expandafter\def\csname pn@hot/qt/bridge/route/complete\endcsname{$+5.2$}
\expandafter\def\csname pn@hot/qt/bridge/route/complete/abs\endcsname{5.2}
\expandafter\def\csname pn@hot/qt/bridge/route/complete/ci\endcsname{[$4.3$, $6.1$]}
\expandafter\def\csname pn@hot/qt/bridge/route/complete/full\endcsname{$+5.2$ [$4.3$, $6.1$]}
\expandafter\def\csname pn@hot/qt/bridge/window/EM\endcsname{$+0.9$}
\expandafter\def\csname pn@hot/qt/bridge/window/EM/abs\endcsname{0.9}
\expandafter\def\csname pn@hot/qt/bridge/window/EM/ci\endcsname{[$-0.1$, $1.9$]}
\expandafter\def\csname pn@hot/qt/bridge/window/EM/full\endcsname{$+0.9$ [$-0.1$, $1.9$]}
\expandafter\def\csname pn@hot/qt/bridge/window/complete\endcsname{$+7.0$}
\expandafter\def\csname pn@hot/qt/bridge/window/complete/abs\endcsname{7.0}
\expandafter\def\csname pn@hot/qt/bridge/window/complete/ci\endcsname{[$6.1$, $8.0$]}
\expandafter\def\csname pn@hot/qt/bridge/window/complete/full\endcsname{$+7.0$ [$6.1$, $8.0$]}
\expandafter\def\csname pn@hot/qt/comparison/detour/EM\endcsname{$-0.6$}
\expandafter\def\csname pn@hot/qt/comparison/detour/EM/abs\endcsname{0.6}
\expandafter\def\csname pn@hot/qt/comparison/detour/EM/ci\endcsname{[$-2.5$, $1.4$]}
\expandafter\def\csname pn@hot/qt/comparison/detour/EM/full\endcsname{$-0.2$ [$-2.1$, $1.7$]}
\expandafter\def\csname pn@hot/qt/comparison/detour/complete\endcsname{$+9.8$}
\expandafter\def\csname pn@hot/qt/comparison/detour/complete/abs\endcsname{9.8}
\expandafter\def\csname pn@hot/qt/comparison/detour/complete/ci\endcsname{[$7.5$, $12.2$]}
\expandafter\def\csname pn@hot/qt/comparison/detour/complete/full\endcsname{$+9.8$ [$7.5$, $12.2$]}
\expandafter\def\csname pn@hot/qt/comparison/n\endcsname{303}
\expandafter\def\csname pn@hot/qt/comparison/route/EM\endcsname{$+0.1$}
\expandafter\def\csname pn@hot/qt/comparison/route/EM/abs\endcsname{0.1}
\expandafter\def\csname pn@hot/qt/comparison/route/EM/ci\endcsname{[$-1.2$, $1.4$]}
\expandafter\def\csname pn@hot/qt/comparison/route/EM/full\endcsname{$+0.1$ [$-1.2$, $1.4$]}
\expandafter\def\csname pn@hot/qt/comparison/route/complete\endcsname{$+0.4$}
\expandafter\def\csname pn@hot/qt/comparison/route/complete/abs\endcsname{0.4}
\expandafter\def\csname pn@hot/qt/comparison/route/complete/ci\endcsname{[$-0.2$, $1.1$]}
\expandafter\def\csname pn@hot/qt/comparison/route/complete/full\endcsname{$+0.4$ [$-0.2$, $1.1$]}
\expandafter\def\csname pn@hot/qt/comparison/window/EM\endcsname{$-0.7$}
\expandafter\def\csname pn@hot/qt/comparison/window/EM/abs\endcsname{0.7}
\expandafter\def\csname pn@hot/qt/comparison/window/EM/ci\endcsname{[$-2.5$, $1.2$]}
\expandafter\def\csname pn@hot/qt/comparison/window/EM/full\endcsname{$-0.7$ [$-2.5$, $1.2$]}
\expandafter\def\csname pn@hot/qt/comparison/window/complete\endcsname{$+9.4$}
\expandafter\def\csname pn@hot/qt/comparison/window/complete/abs\endcsname{9.4}
\expandafter\def\csname pn@hot/qt/comparison/window/complete/ci\endcsname{[$7.2$, $11.8$]}
\expandafter\def\csname pn@hot/qt/comparison/window/complete/full\endcsname{$+9.4$ [$7.2$, $11.8$]}
\expandafter\def\csname pn@hot/randomroute/EM/Decomp\endcsname{$+1.2$}
\expandafter\def\csname pn@hot/randomroute/EM/Decomp/abs\endcsname{1.2}
\expandafter\def\csname pn@hot/randomroute/EM/Decomp/ci\endcsname{[$0.5$, $1.9$]}
\expandafter\def\csname pn@hot/randomroute/EM/Decomp/full\endcsname{$+1.2$ [$0.5$, $1.9$]}
\expandafter\def\csname pn@hot/randomroute/EM/IRCoT\endcsname{$+0.9$}
\expandafter\def\csname pn@hot/randomroute/EM/IRCoT/abs\endcsname{0.9}
\expandafter\def\csname pn@hot/randomroute/EM/IRCoT/ci\endcsname{[$0.0$, $1.8$]}
\expandafter\def\csname pn@hot/randomroute/EM/IRCoT/full\endcsname{$+0.9$ [$0.0$, $1.8$]}
\expandafter\def\csname pn@hot/randomroute/EM/ReAct\endcsname{$+1.9$}
\expandafter\def\csname pn@hot/randomroute/EM/ReAct/abs\endcsname{1.9}
\expandafter\def\csname pn@hot/randomroute/EM/ReAct/ci\endcsname{[$0.8$, $3.0$]}
\expandafter\def\csname pn@hot/randomroute/EM/ReAct/full\endcsname{$+1.9$ [$0.8$, $3.0$]}
\expandafter\def\csname pn@hot/randomroute/EM/SR1\endcsname{$+0.3$}
\expandafter\def\csname pn@hot/randomroute/EM/SR1/abs\endcsname{0.3}
\expandafter\def\csname pn@hot/randomroute/EM/SR1/ci\endcsname{[$-0.3$, $0.9$]}
\expandafter\def\csname pn@hot/randomroute/EM/SR1/full\endcsname{$+0.3$ [$-0.3$, $0.9$]}
\expandafter\def\csname pn@hot/randomroute/EM/ext\endcsname{$+1.0$}
\expandafter\def\csname pn@hot/randomroute/EM/ext/abs\endcsname{1.0}
\expandafter\def\csname pn@hot/randomroute/EM/ext/ci\endcsname{[$0.5$, $1.6$]}
\expandafter\def\csname pn@hot/randomroute/EM/ext/full\endcsname{$+1.0$ [$0.5$, $1.6$]}
\expandafter\def\csname pn@hot/randomroute/EM/nsig\endcsname{2}
\expandafter\def\csname pn@hot/randomroute/EM/pool\endcsname{$+1.1$}
\expandafter\def\csname pn@hot/randomroute/EM/pool/abs\endcsname{0.5}
\expandafter\def\csname pn@hot/randomroute/EM/pool/ci\endcsname{[$0.6$, $1.6$]}
\expandafter\def\csname pn@hot/randomroute/EM/pool/full\endcsname{$+1.1$ [$0.6$, $1.6$]}
\expandafter\def\csname pn@hot/randomroute/EM/range\endcsname{$+0.3$ to $+1.9$}
\expandafter\def\csname pn@hot/randomroute/ans/Decomp\endcsname{$+2.4$}
\expandafter\def\csname pn@hot/randomroute/ans/Decomp/abs\endcsname{2.4}
\expandafter\def\csname pn@hot/randomroute/ans/Decomp/ci\endcsname{[$1.5$, $3.3$]}
\expandafter\def\csname pn@hot/randomroute/ans/Decomp/full\endcsname{$+2.4$ [$1.5$, $3.3$]}
\expandafter\def\csname pn@hot/randomroute/ans/IRCoT\endcsname{$+0.4$}
\expandafter\def\csname pn@hot/randomroute/ans/IRCoT/abs\endcsname{0.4}
\expandafter\def\csname pn@hot/randomroute/ans/IRCoT/ci\endcsname{[$-0.3$, $1.0$]}
\expandafter\def\csname pn@hot/randomroute/ans/IRCoT/full\endcsname{$+0.4$ [$-0.3$, $1.0$]}
\expandafter\def\csname pn@hot/randomroute/ans/ReAct\endcsname{$+3.7$}
\expandafter\def\csname pn@hot/randomroute/ans/ReAct/abs\endcsname{3.7}
\expandafter\def\csname pn@hot/randomroute/ans/ReAct/ci\endcsname{[$2.6$, $4.9$]}
\expandafter\def\csname pn@hot/randomroute/ans/ReAct/full\endcsname{$+3.7$ [$2.6$, $4.9$]}
\expandafter\def\csname pn@hot/randomroute/ans/SR1\endcsname{$+0.5$}
\expandafter\def\csname pn@hot/randomroute/ans/SR1/abs\endcsname{0.5}
\expandafter\def\csname pn@hot/randomroute/ans/SR1/ci\endcsname{[$0.1$, $1.0$]}
\expandafter\def\csname pn@hot/randomroute/ans/SR1/full\endcsname{$+0.5$ [$0.1$, $1.0$]}
\expandafter\def\csname pn@hot/randomroute/ans/ext\endcsname{$+1.5$}
\expandafter\def\csname pn@hot/randomroute/ans/ext/abs\endcsname{1.5}
\expandafter\def\csname pn@hot/randomroute/ans/ext/ci\endcsname{[$1.0$, $2.1$]}
\expandafter\def\csname pn@hot/randomroute/ans/ext/full\endcsname{$+1.5$ [$1.0$, $2.1$]}
\expandafter\def\csname pn@hot/randomroute/ans/nsig\endcsname{3}
\expandafter\def\csname pn@hot/randomroute/ans/pool\endcsname{$+1.8$}
\expandafter\def\csname pn@hot/randomroute/ans/pool/abs\endcsname{1.8}
\expandafter\def\csname pn@hot/randomroute/ans/pool/ci\endcsname{[$1.2$, $2.3$]}
\expandafter\def\csname pn@hot/randomroute/ans/pool/full\endcsname{$+1.8$ [$1.2$, $2.3$]}
\expandafter\def\csname pn@hot/randomroute/ans/range\endcsname{$+0.4$ to $+3.7$}
\expandafter\def\csname pn@hot/randomroute/complete/Decomp\endcsname{$+3.2$}
\expandafter\def\csname pn@hot/randomroute/complete/Decomp/abs\endcsname{3.2}
\expandafter\def\csname pn@hot/randomroute/complete/Decomp/ci\endcsname{[$2.3$, $4.2$]}
\expandafter\def\csname pn@hot/randomroute/complete/Decomp/full\endcsname{$+3.2$ [$2.3$, $4.2$]}
\expandafter\def\csname pn@hot/randomroute/complete/IRCoT\endcsname{$+1.6$}
\expandafter\def\csname pn@hot/randomroute/complete/IRCoT/abs\endcsname{1.6}
\expandafter\def\csname pn@hot/randomroute/complete/IRCoT/ci\endcsname{[$0.8$, $2.4$]}
\expandafter\def\csname pn@hot/randomroute/complete/IRCoT/full\endcsname{$+1.6$ [$0.8$, $2.4$]}
\expandafter\def\csname pn@hot/randomroute/complete/ReAct\endcsname{$+5.9$}
\expandafter\def\csname pn@hot/randomroute/complete/ReAct/abs\endcsname{5.9}
\expandafter\def\csname pn@hot/randomroute/complete/ReAct/ci\endcsname{[$4.6$, $7.2$]}
\expandafter\def\csname pn@hot/randomroute/complete/ReAct/full\endcsname{$+5.9$ [$4.6$, $7.2$]}
\expandafter\def\csname pn@hot/randomroute/complete/SR1\endcsname{$+0.9$}
\expandafter\def\csname pn@hot/randomroute/complete/SR1/abs\endcsname{0.9}
\expandafter\def\csname pn@hot/randomroute/complete/SR1/ci\endcsname{[$0.3$, $1.6$]}
\expandafter\def\csname pn@hot/randomroute/complete/SR1/full\endcsname{$+0.9$ [$0.3$, $1.6$]}
\expandafter\def\csname pn@hot/randomroute/complete/ext\endcsname{$+2.8$}
\expandafter\def\csname pn@hot/randomroute/complete/ext/abs\endcsname{2.8}
\expandafter\def\csname pn@hot/randomroute/complete/ext/ci\endcsname{[$2.2$, $3.5$]}
\expandafter\def\csname pn@hot/randomroute/complete/ext/full\endcsname{$+2.8$ [$2.2$, $3.5$]}
\expandafter\def\csname pn@hot/randomroute/complete/nsig\endcsname{4}
\expandafter\def\csname pn@hot/randomroute/complete/pool\endcsname{$+2.9$}
\expandafter\def\csname pn@hot/randomroute/complete/pool/abs\endcsname{2.9}
\expandafter\def\csname pn@hot/randomroute/complete/pool/ci\endcsname{[$2.3$, $3.5$]}
\expandafter\def\csname pn@hot/randomroute/complete/pool/full\endcsname{$+2.9$ [$2.3$, $3.5$]}
\expandafter\def\csname pn@hot/randomroute/complete/range\endcsname{$+0.9$ to $+5.9$}
\expandafter\def\csname pn@hot/rerankVSrrf/EM/Decomp\endcsname{$+2.6$}
\expandafter\def\csname pn@hot/rerankVSrrf/EM/Decomp/abs\endcsname{2.6}
\expandafter\def\csname pn@hot/rerankVSrrf/EM/Decomp/ci\endcsname{[$1.0$, $4.3$]}
\expandafter\def\csname pn@hot/rerankVSrrf/EM/Decomp/full\endcsname{$+2.6$ [$1.0$, $4.3$]}
\expandafter\def\csname pn@hot/rerankVSrrf/EM/IRCoT\endcsname{$+0.7$}
\expandafter\def\csname pn@hot/rerankVSrrf/EM/IRCoT/abs\endcsname{0.7}
\expandafter\def\csname pn@hot/rerankVSrrf/EM/IRCoT/ci\endcsname{[$-1.1$, $2.5$]}
\expandafter\def\csname pn@hot/rerankVSrrf/EM/IRCoT/full\endcsname{$+0.7$ [$-1.1$, $2.5$]}
\expandafter\def\csname pn@hot/rerankVSrrf/EM/ReAct\endcsname{$+3.0$}
\expandafter\def\csname pn@hot/rerankVSrrf/EM/ReAct/abs\endcsname{3.0}
\expandafter\def\csname pn@hot/rerankVSrrf/EM/ReAct/ci\endcsname{[$1.3$, $4.7$]}
\expandafter\def\csname pn@hot/rerankVSrrf/EM/ReAct/full\endcsname{$+3.0$ [$1.3$, $4.7$]}
\expandafter\def\csname pn@hot/rerankVSrrf/EM/SR1\endcsname{$+0.6$}
\expandafter\def\csname pn@hot/rerankVSrrf/EM/SR1/abs\endcsname{0.6}
\expandafter\def\csname pn@hot/rerankVSrrf/EM/SR1/ci\endcsname{[$-1.1$, $2.3$]}
\expandafter\def\csname pn@hot/rerankVSrrf/EM/SR1/full\endcsname{$+0.6$ [$-1.1$, $2.3$]}
\expandafter\def\csname pn@hot/rerankVSrrf/EM/ext\endcsname{$+1.4$}
\expandafter\def\csname pn@hot/rerankVSrrf/EM/ext/abs\endcsname{1.4}
\expandafter\def\csname pn@hot/rerankVSrrf/EM/ext/ci\endcsname{[$0.3$, $2.6$]}
\expandafter\def\csname pn@hot/rerankVSrrf/EM/ext/full\endcsname{$+1.4$ [$0.3$, $2.6$]}
\expandafter\def\csname pn@hot/rerankVSrrf/EM/nsig\endcsname{2}
\expandafter\def\csname pn@hot/rerankVSrrf/EM/pool\endcsname{$+1.7$}
\expandafter\def\csname pn@hot/rerankVSrrf/EM/pool/abs\endcsname{1.7}
\expandafter\def\csname pn@hot/rerankVSrrf/EM/pool/ci\endcsname{[$0.7$, $2.7$]}
\expandafter\def\csname pn@hot/rerankVSrrf/EM/pool/full\endcsname{$+1.7$ [$0.7$, $2.7$]}
\expandafter\def\csname pn@hot/rerankVSrrf/EM/range\endcsname{$+0.6$ to $+3.0$}
\expandafter\def\csname pn@hot/rerankVSrrf/ans/Decomp\endcsname{$+7.1$}
\expandafter\def\csname pn@hot/rerankVSrrf/ans/Decomp/abs\endcsname{7.1}
\expandafter\def\csname pn@hot/rerankVSrrf/ans/Decomp/ci\endcsname{[$5.6$, $8.6$]}
\expandafter\def\csname pn@hot/rerankVSrrf/ans/Decomp/full\endcsname{$+7.1$ [$5.6$, $8.6$]}
\expandafter\def\csname pn@hot/rerankVSrrf/ans/IRCoT\endcsname{$+5.5$}
\expandafter\def\csname pn@hot/rerankVSrrf/ans/IRCoT/abs\endcsname{5.5}
\expandafter\def\csname pn@hot/rerankVSrrf/ans/IRCoT/ci\endcsname{[$4.1$, $7.0$]}
\expandafter\def\csname pn@hot/rerankVSrrf/ans/IRCoT/full\endcsname{$+5.5$ [$4.1$, $7.0$]}
\expandafter\def\csname pn@hot/rerankVSrrf/ans/ReAct\endcsname{$+8.1$}
\expandafter\def\csname pn@hot/rerankVSrrf/ans/ReAct/abs\endcsname{8.1}
\expandafter\def\csname pn@hot/rerankVSrrf/ans/ReAct/ci\endcsname{[$6.6$, $9.7$]}
\expandafter\def\csname pn@hot/rerankVSrrf/ans/ReAct/full\endcsname{$+8.1$ [$6.6$, $9.7$]}
\expandafter\def\csname pn@hot/rerankVSrrf/ans/SR1\endcsname{$+5.6$}
\expandafter\def\csname pn@hot/rerankVSrrf/ans/SR1/abs\endcsname{5.6}
\expandafter\def\csname pn@hot/rerankVSrrf/ans/SR1/ci\endcsname{[$4.3$, $7.0$]}
\expandafter\def\csname pn@hot/rerankVSrrf/ans/SR1/full\endcsname{$+5.6$ [$4.3$, $7.0$]}
\expandafter\def\csname pn@hot/rerankVSrrf/ans/ext\endcsname{$+6.4$}
\expandafter\def\csname pn@hot/rerankVSrrf/ans/ext/abs\endcsname{6.4}
\expandafter\def\csname pn@hot/rerankVSrrf/ans/ext/ci\endcsname{[$5.5$, $7.4$]}
\expandafter\def\csname pn@hot/rerankVSrrf/ans/ext/full\endcsname{$+6.4$ [$5.5$, $7.4$]}
\expandafter\def\csname pn@hot/rerankVSrrf/ans/nsig\endcsname{4}
\expandafter\def\csname pn@hot/rerankVSrrf/ans/pool\endcsname{$+6.6$}
\expandafter\def\csname pn@hot/rerankVSrrf/ans/pool/abs\endcsname{6.6}
\expandafter\def\csname pn@hot/rerankVSrrf/ans/pool/ci\endcsname{[$5.7$, $7.5$]}
\expandafter\def\csname pn@hot/rerankVSrrf/ans/pool/full\endcsname{$+6.6$ [$5.7$, $7.5$]}
\expandafter\def\csname pn@hot/rerankVSrrf/ans/range\endcsname{$+5.5$ to $+8.1$}
\expandafter\def\csname pn@hot/rerankVSrrf/complete/Decomp\endcsname{$+7.2$}
\expandafter\def\csname pn@hot/rerankVSrrf/complete/Decomp/abs\endcsname{7.2}
\expandafter\def\csname pn@hot/rerankVSrrf/complete/Decomp/ci\endcsname{[$5.4$, $9.0$]}
\expandafter\def\csname pn@hot/rerankVSrrf/complete/Decomp/full\endcsname{$+7.2$ [$5.4$, $9.0$]}
\expandafter\def\csname pn@hot/rerankVSrrf/complete/IRCoT\endcsname{$+15.3$}
\expandafter\def\csname pn@hot/rerankVSrrf/complete/IRCoT/abs\endcsname{15.3}
\expandafter\def\csname pn@hot/rerankVSrrf/complete/IRCoT/ci\endcsname{[$13.0$, $17.7$]}
\expandafter\def\csname pn@hot/rerankVSrrf/complete/IRCoT/full\endcsname{$+15.3$ [$13.0$, $17.7$]}
\expandafter\def\csname pn@hot/rerankVSrrf/complete/ReAct\endcsname{$+13.7$}
\expandafter\def\csname pn@hot/rerankVSrrf/complete/ReAct/abs\endcsname{13.7}
\expandafter\def\csname pn@hot/rerankVSrrf/complete/ReAct/ci\endcsname{[$11.5$, $15.7$]}
\expandafter\def\csname pn@hot/rerankVSrrf/complete/ReAct/full\endcsname{$+13.7$ [$11.5$, $15.7$]}
\expandafter\def\csname pn@hot/rerankVSrrf/complete/SR1\endcsname{$+5.7$}
\expandafter\def\csname pn@hot/rerankVSrrf/complete/SR1/abs\endcsname{5.7}
\expandafter\def\csname pn@hot/rerankVSrrf/complete/SR1/ci\endcsname{[$3.9$, $7.6$]}
\expandafter\def\csname pn@hot/rerankVSrrf/complete/SR1/full\endcsname{$+5.7$ [$3.9$, $7.6$]}
\expandafter\def\csname pn@hot/rerankVSrrf/complete/ext\endcsname{$+11.6$}
\expandafter\def\csname pn@hot/rerankVSrrf/complete/ext/abs\endcsname{11.6}
\expandafter\def\csname pn@hot/rerankVSrrf/complete/ext/ci\endcsname{[$10.2$, $13.0$]}
\expandafter\def\csname pn@hot/rerankVSrrf/complete/ext/full\endcsname{$+11.6$ [$10.2$, $13.0$]}
\expandafter\def\csname pn@hot/rerankVSrrf/complete/nsig\endcsname{4}
\expandafter\def\csname pn@hot/rerankVSrrf/complete/pool\endcsname{$+10.5$}
\expandafter\def\csname pn@hot/rerankVSrrf/complete/pool/abs\endcsname{10.5}
\expandafter\def\csname pn@hot/rerankVSrrf/complete/pool/ci\endcsname{[$9.2$, $11.8$]}
\expandafter\def\csname pn@hot/rerankVSrrf/complete/pool/full\endcsname{$+10.5$ [$9.2$, $11.8$]}
\expandafter\def\csname pn@hot/rerankVSrrf/complete/range\endcsname{$+5.7$ to $+15.3$}
\expandafter\def\csname pn@hot/rerankcalls/mean\endcsname{28.7}
\expandafter\def\csname pn@hot/rerankcalls/range\endcsname{24--36}
\expandafter\def\csname pn@hot/rerankwin/EM/Decomp\endcsname{$+0.7$}
\expandafter\def\csname pn@hot/rerankwin/EM/Decomp/abs\endcsname{0.7}
\expandafter\def\csname pn@hot/rerankwin/EM/Decomp/ci\endcsname{[$-0.6$, $1.9$]}
\expandafter\def\csname pn@hot/rerankwin/EM/Decomp/full\endcsname{$+0.7$ [$-0.6$, $1.9$]}
\expandafter\def\csname pn@hot/rerankwin/EM/IRCoT\endcsname{$+0.7$}
\expandafter\def\csname pn@hot/rerankwin/EM/IRCoT/abs\endcsname{0.7}
\expandafter\def\csname pn@hot/rerankwin/EM/IRCoT/ci\endcsname{[$-0.7$, $2.1$]}
\expandafter\def\csname pn@hot/rerankwin/EM/IRCoT/full\endcsname{$+0.7$ [$-0.7$, $2.1$]}
\expandafter\def\csname pn@hot/rerankwin/EM/ReAct\endcsname{$0.0$}
\expandafter\def\csname pn@hot/rerankwin/EM/ReAct/abs\endcsname{0.0}
\expandafter\def\csname pn@hot/rerankwin/EM/ReAct/ci\endcsname{[$-1.2$, $1.2$]}
\expandafter\def\csname pn@hot/rerankwin/EM/ReAct/full\endcsname{$0.0$ [$-1.2$, $1.2$]}
\expandafter\def\csname pn@hot/rerankwin/EM/SR1\endcsname{$-0.1$}
\expandafter\def\csname pn@hot/rerankwin/EM/SR1/abs\endcsname{0.1}
\expandafter\def\csname pn@hot/rerankwin/EM/SR1/ci\endcsname{[$-1.4$, $1.3$]}
\expandafter\def\csname pn@hot/rerankwin/EM/SR1/full\endcsname{$-0.1$ [$-1.4$, $1.3$]}
\expandafter\def\csname pn@hot/rerankwin/EM/ext\endcsname{$+0.2$}
\expandafter\def\csname pn@hot/rerankwin/EM/ext/abs\endcsname{0.2}
\expandafter\def\csname pn@hot/rerankwin/EM/ext/ci\endcsname{[$-0.7$, $1.1$]}
\expandafter\def\csname pn@hot/rerankwin/EM/ext/full\endcsname{$+0.2$ [$-0.7$, $1.1$]}
\expandafter\def\csname pn@hot/rerankwin/EM/nsig\endcsname{0}
\expandafter\def\csname pn@hot/rerankwin/EM/pool\endcsname{$+0.3$}
\expandafter\def\csname pn@hot/rerankwin/EM/pool/abs\endcsname{0.3}
\expandafter\def\csname pn@hot/rerankwin/EM/pool/ci\endcsname{[$-0.5$, $1.1$]}
\expandafter\def\csname pn@hot/rerankwin/EM/pool/full\endcsname{$+0.3$ [$-0.5$, $1.1$]}
\expandafter\def\csname pn@hot/rerankwin/EM/range\endcsname{$-0.1$ to $+0.7$}
\expandafter\def\csname pn@hot/rerankwin/ans/Decomp\endcsname{$+0.9$}
\expandafter\def\csname pn@hot/rerankwin/ans/Decomp/abs\endcsname{0.9}
\expandafter\def\csname pn@hot/rerankwin/ans/Decomp/ci\endcsname{[$0.3$, $1.6$]}
\expandafter\def\csname pn@hot/rerankwin/ans/Decomp/full\endcsname{$+0.9$ [$0.3$, $1.6$]}
\expandafter\def\csname pn@hot/rerankwin/ans/IRCoT\endcsname{$+0.3$}
\expandafter\def\csname pn@hot/rerankwin/ans/IRCoT/abs\endcsname{0.3}
\expandafter\def\csname pn@hot/rerankwin/ans/IRCoT/ci\endcsname{[$-0.2$, $0.8$]}
\expandafter\def\csname pn@hot/rerankwin/ans/IRCoT/full\endcsname{$+0.3$ [$-0.2$, $0.8$]}
\expandafter\def\csname pn@hot/rerankwin/ans/ReAct\endcsname{$+0.6$}
\expandafter\def\csname pn@hot/rerankwin/ans/ReAct/abs\endcsname{0.6}
\expandafter\def\csname pn@hot/rerankwin/ans/ReAct/ci\endcsname{[$0.0$, $1.3$]}
\expandafter\def\csname pn@hot/rerankwin/ans/ReAct/full\endcsname{$+0.6$ [$0.0$, $1.3$]}
\expandafter\def\csname pn@hot/rerankwin/ans/SR1\endcsname{$+0.3$}
\expandafter\def\csname pn@hot/rerankwin/ans/SR1/abs\endcsname{0.3}
\expandafter\def\csname pn@hot/rerankwin/ans/SR1/ci\endcsname{[$-0.3$, $0.8$]}
\expandafter\def\csname pn@hot/rerankwin/ans/SR1/full\endcsname{$+0.3$ [$-0.3$, $0.8$]}
\expandafter\def\csname pn@hot/rerankwin/ans/ext\endcsname{$+0.4$}
\expandafter\def\csname pn@hot/rerankwin/ans/ext/abs\endcsname{0.4}
\expandafter\def\csname pn@hot/rerankwin/ans/ext/ci\endcsname{[$0.0$, $0.8$]}
\expandafter\def\csname pn@hot/rerankwin/ans/ext/full\endcsname{$+0.4$ [$0.0$, $0.8$]}
\expandafter\def\csname pn@hot/rerankwin/ans/nsig\endcsname{1}
\expandafter\def\csname pn@hot/rerankwin/ans/pool\endcsname{$+0.5$}
\expandafter\def\csname pn@hot/rerankwin/ans/pool/abs\endcsname{0.5}
\expandafter\def\csname pn@hot/rerankwin/ans/pool/ci\endcsname{[$0.2$, $0.9$]}
\expandafter\def\csname pn@hot/rerankwin/ans/pool/full\endcsname{$+0.5$ [$0.2$, $0.9$]}
\expandafter\def\csname pn@hot/rerankwin/ans/range\endcsname{$+0.3$ to $+0.9$}
\expandafter\def\csname pn@hot/rerankwin/complete/Decomp\endcsname{$+3.4$}
\expandafter\def\csname pn@hot/rerankwin/complete/Decomp/abs\endcsname{3.4}
\expandafter\def\csname pn@hot/rerankwin/complete/Decomp/ci\endcsname{[$2.5$, $4.3$]}
\expandafter\def\csname pn@hot/rerankwin/complete/Decomp/full\endcsname{$+3.4$ [$2.5$, $4.3$]}
\expandafter\def\csname pn@hot/rerankwin/complete/IRCoT\endcsname{$+3.5$}
\expandafter\def\csname pn@hot/rerankwin/complete/IRCoT/abs\endcsname{3.5}
\expandafter\def\csname pn@hot/rerankwin/complete/IRCoT/ci\endcsname{[$2.5$, $4.6$]}
\expandafter\def\csname pn@hot/rerankwin/complete/IRCoT/full\endcsname{$+3.5$ [$2.5$, $4.6$]}
\expandafter\def\csname pn@hot/rerankwin/complete/ReAct\endcsname{$+3.7$}
\expandafter\def\csname pn@hot/rerankwin/complete/ReAct/abs\endcsname{3.7}
\expandafter\def\csname pn@hot/rerankwin/complete/ReAct/ci\endcsname{[$2.7$, $4.9$]}
\expandafter\def\csname pn@hot/rerankwin/complete/ReAct/full\endcsname{$+3.7$ [$2.7$, $4.9$]}
\expandafter\def\csname pn@hot/rerankwin/complete/SR1\endcsname{$+2.8$}
\expandafter\def\csname pn@hot/rerankwin/complete/SR1/abs\endcsname{2.8}
\expandafter\def\csname pn@hot/rerankwin/complete/SR1/ci\endcsname{[$1.9$, $3.8$]}
\expandafter\def\csname pn@hot/rerankwin/complete/SR1/full\endcsname{$+2.8$ [$1.9$, $3.8$]}
\expandafter\def\csname pn@hot/rerankwin/complete/ext\endcsname{$+3.4$}
\expandafter\def\csname pn@hot/rerankwin/complete/ext/abs\endcsname{3.4}
\expandafter\def\csname pn@hot/rerankwin/complete/ext/ci\endcsname{[$2.6$, $4.2$]}
\expandafter\def\csname pn@hot/rerankwin/complete/ext/full\endcsname{$+3.4$ [$2.6$, $4.2$]}
\expandafter\def\csname pn@hot/rerankwin/complete/nsig\endcsname{4}
\expandafter\def\csname pn@hot/rerankwin/complete/pool\endcsname{$+3.4$}
\expandafter\def\csname pn@hot/rerankwin/complete/pool/abs\endcsname{3.4}
\expandafter\def\csname pn@hot/rerankwin/complete/pool/ci\endcsname{[$2.7$, $4.1$]}
\expandafter\def\csname pn@hot/rerankwin/complete/pool/full\endcsname{$+3.4$ [$2.7$, $4.1$]}
\expandafter\def\csname pn@hot/rerankwin/complete/range\endcsname{$+2.8$ to $+3.7$}
\expandafter\def\csname pn@hot/rewired/EM/Decomp\endcsname{$-0.1$}
\expandafter\def\csname pn@hot/rewired/EM/Decomp/abs\endcsname{0.1}
\expandafter\def\csname pn@hot/rewired/EM/Decomp/ci\endcsname{[$-1.2$, $0.9$]}
\expandafter\def\csname pn@hot/rewired/EM/Decomp/full\endcsname{$-0.1$ [$-1.2$, $0.9$]}
\expandafter\def\csname pn@hot/rewired/EM/IRCoT\endcsname{$-0.5$}
\expandafter\def\csname pn@hot/rewired/EM/IRCoT/abs\endcsname{0.5}
\expandafter\def\csname pn@hot/rewired/EM/IRCoT/ci\endcsname{[$-1.6$, $0.5$]}
\expandafter\def\csname pn@hot/rewired/EM/IRCoT/full\endcsname{$-0.5$ [$-1.6$, $0.5$]}
\expandafter\def\csname pn@hot/rewired/EM/ReAct\endcsname{$-0.7$}
\expandafter\def\csname pn@hot/rewired/EM/ReAct/abs\endcsname{0.7}
\expandafter\def\csname pn@hot/rewired/EM/ReAct/ci\endcsname{[$-1.9$, $0.3$]}
\expandafter\def\csname pn@hot/rewired/EM/ReAct/full\endcsname{$-0.7$ [$-1.9$, $0.3$]}
\expandafter\def\csname pn@hot/rewired/EM/SR1\endcsname{$-0.1$}
\expandafter\def\csname pn@hot/rewired/EM/SR1/abs\endcsname{0.1}
\expandafter\def\csname pn@hot/rewired/EM/SR1/ci\endcsname{[$-1.2$, $1.0$]}
\expandafter\def\csname pn@hot/rewired/EM/SR1/full\endcsname{$-0.1$ [$-1.2$, $1.0$]}
\expandafter\def\csname pn@hot/rewired/EM/ext\endcsname{$-0.5$}
\expandafter\def\csname pn@hot/rewired/EM/ext/abs\endcsname{0.5}
\expandafter\def\csname pn@hot/rewired/EM/ext/ci\endcsname{[$-1.2$, $0.2$]}
\expandafter\def\csname pn@hot/rewired/EM/ext/full\endcsname{$-0.5$ [$-1.2$, $0.2$]}
\expandafter\def\csname pn@hot/rewired/EM/nsig\endcsname{0}
\expandafter\def\csname pn@hot/rewired/EM/pool\endcsname{$-0.4$}
\expandafter\def\csname pn@hot/rewired/EM/pool/abs\endcsname{0.4}
\expandafter\def\csname pn@hot/rewired/EM/pool/ci\endcsname{[$-1.0$, $0.2$]}
\expandafter\def\csname pn@hot/rewired/EM/pool/full\endcsname{$-0.4$ [$-1.0$, $0.2$]}
\expandafter\def\csname pn@hot/rewired/EM/range\endcsname{$-0.7$ to $-0.1$}
\expandafter\def\csname pn@hot/rewired/ans/Decomp\endcsname{$+1.7$}
\expandafter\def\csname pn@hot/rewired/ans/Decomp/abs\endcsname{1.7}
\expandafter\def\csname pn@hot/rewired/ans/Decomp/ci\endcsname{[$0.8$, $2.6$]}
\expandafter\def\csname pn@hot/rewired/ans/Decomp/full\endcsname{$+1.7$ [$0.8$, $2.6$]}
\expandafter\def\csname pn@hot/rewired/ans/IRCoT\endcsname{$+0.6$}
\expandafter\def\csname pn@hot/rewired/ans/IRCoT/abs\endcsname{0.6}
\expandafter\def\csname pn@hot/rewired/ans/IRCoT/ci\endcsname{[$0.2$, $1.1$]}
\expandafter\def\csname pn@hot/rewired/ans/IRCoT/full\endcsname{$+0.6$ [$0.2$, $1.1$]}
\expandafter\def\csname pn@hot/rewired/ans/ReAct\endcsname{$+1.0$}
\expandafter\def\csname pn@hot/rewired/ans/ReAct/abs\endcsname{1.0}
\expandafter\def\csname pn@hot/rewired/ans/ReAct/ci\endcsname{[$0.4$, $1.6$]}
\expandafter\def\csname pn@hot/rewired/ans/ReAct/full\endcsname{$+1.0$ [$0.4$, $1.6$]}
\expandafter\def\csname pn@hot/rewired/ans/SR1\endcsname{$+0.7$}
\expandafter\def\csname pn@hot/rewired/ans/SR1/abs\endcsname{0.7}
\expandafter\def\csname pn@hot/rewired/ans/SR1/ci\endcsname{[$0.1$, $1.3$]}
\expandafter\def\csname pn@hot/rewired/ans/SR1/full\endcsname{$+0.7$ [$0.1$, $1.3$]}
\expandafter\def\csname pn@hot/rewired/ans/ext\endcsname{$+0.8$}
\expandafter\def\csname pn@hot/rewired/ans/ext/abs\endcsname{0.8}
\expandafter\def\csname pn@hot/rewired/ans/ext/ci\endcsname{[$0.4$, $1.1$]}
\expandafter\def\csname pn@hot/rewired/ans/ext/full\endcsname{$+0.8$ [$0.4$, $1.1$]}
\expandafter\def\csname pn@hot/rewired/ans/nsig\endcsname{4}
\expandafter\def\csname pn@hot/rewired/ans/pool\endcsname{$+1.0$}
\expandafter\def\csname pn@hot/rewired/ans/pool/abs\endcsname{1.0}
\expandafter\def\csname pn@hot/rewired/ans/pool/ci\endcsname{[$0.7$, $1.4$]}
\expandafter\def\csname pn@hot/rewired/ans/pool/full\endcsname{$+1.0$ [$0.7$, $1.4$]}
\expandafter\def\csname pn@hot/rewired/ans/range\endcsname{$+0.6$ to $+1.7$}
\expandafter\def\csname pn@hot/rewired/complete/Decomp\endcsname{$+3.7$}
\expandafter\def\csname pn@hot/rewired/complete/Decomp/abs\endcsname{3.7}
\expandafter\def\csname pn@hot/rewired/complete/Decomp/ci\endcsname{[$2.5$, $4.8$]}
\expandafter\def\csname pn@hot/rewired/complete/Decomp/full\endcsname{$+3.7$ [$2.5$, $4.8$]}
\expandafter\def\csname pn@hot/rewired/complete/IRCoT\endcsname{$+1.2$}
\expandafter\def\csname pn@hot/rewired/complete/IRCoT/abs\endcsname{1.2}
\expandafter\def\csname pn@hot/rewired/complete/IRCoT/ci\endcsname{[$0.5$, $1.9$]}
\expandafter\def\csname pn@hot/rewired/complete/IRCoT/full\endcsname{$+1.2$ [$0.5$, $1.9$]}
\expandafter\def\csname pn@hot/rewired/complete/ReAct\endcsname{$+1.7$}
\expandafter\def\csname pn@hot/rewired/complete/ReAct/abs\endcsname{1.7}
\expandafter\def\csname pn@hot/rewired/complete/ReAct/ci\endcsname{[$0.9$, $2.5$]}
\expandafter\def\csname pn@hot/rewired/complete/ReAct/full\endcsname{$+1.7$ [$0.9$, $2.5$]}
\expandafter\def\csname pn@hot/rewired/complete/SR1\endcsname{$+1.6$}
\expandafter\def\csname pn@hot/rewired/complete/SR1/abs\endcsname{1.6}
\expandafter\def\csname pn@hot/rewired/complete/SR1/ci\endcsname{[$0.9$, $2.3$]}
\expandafter\def\csname pn@hot/rewired/complete/SR1/full\endcsname{$+1.6$ [$0.9$, $2.3$]}
\expandafter\def\csname pn@hot/rewired/complete/ext\endcsname{$+1.5$}
\expandafter\def\csname pn@hot/rewired/complete/ext/abs\endcsname{1.5}
\expandafter\def\csname pn@hot/rewired/complete/ext/ci\endcsname{[$1.1$, $1.9$]}
\expandafter\def\csname pn@hot/rewired/complete/ext/full\endcsname{$+1.5$ [$1.1$, $1.9$]}
\expandafter\def\csname pn@hot/rewired/complete/nsig\endcsname{4}
\expandafter\def\csname pn@hot/rewired/complete/pool\endcsname{$+2.0$}
\expandafter\def\csname pn@hot/rewired/complete/pool/abs\endcsname{2.0}
\expandafter\def\csname pn@hot/rewired/complete/pool/ci\endcsname{[$1.6$, $2.5$]}
\expandafter\def\csname pn@hot/rewired/complete/pool/full\endcsname{$+2.0$ [$1.6$, $2.5$]}
\expandafter\def\csname pn@hot/rewired/complete/range\endcsname{$+1.2$ to $+3.7$}
\expandafter\def\csname pn@hot/rewired10/ans/Decomp\endcsname{$+1.8$}
\expandafter\def\csname pn@hot/rewired10/ans/Decomp/abs\endcsname{1.8}
\expandafter\def\csname pn@hot/rewired10/ans/Decomp/ci\endcsname{[$1.1$, $2.5$]}
\expandafter\def\csname pn@hot/rewired10/ans/Decomp/full\endcsname{$+1.8$ [$1.1$, $2.5$]}
\expandafter\def\csname pn@hot/rewired10/ans/IRCoT\endcsname{$+0.7$}
\expandafter\def\csname pn@hot/rewired10/ans/IRCoT/abs\endcsname{0.7}
\expandafter\def\csname pn@hot/rewired10/ans/IRCoT/ci\endcsname{[$0.3$, $1.0$]}
\expandafter\def\csname pn@hot/rewired10/ans/IRCoT/full\endcsname{$+0.7$ [$0.3$, $1.0$]}
\expandafter\def\csname pn@hot/rewired10/ans/ReAct\endcsname{$+1.0$}
\expandafter\def\csname pn@hot/rewired10/ans/ReAct/abs\endcsname{1.0}
\expandafter\def\csname pn@hot/rewired10/ans/ReAct/ci\endcsname{[$0.5$, $1.5$]}
\expandafter\def\csname pn@hot/rewired10/ans/ReAct/full\endcsname{$+1.0$ [$0.5$, $1.5$]}
\expandafter\def\csname pn@hot/rewired10/ans/SR1\endcsname{$+0.7$}
\expandafter\def\csname pn@hot/rewired10/ans/SR1/abs\endcsname{0.7}
\expandafter\def\csname pn@hot/rewired10/ans/SR1/ci\endcsname{[$0.3$, $1.2$]}
\expandafter\def\csname pn@hot/rewired10/ans/SR1/full\endcsname{$+0.7$ [$0.3$, $1.2$]}
\expandafter\def\csname pn@hot/rewired10/ans/ext\endcsname{$+0.8$}
\expandafter\def\csname pn@hot/rewired10/ans/ext/abs\endcsname{0.8}
\expandafter\def\csname pn@hot/rewired10/ans/ext/ci\endcsname{[$0.5$, $1.1$]}
\expandafter\def\csname pn@hot/rewired10/ans/ext/full\endcsname{$+0.8$ [$0.5$, $1.1$]}
\expandafter\def\csname pn@hot/rewired10/ans/nsig\endcsname{4}
\expandafter\def\csname pn@hot/rewired10/ans/pool\endcsname{$+1.0$}
\expandafter\def\csname pn@hot/rewired10/ans/pool/abs\endcsname{1.0}
\expandafter\def\csname pn@hot/rewired10/ans/pool/ci\endcsname{[$0.7$, $1.3$]}
\expandafter\def\csname pn@hot/rewired10/ans/pool/full\endcsname{$+1.0$ [$0.7$, $1.3$]}
\expandafter\def\csname pn@hot/rewired10/ans/range\endcsname{$+0.7$ to $+1.8$}
\expandafter\def\csname pn@hot/rewired10/complete/Decomp\endcsname{$+3.4$}
\expandafter\def\csname pn@hot/rewired10/complete/Decomp/abs\endcsname{3.4}
\expandafter\def\csname pn@hot/rewired10/complete/Decomp/ci\endcsname{[$2.6$, $4.2$]}
\expandafter\def\csname pn@hot/rewired10/complete/Decomp/full\endcsname{$+3.4$ [$2.6$, $4.2$]}
\expandafter\def\csname pn@hot/rewired10/complete/IRCoT\endcsname{$+1.2$}
\expandafter\def\csname pn@hot/rewired10/complete/IRCoT/abs\endcsname{1.2}
\expandafter\def\csname pn@hot/rewired10/complete/IRCoT/ci\endcsname{[$0.7$, $1.7$]}
\expandafter\def\csname pn@hot/rewired10/complete/IRCoT/full\endcsname{$+1.2$ [$0.7$, $1.7$]}
\expandafter\def\csname pn@hot/rewired10/complete/ReAct\endcsname{$+1.9$}
\expandafter\def\csname pn@hot/rewired10/complete/ReAct/abs\endcsname{1.9}
\expandafter\def\csname pn@hot/rewired10/complete/ReAct/ci\endcsname{[$1.3$, $2.5$]}
\expandafter\def\csname pn@hot/rewired10/complete/ReAct/full\endcsname{$+1.9$ [$1.3$, $2.5$]}
\expandafter\def\csname pn@hot/rewired10/complete/SR1\endcsname{$+1.7$}
\expandafter\def\csname pn@hot/rewired10/complete/SR1/abs\endcsname{1.7}
\expandafter\def\csname pn@hot/rewired10/complete/SR1/ci\endcsname{[$1.1$, $2.2$]}
\expandafter\def\csname pn@hot/rewired10/complete/SR1/full\endcsname{$+1.7$ [$1.1$, $2.2$]}
\expandafter\def\csname pn@hot/rewired10/complete/ext\endcsname{$+1.6$}
\expandafter\def\csname pn@hot/rewired10/complete/ext/abs\endcsname{1.6}
\expandafter\def\csname pn@hot/rewired10/complete/ext/ci\endcsname{[$1.2$, $2.0$]}
\expandafter\def\csname pn@hot/rewired10/complete/ext/full\endcsname{$+1.6$ [$1.2$, $2.0$]}
\expandafter\def\csname pn@hot/rewired10/complete/nsig\endcsname{4}
\expandafter\def\csname pn@hot/rewired10/complete/pool\endcsname{$+2.0$}
\expandafter\def\csname pn@hot/rewired10/complete/pool/abs\endcsname{2.0}
\expandafter\def\csname pn@hot/rewired10/complete/pool/ci\endcsname{[$1.7$, $2.4$]}
\expandafter\def\csname pn@hot/rewired10/complete/pool/full\endcsname{$+2.0$ [$1.7$, $2.4$]}
\expandafter\def\csname pn@hot/rewired10/complete/range\endcsname{$+1.2$ to $+3.4$}
\expandafter\def\csname pn@hot/route/EM/Decomp\endcsname{$+1.2$}
\expandafter\def\csname pn@hot/route/EM/Decomp/abs\endcsname{1.2}
\expandafter\def\csname pn@hot/route/EM/Decomp/ci\endcsname{[$0.4$, $2.1$]}
\expandafter\def\csname pn@hot/route/EM/Decomp/full\endcsname{$+1.2$ [$0.4$, $2.1$]}
\expandafter\def\csname pn@hot/route/EM/IRCoT\endcsname{$+1.1$}
\expandafter\def\csname pn@hot/route/EM/IRCoT/abs\endcsname{1.1}
\expandafter\def\csname pn@hot/route/EM/IRCoT/ci\endcsname{[$0.0$, $2.3$]}
\expandafter\def\csname pn@hot/route/EM/IRCoT/full\endcsname{$+1.1$ [$0.0$, $2.3$]}
\expandafter\def\csname pn@hot/route/EM/IRCoTSR1\endcsname{$+0.6$}
\expandafter\def\csname pn@hot/route/EM/IRCoTSR1/abs\endcsname{0.6}
\expandafter\def\csname pn@hot/route/EM/IRCoTSR1/ci\endcsname{[$0.0$, $1.3$]}
\expandafter\def\csname pn@hot/route/EM/IRCoTSR1/full\endcsname{$+0.6$ [$0.0$, $1.3$]}
\expandafter\def\csname pn@hot/route/EM/ReAct\endcsname{$+3.0$}
\expandafter\def\csname pn@hot/route/EM/ReAct/abs\endcsname{3.0}
\expandafter\def\csname pn@hot/route/EM/ReAct/ci\endcsname{[$1.6$, $4.5$]}
\expandafter\def\csname pn@hot/route/EM/ReAct/full\endcsname{$+3.0$ [$1.6$, $4.5$]}
\expandafter\def\csname pn@hot/route/EM/SR1\endcsname{$+0.1$}
\expandafter\def\csname pn@hot/route/EM/SR1/abs\endcsname{0.1}
\expandafter\def\csname pn@hot/route/EM/SR1/ci\endcsname{[$-0.5$, $0.8$]}
\expandafter\def\csname pn@hot/route/EM/SR1/full\endcsname{$+0.1$ [$-0.5$, $0.8$]}
\expandafter\def\csname pn@hot/route/EM/ext\endcsname{$+1.4$}
\expandafter\def\csname pn@hot/route/EM/ext/abs\endcsname{1.4}
\expandafter\def\csname pn@hot/route/EM/ext/ci\endcsname{[$0.7$, $2.1$]}
\expandafter\def\csname pn@hot/route/EM/ext/full\endcsname{$+1.4$ [$0.7$, $2.1$]}
\expandafter\def\csname pn@hot/route/EM/nsig\endcsname{2}
\expandafter\def\csname pn@hot/route/EM/pool\endcsname{$+1.4$}
\expandafter\def\csname pn@hot/route/EM/pool/abs\endcsname{0.6}
\expandafter\def\csname pn@hot/route/EM/pool/ci\endcsname{[$0.8$, $2.0$]}
\expandafter\def\csname pn@hot/route/EM/pool/full\endcsname{$+1.4$ [$0.8$, $2.0$]}
\expandafter\def\csname pn@hot/route/EM/range\endcsname{$+0.1$ to $+3.0$}
\expandafter\def\csname pn@hot/route/ans/Decomp\endcsname{$+2.8$}
\expandafter\def\csname pn@hot/route/ans/Decomp/abs\endcsname{2.8}
\expandafter\def\csname pn@hot/route/ans/Decomp/ci\endcsname{[$1.9$, $3.8$]}
\expandafter\def\csname pn@hot/route/ans/Decomp/full\endcsname{$+2.8$ [$1.9$, $3.8$]}
\expandafter\def\csname pn@hot/route/ans/IRCoT\endcsname{$+1.2$}
\expandafter\def\csname pn@hot/route/ans/IRCoT/abs\endcsname{1.2}
\expandafter\def\csname pn@hot/route/ans/IRCoT/ci\endcsname{[$0.4$, $2.0$]}
\expandafter\def\csname pn@hot/route/ans/IRCoT/full\endcsname{$+1.2$ [$0.4$, $2.0$]}
\expandafter\def\csname pn@hot/route/ans/ReAct\endcsname{$+5.0$}
\expandafter\def\csname pn@hot/route/ans/ReAct/abs\endcsname{5.0}
\expandafter\def\csname pn@hot/route/ans/ReAct/ci\endcsname{[$3.6$, $6.4$]}
\expandafter\def\csname pn@hot/route/ans/ReAct/full\endcsname{$+5.0$ [$3.6$, $6.4$]}
\expandafter\def\csname pn@hot/route/ans/SR1\endcsname{$+0.4$}
\expandafter\def\csname pn@hot/route/ans/SR1/abs\endcsname{0.4}
\expandafter\def\csname pn@hot/route/ans/SR1/ci\endcsname{[$-0.1$, $1.0$]}
\expandafter\def\csname pn@hot/route/ans/SR1/full\endcsname{$+0.4$ [$-0.1$, $1.0$]}
\expandafter\def\csname pn@hot/route/ans/ext\endcsname{$+2.2$}
\expandafter\def\csname pn@hot/route/ans/ext/abs\endcsname{2.2}
\expandafter\def\csname pn@hot/route/ans/ext/ci\endcsname{[$1.5$, $2.9$]}
\expandafter\def\csname pn@hot/route/ans/ext/full\endcsname{$+2.2$ [$1.5$, $2.9$]}
\expandafter\def\csname pn@hot/route/ans/nsig\endcsname{3}
\expandafter\def\csname pn@hot/route/ans/pool\endcsname{$+2.4$}
\expandafter\def\csname pn@hot/route/ans/pool/abs\endcsname{2.4}
\expandafter\def\csname pn@hot/route/ans/pool/ci\endcsname{[$1.8$, $3.0$]}
\expandafter\def\csname pn@hot/route/ans/pool/full\endcsname{$+2.4$ [$1.8$, $3.0$]}
\expandafter\def\csname pn@hot/route/ans/range\endcsname{$+0.4$ to $+5.0$}
\expandafter\def\csname pn@hot/route/complete/Decomp\endcsname{$+4.2$}
\expandafter\def\csname pn@hot/route/complete/Decomp/abs\endcsname{4.2}
\expandafter\def\csname pn@hot/route/complete/Decomp/ci\endcsname{[$3.2$, $5.3$]}
\expandafter\def\csname pn@hot/route/complete/Decomp/full\endcsname{$+4.2$ [$3.2$, $5.3$]}
\expandafter\def\csname pn@hot/route/complete/IRCoT\endcsname{$+3.0$}
\expandafter\def\csname pn@hot/route/complete/IRCoT/abs\endcsname{3.0}
\expandafter\def\csname pn@hot/route/complete/IRCoT/ci\endcsname{[$1.9$, $4.1$]}
\expandafter\def\csname pn@hot/route/complete/IRCoT/full\endcsname{$+3.0$ [$1.9$, $4.1$]}
\expandafter\def\csname pn@hot/route/complete/ReAct\endcsname{$+8.6$}
\expandafter\def\csname pn@hot/route/complete/ReAct/abs\endcsname{8.6}
\expandafter\def\csname pn@hot/route/complete/ReAct/ci\endcsname{[$7.0$, $10.2$]}
\expandafter\def\csname pn@hot/route/complete/ReAct/full\endcsname{$+8.6$ [$7.0$, $10.2$]}
\expandafter\def\csname pn@hot/route/complete/SR1\endcsname{$+1.2$}
\expandafter\def\csname pn@hot/route/complete/SR1/abs\endcsname{1.2}
\expandafter\def\csname pn@hot/route/complete/SR1/ci\endcsname{[$0.5$, $2.0$]}
\expandafter\def\csname pn@hot/route/complete/SR1/full\endcsname{$+1.2$ [$0.5$, $2.0$]}
\expandafter\def\csname pn@hot/route/complete/ext\endcsname{$+4.3$}
\expandafter\def\csname pn@hot/route/complete/ext/abs\endcsname{4.3}
\expandafter\def\csname pn@hot/route/complete/ext/ci\endcsname{[$3.5$, $5.1$]}
\expandafter\def\csname pn@hot/route/complete/ext/full\endcsname{$+4.3$ [$3.5$, $5.1$]}
\expandafter\def\csname pn@hot/route/complete/nsig\endcsname{4}
\expandafter\def\csname pn@hot/route/complete/pool\endcsname{$+4.3$}
\expandafter\def\csname pn@hot/route/complete/pool/abs\endcsname{4.3}
\expandafter\def\csname pn@hot/route/complete/pool/ci\endcsname{[$3.6$, $5.0$]}
\expandafter\def\csname pn@hot/route/complete/pool/full\endcsname{$+4.3$ [$3.6$, $5.0$]}
\expandafter\def\csname pn@hot/route/complete/range\endcsname{$+1.2$ to $+8.6$}
\expandafter\def\csname pn@hot/routed/Decomp\endcsname{0.41}
\expandafter\def\csname pn@hot/routed/IRCoT\endcsname{1.71}
\expandafter\def\csname pn@hot/routed/ReAct\endcsname{1.52}
\expandafter\def\csname pn@hot/routed/SR1\endcsname{0.32}
\expandafter\def\csname pn@hot/routedshare/Decomp\endcsname{8.2}
\expandafter\def\csname pn@hot/routedshare/IRCoT\endcsname{34.2}
\expandafter\def\csname pn@hot/routedshare/ReAct\endcsname{32.1}
\expandafter\def\csname pn@hot/routedshare/SR1\endcsname{13.5}
\expandafter\def\csname pn@hot/routeonly/EM/Decomp\endcsname{$+0.5$}
\expandafter\def\csname pn@hot/routeonly/EM/Decomp/abs\endcsname{0.5}
\expandafter\def\csname pn@hot/routeonly/EM/Decomp/ci\endcsname{[$-0.2$, $1.3$]}
\expandafter\def\csname pn@hot/routeonly/EM/Decomp/full\endcsname{$+0.5$ [$-0.2$, $1.3$]}
\expandafter\def\csname pn@hot/routeonly/EM/IRCoT\endcsname{$+0.3$}
\expandafter\def\csname pn@hot/routeonly/EM/IRCoT/abs\endcsname{0.3}
\expandafter\def\csname pn@hot/routeonly/EM/IRCoT/ci\endcsname{[$-0.9$, $1.5$]}
\expandafter\def\csname pn@hot/routeonly/EM/IRCoT/full\endcsname{$+0.3$ [$-0.9$, $1.5$]}
\expandafter\def\csname pn@hot/routeonly/EM/ReAct\endcsname{$+0.7$}
\expandafter\def\csname pn@hot/routeonly/EM/ReAct/abs\endcsname{0.7}
\expandafter\def\csname pn@hot/routeonly/EM/ReAct/ci\endcsname{[$-0.6$, $1.9$]}
\expandafter\def\csname pn@hot/routeonly/EM/ReAct/full\endcsname{$+0.7$ [$-0.6$, $1.9$]}
\expandafter\def\csname pn@hot/routeonly/EM/SR1\endcsname{$-0.5$}
\expandafter\def\csname pn@hot/routeonly/EM/SR1/abs\endcsname{0.5}
\expandafter\def\csname pn@hot/routeonly/EM/SR1/ci\endcsname{[$-1.1$, $0.1$]}
\expandafter\def\csname pn@hot/routeonly/EM/SR1/full\endcsname{$-0.5$ [$-1.1$, $0.1$]}
\expandafter\def\csname pn@hot/routeonly/EM/ext\endcsname{$+0.2$}
\expandafter\def\csname pn@hot/routeonly/EM/ext/abs\endcsname{0.2}
\expandafter\def\csname pn@hot/routeonly/EM/ext/ci\endcsname{[$-0.4$, $0.8$]}
\expandafter\def\csname pn@hot/routeonly/EM/ext/full\endcsname{$+0.2$ [$-0.4$, $0.8$]}
\expandafter\def\csname pn@hot/routeonly/EM/nsig\endcsname{0}
\expandafter\def\csname pn@hot/routeonly/EM/pool\endcsname{$+0.3$}
\expandafter\def\csname pn@hot/routeonly/EM/pool/abs\endcsname{0.3}
\expandafter\def\csname pn@hot/routeonly/EM/pool/ci\endcsname{[$-0.3$, $0.8$]}
\expandafter\def\csname pn@hot/routeonly/EM/pool/full\endcsname{$+0.3$ [$-0.3$, $0.8$]}
\expandafter\def\csname pn@hot/routeonly/EM/range\endcsname{$-0.5$ to $+0.7$}
\expandafter\def\csname pn@hot/routeonly/ans/Decomp\endcsname{$+1.8$}
\expandafter\def\csname pn@hot/routeonly/ans/Decomp/abs\endcsname{1.8}
\expandafter\def\csname pn@hot/routeonly/ans/Decomp/ci\endcsname{[$1.1$, $2.6$]}
\expandafter\def\csname pn@hot/routeonly/ans/Decomp/full\endcsname{$+1.8$ [$1.1$, $2.6$]}
\expandafter\def\csname pn@hot/routeonly/ans/IRCoT\endcsname{$+0.6$}
\expandafter\def\csname pn@hot/routeonly/ans/IRCoT/abs\endcsname{0.6}
\expandafter\def\csname pn@hot/routeonly/ans/IRCoT/ci\endcsname{[$-0.2$, $1.5$]}
\expandafter\def\csname pn@hot/routeonly/ans/IRCoT/full\endcsname{$+0.6$ [$-0.2$, $1.5$]}
\expandafter\def\csname pn@hot/routeonly/ans/ReAct\endcsname{$+3.0$}
\expandafter\def\csname pn@hot/routeonly/ans/ReAct/abs\endcsname{3.0}
\expandafter\def\csname pn@hot/routeonly/ans/ReAct/ci\endcsname{[$1.8$, $4.3$]}
\expandafter\def\csname pn@hot/routeonly/ans/ReAct/full\endcsname{$+3.0$ [$1.8$, $4.3$]}
\expandafter\def\csname pn@hot/routeonly/ans/SR1\endcsname{$-0.4$}
\expandafter\def\csname pn@hot/routeonly/ans/SR1/abs\endcsname{0.4}
\expandafter\def\csname pn@hot/routeonly/ans/SR1/ci\endcsname{[$-0.9$, $0.2$]}
\expandafter\def\csname pn@hot/routeonly/ans/SR1/full\endcsname{$-0.4$ [$-0.9$, $0.2$]}
\expandafter\def\csname pn@hot/routeonly/ans/ext\endcsname{$+1.1$}
\expandafter\def\csname pn@hot/routeonly/ans/ext/abs\endcsname{1.1}
\expandafter\def\csname pn@hot/routeonly/ans/ext/ci\endcsname{[$0.5$, $1.7$]}
\expandafter\def\csname pn@hot/routeonly/ans/ext/full\endcsname{$+1.1$ [$0.5$, $1.7$]}
\expandafter\def\csname pn@hot/routeonly/ans/nsig\endcsname{2}
\expandafter\def\csname pn@hot/routeonly/ans/pool\endcsname{$+1.3$}
\expandafter\def\csname pn@hot/routeonly/ans/pool/abs\endcsname{1.3}
\expandafter\def\csname pn@hot/routeonly/ans/pool/ci\endcsname{[$0.8$, $1.8$]}
\expandafter\def\csname pn@hot/routeonly/ans/pool/full\endcsname{$+1.3$ [$0.8$, $1.8$]}
\expandafter\def\csname pn@hot/routeonly/ans/range\endcsname{$-0.4$ to $+3.0$}
\expandafter\def\csname pn@hot/routeonly/complete/Decomp\endcsname{$+2.3$}
\expandafter\def\csname pn@hot/routeonly/complete/Decomp/abs\endcsname{2.3}
\expandafter\def\csname pn@hot/routeonly/complete/Decomp/ci\endcsname{[$1.5$, $3.2$]}
\expandafter\def\csname pn@hot/routeonly/complete/Decomp/full\endcsname{$+2.3$ [$1.5$, $3.2$]}
\expandafter\def\csname pn@hot/routeonly/complete/IRCoT\endcsname{$+0.4$}
\expandafter\def\csname pn@hot/routeonly/complete/IRCoT/abs\endcsname{0.4}
\expandafter\def\csname pn@hot/routeonly/complete/IRCoT/ci\endcsname{[$-0.8$, $1.5$]}
\expandafter\def\csname pn@hot/routeonly/complete/IRCoT/full\endcsname{$+0.4$ [$-0.8$, $1.5$]}
\expandafter\def\csname pn@hot/routeonly/complete/ReAct\endcsname{$+4.7$}
\expandafter\def\csname pn@hot/routeonly/complete/ReAct/abs\endcsname{4.7}
\expandafter\def\csname pn@hot/routeonly/complete/ReAct/ci\endcsname{[$3.1$, $6.3$]}
\expandafter\def\csname pn@hot/routeonly/complete/ReAct/full\endcsname{$+4.7$ [$3.1$, $6.3$]}
\expandafter\def\csname pn@hot/routeonly/complete/SR1\endcsname{$+0.3$}
\expandafter\def\csname pn@hot/routeonly/complete/SR1/abs\endcsname{0.3}
\expandafter\def\csname pn@hot/routeonly/complete/SR1/ci\endcsname{[$-0.5$, $1.0$]}
\expandafter\def\csname pn@hot/routeonly/complete/SR1/full\endcsname{$+0.3$ [$-0.5$, $1.0$]}
\expandafter\def\csname pn@hot/routeonly/complete/ext\endcsname{$+1.8$}
\expandafter\def\csname pn@hot/routeonly/complete/ext/abs\endcsname{1.8}
\expandafter\def\csname pn@hot/routeonly/complete/ext/ci\endcsname{[$1.0$, $2.6$]}
\expandafter\def\csname pn@hot/routeonly/complete/ext/full\endcsname{$+1.8$ [$1.0$, $2.6$]}
\expandafter\def\csname pn@hot/routeonly/complete/nsig\endcsname{2}
\expandafter\def\csname pn@hot/routeonly/complete/pool\endcsname{$+1.9$}
\expandafter\def\csname pn@hot/routeonly/complete/pool/abs\endcsname{1.9}
\expandafter\def\csname pn@hot/routeonly/complete/pool/ci\endcsname{[$1.2$, $2.6$]}
\expandafter\def\csname pn@hot/routeonly/complete/pool/full\endcsname{$+1.9$ [$1.2$, $2.6$]}
\expandafter\def\csname pn@hot/routeonly/complete/range\endcsname{$+0.3$ to $+4.7$}
\expandafter\def\csname pn@hot/rule/concatdd\endcsname{40.0}
\expandafter\def\csname pn@hot/rule/concatnodd\endcsname{40.0}
\expandafter\def\csname pn@hot/rule/coverage\endcsname{65.2}
\expandafter\def\csname pn@hot/rule/g2dd\endcsname{55.9}
\expandafter\def\csname pn@hot/rule/g2nodd\endcsname{55.9}
\expandafter\def\csname pn@hot/rule/rrfdd\endcsname{47.5}
\expandafter\def\csname pn@hot/rule/rrfnodd\endcsname{47.5}
\expandafter\def\csname pn@hot/rule/scoremaxdd\endcsname{48.1}
\expandafter\def\csname pn@hot/rule/scoremaxnodd\endcsname{48.1}
\expandafter\def\csname pn@hot/rule/top2dd\endcsname{46.1}
\expandafter\def\csname pn@hot/rule/top2nodd\endcsname{46.1}
\expandafter\def\csname pn@hot/rule/top3dd\endcsname{51.9}
\expandafter\def\csname pn@hot/rule/top3nodd\endcsname{51.9}
\expandafter\def\csname pn@hot/simgraph/EM/Decomp\endcsname{$+1.8$}
\expandafter\def\csname pn@hot/simgraph/EM/Decomp/abs\endcsname{1.8}
\expandafter\def\csname pn@hot/simgraph/EM/Decomp/ci\endcsname{[$0.5$, $3.1$]}
\expandafter\def\csname pn@hot/simgraph/EM/Decomp/full\endcsname{$+1.8$ [$0.5$, $3.1$]}
\expandafter\def\csname pn@hot/simgraph/EM/IRCoT\endcsname{$+1.6$}
\expandafter\def\csname pn@hot/simgraph/EM/IRCoT/abs\endcsname{1.6}
\expandafter\def\csname pn@hot/simgraph/EM/IRCoT/ci\endcsname{[$0.4$, $2.8$]}
\expandafter\def\csname pn@hot/simgraph/EM/IRCoT/full\endcsname{$+1.6$ [$0.4$, $2.8$]}
\expandafter\def\csname pn@hot/simgraph/EM/ReAct\endcsname{$+0.1$}
\expandafter\def\csname pn@hot/simgraph/EM/ReAct/abs\endcsname{0.1}
\expandafter\def\csname pn@hot/simgraph/EM/ReAct/ci\endcsname{[$-1.2$, $1.4$]}
\expandafter\def\csname pn@hot/simgraph/EM/ReAct/full\endcsname{$+0.1$ [$-1.2$, $1.4$]}
\expandafter\def\csname pn@hot/simgraph/EM/SR1\endcsname{$+0.8$}
\expandafter\def\csname pn@hot/simgraph/EM/SR1/abs\endcsname{0.8}
\expandafter\def\csname pn@hot/simgraph/EM/SR1/ci\endcsname{[$-0.6$, $2.2$]}
\expandafter\def\csname pn@hot/simgraph/EM/SR1/full\endcsname{$+0.8$ [$-0.6$, $2.2$]}
\expandafter\def\csname pn@hot/simgraph/EM/ext\endcsname{$+0.8$}
\expandafter\def\csname pn@hot/simgraph/EM/ext/abs\endcsname{0.8}
\expandafter\def\csname pn@hot/simgraph/EM/ext/ci\endcsname{[$0.0$, $1.7$]}
\expandafter\def\csname pn@hot/simgraph/EM/ext/full\endcsname{$+0.8$ [$0.0$, $1.7$]}
\expandafter\def\csname pn@hot/simgraph/EM/nsig\endcsname{2}
\expandafter\def\csname pn@hot/simgraph/EM/pool\endcsname{$+1.1$}
\expandafter\def\csname pn@hot/simgraph/EM/pool/abs\endcsname{1.1}
\expandafter\def\csname pn@hot/simgraph/EM/pool/ci\endcsname{[$0.3$, $1.9$]}
\expandafter\def\csname pn@hot/simgraph/EM/pool/full\endcsname{$+1.1$ [$0.3$, $1.9$]}
\expandafter\def\csname pn@hot/simgraph/EM/range\endcsname{$+0.1$ to $+1.8$}
\expandafter\def\csname pn@hot/simgraph/ans/Decomp\endcsname{$+4.0$}
\expandafter\def\csname pn@hot/simgraph/ans/Decomp/abs\endcsname{4.0}
\expandafter\def\csname pn@hot/simgraph/ans/Decomp/ci\endcsname{[$3.0$, $5.1$]}
\expandafter\def\csname pn@hot/simgraph/ans/Decomp/full\endcsname{$+4.0$ [$3.0$, $5.1$]}
\expandafter\def\csname pn@hot/simgraph/ans/IRCoT\endcsname{$+0.8$}
\expandafter\def\csname pn@hot/simgraph/ans/IRCoT/abs\endcsname{0.8}
\expandafter\def\csname pn@hot/simgraph/ans/IRCoT/ci\endcsname{[$0.3$, $1.4$]}
\expandafter\def\csname pn@hot/simgraph/ans/IRCoT/full\endcsname{$+0.8$ [$0.3$, $1.4$]}
\expandafter\def\csname pn@hot/simgraph/ans/ReAct\endcsname{$+1.9$}
\expandafter\def\csname pn@hot/simgraph/ans/ReAct/abs\endcsname{1.9}
\expandafter\def\csname pn@hot/simgraph/ans/ReAct/ci\endcsname{[$1.1$, $2.7$]}
\expandafter\def\csname pn@hot/simgraph/ans/ReAct/full\endcsname{$+1.9$ [$1.1$, $2.7$]}
\expandafter\def\csname pn@hot/simgraph/ans/SR1\endcsname{$+1.9$}
\expandafter\def\csname pn@hot/simgraph/ans/SR1/abs\endcsname{1.9}
\expandafter\def\csname pn@hot/simgraph/ans/SR1/ci\endcsname{[$1.1$, $2.7$]}
\expandafter\def\csname pn@hot/simgraph/ans/SR1/full\endcsname{$+1.9$ [$1.1$, $2.7$]}
\expandafter\def\csname pn@hot/simgraph/ans/ext\endcsname{$+1.6$}
\expandafter\def\csname pn@hot/simgraph/ans/ext/abs\endcsname{1.6}
\expandafter\def\csname pn@hot/simgraph/ans/ext/ci\endcsname{[$1.1$, $2.0$]}
\expandafter\def\csname pn@hot/simgraph/ans/ext/full\endcsname{$+1.6$ [$1.1$, $2.0$]}
\expandafter\def\csname pn@hot/simgraph/ans/nsig\endcsname{4}
\expandafter\def\csname pn@hot/simgraph/ans/pool\endcsname{$+2.2$}
\expandafter\def\csname pn@hot/simgraph/ans/pool/abs\endcsname{2.2}
\expandafter\def\csname pn@hot/simgraph/ans/pool/ci\endcsname{[$1.7$, $2.7$]}
\expandafter\def\csname pn@hot/simgraph/ans/pool/full\endcsname{$+2.2$ [$1.7$, $2.7$]}
\expandafter\def\csname pn@hot/simgraph/ans/range\endcsname{$+0.8$ to $+4.0$}
\expandafter\def\csname pn@hot/simgraph/complete/Decomp\endcsname{$+6.0$}
\expandafter\def\csname pn@hot/simgraph/complete/Decomp/abs\endcsname{6.0}
\expandafter\def\csname pn@hot/simgraph/complete/Decomp/ci\endcsname{[$4.7$, $7.3$]}
\expandafter\def\csname pn@hot/simgraph/complete/Decomp/full\endcsname{$+6.0$ [$4.7$, $7.3$]}
\expandafter\def\csname pn@hot/simgraph/complete/IRCoT\endcsname{$+2.3$}
\expandafter\def\csname pn@hot/simgraph/complete/IRCoT/abs\endcsname{2.3}
\expandafter\def\csname pn@hot/simgraph/complete/IRCoT/ci\endcsname{[$1.5$, $3.1$]}
\expandafter\def\csname pn@hot/simgraph/complete/IRCoT/full\endcsname{$+2.3$ [$1.5$, $3.1$]}
\expandafter\def\csname pn@hot/simgraph/complete/ReAct\endcsname{$+3.7$}
\expandafter\def\csname pn@hot/simgraph/complete/ReAct/abs\endcsname{3.7}
\expandafter\def\csname pn@hot/simgraph/complete/ReAct/ci\endcsname{[$2.8$, $4.7$]}
\expandafter\def\csname pn@hot/simgraph/complete/ReAct/full\endcsname{$+3.7$ [$2.8$, $4.7$]}
\expandafter\def\csname pn@hot/simgraph/complete/SR1\endcsname{$+3.9$}
\expandafter\def\csname pn@hot/simgraph/complete/SR1/abs\endcsname{3.9}
\expandafter\def\csname pn@hot/simgraph/complete/SR1/ci\endcsname{[$2.9$, $5.0$]}
\expandafter\def\csname pn@hot/simgraph/complete/SR1/full\endcsname{$+3.9$ [$2.9$, $5.0$]}
\expandafter\def\csname pn@hot/simgraph/complete/ext\endcsname{$+3.3$}
\expandafter\def\csname pn@hot/simgraph/complete/ext/abs\endcsname{3.3}
\expandafter\def\csname pn@hot/simgraph/complete/ext/ci\endcsname{[$2.7$, $4.0$]}
\expandafter\def\csname pn@hot/simgraph/complete/ext/full\endcsname{$+3.3$ [$2.7$, $4.0$]}
\expandafter\def\csname pn@hot/simgraph/complete/nsig\endcsname{4}
\expandafter\def\csname pn@hot/simgraph/complete/pool\endcsname{$+4.0$}
\expandafter\def\csname pn@hot/simgraph/complete/pool/abs\endcsname{4.0}
\expandafter\def\csname pn@hot/simgraph/complete/pool/ci\endcsname{[$3.3$, $4.7$]}
\expandafter\def\csname pn@hot/simgraph/complete/pool/full\endcsname{$+4.0$ [$3.3$, $4.7$]}
\expandafter\def\csname pn@hot/simgraph/complete/range\endcsname{$+2.3$ to $+6.0$}
\expandafter\def\csname pn@hot/sr1native\endcsname{44.8}
\expandafter\def\csname pn@hot/sr1own/detour\endcsname{46.1}
\expandafter\def\csname pn@hot/sr1own/detourVSinterleave\endcsname{$+1.8$}
\expandafter\def\csname pn@hot/sr1own/detourVSinterleave/abs\endcsname{1.8}
\expandafter\def\csname pn@hot/sr1own/detourVSinterleave/ci\endcsname{[$0.0$, $3.7$]}
\expandafter\def\csname pn@hot/sr1own/detourVSinterleave/full\endcsname{$+1.8$ [$0.0$, $3.7$]}
\expandafter\def\csname pn@hot/sr1own/detourVSnative\endcsname{$+1.3$}
\expandafter\def\csname pn@hot/sr1own/detourVSnative/abs\endcsname{1.3}
\expandafter\def\csname pn@hot/sr1own/detourVSnative/ci\endcsname{[$-0.8$, $3.4$]}
\expandafter\def\csname pn@hot/sr1own/detourVSnative/full\endcsname{$+1.3$ [$-0.8$, $3.4$]}
\expandafter\def\csname pn@hot/sr1own/interleave\endcsname{44.3}
\expandafter\def\csname pn@hot/sr1own/interleaveVSnative\endcsname{$-0.5$}
\expandafter\def\csname pn@hot/sr1own/interleaveVSnative/abs\endcsname{0.5}
\expandafter\def\csname pn@hot/sr1own/interleaveVSnative/ci\endcsname{[$-2.5$, $1.5$]}
\expandafter\def\csname pn@hot/sr1own/interleaveVSnative/full\endcsname{$-0.5$ [$-2.5$, $1.5$]}
\expandafter\def\csname pn@hot/sr1own/native\endcsname{44.8}
\expandafter\def\csname pn@hot/sr1own/rrf\endcsname{42.1}
\expandafter\def\csname pn@hot/sr1own/rrfVSnative\endcsname{$-2.7$}
\expandafter\def\csname pn@hot/sr1own/rrfVSnative/abs\endcsname{2.7}
\expandafter\def\csname pn@hot/sr1own/rrfVSnative/ci\endcsname{[$-4.7$, $-0.6$]}
\expandafter\def\csname pn@hot/sr1own/rrfVSnative/full\endcsname{$-2.7$ [$-4.7$, $-0.6$]}
\expandafter\def\csname pn@hot/sr1own/window\endcsname{45.7}
\expandafter\def\csname pn@hot/sr1own/windowVSinterleave\endcsname{$+1.5$}
\expandafter\def\csname pn@hot/sr1own/windowVSinterleave/abs\endcsname{1.5}
\expandafter\def\csname pn@hot/sr1own/windowVSinterleave/ci\endcsname{[$-0.3$, $3.3$]}
\expandafter\def\csname pn@hot/sr1own/windowVSinterleave/full\endcsname{$+1.5$ [$-0.3$, $3.3$]}
\expandafter\def\csname pn@hot/sr1own/windowVSnative\endcsname{$+0.9$}
\expandafter\def\csname pn@hot/sr1own/windowVSnative/abs\endcsname{0.9}
\expandafter\def\csname pn@hot/sr1own/windowVSnative/ci\endcsname{[$-1.1$, $3.1$]}
\expandafter\def\csname pn@hot/sr1own/windowVSnative/full\endcsname{$+0.9$ [$-1.1$, $3.1$]}
\expandafter\def\csname pn@hot/titleroute/EM/Decomp\endcsname{$+0.6$}
\expandafter\def\csname pn@hot/titleroute/EM/Decomp/abs\endcsname{0.6}
\expandafter\def\csname pn@hot/titleroute/EM/Decomp/ci\endcsname{[$-0.2$, $1.4$]}
\expandafter\def\csname pn@hot/titleroute/EM/Decomp/full\endcsname{$+0.6$ [$-0.2$, $1.4$]}
\expandafter\def\csname pn@hot/titleroute/EM/IRCoT\endcsname{$-0.3$}
\expandafter\def\csname pn@hot/titleroute/EM/IRCoT/abs\endcsname{0.3}
\expandafter\def\csname pn@hot/titleroute/EM/IRCoT/ci\endcsname{[$-1.4$, $0.8$]}
\expandafter\def\csname pn@hot/titleroute/EM/IRCoT/full\endcsname{$-0.3$ [$-1.4$, $0.8$]}
\expandafter\def\csname pn@hot/titleroute/EM/ReAct\endcsname{$+0.2$}
\expandafter\def\csname pn@hot/titleroute/EM/ReAct/abs\endcsname{0.2}
\expandafter\def\csname pn@hot/titleroute/EM/ReAct/ci\endcsname{[$-1.1$, $1.5$]}
\expandafter\def\csname pn@hot/titleroute/EM/ReAct/full\endcsname{$+0.2$ [$-1.1$, $1.5$]}
\expandafter\def\csname pn@hot/titleroute/EM/SR1\endcsname{$+0.2$}
\expandafter\def\csname pn@hot/titleroute/EM/SR1/abs\endcsname{0.2}
\expandafter\def\csname pn@hot/titleroute/EM/SR1/ci\endcsname{[$-0.5$, $0.9$]}
\expandafter\def\csname pn@hot/titleroute/EM/SR1/full\endcsname{$+0.2$ [$-0.5$, $0.9$]}
\expandafter\def\csname pn@hot/titleroute/EM/ext\endcsname{$0.0$}
\expandafter\def\csname pn@hot/titleroute/EM/ext/abs\endcsname{0.0}
\expandafter\def\csname pn@hot/titleroute/EM/ext/ci\endcsname{[$-0.6$, $0.7$]}
\expandafter\def\csname pn@hot/titleroute/EM/ext/full\endcsname{$0.0$ [$-0.6$, $0.7$]}
\expandafter\def\csname pn@hot/titleroute/EM/nsig\endcsname{0}
\expandafter\def\csname pn@hot/titleroute/EM/pool\endcsname{$+0.2$}
\expandafter\def\csname pn@hot/titleroute/EM/pool/abs\endcsname{0.2}
\expandafter\def\csname pn@hot/titleroute/EM/pool/ci\endcsname{[$-0.4$, $0.7$]}
\expandafter\def\csname pn@hot/titleroute/EM/pool/full\endcsname{$+0.2$ [$-0.4$, $0.7$]}
\expandafter\def\csname pn@hot/titleroute/EM/range\endcsname{$-0.3$ to $+0.6$}
\expandafter\def\csname pn@hot/titleroute/ans/Decomp\endcsname{$+0.9$}
\expandafter\def\csname pn@hot/titleroute/ans/Decomp/abs\endcsname{0.9}
\expandafter\def\csname pn@hot/titleroute/ans/Decomp/ci\endcsname{[$0.1$, $1.8$]}
\expandafter\def\csname pn@hot/titleroute/ans/Decomp/full\endcsname{$+0.9$ [$0.1$, $1.8$]}
\expandafter\def\csname pn@hot/titleroute/ans/IRCoT\endcsname{$-0.4$}
\expandafter\def\csname pn@hot/titleroute/ans/IRCoT/abs\endcsname{0.4}
\expandafter\def\csname pn@hot/titleroute/ans/IRCoT/ci\endcsname{[$-1.2$, $0.4$]}
\expandafter\def\csname pn@hot/titleroute/ans/IRCoT/full\endcsname{$-0.4$ [$-1.2$, $0.4$]}
\expandafter\def\csname pn@hot/titleroute/ans/ReAct\endcsname{$+0.6$}
\expandafter\def\csname pn@hot/titleroute/ans/ReAct/abs\endcsname{0.6}
\expandafter\def\csname pn@hot/titleroute/ans/ReAct/ci\endcsname{[$-0.5$, $1.8$]}
\expandafter\def\csname pn@hot/titleroute/ans/ReAct/full\endcsname{$+0.6$ [$-0.5$, $1.8$]}
\expandafter\def\csname pn@hot/titleroute/ans/SR1\endcsname{$+0.4$}
\expandafter\def\csname pn@hot/titleroute/ans/SR1/abs\endcsname{0.4}
\expandafter\def\csname pn@hot/titleroute/ans/SR1/ci\endcsname{[$-0.1$, $0.9$]}
\expandafter\def\csname pn@hot/titleroute/ans/SR1/full\endcsname{$+0.4$ [$-0.1$, $0.9$]}
\expandafter\def\csname pn@hot/titleroute/ans/ext\endcsname{$+0.2$}
\expandafter\def\csname pn@hot/titleroute/ans/ext/abs\endcsname{0.2}
\expandafter\def\csname pn@hot/titleroute/ans/ext/ci\endcsname{[$-0.4$, $0.8$]}
\expandafter\def\csname pn@hot/titleroute/ans/ext/full\endcsname{$+0.2$ [$-0.4$, $0.8$]}
\expandafter\def\csname pn@hot/titleroute/ans/nsig\endcsname{0}
\expandafter\def\csname pn@hot/titleroute/ans/pool\endcsname{$+0.4$}
\expandafter\def\csname pn@hot/titleroute/ans/pool/abs\endcsname{0.4}
\expandafter\def\csname pn@hot/titleroute/ans/pool/ci\endcsname{[$-0.1$, $0.9$]}
\expandafter\def\csname pn@hot/titleroute/ans/pool/full\endcsname{$+0.4$ [$-0.1$, $0.9$]}
\expandafter\def\csname pn@hot/titleroute/ans/range\endcsname{$-0.4$ to $+0.9$}
\expandafter\def\csname pn@hot/titleroute/complete/Decomp\endcsname{$+1.3$}
\expandafter\def\csname pn@hot/titleroute/complete/Decomp/abs\endcsname{1.3}
\expandafter\def\csname pn@hot/titleroute/complete/Decomp/ci\endcsname{[$0.4$, $2.2$]}
\expandafter\def\csname pn@hot/titleroute/complete/Decomp/full\endcsname{$+1.3$ [$0.4$, $2.2$]}
\expandafter\def\csname pn@hot/titleroute/complete/IRCoT\endcsname{$+0.3$}
\expandafter\def\csname pn@hot/titleroute/complete/IRCoT/abs\endcsname{0.3}
\expandafter\def\csname pn@hot/titleroute/complete/IRCoT/ci\endcsname{[$-0.7$, $1.3$]}
\expandafter\def\csname pn@hot/titleroute/complete/IRCoT/full\endcsname{$+0.3$ [$-0.7$, $1.3$]}
\expandafter\def\csname pn@hot/titleroute/complete/ReAct\endcsname{$+1.7$}
\expandafter\def\csname pn@hot/titleroute/complete/ReAct/abs\endcsname{1.7}
\expandafter\def\csname pn@hot/titleroute/complete/ReAct/ci\endcsname{[$0.3$, $3.1$]}
\expandafter\def\csname pn@hot/titleroute/complete/ReAct/full\endcsname{$+1.7$ [$0.3$, $3.1$]}
\expandafter\def\csname pn@hot/titleroute/complete/SR1\endcsname{$+0.6$}
\expandafter\def\csname pn@hot/titleroute/complete/SR1/abs\endcsname{0.6}
\expandafter\def\csname pn@hot/titleroute/complete/SR1/ci\endcsname{[$-0.1$, $1.3$]}
\expandafter\def\csname pn@hot/titleroute/complete/SR1/full\endcsname{$+0.6$ [$-0.1$, $1.3$]}
\expandafter\def\csname pn@hot/titleroute/complete/ext\endcsname{$+0.9$}
\expandafter\def\csname pn@hot/titleroute/complete/ext/abs\endcsname{0.9}
\expandafter\def\csname pn@hot/titleroute/complete/ext/ci\endcsname{[$0.2$, $1.6$]}
\expandafter\def\csname pn@hot/titleroute/complete/ext/full\endcsname{$+0.9$ [$0.2$, $1.6$]}
\expandafter\def\csname pn@hot/titleroute/complete/nsig\endcsname{2}
\expandafter\def\csname pn@hot/titleroute/complete/pool\endcsname{$+1.0$}
\expandafter\def\csname pn@hot/titleroute/complete/pool/abs\endcsname{1.0}
\expandafter\def\csname pn@hot/titleroute/complete/pool/ci\endcsname{[$0.4$, $1.6$]}
\expandafter\def\csname pn@hot/titleroute/complete/pool/full\endcsname{$+1.0$ [$0.4$, $1.6$]}
\expandafter\def\csname pn@hot/titleroute/complete/range\endcsname{$+0.3$ to $+1.7$}
\expandafter\def\csname pn@hot/vsrrf/EM/Decomp\endcsname{$+3.3$}
\expandafter\def\csname pn@hot/vsrrf/EM/Decomp/abs\endcsname{3.3}
\expandafter\def\csname pn@hot/vsrrf/EM/Decomp/ci\endcsname{[$1.7$, $4.9$]}
\expandafter\def\csname pn@hot/vsrrf/EM/Decomp/full\endcsname{$+3.3$ [$1.7$, $4.9$]}
\expandafter\def\csname pn@hot/vsrrf/EM/IRCoT\endcsname{$+4.3$}
\expandafter\def\csname pn@hot/vsrrf/EM/IRCoT/abs\endcsname{4.3}
\expandafter\def\csname pn@hot/vsrrf/EM/IRCoT/ci\endcsname{[$2.5$, $6.0$]}
\expandafter\def\csname pn@hot/vsrrf/EM/IRCoT/full\endcsname{$+4.3$ [$2.5$, $6.0$]}
\expandafter\def\csname pn@hot/vsrrf/EM/ReAct\endcsname{$+5.2$}
\expandafter\def\csname pn@hot/vsrrf/EM/ReAct/abs\endcsname{5.2}
\expandafter\def\csname pn@hot/vsrrf/EM/ReAct/ci\endcsname{[$3.3$, $7.1$]}
\expandafter\def\csname pn@hot/vsrrf/EM/ReAct/full\endcsname{$+5.2$ [$3.3$, $7.1$]}
\expandafter\def\csname pn@hot/vsrrf/EM/SR1\endcsname{$+1.3$}
\expandafter\def\csname pn@hot/vsrrf/EM/SR1/abs\endcsname{1.3}
\expandafter\def\csname pn@hot/vsrrf/EM/SR1/ci\endcsname{[$-0.3$, $2.9$]}
\expandafter\def\csname pn@hot/vsrrf/EM/SR1/full\endcsname{$+1.3$ [$-0.3$, $2.9$]}
\expandafter\def\csname pn@hot/vsrrf/EM/ext\endcsname{$+3.6$}
\expandafter\def\csname pn@hot/vsrrf/EM/ext/abs\endcsname{3.6}
\expandafter\def\csname pn@hot/vsrrf/EM/ext/ci\endcsname{[$2.5$, $4.7$]}
\expandafter\def\csname pn@hot/vsrrf/EM/ext/full\endcsname{$+3.6$ [$2.5$, $4.7$]}
\expandafter\def\csname pn@hot/vsrrf/EM/nsig\endcsname{3}
\expandafter\def\csname pn@hot/vsrrf/EM/pool\endcsname{$+3.5$}
\expandafter\def\csname pn@hot/vsrrf/EM/pool/abs\endcsname{3.5}
\expandafter\def\csname pn@hot/vsrrf/EM/pool/ci\endcsname{[$2.5$, $4.5$]}
\expandafter\def\csname pn@hot/vsrrf/EM/pool/full\endcsname{$+3.5$ [$2.5$, $4.5$]}
\expandafter\def\csname pn@hot/vsrrf/EM/range\endcsname{$+1.3$ to $+5.2$}
\expandafter\def\csname pn@hot/vsrrf/ans/Decomp\endcsname{$+8.0$}
\expandafter\def\csname pn@hot/vsrrf/ans/Decomp/abs\endcsname{8.0}
\expandafter\def\csname pn@hot/vsrrf/ans/Decomp/ci\endcsname{[$6.4$, $9.7$]}
\expandafter\def\csname pn@hot/vsrrf/ans/Decomp/full\endcsname{$+8.0$ [$6.4$, $9.7$]}
\expandafter\def\csname pn@hot/vsrrf/ans/IRCoT\endcsname{$+7.2$}
\expandafter\def\csname pn@hot/vsrrf/ans/IRCoT/abs\endcsname{7.2}
\expandafter\def\csname pn@hot/vsrrf/ans/IRCoT/ci\endcsname{[$5.7$, $8.7$]}
\expandafter\def\csname pn@hot/vsrrf/ans/IRCoT/full\endcsname{$+7.2$ [$5.7$, $8.7$]}
\expandafter\def\csname pn@hot/vsrrf/ans/ReAct\endcsname{$+12.3$}
\expandafter\def\csname pn@hot/vsrrf/ans/ReAct/abs\endcsname{12.3}
\expandafter\def\csname pn@hot/vsrrf/ans/ReAct/ci\endcsname{[$10.4$, $14.2$]}
\expandafter\def\csname pn@hot/vsrrf/ans/ReAct/full\endcsname{$+12.3$ [$10.4$, $14.2$]}
\expandafter\def\csname pn@hot/vsrrf/ans/SR1\endcsname{$+5.4$}
\expandafter\def\csname pn@hot/vsrrf/ans/SR1/abs\endcsname{5.4}
\expandafter\def\csname pn@hot/vsrrf/ans/SR1/ci\endcsname{[$4.1$, $6.8$]}
\expandafter\def\csname pn@hot/vsrrf/ans/SR1/full\endcsname{$+5.4$ [$4.1$, $6.8$]}
\expandafter\def\csname pn@hot/vsrrf/ans/ext\endcsname{$+8.3$}
\expandafter\def\csname pn@hot/vsrrf/ans/ext/abs\endcsname{8.3}
\expandafter\def\csname pn@hot/vsrrf/ans/ext/ci\endcsname{[$7.3$, $9.3$]}
\expandafter\def\csname pn@hot/vsrrf/ans/ext/full\endcsname{$+8.3$ [$7.3$, $9.3$]}
\expandafter\def\csname pn@hot/vsrrf/ans/nsig\endcsname{4}
\expandafter\def\csname pn@hot/vsrrf/ans/pool\endcsname{$+8.2$}
\expandafter\def\csname pn@hot/vsrrf/ans/pool/abs\endcsname{8.2}
\expandafter\def\csname pn@hot/vsrrf/ans/pool/ci\endcsname{[$7.3$, $9.2$]}
\expandafter\def\csname pn@hot/vsrrf/ans/pool/full\endcsname{$+8.2$ [$7.3$, $9.2$]}
\expandafter\def\csname pn@hot/vsrrf/ans/range\endcsname{$+5.4$ to $+12.3$}
\expandafter\def\csname pn@hot/vsrrf/complete/Decomp\endcsname{$+15.5$}
\expandafter\def\csname pn@hot/vsrrf/complete/Decomp/abs\endcsname{15.5}
\expandafter\def\csname pn@hot/vsrrf/complete/Decomp/ci\endcsname{[$13.5$, $17.4$]}
\expandafter\def\csname pn@hot/vsrrf/complete/Decomp/full\endcsname{$+15.5$ [$13.5$, $17.4$]}
\expandafter\def\csname pn@hot/vsrrf/complete/IRCoT\endcsname{$+25.7$}
\expandafter\def\csname pn@hot/vsrrf/complete/IRCoT/abs\endcsname{25.7}
\expandafter\def\csname pn@hot/vsrrf/complete/IRCoT/ci\endcsname{[$23.4$, $28.1$]}
\expandafter\def\csname pn@hot/vsrrf/complete/IRCoT/full\endcsname{$+25.7$ [$23.4$, $28.1$]}
\expandafter\def\csname pn@hot/vsrrf/complete/ReAct\endcsname{$+27.9$}
\expandafter\def\csname pn@hot/vsrrf/complete/ReAct/abs\endcsname{27.9}
\expandafter\def\csname pn@hot/vsrrf/complete/ReAct/ci\endcsname{[$25.6$, $30.2$]}
\expandafter\def\csname pn@hot/vsrrf/complete/ReAct/full\endcsname{$+27.9$ [$25.6$, $30.2$]}
\expandafter\def\csname pn@hot/vsrrf/complete/SR1\endcsname{$+11.5$}
\expandafter\def\csname pn@hot/vsrrf/complete/SR1/abs\endcsname{11.5}
\expandafter\def\csname pn@hot/vsrrf/complete/SR1/ci\endcsname{[$9.8$, $13.2$]}
\expandafter\def\csname pn@hot/vsrrf/complete/SR1/full\endcsname{$+11.5$ [$9.8$, $13.2$]}
\expandafter\def\csname pn@hot/vsrrf/complete/ext\endcsname{$+21.7$}
\expandafter\def\csname pn@hot/vsrrf/complete/ext/abs\endcsname{21.7}
\expandafter\def\csname pn@hot/vsrrf/complete/ext/ci\endcsname{[$20.3$, $23.1$]}
\expandafter\def\csname pn@hot/vsrrf/complete/ext/full\endcsname{$+21.7$ [$20.3$, $23.1$]}
\expandafter\def\csname pn@hot/vsrrf/complete/nsig\endcsname{4}
\expandafter\def\csname pn@hot/vsrrf/complete/pool\endcsname{$+20.1$}
\expandafter\def\csname pn@hot/vsrrf/complete/pool/abs\endcsname{20.1}
\expandafter\def\csname pn@hot/vsrrf/complete/pool/ci\endcsname{[$18.9$, $21.4$]}
\expandafter\def\csname pn@hot/vsrrf/complete/pool/full\endcsname{$+20.1$ [$18.9$, $21.4$]}
\expandafter\def\csname pn@hot/vsrrf/complete/range\endcsname{$+11.5$ to $+27.9$}
\expandafter\def\csname pn@hot/winVSrerank/EM/Decomp\endcsname{$-0.5$}
\expandafter\def\csname pn@hot/winVSrerank/EM/Decomp/abs\endcsname{0.5}
\expandafter\def\csname pn@hot/winVSrerank/EM/Decomp/ci\endcsname{[$-2.1$, $1.1$]}
\expandafter\def\csname pn@hot/winVSrerank/EM/Decomp/full\endcsname{$-0.5$ [$-2.1$, $1.1$]}
\expandafter\def\csname pn@hot/winVSrerank/EM/IRCoT\endcsname{$+2.5$}
\expandafter\def\csname pn@hot/winVSrerank/EM/IRCoT/abs\endcsname{2.5}
\expandafter\def\csname pn@hot/winVSrerank/EM/IRCoT/ci\endcsname{[$0.9$, $4.1$]}
\expandafter\def\csname pn@hot/winVSrerank/EM/IRCoT/full\endcsname{$+2.5$ [$0.9$, $4.1$]}
\expandafter\def\csname pn@hot/winVSrerank/EM/ReAct\endcsname{$-0.8$}
\expandafter\def\csname pn@hot/winVSrerank/EM/ReAct/abs\endcsname{0.8}
\expandafter\def\csname pn@hot/winVSrerank/EM/ReAct/ci\endcsname{[$-2.4$, $0.7$]}
\expandafter\def\csname pn@hot/winVSrerank/EM/ReAct/full\endcsname{$-0.8$ [$-2.4$, $0.7$]}
\expandafter\def\csname pn@hot/winVSrerank/EM/SR1\endcsname{$+0.5$}
\expandafter\def\csname pn@hot/winVSrerank/EM/SR1/abs\endcsname{0.5}
\expandafter\def\csname pn@hot/winVSrerank/EM/SR1/ci\endcsname{[$-1.1$, $2.2$]}
\expandafter\def\csname pn@hot/winVSrerank/EM/SR1/full\endcsname{$+0.5$ [$-1.1$, $2.2$]}
\expandafter\def\csname pn@hot/winVSrerank/EM/ext\endcsname{$+0.7$}
\expandafter\def\csname pn@hot/winVSrerank/EM/ext/abs\endcsname{0.7}
\expandafter\def\csname pn@hot/winVSrerank/EM/ext/ci\endcsname{[$-0.4$, $1.8$]}
\expandafter\def\csname pn@hot/winVSrerank/EM/ext/full\endcsname{$+0.7$ [$-0.4$, $1.8$]}
\expandafter\def\csname pn@hot/winVSrerank/EM/nsig\endcsname{1}
\expandafter\def\csname pn@hot/winVSrerank/EM/pool\endcsname{$+0.4$}
\expandafter\def\csname pn@hot/winVSrerank/EM/pool/abs\endcsname{0.4}
\expandafter\def\csname pn@hot/winVSrerank/EM/pool/ci\endcsname{[$-0.6$, $1.5$]}
\expandafter\def\csname pn@hot/winVSrerank/EM/pool/full\endcsname{$+0.4$ [$-0.6$, $1.5$]}
\expandafter\def\csname pn@hot/winVSrerank/EM/range\endcsname{$-0.8$ to $+2.5$}
\expandafter\def\csname pn@hot/winVSrerank/ans/Decomp\endcsname{$-1.9$}
\expandafter\def\csname pn@hot/winVSrerank/ans/Decomp/abs\endcsname{1.9}
\expandafter\def\csname pn@hot/winVSrerank/ans/Decomp/ci\endcsname{[$-3.1$, $-0.7$]}
\expandafter\def\csname pn@hot/winVSrerank/ans/Decomp/full\endcsname{$-1.9$ [$-3.1$, $-0.7$]}
\expandafter\def\csname pn@hot/winVSrerank/ans/IRCoT\endcsname{$+0.5$}
\expandafter\def\csname pn@hot/winVSrerank/ans/IRCoT/abs\endcsname{0.5}
\expandafter\def\csname pn@hot/winVSrerank/ans/IRCoT/ci\endcsname{[$-0.3$, $1.3$]}
\expandafter\def\csname pn@hot/winVSrerank/ans/IRCoT/full\endcsname{$+0.5$ [$-0.3$, $1.3$]}
\expandafter\def\csname pn@hot/winVSrerank/ans/ReAct\endcsname{$-0.8$}
\expandafter\def\csname pn@hot/winVSrerank/ans/ReAct/abs\endcsname{0.8}
\expandafter\def\csname pn@hot/winVSrerank/ans/ReAct/ci\endcsname{[$-1.9$, $0.1$]}
\expandafter\def\csname pn@hot/winVSrerank/ans/ReAct/full\endcsname{$-0.8$ [$-1.9$, $0.1$]}
\expandafter\def\csname pn@hot/winVSrerank/ans/SR1\endcsname{$-0.6$}
\expandafter\def\csname pn@hot/winVSrerank/ans/SR1/abs\endcsname{0.6}
\expandafter\def\csname pn@hot/winVSrerank/ans/SR1/ci\endcsname{[$-1.5$, $0.2$]}
\expandafter\def\csname pn@hot/winVSrerank/ans/SR1/full\endcsname{$-0.6$ [$-1.5$, $0.2$]}
\expandafter\def\csname pn@hot/winVSrerank/ans/ext\endcsname{$-0.3$}
\expandafter\def\csname pn@hot/winVSrerank/ans/ext/abs\endcsname{0.3}
\expandafter\def\csname pn@hot/winVSrerank/ans/ext/ci\endcsname{[$-1.0$, $0.3$]}
\expandafter\def\csname pn@hot/winVSrerank/ans/ext/full\endcsname{$-0.3$ [$-1.0$, $0.3$]}
\expandafter\def\csname pn@hot/winVSrerank/ans/nsig\endcsname{1}
\expandafter\def\csname pn@hot/winVSrerank/ans/pool\endcsname{$-0.7$}
\expandafter\def\csname pn@hot/winVSrerank/ans/pool/abs\endcsname{0.7}
\expandafter\def\csname pn@hot/winVSrerank/ans/pool/ci\endcsname{[$-1.4$, $-0.1$]}
\expandafter\def\csname pn@hot/winVSrerank/ans/pool/full\endcsname{$-0.7$ [$-1.4$, $-0.1$]}
\expandafter\def\csname pn@hot/winVSrerank/ans/range\endcsname{$-1.9$ to $+0.5$}
\expandafter\def\csname pn@hot/winVSrerank/complete/Decomp\endcsname{$+4.1$}
\expandafter\def\csname pn@hot/winVSrerank/complete/Decomp/abs\endcsname{4.1}
\expandafter\def\csname pn@hot/winVSrerank/complete/Decomp/ci\endcsname{[$2.5$, $5.6$]}
\expandafter\def\csname pn@hot/winVSrerank/complete/Decomp/full\endcsname{$+4.1$ [$2.5$, $5.6$]}
\expandafter\def\csname pn@hot/winVSrerank/complete/IRCoT\endcsname{$+7.4$}
\expandafter\def\csname pn@hot/winVSrerank/complete/IRCoT/abs\endcsname{7.4}
\expandafter\def\csname pn@hot/winVSrerank/complete/IRCoT/ci\endcsname{[$6.0$, $8.9$]}
\expandafter\def\csname pn@hot/winVSrerank/complete/IRCoT/full\endcsname{$+7.4$ [$6.0$, $8.9$]}
\expandafter\def\csname pn@hot/winVSrerank/complete/ReAct\endcsname{$+5.6$}
\expandafter\def\csname pn@hot/winVSrerank/complete/ReAct/abs\endcsname{5.6}
\expandafter\def\csname pn@hot/winVSrerank/complete/ReAct/ci\endcsname{[$4.3$, $7.0$]}
\expandafter\def\csname pn@hot/winVSrerank/complete/ReAct/full\endcsname{$+5.6$ [$4.3$, $7.0$]}
\expandafter\def\csname pn@hot/winVSrerank/complete/SR1\endcsname{$+4.5$}
\expandafter\def\csname pn@hot/winVSrerank/complete/SR1/abs\endcsname{4.5}
\expandafter\def\csname pn@hot/winVSrerank/complete/SR1/ci\endcsname{[$3.2$, $5.9$]}
\expandafter\def\csname pn@hot/winVSrerank/complete/SR1/full\endcsname{$+4.5$ [$3.2$, $5.9$]}
\expandafter\def\csname pn@hot/winVSrerank/complete/ext\endcsname{$+5.8$}
\expandafter\def\csname pn@hot/winVSrerank/complete/ext/abs\endcsname{5.8}
\expandafter\def\csname pn@hot/winVSrerank/complete/ext/ci\endcsname{[$4.8$, $7.0$]}
\expandafter\def\csname pn@hot/winVSrerank/complete/ext/full\endcsname{$+5.8$ [$4.8$, $7.0$]}
\expandafter\def\csname pn@hot/winVSrerank/complete/nsig\endcsname{4}
\expandafter\def\csname pn@hot/winVSrerank/complete/pool\endcsname{$+5.4$}
\expandafter\def\csname pn@hot/winVSrerank/complete/pool/abs\endcsname{5.4}
\expandafter\def\csname pn@hot/winVSrerank/complete/pool/ci\endcsname{[$4.4$, $6.5$]}
\expandafter\def\csname pn@hot/winVSrerank/complete/pool/full\endcsname{$+5.4$ [$4.4$, $6.5$]}
\expandafter\def\csname pn@hot/winVSrerank/complete/range\endcsname{$+4.1$ to $+7.4$}
\expandafter\def\csname pn@hot/wincpu/range\endcsname{0.3--0.4}
\expandafter\def\csname pn@hot/window/EM/Decomp\endcsname{$+0.6$}
\expandafter\def\csname pn@hot/window/EM/Decomp/abs\endcsname{0.6}
\expandafter\def\csname pn@hot/window/EM/Decomp/ci\endcsname{[$-0.8$, $2.1$]}
\expandafter\def\csname pn@hot/window/EM/Decomp/full\endcsname{$+0.6$ [$-0.8$, $2.1$]}
\expandafter\def\csname pn@hot/window/EM/IRCoT\endcsname{$+0.8$}
\expandafter\def\csname pn@hot/window/EM/IRCoT/abs\endcsname{0.8}
\expandafter\def\csname pn@hot/window/EM/IRCoT/ci\endcsname{[$-0.6$, $2.2$]}
\expandafter\def\csname pn@hot/window/EM/IRCoT/full\endcsname{$+0.8$ [$-0.6$, $2.2$]}
\expandafter\def\csname pn@hot/window/EM/ReAct\endcsname{$+0.4$}
\expandafter\def\csname pn@hot/window/EM/ReAct/abs\endcsname{0.4}
\expandafter\def\csname pn@hot/window/EM/ReAct/ci\endcsname{[$-1.0$, $1.7$]}
\expandafter\def\csname pn@hot/window/EM/ReAct/full\endcsname{$+0.4$ [$-1.0$, $1.7$]}
\expandafter\def\csname pn@hot/window/EM/SR1\endcsname{$+0.5$}
\expandafter\def\csname pn@hot/window/EM/SR1/abs\endcsname{0.5}
\expandafter\def\csname pn@hot/window/EM/SR1/ci\endcsname{[$-1.0$, $1.9$]}
\expandafter\def\csname pn@hot/window/EM/SR1/full\endcsname{$+0.5$ [$-1.0$, $1.9$]}
\expandafter\def\csname pn@hot/window/EM/ext\endcsname{$+0.6$}
\expandafter\def\csname pn@hot/window/EM/ext/abs\endcsname{0.6}
\expandafter\def\csname pn@hot/window/EM/ext/ci\endcsname{[$-0.4$, $1.5$]}
\expandafter\def\csname pn@hot/window/EM/ext/full\endcsname{$+0.6$ [$-0.4$, $1.5$]}
\expandafter\def\csname pn@hot/window/EM/nsig\endcsname{0}
\expandafter\def\csname pn@hot/window/EM/pool\endcsname{$+0.6$}
\expandafter\def\csname pn@hot/window/EM/pool/abs\endcsname{0.6}
\expandafter\def\csname pn@hot/window/EM/pool/ci\endcsname{[$-0.3$, $1.4$]}
\expandafter\def\csname pn@hot/window/EM/pool/full\endcsname{$+0.6$ [$-0.3$, $1.4$]}
\expandafter\def\csname pn@hot/window/EM/range\endcsname{$+0.4$ to $+0.8$}
\expandafter\def\csname pn@hot/window/ans/Decomp\endcsname{$+4.5$}
\expandafter\def\csname pn@hot/window/ans/Decomp/abs\endcsname{4.5}
\expandafter\def\csname pn@hot/window/ans/Decomp/ci\endcsname{[$3.3$, $5.8$]}
\expandafter\def\csname pn@hot/window/ans/Decomp/full\endcsname{$+4.5$ [$3.3$, $5.8$]}
\expandafter\def\csname pn@hot/window/ans/IRCoT\endcsname{$+1.1$}
\expandafter\def\csname pn@hot/window/ans/IRCoT/abs\endcsname{1.1}
\expandafter\def\csname pn@hot/window/ans/IRCoT/ci\endcsname{[$0.5$, $1.7$]}
\expandafter\def\csname pn@hot/window/ans/IRCoT/full\endcsname{$+1.1$ [$0.5$, $1.7$]}
\expandafter\def\csname pn@hot/window/ans/ReAct\endcsname{$+2.9$}
\expandafter\def\csname pn@hot/window/ans/ReAct/abs\endcsname{2.9}
\expandafter\def\csname pn@hot/window/ans/ReAct/ci\endcsname{[$1.9$, $3.9$]}
\expandafter\def\csname pn@hot/window/ans/ReAct/full\endcsname{$+2.9$ [$1.9$, $3.9$]}
\expandafter\def\csname pn@hot/window/ans/SR1\endcsname{$+2.7$}
\expandafter\def\csname pn@hot/window/ans/SR1/abs\endcsname{2.7}
\expandafter\def\csname pn@hot/window/ans/SR1/ci\endcsname{[$1.8$, $3.7$]}
\expandafter\def\csname pn@hot/window/ans/SR1/full\endcsname{$+2.7$ [$1.8$, $3.7$]}
\expandafter\def\csname pn@hot/window/ans/ext\endcsname{$+2.2$}
\expandafter\def\csname pn@hot/window/ans/ext/abs\endcsname{2.2}
\expandafter\def\csname pn@hot/window/ans/ext/ci\endcsname{[$1.7$, $2.8$]}
\expandafter\def\csname pn@hot/window/ans/ext/full\endcsname{$+2.2$ [$1.7$, $2.8$]}
\expandafter\def\csname pn@hot/window/ans/nsig\endcsname{4}
\expandafter\def\csname pn@hot/window/ans/pool\endcsname{$+2.8$}
\expandafter\def\csname pn@hot/window/ans/pool/abs\endcsname{2.8}
\expandafter\def\csname pn@hot/window/ans/pool/ci\endcsname{[$2.2$, $3.4$]}
\expandafter\def\csname pn@hot/window/ans/pool/full\endcsname{$+2.8$ [$2.2$, $3.4$]}
\expandafter\def\csname pn@hot/window/ans/range\endcsname{$+1.1$ to $+4.5$}
\expandafter\def\csname pn@hot/window/complete/Decomp\endcsname{$+9.7$}
\expandafter\def\csname pn@hot/window/complete/Decomp/abs\endcsname{9.7}
\expandafter\def\csname pn@hot/window/complete/Decomp/ci\endcsname{[$8.1$, $11.4$]}
\expandafter\def\csname pn@hot/window/complete/Decomp/full\endcsname{$+9.7$ [$8.1$, $11.4$]}
\expandafter\def\csname pn@hot/window/complete/IRCoT\endcsname{$+5.4$}
\expandafter\def\csname pn@hot/window/complete/IRCoT/abs\endcsname{5.4}
\expandafter\def\csname pn@hot/window/complete/IRCoT/ci\endcsname{[$4.2$, $6.6$]}
\expandafter\def\csname pn@hot/window/complete/IRCoT/full\endcsname{$+5.4$ [$4.2$, $6.6$]}
\expandafter\def\csname pn@hot/window/complete/ReAct\endcsname{$+7.5$}
\expandafter\def\csname pn@hot/window/complete/ReAct/abs\endcsname{7.5}
\expandafter\def\csname pn@hot/window/complete/ReAct/ci\endcsname{[$6.2$, $8.9$]}
\expandafter\def\csname pn@hot/window/complete/ReAct/full\endcsname{$+7.5$ [$6.2$, $8.9$]}
\expandafter\def\csname pn@hot/window/complete/SR1\endcsname{$+7.4$}
\expandafter\def\csname pn@hot/window/complete/SR1/abs\endcsname{7.4}
\expandafter\def\csname pn@hot/window/complete/SR1/ci\endcsname{[$6.1$, $8.8$]}
\expandafter\def\csname pn@hot/window/complete/SR1/full\endcsname{$+7.4$ [$6.1$, $8.8$]}
\expandafter\def\csname pn@hot/window/complete/ext\endcsname{$+6.8$}
\expandafter\def\csname pn@hot/window/complete/ext/abs\endcsname{6.8}
\expandafter\def\csname pn@hot/window/complete/ext/ci\endcsname{[$5.9$, $7.7$]}
\expandafter\def\csname pn@hot/window/complete/ext/full\endcsname{$+6.8$ [$5.9$, $7.7$]}
\expandafter\def\csname pn@hot/window/complete/nsig\endcsname{4}
\expandafter\def\csname pn@hot/window/complete/pool\endcsname{$+7.5$}
\expandafter\def\csname pn@hot/window/complete/pool/abs\endcsname{7.5}
\expandafter\def\csname pn@hot/window/complete/pool/ci\endcsname{[$6.6$, $8.4$]}
\expandafter\def\csname pn@hot/window/complete/pool/full\endcsname{$+7.5$ [$6.6$, $8.4$]}
\expandafter\def\csname pn@hot/window/complete/range\endcsname{$+5.4$ to $+9.7$}
\expandafter\def\csname pn@pre/hot/Decomp/T1/interleave\endcsname{40.0}
\expandafter\def\csname pn@pre/hot/Decomp/T1/rrf\endcsname{40.0}
\expandafter\def\csname pn@pre/hot/Decomp/T1/window\endcsname{40.0}
\expandafter\def\csname pn@pre/hot/Decomp/T2/interleave\endcsname{41.1}
\expandafter\def\csname pn@pre/hot/Decomp/T2/rrf\endcsname{41.2}
\expandafter\def\csname pn@pre/hot/Decomp/T2/window\endcsname{44.7}
\expandafter\def\csname pn@pre/hot/Decomp/T3/interleave\endcsname{42.0}
\expandafter\def\csname pn@pre/hot/Decomp/T3/rrf\endcsname{42.2}
\expandafter\def\csname pn@pre/hot/Decomp/T3/window\endcsname{48.5}
\expandafter\def\csname pn@pre/hot/Decomp/T4/interleave\endcsname{42.7}
\expandafter\def\csname pn@pre/hot/Decomp/T4/rrf\endcsname{41.3}
\expandafter\def\csname pn@pre/hot/Decomp/T4/window\endcsname{49.7}
\expandafter\def\csname pn@pre/hot/Decomp/T5/interleave\endcsname{43.1}
\expandafter\def\csname pn@pre/hot/Decomp/T5/rrf\endcsname{41.2}
\expandafter\def\csname pn@pre/hot/Decomp/T5/window\endcsname{51.3}
\expandafter\def\csname pn@pre/hot/Decomp/T6/interleave\endcsname{42.7}
\expandafter\def\csname pn@pre/hot/Decomp/T6/rrf\endcsname{41.2}
\expandafter\def\csname pn@pre/hot/Decomp/T6/window\endcsname{52.5}
\expandafter\def\csname pn@pre/hot/Decomp/drop\endcsname{1.0}
\expandafter\def\csname pn@pre/hot/Decomp/drop/ci\endcsname{[$-0.4$, $2.4$]}
\expandafter\def\csname pn@pre/hot/Decomp/gap6\endcsname{9.7}
\expandafter\def\csname pn@pre/hot/Decomp/peak\endcsname{42.2}
\expandafter\def\csname pn@pre/hot/Decomp/peakT\endcsname{3}
\expandafter\def\csname pn@pre/hot/Decomp/windowmono\endcsname{yes}
\expandafter\def\csname pn@pre/hot/IRCoT/T1/interleave\endcsname{40.0}
\expandafter\def\csname pn@pre/hot/IRCoT/T1/rrf\endcsname{40.0}
\expandafter\def\csname pn@pre/hot/IRCoT/T1/window\endcsname{40.0}
\expandafter\def\csname pn@pre/hot/IRCoT/T2/interleave\endcsname{62.3}
\expandafter\def\csname pn@pre/hot/IRCoT/T2/rrf\endcsname{62.3}
\expandafter\def\csname pn@pre/hot/IRCoT/T2/window\endcsname{67.4}
\expandafter\def\csname pn@pre/hot/IRCoT/T3/interleave\endcsname{66.3}
\expandafter\def\csname pn@pre/hot/IRCoT/T3/rrf\endcsname{65.5}
\expandafter\def\csname pn@pre/hot/IRCoT/T3/window\endcsname{71.0}
\expandafter\def\csname pn@pre/hot/IRCoT/T4/interleave\endcsname{67.2}
\expandafter\def\csname pn@pre/hot/IRCoT/T4/rrf\endcsname{62.7}
\expandafter\def\csname pn@pre/hot/IRCoT/T4/window\endcsname{72.3}
\expandafter\def\csname pn@pre/hot/IRCoT/T5/interleave\endcsname{68.1}
\expandafter\def\csname pn@pre/hot/IRCoT/T5/rrf\endcsname{57.6}
\expandafter\def\csname pn@pre/hot/IRCoT/T5/window\endcsname{73.0}
\expandafter\def\csname pn@pre/hot/IRCoT/T6/interleave\endcsname{68.3}
\expandafter\def\csname pn@pre/hot/IRCoT/T6/rrf\endcsname{50.9}
\expandafter\def\csname pn@pre/hot/IRCoT/T6/window\endcsname{73.7}
\expandafter\def\csname pn@pre/hot/IRCoT/drop\endcsname{14.5}
\expandafter\def\csname pn@pre/hot/IRCoT/drop/ci\endcsname{[$12.3$, $16.6$]}
\expandafter\def\csname pn@pre/hot/IRCoT/gap6\endcsname{5.4}
\expandafter\def\csname pn@pre/hot/IRCoT/peak\endcsname{65.5}
\expandafter\def\csname pn@pre/hot/IRCoT/peakT\endcsname{3}
\expandafter\def\csname pn@pre/hot/IRCoT/windowmono\endcsname{yes}
\expandafter\def\csname pn@pre/hot/ReAct/T1/interleave\endcsname{40.0}
\expandafter\def\csname pn@pre/hot/ReAct/T1/rrf\endcsname{40.0}
\expandafter\def\csname pn@pre/hot/ReAct/T1/window\endcsname{40.0}
\expandafter\def\csname pn@pre/hot/ReAct/T2/interleave\endcsname{48.1}
\expandafter\def\csname pn@pre/hot/ReAct/T2/rrf\endcsname{48.1}
\expandafter\def\csname pn@pre/hot/ReAct/T2/window\endcsname{52.3}
\expandafter\def\csname pn@pre/hot/ReAct/T3/interleave\endcsname{49.7}
\expandafter\def\csname pn@pre/hot/ReAct/T3/rrf\endcsname{45.9}
\expandafter\def\csname pn@pre/hot/ReAct/T3/window\endcsname{55.7}
\expandafter\def\csname pn@pre/hot/ReAct/T4/interleave\endcsname{50.1}
\expandafter\def\csname pn@pre/hot/ReAct/T4/rrf\endcsname{43.3}
\expandafter\def\csname pn@pre/hot/ReAct/T4/window\endcsname{57.7}
\expandafter\def\csname pn@pre/hot/ReAct/T5/interleave\endcsname{50.9}
\expandafter\def\csname pn@pre/hot/ReAct/T5/rrf\endcsname{40.7}
\expandafter\def\csname pn@pre/hot/ReAct/T5/window\endcsname{58.3}
\expandafter\def\csname pn@pre/hot/ReAct/T6/interleave\endcsname{51.3}
\expandafter\def\csname pn@pre/hot/ReAct/T6/rrf\endcsname{39.6}
\expandafter\def\csname pn@pre/hot/ReAct/T6/window\endcsname{58.9}
\expandafter\def\csname pn@pre/hot/ReAct/drop\endcsname{5.6}
\expandafter\def\csname pn@pre/hot/ReAct/gap6\endcsname{7.5}
\expandafter\def\csname pn@pre/hot/ReAct/peak\endcsname{48.1}
\expandafter\def\csname pn@pre/hot/ReAct/peakT\endcsname{2}
\expandafter\def\csname pn@pre/hot/ReAct/windowmono\endcsname{yes}
\expandafter\def\csname pn@pre/hot/SR1/T1/interleave\endcsname{40.0}
\expandafter\def\csname pn@pre/hot/SR1/T1/rrf\endcsname{40.0}
\expandafter\def\csname pn@pre/hot/SR1/T1/window\endcsname{40.0}
\expandafter\def\csname pn@pre/hot/SR1/T2/interleave\endcsname{41.6}
\expandafter\def\csname pn@pre/hot/SR1/T2/rrf\endcsname{41.3}
\expandafter\def\csname pn@pre/hot/SR1/T2/window\endcsname{44.1}
\expandafter\def\csname pn@pre/hot/SR1/T3/interleave\endcsname{57.6}
\expandafter\def\csname pn@pre/hot/SR1/T3/rrf\endcsname{55.5}
\expandafter\def\csname pn@pre/hot/SR1/T3/window\endcsname{64.5}
\expandafter\def\csname pn@pre/hot/SR1/T4/interleave\endcsname{60.6}
\expandafter\def\csname pn@pre/hot/SR1/T4/rrf\endcsname{58.3}
\expandafter\def\csname pn@pre/hot/SR1/T4/window\endcsname{67.7}
\expandafter\def\csname pn@pre/hot/SR1/T5/interleave\endcsname{61.3}
\expandafter\def\csname pn@pre/hot/SR1/T5/rrf\endcsname{58.3}
\expandafter\def\csname pn@pre/hot/SR1/T5/window\endcsname{68.3}
\expandafter\def\csname pn@pre/hot/SR1/T6/interleave\endcsname{61.3}
\expandafter\def\csname pn@pre/hot/SR1/T6/rrf\endcsname{58.4}
\expandafter\def\csname pn@pre/hot/SR1/T6/window\endcsname{68.7}
\expandafter\def\csname pn@pre/hot/SR1/drop\endcsname{0.0}
\expandafter\def\csname pn@pre/hot/SR1/gap6\endcsname{7.4}
\expandafter\def\csname pn@pre/hot/SR1/peak\endcsname{58.4}
\expandafter\def\csname pn@pre/hot/SR1/peakT\endcsname{6}
\expandafter\def\csname pn@pre/hot/SR1/windowmono\endcsname{yes}
\expandafter\def\csname pn@pre/wiki/Decomp/T1/interleave\endcsname{30.1}
\expandafter\def\csname pn@pre/wiki/Decomp/T1/rrf\endcsname{30.1}
\expandafter\def\csname pn@pre/wiki/Decomp/T1/window\endcsname{30.1}
\expandafter\def\csname pn@pre/wiki/Decomp/T2/interleave\endcsname{31.5}
\expandafter\def\csname pn@pre/wiki/Decomp/T2/rrf\endcsname{30.9}
\expandafter\def\csname pn@pre/wiki/Decomp/T2/window\endcsname{34.0}
\expandafter\def\csname pn@pre/wiki/Decomp/T3/interleave\endcsname{32.5}
\expandafter\def\csname pn@pre/wiki/Decomp/T3/rrf\endcsname{32.2}
\expandafter\def\csname pn@pre/wiki/Decomp/T3/window\endcsname{36.3}
\expandafter\def\csname pn@pre/wiki/Decomp/T4/interleave\endcsname{32.6}
\expandafter\def\csname pn@pre/wiki/Decomp/T4/rrf\endcsname{32.9}
\expandafter\def\csname pn@pre/wiki/Decomp/T4/window\endcsname{36.9}
\expandafter\def\csname pn@pre/wiki/Decomp/T5/interleave\endcsname{32.9}
\expandafter\def\csname pn@pre/wiki/Decomp/T5/rrf\endcsname{33.4}
\expandafter\def\csname pn@pre/wiki/Decomp/T5/window\endcsname{37.6}
\expandafter\def\csname pn@pre/wiki/Decomp/T6/interleave\endcsname{32.7}
\expandafter\def\csname pn@pre/wiki/Decomp/T6/rrf\endcsname{33.8}
\expandafter\def\csname pn@pre/wiki/Decomp/T6/window\endcsname{37.8}
\expandafter\def\csname pn@pre/wiki/Decomp/drop\endcsname{0.0}
\expandafter\def\csname pn@pre/wiki/Decomp/gap6\endcsname{5.1}
\expandafter\def\csname pn@pre/wiki/Decomp/peak\endcsname{33.8}
\expandafter\def\csname pn@pre/wiki/Decomp/peakT\endcsname{6}
\expandafter\def\csname pn@pre/wiki/Decomp/windowmono\endcsname{yes}
\expandafter\def\csname pn@pre/wiki/IRCoT/T1/interleave\endcsname{30.1}
\expandafter\def\csname pn@pre/wiki/IRCoT/T1/rrf\endcsname{30.1}
\expandafter\def\csname pn@pre/wiki/IRCoT/T1/window\endcsname{30.1}
\expandafter\def\csname pn@pre/wiki/IRCoT/T2/interleave\endcsname{59.1}
\expandafter\def\csname pn@pre/wiki/IRCoT/T2/rrf\endcsname{57.7}
\expandafter\def\csname pn@pre/wiki/IRCoT/T2/window\endcsname{62.8}
\expandafter\def\csname pn@pre/wiki/IRCoT/T3/interleave\endcsname{64.9}
\expandafter\def\csname pn@pre/wiki/IRCoT/T3/rrf\endcsname{63.5}
\expandafter\def\csname pn@pre/wiki/IRCoT/T3/window\endcsname{68.5}
\expandafter\def\csname pn@pre/wiki/IRCoT/T4/interleave\endcsname{66.0}
\expandafter\def\csname pn@pre/wiki/IRCoT/T4/rrf\endcsname{63.2}
\expandafter\def\csname pn@pre/wiki/IRCoT/T4/window\endcsname{69.9}
\expandafter\def\csname pn@pre/wiki/IRCoT/T5/interleave\endcsname{66.7}
\expandafter\def\csname pn@pre/wiki/IRCoT/T5/rrf\endcsname{57.7}
\expandafter\def\csname pn@pre/wiki/IRCoT/T5/window\endcsname{70.3}
\expandafter\def\csname pn@pre/wiki/IRCoT/T6/interleave\endcsname{66.8}
\expandafter\def\csname pn@pre/wiki/IRCoT/T6/rrf\endcsname{53.1}
\expandafter\def\csname pn@pre/wiki/IRCoT/T6/window\endcsname{70.6}
\expandafter\def\csname pn@pre/wiki/IRCoT/drop\endcsname{10.9}
\expandafter\def\csname pn@pre/wiki/IRCoT/drop/ci\endcsname{[$8.9$, $13.0$]}
\expandafter\def\csname pn@pre/wiki/IRCoT/gap6\endcsname{3.8}
\expandafter\def\csname pn@pre/wiki/IRCoT/peak\endcsname{64.0}
\expandafter\def\csname pn@pre/wiki/IRCoT/peakT\endcsname{3}
\expandafter\def\csname pn@pre/wiki/IRCoT/windowmono\endcsname{yes}
\expandafter\def\csname pn@pre/wiki/ReAct/T1/interleave\endcsname{30.1}
\expandafter\def\csname pn@pre/wiki/ReAct/T1/rrf\endcsname{30.1}
\expandafter\def\csname pn@pre/wiki/ReAct/T1/window\endcsname{30.1}
\expandafter\def\csname pn@pre/wiki/ReAct/T2/interleave\endcsname{46.8}
\expandafter\def\csname pn@pre/wiki/ReAct/T2/rrf\endcsname{46.2}
\expandafter\def\csname pn@pre/wiki/ReAct/T2/window\endcsname{49.3}
\expandafter\def\csname pn@pre/wiki/ReAct/T3/interleave\endcsname{50.7}
\expandafter\def\csname pn@pre/wiki/ReAct/T3/rrf\endcsname{47.1}
\expandafter\def\csname pn@pre/wiki/ReAct/T3/window\endcsname{54.3}
\expandafter\def\csname pn@pre/wiki/ReAct/T4/interleave\endcsname{52.3}
\expandafter\def\csname pn@pre/wiki/ReAct/T4/rrf\endcsname{46.4}
\expandafter\def\csname pn@pre/wiki/ReAct/T4/window\endcsname{57.2}
\expandafter\def\csname pn@pre/wiki/ReAct/T5/interleave\endcsname{53.1}
\expandafter\def\csname pn@pre/wiki/ReAct/T5/rrf\endcsname{45.1}
\expandafter\def\csname pn@pre/wiki/ReAct/T5/window\endcsname{58.9}
\expandafter\def\csname pn@pre/wiki/ReAct/T6/interleave\endcsname{53.9}
\expandafter\def\csname pn@pre/wiki/ReAct/T6/rrf\endcsname{43.9}
\expandafter\def\csname pn@pre/wiki/ReAct/T6/window\endcsname{60.1}
\expandafter\def\csname pn@pre/wiki/ReAct/drop\endcsname{3.3}
\expandafter\def\csname pn@pre/wiki/ReAct/drop/ci\endcsname{[$1.7$, $4.9$]}
\expandafter\def\csname pn@pre/wiki/ReAct/gap6\endcsname{6.3}
\expandafter\def\csname pn@pre/wiki/ReAct/peak\endcsname{47.1}
\expandafter\def\csname pn@pre/wiki/ReAct/peakT\endcsname{3}
\expandafter\def\csname pn@pre/wiki/ReAct/windowmono\endcsname{yes}
\expandafter\def\csname pn@pre/wiki/SR1/T1/interleave\endcsname{30.1}
\expandafter\def\csname pn@pre/wiki/SR1/T1/rrf\endcsname{30.1}
\expandafter\def\csname pn@pre/wiki/SR1/T1/window\endcsname{30.1}
\expandafter\def\csname pn@pre/wiki/SR1/T2/interleave\endcsname{32.0}
\expandafter\def\csname pn@pre/wiki/SR1/T2/rrf\endcsname{31.4}
\expandafter\def\csname pn@pre/wiki/SR1/T2/window\endcsname{33.5}
\expandafter\def\csname pn@pre/wiki/SR1/T3/interleave\endcsname{58.2}
\expandafter\def\csname pn@pre/wiki/SR1/T3/rrf\endcsname{53.5}
\expandafter\def\csname pn@pre/wiki/SR1/T3/window\endcsname{62.3}
\expandafter\def\csname pn@pre/wiki/SR1/T4/interleave\endcsname{62.0}
\expandafter\def\csname pn@pre/wiki/SR1/T4/rrf\endcsname{56.8}
\expandafter\def\csname pn@pre/wiki/SR1/T4/window\endcsname{67.1}
\expandafter\def\csname pn@pre/wiki/SR1/T5/interleave\endcsname{69.5}
\expandafter\def\csname pn@pre/wiki/SR1/T5/rrf\endcsname{62.3}
\expandafter\def\csname pn@pre/wiki/SR1/T5/window\endcsname{75.3}
\expandafter\def\csname pn@pre/wiki/SR1/T6/interleave\endcsname{70.3}
\expandafter\def\csname pn@pre/wiki/SR1/T6/rrf\endcsname{62.5}
\expandafter\def\csname pn@pre/wiki/SR1/T6/window\endcsname{76.6}
\expandafter\def\csname pn@pre/wiki/SR1/drop\endcsname{0.0}
\expandafter\def\csname pn@pre/wiki/SR1/gap6\endcsname{6.3}
\expandafter\def\csname pn@pre/wiki/SR1/peak\endcsname{62.5}
\expandafter\def\csname pn@pre/wiki/SR1/peakT\endcsname{6}
\expandafter\def\csname pn@pre/wiki/SR1/windowmono\endcsname{yes}
\expandafter\def\csname pn@rev/x1/G0/hit\endcsname{71}
\expandafter\def\csname pn@rev/x1/G0/recall\endcsname{16.3}
\expandafter\def\csname pn@rev/x1/G1/hit\endcsname{3}
\expandafter\def\csname pn@rev/x1/G1/recall\endcsname{0.3}
\expandafter\def\csname pn@rev/x1/G2/hit\endcsname{5}
\expandafter\def\csname pn@rev/x1/G2/recall\endcsname{0.6}
\expandafter\def\csname pn@rev/x1/G3/hit\endcsname{23}
\expandafter\def\csname pn@rev/x1/G3/recall\endcsname{2.5}
\expandafter\def\csname pn@rev/x1/G3t-G2/hit\endcsname{$+15.0$ [$10.3$, $19.7$]}
\expandafter\def\csname pn@rev/x1/G3t-G2/recall\endcsname{$+1.8$ [$1.3$, $2.4$]}
\expandafter\def\csname pn@rev/x1/G3t/hit\endcsname{20}
\expandafter\def\csname pn@rev/x1/G3t/recall\endcsname{2.4}
\expandafter\def\csname pn@rev/x1/G5/hit\endcsname{80}
\expandafter\def\csname pn@rev/x1/G5/recall\endcsname{23.6}
\expandafter\def\csname pn@rev/x1/G5t-G0/hit\endcsname{$+3.7$ [$-2.0$, $9.3$]}
\expandafter\def\csname pn@rev/x1/G5t-G0/recall\endcsname{$+3.7$ [$1.7$, $5.6$]}
\expandafter\def\csname pn@rev/x1/G5t-G3t/hit\endcsname{$+54.3$ [$48.3$, $60.3$]}
\expandafter\def\csname pn@rev/x1/G5t-G3t/recall\endcsname{$+17.6$ [$15.7$, $19.5$]}
\expandafter\def\csname pn@rev/x1/G5t-G5/hit\endcsname{$-5.7$ [$-9.7$, $-1.7$]}
\expandafter\def\csname pn@rev/x1/G5t-G5/recall\endcsname{$-3.6$ [$-5.0$, $-2.3$]}
\expandafter\def\csname pn@rev/x1/G5t-G8/hit\endcsname{$-2.7$ [$-6.3$, $1.0$]}
\expandafter\def\csname pn@rev/x1/G5t-G8/recall\endcsname{$-1.8$ [$-2.9$, $-0.8$]}
\expandafter\def\csname pn@rev/x1/G5t/hit\endcsname{75}
\expandafter\def\csname pn@rev/x1/G5t/recall\endcsname{20.0}
\expandafter\def\csname pn@rev/x1/G6/hit\endcsname{80}
\expandafter\def\csname pn@rev/x1/G6/recall\endcsname{24.1}
\expandafter\def\csname pn@rev/x1/G6t-G0/hit\endcsname{$+5.7$ [$0.3$, $11.0$]}
\expandafter\def\csname pn@rev/x1/G6t-G0/recall\endcsname{$+5.3$ [$3.3$, $7.2$]}
\expandafter\def\csname pn@rev/x1/G6t-G5t/hit\endcsname{$+2.0$ [$-1.7$, $5.7$]}
\expandafter\def\csname pn@rev/x1/G6t-G5t/recall\endcsname{$+1.6$ [$0.7$, $2.6$]}
\expandafter\def\csname pn@rev/x1/G6t-G6/hit\endcsname{$-3.7$ [$-7.7$, $0.3$]}
\expandafter\def\csname pn@rev/x1/G6t-G6/recall\endcsname{$-2.6$ [$-3.9$, $-1.3$]}
\expandafter\def\csname pn@rev/x1/G6t-G9/hit\endcsname{$-1.3$ [$-5.3$, $2.3$]}
\expandafter\def\csname pn@rev/x1/G6t-G9/recall\endcsname{$-1.3$ [$-2.7$, $0.0$]}
\expandafter\def\csname pn@rev/x1/G6t/hit\endcsname{77}
\expandafter\def\csname pn@rev/x1/G6t/recall\endcsname{21.6}
\expandafter\def\csname pn@rev/x1/G7/hit\endcsname{19}
\expandafter\def\csname pn@rev/x1/G7/recall\endcsname{4.5}
\expandafter\def\csname pn@rev/x1/G8-G0/hit\endcsname{$+6.3$ [$0.7$, $12.3$]}
\expandafter\def\csname pn@rev/x1/G8-G0/recall\endcsname{$+5.5$ [$3.5$, $7.5$]}
\expandafter\def\csname pn@rev/x1/G8/hit\endcsname{77}
\expandafter\def\csname pn@rev/x1/G8/recall\endcsname{21.8}
\expandafter\def\csname pn@rev/x1/G9-G0/hit\endcsname{$+7.0$ [$1.7$, $12.3$]}
\expandafter\def\csname pn@rev/x1/G9-G0/recall\endcsname{$+6.6$ [$4.6$, $8.5$]}
\expandafter\def\csname pn@rev/x1/G9-G8/hit\endcsname{$+0.7$ [$-3.0$, $4.3$]}
\expandafter\def\csname pn@rev/x1/G9-G8/recall\endcsname{$+1.1$ [$-0.0$, $2.2$]}
\expandafter\def\csname pn@rev/x1/G9/hit\endcsname{78}
\expandafter\def\csname pn@rev/x1/G9/recall\endcsname{22.9}
\expandafter\def\csname pn@rev/x2b/hot/G2+dd\endcsname{46.0}
\expandafter\def\csname pn@rev/x2b/hot/detour\endcsname{47.9}
\expandafter\def\csname pn@rev/x2b/hot/detour-G2+dd\endcsname{$+1.9$ [$0.3$, $3.5$]}
\expandafter\def\csname pn@rev/x2b/hot/detour-native\endcsname{$+4.7$ [$2.9$, $6.4$]}
\expandafter\def\csname pn@rev/x2b/hot/native\endcsname{43.3}
\expandafter\def\csname pn@rev/x2b/wiki/G2+dd\endcsname{40.0}
\expandafter\def\csname pn@rev/x2b/wiki/detour\endcsname{42.6}
\expandafter\def\csname pn@rev/x2b/wiki/detour-G2+dd\endcsname{$+2.6$ [$0.6$, $4.6$]}
\expandafter\def\csname pn@rev/x2b/wiki/detour-native\endcsname{$+4.9$ [$2.9$, $7.1$]}
\expandafter\def\csname pn@rev/x2b/wiki/native\endcsname{37.7}
\expandafter\def\csname pn@rev/x3/wiki/theta/EMmax\endcsname{35.7}
\expandafter\def\csname pn@rev/x3/wiki/theta/EMmin\endcsname{34.0}
\expandafter\def\csname pn@rev/x3/wiki/theta0.6/EMgain\endcsname{$+0.8$ [$0.3$, $1.3$]}
\expandafter\def\csname pn@rev/x3/wiki/theta0.6/rerouted\endcsname{41}
\expandafter\def\csname pn@rev/x3/wiki/theta0.7/rerouted\endcsname{31}
\expandafter\def\csname pn@rev/x3/wiki/theta0.8/rerouted\endcsname{25}
\expandafter\def\csname pn@rev/x3/wiki/theta0.9/rerouted\endcsname{16}
\expandafter\def\csname pn@rev/x4/hot/n10/delta\endcsname{$+$3.0}
\expandafter\def\csname pn@rev/x4/hot/n20/delta\endcsname{$+$2.8}
\expandafter\def\csname pn@rev/x4/hot/n5/delta\endcsname{$+$6.2}
\expandafter\def\csname pn@rev/x4/wiki/n10/delta\endcsname{$+$5.3}
\expandafter\def\csname pn@rev/x4/wiki/n20/delta\endcsname{$+$4.2}
\expandafter\def\csname pn@rev/x4/wiki/n5/delta\endcsname{$+$5.2}
\expandafter\def\csname pn@rev/x5/hot/EM\endcsname{$+0.0$ [$-1.3$, $1.3$]}
\expandafter\def\csname pn@rev/x5/hot/complete\endcsname{+4.1}
\expandafter\def\csname pn@rev/x5/wiki/EM\endcsname{$-0.2$ [$-1.9$, $1.5$]}
\expandafter\def\csname pn@rev/x5/wiki/complete\endcsname{+2.9}
\expandafter\def\csname pn@rev/x6/curve/n1000/complete\endcsname{$-$0.2}
\expandafter\def\csname pn@rev/x6/transfer/hot/complete\endcsname{$+$0.1}
\expandafter\def\csname pn@rev/x6/transfer/wiki/complete\endcsname{$-$1.1}
\expandafter\def\csname pn@rev/x7/hot/QR/EM\endcsname{40.7}
\expandafter\def\csname pn@rev/x7/hot/QRDW/EM\endcsname{40.7}
\expandafter\def\csname pn@rev/x7/hot/detourVSqr\endcsname{$+0.2$ [$-0.8$, $1.2$]}
\expandafter\def\csname pn@rev/x7/hot/detourVSqr/complete\endcsname{+3.7}
\expandafter\def\csname pn@rev/x7/passages\endcsname{28}
\expandafter\def\csname pn@rev/x7/wiki/QR/EM\endcsname{32.7}
\expandafter\def\csname pn@rev/x7/wiki/QRDW/EM\endcsname{33.1}
\expandafter\def\csname pn@rev/x7/wiki/detourVSqr\endcsname{$+2.6$ [$1.3$, $3.9$]}
\expandafter\def\csname pn@rev/x7/wiki/detourVSqr/complete\endcsname{+14.0}
\expandafter\def\csname pn@rev/x8/C2\endcsname{$+1.6$ [$0.7$, $2.4$]}
\expandafter\def\csname pn@rev/x8/C3\endcsname{$+1.4$ [$0.3$, $2.4$]}
\expandafter\def\csname pn@rev/x8/hot/C1\endcsname{$+1.4$ [$0.4$, $2.3$]}
\expandafter\def\csname pn@rev/x8/wiki/C1\endcsname{$+3.9$ [$2.8$, $5.0$]}
\expandafter\def\csname pn@wiki/EM/Decomp/detour\endcsname{30.7}
\expandafter\def\csname pn@wiki/EM/Decomp/droprepeats\endcsname{28.4}
\expandafter\def\csname pn@wiki/EM/Decomp/interleave\endcsname{27.7}
\expandafter\def\csname pn@wiki/EM/Decomp/named\endcsname{28.4}
\expandafter\def\csname pn@wiki/EM/Decomp/novelquery\endcsname{28.3}
\expandafter\def\csname pn@wiki/EM/Decomp/randomroute\endcsname{28.9}
\expandafter\def\csname pn@wiki/EM/Decomp/reranker\endcsname{27.5}
\expandafter\def\csname pn@wiki/EM/Decomp/rerankertops\endcsname{27.7}
\expandafter\def\csname pn@wiki/EM/Decomp/rerankerwin\endcsname{27.9}
\expandafter\def\csname pn@wiki/EM/Decomp/rewired\endcsname{27.7}
\expandafter\def\csname pn@wiki/EM/Decomp/routeonly\endcsname{29.0}
\expandafter\def\csname pn@wiki/EM/Decomp/rrf\endcsname{24.7}
\expandafter\def\csname pn@wiki/EM/Decomp/simgraph\endcsname{27.9}
\expandafter\def\csname pn@wiki/EM/Decomp/titleroute\endcsname{29.6}
\expandafter\def\csname pn@wiki/EM/Decomp/window\endcsname{28.5}
\expandafter\def\csname pn@wiki/EM/IRCoT/detour\endcsname{36.1}
\expandafter\def\csname pn@wiki/EM/IRCoT/droprepeats\endcsname{35.9}
\expandafter\def\csname pn@wiki/EM/IRCoT/interleave\endcsname{34.2}
\expandafter\def\csname pn@wiki/EM/IRCoT/named\endcsname{35.6}
\expandafter\def\csname pn@wiki/EM/IRCoT/novelquery\endcsname{36.8}
\expandafter\def\csname pn@wiki/EM/IRCoT/randomroute\endcsname{35.1}
\expandafter\def\csname pn@wiki/EM/IRCoT/reranker\endcsname{34.5}
\expandafter\def\csname pn@wiki/EM/IRCoT/rerankertops\endcsname{34.9}
\expandafter\def\csname pn@wiki/EM/IRCoT/rerankerwin\endcsname{35.9}
\expandafter\def\csname pn@wiki/EM/IRCoT/rewired\endcsname{35.4}
\expandafter\def\csname pn@wiki/EM/IRCoT/routeonly\endcsname{35.7}
\expandafter\def\csname pn@wiki/EM/IRCoT/rrf\endcsname{32.3}
\expandafter\def\csname pn@wiki/EM/IRCoT/simgraph\endcsname{35.7}
\expandafter\def\csname pn@wiki/EM/IRCoT/titleroute\endcsname{36.3}
\expandafter\def\csname pn@wiki/EM/IRCoT/window\endcsname{36.2}
\expandafter\def\csname pn@wiki/EM/ReAct/detour\endcsname{36.1}
\expandafter\def\csname pn@wiki/EM/ReAct/droprepeats\endcsname{32.5}
\expandafter\def\csname pn@wiki/EM/ReAct/interleave\endcsname{29.2}
\expandafter\def\csname pn@wiki/EM/ReAct/named\endcsname{32.1}
\expandafter\def\csname pn@wiki/EM/ReAct/novelquery\endcsname{33.3}
\expandafter\def\csname pn@wiki/EM/ReAct/randomroute\endcsname{34.1}
\expandafter\def\csname pn@wiki/EM/ReAct/reranker\endcsname{33.3}
\expandafter\def\csname pn@wiki/EM/ReAct/rerankertops\endcsname{33.3}
\expandafter\def\csname pn@wiki/EM/ReAct/rerankerwin\endcsname{34.4}
\expandafter\def\csname pn@wiki/EM/ReAct/rewired\endcsname{32.3}
\expandafter\def\csname pn@wiki/EM/ReAct/routeonly\endcsname{32.0}
\expandafter\def\csname pn@wiki/EM/ReAct/rrf\endcsname{28.1}
\expandafter\def\csname pn@wiki/EM/ReAct/simgraph\endcsname{32.1}
\expandafter\def\csname pn@wiki/EM/ReAct/titleroute\endcsname{34.7}
\expandafter\def\csname pn@wiki/EM/ReAct/window\endcsname{32.3}
\expandafter\def\csname pn@wiki/EM/SR1/detour\endcsname{36.5}
\expandafter\def\csname pn@wiki/EM/SR1/droprepeats\endcsname{36.2}
\expandafter\def\csname pn@wiki/EM/SR1/interleave\endcsname{34.1}
\expandafter\def\csname pn@wiki/EM/SR1/named\endcsname{35.3}
\expandafter\def\csname pn@wiki/EM/SR1/novelquery\endcsname{36.3}
\expandafter\def\csname pn@wiki/EM/SR1/randomroute\endcsname{36.4}
\expandafter\def\csname pn@wiki/EM/SR1/reranker\endcsname{34.1}
\expandafter\def\csname pn@wiki/EM/SR1/rerankertops\endcsname{33.8}
\expandafter\def\csname pn@wiki/EM/SR1/rerankerwin\endcsname{35.2}
\expandafter\def\csname pn@wiki/EM/SR1/rewired\endcsname{34.8}
\expandafter\def\csname pn@wiki/EM/SR1/routeonly\endcsname{34.3}
\expandafter\def\csname pn@wiki/EM/SR1/rrf\endcsname{33.9}
\expandafter\def\csname pn@wiki/EM/SR1/simgraph\endcsname{34.5}
\expandafter\def\csname pn@wiki/EM/SR1/titleroute\endcsname{36.7}
\expandafter\def\csname pn@wiki/EM/SR1/window\endcsname{36.3}
\expandafter\def\csname pn@wiki/EM/pool/detour\endcsname{34.9}
\expandafter\def\csname pn@wiki/EM/pool/droprepeats\endcsname{33.3}
\expandafter\def\csname pn@wiki/EM/pool/interleave\endcsname{31.3}
\expandafter\def\csname pn@wiki/EM/pool/named\endcsname{32.9}
\expandafter\def\csname pn@wiki/EM/pool/novelquery\endcsname{33.7}
\expandafter\def\csname pn@wiki/EM/pool/randomroute\endcsname{33.6}
\expandafter\def\csname pn@wiki/EM/pool/reranker\endcsname{32.4}
\expandafter\def\csname pn@wiki/EM/pool/rerankertops\endcsname{32.4}
\expandafter\def\csname pn@wiki/EM/pool/rerankerwin\endcsname{33.4}
\expandafter\def\csname pn@wiki/EM/pool/rewired\endcsname{32.6}
\expandafter\def\csname pn@wiki/EM/pool/routeonly\endcsname{32.7}
\expandafter\def\csname pn@wiki/EM/pool/rrf\endcsname{29.7}
\expandafter\def\csname pn@wiki/EM/pool/simgraph\endcsname{32.6}
\expandafter\def\csname pn@wiki/EM/pool/titleroute\endcsname{34.3}
\expandafter\def\csname pn@wiki/EM/pool/window\endcsname{33.3}
\expandafter\def\csname pn@wiki/calls/Decomp\endcsname{6.00}
\expandafter\def\csname pn@wiki/calls/IRCoT\endcsname{6.00}
\expandafter\def\csname pn@wiki/calls/ReAct\endcsname{5.64}
\expandafter\def\csname pn@wiki/calls/SR1\endcsname{3.81}
\expandafter\def\csname pn@wiki/complete/Decomp/detour\endcsname{49.1}
\expandafter\def\csname pn@wiki/complete/Decomp/droprepeats\endcsname{37.7}
\expandafter\def\csname pn@wiki/complete/Decomp/interleave\endcsname{32.7}
\expandafter\def\csname pn@wiki/complete/Decomp/named\endcsname{34.5}
\expandafter\def\csname pn@wiki/complete/Decomp/novelquery\endcsname{38.9}
\expandafter\def\csname pn@wiki/complete/Decomp/randomroute\endcsname{40.3}
\expandafter\def\csname pn@wiki/complete/Decomp/reranker\endcsname{34.2}
\expandafter\def\csname pn@wiki/complete/Decomp/rerankertops\endcsname{35.0}
\expandafter\def\csname pn@wiki/complete/Decomp/rerankerwin\endcsname{37.8}
\expandafter\def\csname pn@wiki/complete/Decomp/rewired\endcsname{36.9}
\expandafter\def\csname pn@wiki/complete/Decomp/routeonly\endcsname{40.5}
\expandafter\def\csname pn@wiki/complete/Decomp/rrf\endcsname{33.8}
\expandafter\def\csname pn@wiki/complete/Decomp/rrfnorepeat\endcsname{34.0}
\expandafter\def\csname pn@wiki/complete/Decomp/rrfskip\endcsname{34.6}
\expandafter\def\csname pn@wiki/complete/Decomp/simgraph\endcsname{35.2}
\expandafter\def\csname pn@wiki/complete/Decomp/titleroute\endcsname{43.4}
\expandafter\def\csname pn@wiki/complete/Decomp/window\endcsname{37.8}
\expandafter\def\csname pn@wiki/complete/IRCoT/detour\endcsname{75.1}
\expandafter\def\csname pn@wiki/complete/IRCoT/droprepeats\endcsname{70.5}
\expandafter\def\csname pn@wiki/complete/IRCoT/interleave\endcsname{66.8}
\expandafter\def\csname pn@wiki/complete/IRCoT/named\endcsname{68.3}
\expandafter\def\csname pn@wiki/complete/IRCoT/novelquery\endcsname{71.7}
\expandafter\def\csname pn@wiki/complete/IRCoT/randomroute\endcsname{72.5}
\expandafter\def\csname pn@wiki/complete/IRCoT/reranker\endcsname{57.7}
\expandafter\def\csname pn@wiki/complete/IRCoT/rerankertops\endcsname{64.8}
\expandafter\def\csname pn@wiki/complete/IRCoT/rerankerwin\endcsname{70.5}
\expandafter\def\csname pn@wiki/complete/IRCoT/rewired\endcsname{69.9}
\expandafter\def\csname pn@wiki/complete/IRCoT/routeonly\endcsname{69.1}
\expandafter\def\csname pn@wiki/complete/IRCoT/rrf\endcsname{52.3}
\expandafter\def\csname pn@wiki/complete/IRCoT/rrfnorepeat\endcsname{64.3}
\expandafter\def\csname pn@wiki/complete/IRCoT/rrfskip\endcsname{60.3}
\expandafter\def\csname pn@wiki/complete/IRCoT/simgraph\endcsname{68.8}
\expandafter\def\csname pn@wiki/complete/IRCoT/titleroute\endcsname{74.5}
\expandafter\def\csname pn@wiki/complete/IRCoT/window\endcsname{70.6}
\expandafter\def\csname pn@wiki/complete/ReAct/detour\endcsname{71.0}
\expandafter\def\csname pn@wiki/complete/ReAct/droprepeats\endcsname{59.6}
\expandafter\def\csname pn@wiki/complete/ReAct/interleave\endcsname{53.9}
\expandafter\def\csname pn@wiki/complete/ReAct/named\endcsname{56.1}
\expandafter\def\csname pn@wiki/complete/ReAct/novelquery\endcsname{62.5}
\expandafter\def\csname pn@wiki/complete/ReAct/randomroute\endcsname{64.2}
\expandafter\def\csname pn@wiki/complete/ReAct/reranker\endcsname{52.2}
\expandafter\def\csname pn@wiki/complete/ReAct/rerankertops\endcsname{55.5}
\expandafter\def\csname pn@wiki/complete/ReAct/rerankerwin\endcsname{60.5}
\expandafter\def\csname pn@wiki/complete/ReAct/rewired\endcsname{59.0}
\expandafter\def\csname pn@wiki/complete/ReAct/routeonly\endcsname{60.9}
\expandafter\def\csname pn@wiki/complete/ReAct/rrf\endcsname{43.9}
\expandafter\def\csname pn@wiki/complete/ReAct/rrfnorepeat\endcsname{49.5}
\expandafter\def\csname pn@wiki/complete/ReAct/rrfskip\endcsname{48.5}
\expandafter\def\csname pn@wiki/complete/ReAct/simgraph\endcsname{56.3}
\expandafter\def\csname pn@wiki/complete/ReAct/titleroute\endcsname{67.9}
\expandafter\def\csname pn@wiki/complete/ReAct/window\endcsname{60.1}
\expandafter\def\csname pn@wiki/complete/SR1/detour\endcsname{78.1}
\expandafter\def\csname pn@wiki/complete/SR1/droprepeats\endcsname{76.2}
\expandafter\def\csname pn@wiki/complete/SR1/interleave\endcsname{70.3}
\expandafter\def\csname pn@wiki/complete/SR1/named\endcsname{71.9}
\expandafter\def\csname pn@wiki/complete/SR1/novelquery\endcsname{77.8}
\expandafter\def\csname pn@wiki/complete/SR1/randomroute\endcsname{76.9}
\expandafter\def\csname pn@wiki/complete/SR1/reranker\endcsname{63.5}
\expandafter\def\csname pn@wiki/complete/SR1/rerankertops\endcsname{69.4}
\expandafter\def\csname pn@wiki/complete/SR1/rerankerwin\endcsname{75.9}
\expandafter\def\csname pn@wiki/complete/SR1/rewired\endcsname{75.1}
\expandafter\def\csname pn@wiki/complete/SR1/routeonly\endcsname{71.0}
\expandafter\def\csname pn@wiki/complete/SR1/rrf\endcsname{62.5}
\expandafter\def\csname pn@wiki/complete/SR1/rrfnorepeat\endcsname{64.1}
\expandafter\def\csname pn@wiki/complete/SR1/rrfskip\endcsname{67.7}
\expandafter\def\csname pn@wiki/complete/SR1/simgraph\endcsname{71.0}
\expandafter\def\csname pn@wiki/complete/SR1/titleroute\endcsname{77.6}
\expandafter\def\csname pn@wiki/complete/SR1/window\endcsname{76.6}
\expandafter\def\csname pn@wiki/complete/pool/detour\endcsname{68.3}
\expandafter\def\csname pn@wiki/complete/pool/droprepeats\endcsname{61.0}
\expandafter\def\csname pn@wiki/complete/pool/interleave\endcsname{55.9}
\expandafter\def\csname pn@wiki/complete/pool/named\endcsname{57.7}
\expandafter\def\csname pn@wiki/complete/pool/novelquery\endcsname{62.7}
\expandafter\def\csname pn@wiki/complete/pool/randomroute\endcsname{63.5}
\expandafter\def\csname pn@wiki/complete/pool/reranker\endcsname{51.9}
\expandafter\def\csname pn@wiki/complete/pool/rerankertops\endcsname{56.2}
\expandafter\def\csname pn@wiki/complete/pool/rerankerwin\endcsname{61.2}
\expandafter\def\csname pn@wiki/complete/pool/rewired\endcsname{60.2}
\expandafter\def\csname pn@wiki/complete/pool/routeonly\endcsname{60.4}
\expandafter\def\csname pn@wiki/complete/pool/rrf\endcsname{48.1}
\expandafter\def\csname pn@wiki/complete/pool/rrfnorepeat\endcsname{53.0}
\expandafter\def\csname pn@wiki/complete/pool/rrfskip\endcsname{52.8}
\expandafter\def\csname pn@wiki/complete/pool/simgraph\endcsname{57.8}
\expandafter\def\csname pn@wiki/complete/pool/titleroute\endcsname{65.9}
\expandafter\def\csname pn@wiki/complete/pool/window\endcsname{61.3}
\expandafter\def\csname pn@wiki/compose/EM/Decomp\endcsname{$+1.7$}
\expandafter\def\csname pn@wiki/compose/EM/Decomp/abs\endcsname{1.7}
\expandafter\def\csname pn@wiki/compose/EM/Decomp/ci\endcsname{[$-0.1$, $3.4$]}
\expandafter\def\csname pn@wiki/compose/EM/Decomp/full\endcsname{$+1.7$ [$-0.1$, $3.4$]}
\expandafter\def\csname pn@wiki/compose/EM/IRCoT\endcsname{$+0.5$}
\expandafter\def\csname pn@wiki/compose/EM/IRCoT/abs\endcsname{0.5}
\expandafter\def\csname pn@wiki/compose/EM/IRCoT/ci\endcsname{[$-1.4$, $2.3$]}
\expandafter\def\csname pn@wiki/compose/EM/IRCoT/full\endcsname{$+0.5$ [$-1.4$, $2.3$]}
\expandafter\def\csname pn@wiki/compose/EM/ReAct\endcsname{$+4.1$}
\expandafter\def\csname pn@wiki/compose/EM/ReAct/abs\endcsname{4.1}
\expandafter\def\csname pn@wiki/compose/EM/ReAct/ci\endcsname{[$2.3$, $5.9$]}
\expandafter\def\csname pn@wiki/compose/EM/ReAct/full\endcsname{$+4.1$ [$2.3$, $5.9$]}
\expandafter\def\csname pn@wiki/compose/EM/SR1\endcsname{$+2.2$}
\expandafter\def\csname pn@wiki/compose/EM/SR1/abs\endcsname{2.2}
\expandafter\def\csname pn@wiki/compose/EM/SR1/ci\endcsname{[$0.5$, $3.9$]}
\expandafter\def\csname pn@wiki/compose/EM/SR1/full\endcsname{$+2.2$ [$0.5$, $3.9$]}
\expandafter\def\csname pn@wiki/compose/EM/ext\endcsname{$+2.3$}
\expandafter\def\csname pn@wiki/compose/EM/ext/abs\endcsname{2.3}
\expandafter\def\csname pn@wiki/compose/EM/ext/ci\endcsname{[$1.1$, $3.5$]}
\expandafter\def\csname pn@wiki/compose/EM/ext/full\endcsname{$+2.3$ [$1.1$, $3.5$]}
\expandafter\def\csname pn@wiki/compose/EM/nsig\endcsname{2}
\expandafter\def\csname pn@wiki/compose/EM/pool\endcsname{$+2.1$}
\expandafter\def\csname pn@wiki/compose/EM/pool/abs\endcsname{3.2}
\expandafter\def\csname pn@wiki/compose/EM/pool/ci\endcsname{[$1.1$, $3.2$]}
\expandafter\def\csname pn@wiki/compose/EM/pool/full\endcsname{$+2.1$ [$1.1$, $3.2$]}
\expandafter\def\csname pn@wiki/compose/EM/range\endcsname{$+0.5$ to $+4.1$}
\expandafter\def\csname pn@wiki/compose/ans/Decomp\endcsname{$+5.6$}
\expandafter\def\csname pn@wiki/compose/ans/Decomp/abs\endcsname{5.6}
\expandafter\def\csname pn@wiki/compose/ans/Decomp/ci\endcsname{[$4.4$, $7.0$]}
\expandafter\def\csname pn@wiki/compose/ans/Decomp/full\endcsname{$+5.6$ [$4.4$, $7.0$]}
\expandafter\def\csname pn@wiki/compose/ans/IRCoT\endcsname{$+2.9$}
\expandafter\def\csname pn@wiki/compose/ans/IRCoT/abs\endcsname{2.9}
\expandafter\def\csname pn@wiki/compose/ans/IRCoT/ci\endcsname{[$1.9$, $3.9$]}
\expandafter\def\csname pn@wiki/compose/ans/IRCoT/full\endcsname{$+2.9$ [$1.9$, $3.9$]}
\expandafter\def\csname pn@wiki/compose/ans/ReAct\endcsname{$+5.8$}
\expandafter\def\csname pn@wiki/compose/ans/ReAct/abs\endcsname{5.8}
\expandafter\def\csname pn@wiki/compose/ans/ReAct/ci\endcsname{[$4.4$, $7.2$]}
\expandafter\def\csname pn@wiki/compose/ans/ReAct/full\endcsname{$+5.8$ [$4.4$, $7.2$]}
\expandafter\def\csname pn@wiki/compose/ans/SR1\endcsname{$+2.3$}
\expandafter\def\csname pn@wiki/compose/ans/SR1/abs\endcsname{2.3}
\expandafter\def\csname pn@wiki/compose/ans/SR1/ci\endcsname{[$1.3$, $3.3$]}
\expandafter\def\csname pn@wiki/compose/ans/SR1/full\endcsname{$+2.3$ [$1.3$, $3.3$]}
\expandafter\def\csname pn@wiki/compose/ans/ext\endcsname{$+3.6$}
\expandafter\def\csname pn@wiki/compose/ans/ext/abs\endcsname{3.6}
\expandafter\def\csname pn@wiki/compose/ans/ext/ci\endcsname{[$2.9$, $4.4$]}
\expandafter\def\csname pn@wiki/compose/ans/ext/full\endcsname{$+3.6$ [$2.9$, $4.4$]}
\expandafter\def\csname pn@wiki/compose/ans/nsig\endcsname{4}
\expandafter\def\csname pn@wiki/compose/ans/pool\endcsname{$+4.1$}
\expandafter\def\csname pn@wiki/compose/ans/pool/abs\endcsname{4.1}
\expandafter\def\csname pn@wiki/compose/ans/pool/ci\endcsname{[$3.4$, $4.9$]}
\expandafter\def\csname pn@wiki/compose/ans/pool/full\endcsname{$+4.1$ [$3.4$, $4.9$]}
\expandafter\def\csname pn@wiki/compose/ans/range\endcsname{$+2.3$ to $+5.8$}
\expandafter\def\csname pn@wiki/compose/complete/Decomp\endcsname{$+8.6$}
\expandafter\def\csname pn@wiki/compose/complete/Decomp/abs\endcsname{8.6}
\expandafter\def\csname pn@wiki/compose/complete/Decomp/ci\endcsname{[$7.3$, $10.0$]}
\expandafter\def\csname pn@wiki/compose/complete/Decomp/full\endcsname{$+8.6$ [$7.3$, $10.0$]}
\expandafter\def\csname pn@wiki/compose/complete/IRCoT\endcsname{$+5.9$}
\expandafter\def\csname pn@wiki/compose/complete/IRCoT/abs\endcsname{5.9}
\expandafter\def\csname pn@wiki/compose/complete/IRCoT/ci\endcsname{[$4.7$, $7.3$]}
\expandafter\def\csname pn@wiki/compose/complete/IRCoT/full\endcsname{$+5.9$ [$4.7$, $7.3$]}
\expandafter\def\csname pn@wiki/compose/complete/ReAct\endcsname{$+10.1$}
\expandafter\def\csname pn@wiki/compose/complete/ReAct/abs\endcsname{10.1}
\expandafter\def\csname pn@wiki/compose/complete/ReAct/ci\endcsname{[$8.5$, $11.7$]}
\expandafter\def\csname pn@wiki/compose/complete/ReAct/full\endcsname{$+10.1$ [$8.5$, $11.7$]}
\expandafter\def\csname pn@wiki/compose/complete/SR1\endcsname{$+7.1$}
\expandafter\def\csname pn@wiki/compose/complete/SR1/abs\endcsname{7.1}
\expandafter\def\csname pn@wiki/compose/complete/SR1/ci\endcsname{[$5.7$, $8.5$]}
\expandafter\def\csname pn@wiki/compose/complete/SR1/full\endcsname{$+7.1$ [$5.7$, $8.5$]}
\expandafter\def\csname pn@wiki/compose/complete/ext\endcsname{$+7.7$}
\expandafter\def\csname pn@wiki/compose/complete/ext/abs\endcsname{7.7}
\expandafter\def\csname pn@wiki/compose/complete/ext/ci\endcsname{[$6.7$, $8.7$]}
\expandafter\def\csname pn@wiki/compose/complete/ext/full\endcsname{$+7.7$ [$6.7$, $8.7$]}
\expandafter\def\csname pn@wiki/compose/complete/nsig\endcsname{4}
\expandafter\def\csname pn@wiki/compose/complete/pool\endcsname{$+7.9$}
\expandafter\def\csname pn@wiki/compose/complete/pool/abs\endcsname{7.9}
\expandafter\def\csname pn@wiki/compose/complete/pool/ci\endcsname{[$7.0$, $8.8$]}
\expandafter\def\csname pn@wiki/compose/complete/pool/full\endcsname{$+7.9$ [$7.0$, $8.8$]}
\expandafter\def\csname pn@wiki/compose/complete/range\endcsname{$+5.9$ to $+10.1$}
\expandafter\def\csname pn@wiki/cpu/range\endcsname{1.1--12.1}
\expandafter\def\csname pn@wiki/depth/EM/IRCoT\endcsname{$+0.9$}
\expandafter\def\csname pn@wiki/depth/EM/IRCoT/abs\endcsname{0.9}
\expandafter\def\csname pn@wiki/depth/EM/IRCoT/ci\endcsname{[$-1.1$, $2.8$]}
\expandafter\def\csname pn@wiki/depth/EM/IRCoT/full\endcsname{$+0.9$ [$-1.1$, $2.8$]}
\expandafter\def\csname pn@wiki/depth/EM/nsig\endcsname{0}
\expandafter\def\csname pn@wiki/depth/ans/Decomp\endcsname{$+4.8$}
\expandafter\def\csname pn@wiki/depth/ans/Decomp/abs\endcsname{4.8}
\expandafter\def\csname pn@wiki/depth/ans/Decomp/ci\endcsname{[$3.7$, $6.0$]}
\expandafter\def\csname pn@wiki/depth/ans/Decomp/full\endcsname{$+4.8$ [$3.7$, $6.0$]}
\expandafter\def\csname pn@wiki/depth/ans/IRCoT\endcsname{$+11.3$}
\expandafter\def\csname pn@wiki/depth/ans/IRCoT/abs\endcsname{11.3}
\expandafter\def\csname pn@wiki/depth/ans/IRCoT/ci\endcsname{[$9.5$, $13.0$]}
\expandafter\def\csname pn@wiki/depth/ans/IRCoT/full\endcsname{$+11.3$ [$9.5$, $13.0$]}
\expandafter\def\csname pn@wiki/depth/ans/ReAct\endcsname{$+12.4$}
\expandafter\def\csname pn@wiki/depth/ans/ReAct/abs\endcsname{12.4}
\expandafter\def\csname pn@wiki/depth/ans/ReAct/ci\endcsname{[$10.6$, $14.2$]}
\expandafter\def\csname pn@wiki/depth/ans/ReAct/full\endcsname{$+12.4$ [$10.6$, $14.2$]}
\expandafter\def\csname pn@wiki/depth/ans/SR1\endcsname{$+7.4$}
\expandafter\def\csname pn@wiki/depth/ans/SR1/abs\endcsname{7.4}
\expandafter\def\csname pn@wiki/depth/ans/SR1/ci\endcsname{[$6.1$, $8.9$]}
\expandafter\def\csname pn@wiki/depth/ans/SR1/full\endcsname{$+7.4$ [$6.1$, $8.9$]}
\expandafter\def\csname pn@wiki/depth/ans/ext\endcsname{$+10.4$}
\expandafter\def\csname pn@wiki/depth/ans/ext/abs\endcsname{10.4}
\expandafter\def\csname pn@wiki/depth/ans/ext/ci\endcsname{[$9.4$, $11.4$]}
\expandafter\def\csname pn@wiki/depth/ans/ext/full\endcsname{$+10.4$ [$9.4$, $11.4$]}
\expandafter\def\csname pn@wiki/depth/ans/nsig\endcsname{4}
\expandafter\def\csname pn@wiki/depth/ans/pool\endcsname{$+9.0$}
\expandafter\def\csname pn@wiki/depth/ans/pool/abs\endcsname{9.0}
\expandafter\def\csname pn@wiki/depth/ans/pool/ci\endcsname{[$8.1$, $9.8$]}
\expandafter\def\csname pn@wiki/depth/ans/pool/full\endcsname{$+9.0$ [$8.1$, $9.8$]}
\expandafter\def\csname pn@wiki/depth/ans/range\endcsname{$+4.8$ to $+12.4$}
\expandafter\def\csname pn@wiki/depth/complete/Decomp\endcsname{$+3.1$}
\expandafter\def\csname pn@wiki/depth/complete/Decomp/abs\endcsname{3.1}
\expandafter\def\csname pn@wiki/depth/complete/Decomp/ci\endcsname{[$2.3$, $3.9$]}
\expandafter\def\csname pn@wiki/depth/complete/Decomp/full\endcsname{$+3.1$ [$2.3$, $3.9$]}
\expandafter\def\csname pn@wiki/depth/complete/IRCoT\endcsname{$+18.4$}
\expandafter\def\csname pn@wiki/depth/complete/IRCoT/abs\endcsname{18.4}
\expandafter\def\csname pn@wiki/depth/complete/IRCoT/ci\endcsname{[$16.5$, $20.3$]}
\expandafter\def\csname pn@wiki/depth/complete/IRCoT/full\endcsname{$+18.4$ [$16.5$, $20.3$]}
\expandafter\def\csname pn@wiki/depth/complete/ReAct\endcsname{$+14.1$}
\expandafter\def\csname pn@wiki/depth/complete/ReAct/abs\endcsname{14.1}
\expandafter\def\csname pn@wiki/depth/complete/ReAct/ci\endcsname{[$12.4$, $15.9$]}
\expandafter\def\csname pn@wiki/depth/complete/ReAct/full\endcsname{$+14.1$ [$12.4$, $15.9$]}
\expandafter\def\csname pn@wiki/depth/complete/SR1\endcsname{$+13.3$}
\expandafter\def\csname pn@wiki/depth/complete/SR1/abs\endcsname{13.3}
\expandafter\def\csname pn@wiki/depth/complete/SR1/ci\endcsname{[$11.5$, $15.0$]}
\expandafter\def\csname pn@wiki/depth/complete/SR1/full\endcsname{$+13.3$ [$11.5$, $15.0$]}
\expandafter\def\csname pn@wiki/depth/complete/ext\endcsname{$+15.3$}
\expandafter\def\csname pn@wiki/depth/complete/ext/abs\endcsname{15.3}
\expandafter\def\csname pn@wiki/depth/complete/ext/ci\endcsname{[$14.1$, $16.4$]}
\expandafter\def\csname pn@wiki/depth/complete/ext/full\endcsname{$+15.3$ [$14.1$, $16.4$]}
\expandafter\def\csname pn@wiki/depth/complete/nsig\endcsname{4}
\expandafter\def\csname pn@wiki/depth/complete/pool\endcsname{$+12.2$}
\expandafter\def\csname pn@wiki/depth/complete/pool/abs\endcsname{12.2}
\expandafter\def\csname pn@wiki/depth/complete/pool/ci\endcsname{[$11.3$, $13.1$]}
\expandafter\def\csname pn@wiki/depth/complete/pool/full\endcsname{$+12.2$ [$11.3$, $13.1$]}
\expandafter\def\csname pn@wiki/depth/complete/range\endcsname{$+3.1$ to $+18.4$}
\expandafter\def\csname pn@wiki/detour/EM/Decomp\endcsname{$+2.9$}
\expandafter\def\csname pn@wiki/detour/EM/Decomp/abs\endcsname{2.9}
\expandafter\def\csname pn@wiki/detour/EM/Decomp/ci\endcsname{[$1.1$, $4.7$]}
\expandafter\def\csname pn@wiki/detour/EM/Decomp/full\endcsname{$+2.9$ [$1.1$, $4.7$]}
\expandafter\def\csname pn@wiki/detour/EM/IRCoT\endcsname{$+4.3$}
\expandafter\def\csname pn@wiki/detour/EM/IRCoT/abs\endcsname{1.9}
\expandafter\def\csname pn@wiki/detour/EM/IRCoT/ci\endcsname{[$0.0$, $3.9$]}
\expandafter\def\csname pn@wiki/detour/EM/IRCoT/full\endcsname{$+1.9$ [$0.0$, $3.9$]}
\expandafter\def\csname pn@wiki/detour/EM/ReAct\endcsname{$+3.9$}
\expandafter\def\csname pn@wiki/detour/EM/ReAct/abs\endcsname{6.9}
\expandafter\def\csname pn@wiki/detour/EM/ReAct/ci\endcsname{[$5.1$, $8.9$]}
\expandafter\def\csname pn@wiki/detour/EM/ReAct/full\endcsname{$+6.9$ [$5.1$, $8.9$]}
\expandafter\def\csname pn@wiki/detour/EM/SR1\endcsname{$+2.3$}
\expandafter\def\csname pn@wiki/detour/EM/SR1/abs\endcsname{2.3}
\expandafter\def\csname pn@wiki/detour/EM/SR1/ci\endcsname{[$0.6$, $4.1$]}
\expandafter\def\csname pn@wiki/detour/EM/SR1/full\endcsname{$+2.3$ [$0.6$, $4.1$]}
\expandafter\def\csname pn@wiki/detour/EM/ext\endcsname{$+3.7$}
\expandafter\def\csname pn@wiki/detour/EM/ext/abs\endcsname{3.7}
\expandafter\def\csname pn@wiki/detour/EM/ext/ci\endcsname{[$2.5$, $5.0$]}
\expandafter\def\csname pn@wiki/detour/EM/ext/full\endcsname{$+3.7$ [$2.5$, $5.0$]}
\expandafter\def\csname pn@wiki/detour/EM/nsig\endcsname{3}
\expandafter\def\csname pn@wiki/detour/EM/pool\endcsname{$+3.5$}
\expandafter\def\csname pn@wiki/detour/EM/pool/abs\endcsname{4.0}
\expandafter\def\csname pn@wiki/detour/EM/pool/ci\endcsname{[$2.4$, $4.7$]}
\expandafter\def\csname pn@wiki/detour/EM/pool/full\endcsname{$+4.0$ [$2.8$, $5.1$]}
\expandafter\def\csname pn@wiki/detour/EM/range\endcsname{$+1.9$ to $+6.9$}
\expandafter\def\csname pn@wiki/detour/ans/Decomp\endcsname{$+12.3$}
\expandafter\def\csname pn@wiki/detour/ans/Decomp/abs\endcsname{12.3}
\expandafter\def\csname pn@wiki/detour/ans/Decomp/ci\endcsname{[$10.5$, $14.2$]}
\expandafter\def\csname pn@wiki/detour/ans/Decomp/full\endcsname{$+12.3$ [$10.5$, $14.2$]}
\expandafter\def\csname pn@wiki/detour/ans/IRCoT\endcsname{$+3.9$}
\expandafter\def\csname pn@wiki/detour/ans/IRCoT/abs\endcsname{3.9}
\expandafter\def\csname pn@wiki/detour/ans/IRCoT/ci\endcsname{[$2.7$, $5.2$]}
\expandafter\def\csname pn@wiki/detour/ans/IRCoT/full\endcsname{$+3.9$ [$2.7$, $5.2$]}
\expandafter\def\csname pn@wiki/detour/ans/ReAct\endcsname{$+11.3$}
\expandafter\def\csname pn@wiki/detour/ans/ReAct/abs\endcsname{11.3}
\expandafter\def\csname pn@wiki/detour/ans/ReAct/ci\endcsname{[$9.4$, $13.2$]}
\expandafter\def\csname pn@wiki/detour/ans/ReAct/full\endcsname{$+11.3$ [$9.4$, $13.2$]}
\expandafter\def\csname pn@wiki/detour/ans/SR1\endcsname{$+3.2$}
\expandafter\def\csname pn@wiki/detour/ans/SR1/abs\endcsname{3.2}
\expandafter\def\csname pn@wiki/detour/ans/SR1/ci\endcsname{[$2.0$, $4.4$]}
\expandafter\def\csname pn@wiki/detour/ans/SR1/full\endcsname{$+3.2$ [$2.0$, $4.4$]}
\expandafter\def\csname pn@wiki/detour/ans/ext\endcsname{$+6.1$}
\expandafter\def\csname pn@wiki/detour/ans/ext/abs\endcsname{6.1}
\expandafter\def\csname pn@wiki/detour/ans/ext/ci\endcsname{[$5.1$, $7.1$]}
\expandafter\def\csname pn@wiki/detour/ans/ext/full\endcsname{$+6.1$ [$5.1$, $7.1$]}
\expandafter\def\csname pn@wiki/detour/ans/nsig\endcsname{4}
\expandafter\def\csname pn@wiki/detour/ans/pool\endcsname{$+7.7$}
\expandafter\def\csname pn@wiki/detour/ans/pool/abs\endcsname{7.7}
\expandafter\def\csname pn@wiki/detour/ans/pool/ci\endcsname{[$6.7$, $8.7$]}
\expandafter\def\csname pn@wiki/detour/ans/pool/full\endcsname{$+7.7$ [$6.7$, $8.7$]}
\expandafter\def\csname pn@wiki/detour/ans/range\endcsname{$+3.2$ to $+12.3$}
\expandafter\def\csname pn@wiki/detour/complete/Decomp\endcsname{$+16.4$}
\expandafter\def\csname pn@wiki/detour/complete/Decomp/abs\endcsname{16.4}
\expandafter\def\csname pn@wiki/detour/complete/Decomp/ci\endcsname{[$14.6$, $18.3$]}
\expandafter\def\csname pn@wiki/detour/complete/Decomp/full\endcsname{$+16.4$ [$14.6$, $18.3$]}
\expandafter\def\csname pn@wiki/detour/complete/IRCoT\endcsname{$+8.3$}
\expandafter\def\csname pn@wiki/detour/complete/IRCoT/abs\endcsname{8.3}
\expandafter\def\csname pn@wiki/detour/complete/IRCoT/ci\endcsname{[$6.7$, $9.8$]}
\expandafter\def\csname pn@wiki/detour/complete/IRCoT/full\endcsname{$+8.3$ [$6.7$, $9.8$]}
\expandafter\def\csname pn@wiki/detour/complete/ReAct\endcsname{$+17.1$}
\expandafter\def\csname pn@wiki/detour/complete/ReAct/abs\endcsname{17.1}
\expandafter\def\csname pn@wiki/detour/complete/ReAct/ci\endcsname{[$15.1$, $19.2$]}
\expandafter\def\csname pn@wiki/detour/complete/ReAct/full\endcsname{$+17.1$ [$15.1$, $19.2$]}
\expandafter\def\csname pn@wiki/detour/complete/SR1\endcsname{$+7.7$}
\expandafter\def\csname pn@wiki/detour/complete/SR1/abs\endcsname{7.7}
\expandafter\def\csname pn@wiki/detour/complete/SR1/ci\endcsname{[$6.2$, $9.3$]}
\expandafter\def\csname pn@wiki/detour/complete/SR1/full\endcsname{$+7.7$ [$6.2$, $9.3$]}
\expandafter\def\csname pn@wiki/detour/complete/ext\endcsname{$+11.0$}
\expandafter\def\csname pn@wiki/detour/complete/ext/abs\endcsname{11.0}
\expandafter\def\csname pn@wiki/detour/complete/ext/ci\endcsname{[$9.9$, $12.2$]}
\expandafter\def\csname pn@wiki/detour/complete/ext/full\endcsname{$+11.0$ [$9.9$, $12.2$]}
\expandafter\def\csname pn@wiki/detour/complete/nsig\endcsname{4}
\expandafter\def\csname pn@wiki/detour/complete/pool\endcsname{$+12.4$}
\expandafter\def\csname pn@wiki/detour/complete/pool/abs\endcsname{12.4}
\expandafter\def\csname pn@wiki/detour/complete/pool/ci\endcsname{[$11.3$, $13.5$]}
\expandafter\def\csname pn@wiki/detour/complete/pool/full\endcsname{$+12.4$ [$11.3$, $13.5$]}
\expandafter\def\csname pn@wiki/detour/complete/range\endcsname{$+7.7$ to $+17.1$}
\expandafter\def\csname pn@wiki/detourvsreranker/EM/pool\endcsname{$+2.5$}
\expandafter\def\csname pn@wiki/detourvsreranker/EM/pool/abs\endcsname{2.5}
\expandafter\def\csname pn@wiki/detourvsreranker/EM/pool/ci\endcsname{[$1.1$, $3.9$]}
\expandafter\def\csname pn@wiki/detourvsreranker/EM/pool/full\endcsname{$+3.1$ [$1.6$, $4.6$]}
\expandafter\def\csname pn@wiki/detourvsrerankerwin/EM/pool\endcsname{$+1.5$}
\expandafter\def\csname pn@wiki/detourvsrerankerwin/EM/pool/abs\endcsname{1.5}
\expandafter\def\csname pn@wiki/detourvsrerankerwin/EM/pool/ci\endcsname{[$0.3$, $2.7$]}
\expandafter\def\csname pn@wiki/detourvsrerankerwin/EM/pool/full\endcsname{$+1.5$ [$0.3$, $2.7$]}
\expandafter\def\csname pn@wiki/droprepeats/EM/Decomp\endcsname{$-0.1$}
\expandafter\def\csname pn@wiki/droprepeats/EM/Decomp/abs\endcsname{0.1}
\expandafter\def\csname pn@wiki/droprepeats/EM/Decomp/ci\endcsname{[$-0.5$, $0.3$]}
\expandafter\def\csname pn@wiki/droprepeats/EM/Decomp/full\endcsname{$-0.1$ [$-0.5$, $0.3$]}
\expandafter\def\csname pn@wiki/droprepeats/EM/IRCoT\endcsname{$-0.3$}
\expandafter\def\csname pn@wiki/droprepeats/EM/IRCoT/abs\endcsname{0.3}
\expandafter\def\csname pn@wiki/droprepeats/EM/IRCoT/ci\endcsname{[$-0.9$, $0.3$]}
\expandafter\def\csname pn@wiki/droprepeats/EM/IRCoT/full\endcsname{$-0.3$ [$-0.9$, $0.3$]}
\expandafter\def\csname pn@wiki/droprepeats/EM/ReAct\endcsname{$+0.3$}
\expandafter\def\csname pn@wiki/droprepeats/EM/ReAct/abs\endcsname{0.3}
\expandafter\def\csname pn@wiki/droprepeats/EM/ReAct/ci\endcsname{[$-0.3$, $0.9$]}
\expandafter\def\csname pn@wiki/droprepeats/EM/ReAct/full\endcsname{$+0.3$ [$-0.3$, $0.9$]}
\expandafter\def\csname pn@wiki/droprepeats/EM/SR1\endcsname{$-0.1$}
\expandafter\def\csname pn@wiki/droprepeats/EM/SR1/abs\endcsname{0.1}
\expandafter\def\csname pn@wiki/droprepeats/EM/SR1/ci\endcsname{[$-0.4$, $0.3$]}
\expandafter\def\csname pn@wiki/droprepeats/EM/SR1/full\endcsname{$-0.1$ [$-0.4$, $0.3$]}
\expandafter\def\csname pn@wiki/droprepeats/EM/ext\endcsname{$0.0$}
\expandafter\def\csname pn@wiki/droprepeats/EM/ext/abs\endcsname{0.0}
\expandafter\def\csname pn@wiki/droprepeats/EM/ext/ci\endcsname{[$-0.3$, $0.3$]}
\expandafter\def\csname pn@wiki/droprepeats/EM/ext/full\endcsname{$0.0$ [$-0.3$, $0.3$]}
\expandafter\def\csname pn@wiki/droprepeats/EM/nsig\endcsname{0}
\expandafter\def\csname pn@wiki/droprepeats/EM/pool\endcsname{$-0.1$}
\expandafter\def\csname pn@wiki/droprepeats/EM/pool/abs\endcsname{0.1}
\expandafter\def\csname pn@wiki/droprepeats/EM/pool/ci\endcsname{[$-0.3$, $0.2$]}
\expandafter\def\csname pn@wiki/droprepeats/EM/pool/full\endcsname{$-0.1$ [$-0.3$, $0.2$]}
\expandafter\def\csname pn@wiki/droprepeats/EM/range\endcsname{$-0.3$ to $+0.3$}
\expandafter\def\csname pn@wiki/droprepeats/ans/Decomp\endcsname{$+0.1$}
\expandafter\def\csname pn@wiki/droprepeats/ans/Decomp/abs\endcsname{0.1}
\expandafter\def\csname pn@wiki/droprepeats/ans/Decomp/ci\endcsname{[$0.0$, $0.2$]}
\expandafter\def\csname pn@wiki/droprepeats/ans/Decomp/full\endcsname{$+0.1$ [$0.0$, $0.2$]}
\expandafter\def\csname pn@wiki/droprepeats/ans/IRCoT\endcsname{$-0.1$}
\expandafter\def\csname pn@wiki/droprepeats/ans/IRCoT/abs\endcsname{0.1}
\expandafter\def\csname pn@wiki/droprepeats/ans/IRCoT/ci\endcsname{[$-0.2$, $0.0$]}
\expandafter\def\csname pn@wiki/droprepeats/ans/IRCoT/full\endcsname{$-0.1$ [$-0.2$, $0.0$]}
\expandafter\def\csname pn@wiki/droprepeats/ans/ReAct\endcsname{$-0.2$}
\expandafter\def\csname pn@wiki/droprepeats/ans/ReAct/abs\endcsname{0.2}
\expandafter\def\csname pn@wiki/droprepeats/ans/ReAct/ci\endcsname{[$-0.5$, $0.2$]}
\expandafter\def\csname pn@wiki/droprepeats/ans/ReAct/full\endcsname{$-0.2$ [$-0.5$, $0.2$]}
\expandafter\def\csname pn@wiki/droprepeats/ans/SR1\endcsname{$-0.1$}
\expandafter\def\csname pn@wiki/droprepeats/ans/SR1/abs\endcsname{0.1}
\expandafter\def\csname pn@wiki/droprepeats/ans/SR1/ci\endcsname{[$-0.5$, $0.3$]}
\expandafter\def\csname pn@wiki/droprepeats/ans/SR1/full\endcsname{$-0.1$ [$-0.5$, $0.3$]}
\expandafter\def\csname pn@wiki/droprepeats/ans/ext\endcsname{$-0.1$}
\expandafter\def\csname pn@wiki/droprepeats/ans/ext/abs\endcsname{0.1}
\expandafter\def\csname pn@wiki/droprepeats/ans/ext/ci\endcsname{[$-0.3$, $0.1$]}
\expandafter\def\csname pn@wiki/droprepeats/ans/ext/full\endcsname{$-0.1$ [$-0.3$, $0.1$]}
\expandafter\def\csname pn@wiki/droprepeats/ans/nsig\endcsname{0}
\expandafter\def\csname pn@wiki/droprepeats/ans/pool\endcsname{$-0.1$}
\expandafter\def\csname pn@wiki/droprepeats/ans/pool/abs\endcsname{0.1}
\expandafter\def\csname pn@wiki/droprepeats/ans/pool/ci\endcsname{[$-0.2$, $0.1$]}
\expandafter\def\csname pn@wiki/droprepeats/ans/pool/full\endcsname{$-0.1$ [$-0.2$, $0.1$]}
\expandafter\def\csname pn@wiki/droprepeats/ans/range\endcsname{$-0.2$ to $+0.1$}
\expandafter\def\csname pn@wiki/droprepeats/complete/Decomp\endcsname{$-0.1$}
\expandafter\def\csname pn@wiki/droprepeats/complete/Decomp/abs\endcsname{0.1}
\expandafter\def\csname pn@wiki/droprepeats/complete/Decomp/ci\endcsname{[$-0.2$, $0.0$]}
\expandafter\def\csname pn@wiki/droprepeats/complete/Decomp/full\endcsname{$-0.1$ [$-0.2$, $0.0$]}
\expandafter\def\csname pn@wiki/droprepeats/complete/IRCoT\endcsname{$-0.1$}
\expandafter\def\csname pn@wiki/droprepeats/complete/IRCoT/abs\endcsname{0.1}
\expandafter\def\csname pn@wiki/droprepeats/complete/IRCoT/ci\endcsname{[$-0.3$, $0.1$]}
\expandafter\def\csname pn@wiki/droprepeats/complete/IRCoT/full\endcsname{$-0.1$ [$-0.3$, $0.1$]}
\expandafter\def\csname pn@wiki/droprepeats/complete/ReAct\endcsname{$-0.5$}
\expandafter\def\csname pn@wiki/droprepeats/complete/ReAct/abs\endcsname{0.5}
\expandafter\def\csname pn@wiki/droprepeats/complete/ReAct/ci\endcsname{[$-0.9$, $-0.2$]}
\expandafter\def\csname pn@wiki/droprepeats/complete/ReAct/full\endcsname{$-0.5$ [$-0.9$, $-0.2$]}
\expandafter\def\csname pn@wiki/droprepeats/complete/SR1\endcsname{$-0.4$}
\expandafter\def\csname pn@wiki/droprepeats/complete/SR1/abs\endcsname{0.4}
\expandafter\def\csname pn@wiki/droprepeats/complete/SR1/ci\endcsname{[$-0.8$, $0.0$]}
\expandafter\def\csname pn@wiki/droprepeats/complete/SR1/full\endcsname{$-0.4$ [$-0.8$, $0.0$]}
\expandafter\def\csname pn@wiki/droprepeats/complete/ext\endcsname{$-0.3$}
\expandafter\def\csname pn@wiki/droprepeats/complete/ext/abs\endcsname{0.3}
\expandafter\def\csname pn@wiki/droprepeats/complete/ext/ci\endcsname{[$-0.5$, $-0.1$]}
\expandafter\def\csname pn@wiki/droprepeats/complete/ext/full\endcsname{$-0.3$ [$-0.5$, $-0.1$]}
\expandafter\def\csname pn@wiki/droprepeats/complete/nsig\endcsname{1}
\expandafter\def\csname pn@wiki/droprepeats/complete/pool\endcsname{$-0.3$}
\expandafter\def\csname pn@wiki/droprepeats/complete/pool/abs\endcsname{0.3}
\expandafter\def\csname pn@wiki/droprepeats/complete/pool/ci\endcsname{[$-0.5$, $-0.1$]}
\expandafter\def\csname pn@wiki/droprepeats/complete/pool/full\endcsname{$-0.3$ [$-0.5$, $-0.1$]}
\expandafter\def\csname pn@wiki/droprepeats/complete/range\endcsname{$-0.5$ to $-0.1$}
\expandafter\def\csname pn@wiki/flag/Decomp\endcsname{13.0}
\expandafter\def\csname pn@wiki/flag/IRCoT\endcsname{38.1}
\expandafter\def\csname pn@wiki/flag/ReAct\endcsname{34.7}
\expandafter\def\csname pn@wiki/flag/SR1\endcsname{15.1}
\expandafter\def\csname pn@wiki/goldlink\endcsname{66}
\expandafter\def\csname pn@wiki/interVSrrf/EM/Decomp\endcsname{$+3.1$}
\expandafter\def\csname pn@wiki/interVSrrf/EM/Decomp/abs\endcsname{3.1}
\expandafter\def\csname pn@wiki/interVSrrf/EM/Decomp/ci\endcsname{[$1.3$, $4.9$]}
\expandafter\def\csname pn@wiki/interVSrrf/EM/Decomp/full\endcsname{$+3.1$ [$1.3$, $4.9$]}
\expandafter\def\csname pn@wiki/interVSrrf/EM/IRCoT\endcsname{$+1.9$}
\expandafter\def\csname pn@wiki/interVSrrf/EM/IRCoT/abs\endcsname{1.9}
\expandafter\def\csname pn@wiki/interVSrrf/EM/IRCoT/ci\endcsname{[$-0.3$, $4.2$]}
\expandafter\def\csname pn@wiki/interVSrrf/EM/IRCoT/full\endcsname{$+1.9$ [$-0.3$, $4.2$]}
\expandafter\def\csname pn@wiki/interVSrrf/EM/ReAct\endcsname{$+1.1$}
\expandafter\def\csname pn@wiki/interVSrrf/EM/ReAct/abs\endcsname{1.1}
\expandafter\def\csname pn@wiki/interVSrrf/EM/ReAct/ci\endcsname{[$-0.9$, $3.1$]}
\expandafter\def\csname pn@wiki/interVSrrf/EM/ReAct/full\endcsname{$+1.1$ [$-0.9$, $3.1$]}
\expandafter\def\csname pn@wiki/interVSrrf/EM/SR1\endcsname{$+0.3$}
\expandafter\def\csname pn@wiki/interVSrrf/EM/SR1/abs\endcsname{0.3}
\expandafter\def\csname pn@wiki/interVSrrf/EM/SR1/ci\endcsname{[$-1.6$, $2.1$]}
\expandafter\def\csname pn@wiki/interVSrrf/EM/SR1/full\endcsname{$+0.3$ [$-1.6$, $2.1$]}
\expandafter\def\csname pn@wiki/interVSrrf/EM/ext\endcsname{$+1.1$}
\expandafter\def\csname pn@wiki/interVSrrf/EM/ext/abs\endcsname{1.1}
\expandafter\def\csname pn@wiki/interVSrrf/EM/ext/ci\endcsname{[$-0.2$, $2.3$]}
\expandafter\def\csname pn@wiki/interVSrrf/EM/ext/full\endcsname{$+1.1$ [$-0.2$, $2.3$]}
\expandafter\def\csname pn@wiki/interVSrrf/EM/nsig\endcsname{1}
\expandafter\def\csname pn@wiki/interVSrrf/EM/pool\endcsname{$+1.6$}
\expandafter\def\csname pn@wiki/interVSrrf/EM/pool/abs\endcsname{1.6}
\expandafter\def\csname pn@wiki/interVSrrf/EM/pool/ci\endcsname{[$0.5$, $2.7$]}
\expandafter\def\csname pn@wiki/interVSrrf/EM/pool/full\endcsname{$+1.6$ [$0.5$, $2.7$]}
\expandafter\def\csname pn@wiki/interVSrrf/EM/range\endcsname{$+0.3$ to $+3.1$}
\expandafter\def\csname pn@wiki/interVSrrf/ans/Decomp\endcsname{$-0.7$}
\expandafter\def\csname pn@wiki/interVSrrf/ans/Decomp/abs\endcsname{0.7}
\expandafter\def\csname pn@wiki/interVSrrf/ans/Decomp/ci\endcsname{[$-2.0$, $0.6$]}
\expandafter\def\csname pn@wiki/interVSrrf/ans/Decomp/full\endcsname{$-0.7$ [$-2.0$, $0.6$]}
\expandafter\def\csname pn@wiki/interVSrrf/ans/IRCoT\endcsname{$+8.3$}
\expandafter\def\csname pn@wiki/interVSrrf/ans/IRCoT/abs\endcsname{8.3}
\expandafter\def\csname pn@wiki/interVSrrf/ans/IRCoT/ci\endcsname{[$6.5$, $10.0$]}
\expandafter\def\csname pn@wiki/interVSrrf/ans/IRCoT/full\endcsname{$+8.3$ [$6.5$, $10.0$]}
\expandafter\def\csname pn@wiki/interVSrrf/ans/ReAct\endcsname{$+8.3$}
\expandafter\def\csname pn@wiki/interVSrrf/ans/ReAct/abs\endcsname{8.3}
\expandafter\def\csname pn@wiki/interVSrrf/ans/ReAct/ci\endcsname{[$6.5$, $10.1$]}
\expandafter\def\csname pn@wiki/interVSrrf/ans/ReAct/full\endcsname{$+8.3$ [$6.5$, $10.1$]}
\expandafter\def\csname pn@wiki/interVSrrf/ans/SR1\endcsname{$+4.3$}
\expandafter\def\csname pn@wiki/interVSrrf/ans/SR1/abs\endcsname{4.3}
\expandafter\def\csname pn@wiki/interVSrrf/ans/SR1/ci\endcsname{[$2.8$, $5.8$]}
\expandafter\def\csname pn@wiki/interVSrrf/ans/SR1/full\endcsname{$+4.3$ [$2.8$, $5.8$]}
\expandafter\def\csname pn@wiki/interVSrrf/ans/ext\endcsname{$+6.9$}
\expandafter\def\csname pn@wiki/interVSrrf/ans/ext/abs\endcsname{6.9}
\expandafter\def\csname pn@wiki/interVSrrf/ans/ext/ci\endcsname{[$5.9$, $8.0$]}
\expandafter\def\csname pn@wiki/interVSrrf/ans/ext/full\endcsname{$+6.9$ [$5.9$, $8.0$]}
\expandafter\def\csname pn@wiki/interVSrrf/ans/nsig\endcsname{3}
\expandafter\def\csname pn@wiki/interVSrrf/ans/pool\endcsname{$+5.0$}
\expandafter\def\csname pn@wiki/interVSrrf/ans/pool/abs\endcsname{5.0}
\expandafter\def\csname pn@wiki/interVSrrf/ans/pool/ci\endcsname{[$4.2$, $5.9$]}
\expandafter\def\csname pn@wiki/interVSrrf/ans/pool/full\endcsname{$+5.0$ [$4.2$, $5.9$]}
\expandafter\def\csname pn@wiki/interVSrrf/ans/range\endcsname{$-0.7$ to $+8.3$}
\expandafter\def\csname pn@wiki/interVSrrf/complete/Decomp\endcsname{$-1.1$}
\expandafter\def\csname pn@wiki/interVSrrf/complete/Decomp/abs\endcsname{1.1}
\expandafter\def\csname pn@wiki/interVSrrf/complete/Decomp/ci\endcsname{[$-2.3$, $-0.1$]}
\expandafter\def\csname pn@wiki/interVSrrf/complete/Decomp/full\endcsname{$-1.1$ [$-2.3$, $-0.1$]}
\expandafter\def\csname pn@wiki/interVSrrf/complete/IRCoT\endcsname{$+14.5$}
\expandafter\def\csname pn@wiki/interVSrrf/complete/IRCoT/abs\endcsname{14.5}
\expandafter\def\csname pn@wiki/interVSrrf/complete/IRCoT/ci\endcsname{[$12.6$, $16.4$]}
\expandafter\def\csname pn@wiki/interVSrrf/complete/IRCoT/full\endcsname{$+14.5$ [$12.6$, $16.4$]}
\expandafter\def\csname pn@wiki/interVSrrf/complete/ReAct\endcsname{$+10.0$}
\expandafter\def\csname pn@wiki/interVSrrf/complete/ReAct/abs\endcsname{10.0}
\expandafter\def\csname pn@wiki/interVSrrf/complete/ReAct/ci\endcsname{[$8.3$, $11.8$]}
\expandafter\def\csname pn@wiki/interVSrrf/complete/ReAct/full\endcsname{$+10.0$ [$8.3$, $11.8$]}
\expandafter\def\csname pn@wiki/interVSrrf/complete/SR1\endcsname{$+7.8$}
\expandafter\def\csname pn@wiki/interVSrrf/complete/SR1/abs\endcsname{7.8}
\expandafter\def\csname pn@wiki/interVSrrf/complete/SR1/ci\endcsname{[$6.2$, $9.5$]}
\expandafter\def\csname pn@wiki/interVSrrf/complete/SR1/full\endcsname{$+7.8$ [$6.2$, $9.5$]}
\expandafter\def\csname pn@wiki/interVSrrf/complete/ext\endcsname{$+10.8$}
\expandafter\def\csname pn@wiki/interVSrrf/complete/ext/abs\endcsname{10.8}
\expandafter\def\csname pn@wiki/interVSrrf/complete/ext/ci\endcsname{[$9.6$, $11.9$]}
\expandafter\def\csname pn@wiki/interVSrrf/complete/ext/full\endcsname{$+10.8$ [$9.6$, $11.9$]}
\expandafter\def\csname pn@wiki/interVSrrf/complete/nsig\endcsname{4}
\expandafter\def\csname pn@wiki/interVSrrf/complete/pool\endcsname{$+7.8$}
\expandafter\def\csname pn@wiki/interVSrrf/complete/pool/abs\endcsname{7.8}
\expandafter\def\csname pn@wiki/interVSrrf/complete/pool/ci\endcsname{[$6.9$, $8.7$]}
\expandafter\def\csname pn@wiki/interVSrrf/complete/pool/full\endcsname{$+7.8$ [$6.9$, $8.7$]}
\expandafter\def\csname pn@wiki/interVSrrf/complete/range\endcsname{$-1.1$ to $+14.5$}
\expandafter\def\csname pn@wiki/n\endcsname{1{,}500}
\expandafter\def\csname pn@wiki/named/EM/Decomp\endcsname{$+0.1$}
\expandafter\def\csname pn@wiki/named/EM/Decomp/abs\endcsname{0.1}
\expandafter\def\csname pn@wiki/named/EM/Decomp/ci\endcsname{[$-1.2$, $1.5$]}
\expandafter\def\csname pn@wiki/named/EM/Decomp/full\endcsname{$+0.1$ [$-1.2$, $1.5$]}
\expandafter\def\csname pn@wiki/named/EM/IRCoT\endcsname{$+0.6$}
\expandafter\def\csname pn@wiki/named/EM/IRCoT/abs\endcsname{0.6}
\expandafter\def\csname pn@wiki/named/EM/IRCoT/ci\endcsname{[$-0.9$, $2.1$]}
\expandafter\def\csname pn@wiki/named/EM/IRCoT/full\endcsname{$+0.6$ [$-0.9$, $2.1$]}
\expandafter\def\csname pn@wiki/named/EM/ReAct\endcsname{$+0.2$}
\expandafter\def\csname pn@wiki/named/EM/ReAct/abs\endcsname{0.2}
\expandafter\def\csname pn@wiki/named/EM/ReAct/ci\endcsname{[$-1.1$, $1.5$]}
\expandafter\def\csname pn@wiki/named/EM/ReAct/full\endcsname{$+0.2$ [$-1.1$, $1.5$]}
\expandafter\def\csname pn@wiki/named/EM/SR1\endcsname{$+0.9$}
\expandafter\def\csname pn@wiki/named/EM/SR1/abs\endcsname{0.9}
\expandafter\def\csname pn@wiki/named/EM/SR1/ci\endcsname{[$-0.6$, $2.5$]}
\expandafter\def\csname pn@wiki/named/EM/SR1/full\endcsname{$+0.9$ [$-0.6$, $2.5$]}
\expandafter\def\csname pn@wiki/named/EM/ext\endcsname{$+0.6$}
\expandafter\def\csname pn@wiki/named/EM/ext/abs\endcsname{0.6}
\expandafter\def\csname pn@wiki/named/EM/ext/ci\endcsname{[$-0.4$, $1.6$]}
\expandafter\def\csname pn@wiki/named/EM/ext/full\endcsname{$+0.6$ [$-0.4$, $1.6$]}
\expandafter\def\csname pn@wiki/named/EM/nsig\endcsname{0}
\expandafter\def\csname pn@wiki/named/EM/pool\endcsname{$+0.5$}
\expandafter\def\csname pn@wiki/named/EM/pool/abs\endcsname{0.5}
\expandafter\def\csname pn@wiki/named/EM/pool/ci\endcsname{[$-0.4$, $1.3$]}
\expandafter\def\csname pn@wiki/named/EM/pool/full\endcsname{$+0.5$ [$-0.4$, $1.3$]}
\expandafter\def\csname pn@wiki/named/EM/range\endcsname{$+0.1$ to $+0.9$}
\expandafter\def\csname pn@wiki/named/ans/Decomp\endcsname{$+2.5$}
\expandafter\def\csname pn@wiki/named/ans/Decomp/abs\endcsname{2.5}
\expandafter\def\csname pn@wiki/named/ans/Decomp/ci\endcsname{[$1.6$, $3.4$]}
\expandafter\def\csname pn@wiki/named/ans/Decomp/full\endcsname{$+2.5$ [$1.6$, $3.4$]}
\expandafter\def\csname pn@wiki/named/ans/IRCoT\endcsname{$+1.1$}
\expandafter\def\csname pn@wiki/named/ans/IRCoT/abs\endcsname{1.1}
\expandafter\def\csname pn@wiki/named/ans/IRCoT/ci\endcsname{[$0.5$, $1.7$]}
\expandafter\def\csname pn@wiki/named/ans/IRCoT/full\endcsname{$+1.1$ [$0.5$, $1.7$]}
\expandafter\def\csname pn@wiki/named/ans/ReAct\endcsname{$+1.6$}
\expandafter\def\csname pn@wiki/named/ans/ReAct/abs\endcsname{1.6}
\expandafter\def\csname pn@wiki/named/ans/ReAct/ci\endcsname{[$0.8$, $2.4$]}
\expandafter\def\csname pn@wiki/named/ans/ReAct/full\endcsname{$+1.6$ [$0.8$, $2.4$]}
\expandafter\def\csname pn@wiki/named/ans/SR1\endcsname{$+1.0$}
\expandafter\def\csname pn@wiki/named/ans/SR1/abs\endcsname{1.0}
\expandafter\def\csname pn@wiki/named/ans/SR1/ci\endcsname{[$0.2$, $1.8$]}
\expandafter\def\csname pn@wiki/named/ans/SR1/full\endcsname{$+1.0$ [$0.2$, $1.8$]}
\expandafter\def\csname pn@wiki/named/ans/ext\endcsname{$+1.2$}
\expandafter\def\csname pn@wiki/named/ans/ext/abs\endcsname{1.2}
\expandafter\def\csname pn@wiki/named/ans/ext/ci\endcsname{[$0.7$, $1.7$]}
\expandafter\def\csname pn@wiki/named/ans/ext/full\endcsname{$+1.2$ [$0.7$, $1.7$]}
\expandafter\def\csname pn@wiki/named/ans/nsig\endcsname{4}
\expandafter\def\csname pn@wiki/named/ans/pool\endcsname{$+1.5$}
\expandafter\def\csname pn@wiki/named/ans/pool/abs\endcsname{1.5}
\expandafter\def\csname pn@wiki/named/ans/pool/ci\endcsname{[$1.1$, $2.0$]}
\expandafter\def\csname pn@wiki/named/ans/pool/full\endcsname{$+1.5$ [$1.1$, $2.0$]}
\expandafter\def\csname pn@wiki/named/ans/range\endcsname{$+1.0$ to $+2.5$}
\expandafter\def\csname pn@wiki/named/complete/Decomp\endcsname{$+3.3$}
\expandafter\def\csname pn@wiki/named/complete/Decomp/abs\endcsname{3.3}
\expandafter\def\csname pn@wiki/named/complete/Decomp/ci\endcsname{[$2.4$, $4.2$]}
\expandafter\def\csname pn@wiki/named/complete/Decomp/full\endcsname{$+3.3$ [$2.4$, $4.2$]}
\expandafter\def\csname pn@wiki/named/complete/IRCoT\endcsname{$+2.3$}
\expandafter\def\csname pn@wiki/named/complete/IRCoT/abs\endcsname{2.3}
\expandafter\def\csname pn@wiki/named/complete/IRCoT/ci\endcsname{[$1.5$, $3.1$]}
\expandafter\def\csname pn@wiki/named/complete/IRCoT/full\endcsname{$+2.3$ [$1.5$, $3.1$]}
\expandafter\def\csname pn@wiki/named/complete/ReAct\endcsname{$+4.0$}
\expandafter\def\csname pn@wiki/named/complete/ReAct/abs\endcsname{4.0}
\expandafter\def\csname pn@wiki/named/complete/ReAct/ci\endcsname{[$2.9$, $5.1$]}
\expandafter\def\csname pn@wiki/named/complete/ReAct/full\endcsname{$+4.0$ [$2.9$, $5.1$]}
\expandafter\def\csname pn@wiki/named/complete/SR1\endcsname{$+4.7$}
\expandafter\def\csname pn@wiki/named/complete/SR1/abs\endcsname{4.7}
\expandafter\def\csname pn@wiki/named/complete/SR1/ci\endcsname{[$3.5$, $5.9$]}
\expandafter\def\csname pn@wiki/named/complete/SR1/full\endcsname{$+4.7$ [$3.5$, $5.9$]}
\expandafter\def\csname pn@wiki/named/complete/ext\endcsname{$+3.6$}
\expandafter\def\csname pn@wiki/named/complete/ext/abs\endcsname{3.6}
\expandafter\def\csname pn@wiki/named/complete/ext/ci\endcsname{[$3.0$, $4.4$]}
\expandafter\def\csname pn@wiki/named/complete/ext/full\endcsname{$+3.6$ [$3.0$, $4.4$]}
\expandafter\def\csname pn@wiki/named/complete/nsig\endcsname{4}
\expandafter\def\csname pn@wiki/named/complete/pool\endcsname{$+3.6$}
\expandafter\def\csname pn@wiki/named/complete/pool/abs\endcsname{3.6}
\expandafter\def\csname pn@wiki/named/complete/pool/ci\endcsname{[$3.0$, $4.2$]}
\expandafter\def\csname pn@wiki/named/complete/pool/full\endcsname{$+3.6$ [$3.0$, $4.2$]}
\expandafter\def\csname pn@wiki/named/complete/range\endcsname{$+2.3$ to $+4.7$}
\expandafter\def\csname pn@wiki/novelquery/EM/Decomp\endcsname{$+2.4$}
\expandafter\def\csname pn@wiki/novelquery/EM/Decomp/abs\endcsname{2.4}
\expandafter\def\csname pn@wiki/novelquery/EM/Decomp/ci\endcsname{[$1.1$, $3.7$]}
\expandafter\def\csname pn@wiki/novelquery/EM/Decomp/full\endcsname{$+2.4$ [$1.1$, $3.7$]}
\expandafter\def\csname pn@wiki/novelquery/EM/IRCoT\endcsname{$-0.7$}
\expandafter\def\csname pn@wiki/novelquery/EM/IRCoT/abs\endcsname{0.7}
\expandafter\def\csname pn@wiki/novelquery/EM/IRCoT/ci\endcsname{[$-2.5$, $1.1$]}
\expandafter\def\csname pn@wiki/novelquery/EM/IRCoT/full\endcsname{$-0.7$ [$-2.5$, $1.1$]}
\expandafter\def\csname pn@wiki/novelquery/EM/ReAct\endcsname{$+2.8$}
\expandafter\def\csname pn@wiki/novelquery/EM/ReAct/abs\endcsname{2.8}
\expandafter\def\csname pn@wiki/novelquery/EM/ReAct/ci\endcsname{[$1.2$, $4.4$]}
\expandafter\def\csname pn@wiki/novelquery/EM/ReAct/full\endcsname{$+2.8$ [$1.2$, $4.4$]}
\expandafter\def\csname pn@wiki/novelquery/EM/SR1\endcsname{$+0.2$}
\expandafter\def\csname pn@wiki/novelquery/EM/SR1/abs\endcsname{0.2}
\expandafter\def\csname pn@wiki/novelquery/EM/SR1/ci\endcsname{[$-0.6$, $1.0$]}
\expandafter\def\csname pn@wiki/novelquery/EM/SR1/full\endcsname{$+0.2$ [$-0.6$, $1.0$]}
\expandafter\def\csname pn@wiki/novelquery/EM/ext\endcsname{$+0.8$}
\expandafter\def\csname pn@wiki/novelquery/EM/ext/abs\endcsname{0.8}
\expandafter\def\csname pn@wiki/novelquery/EM/ext/ci\endcsname{[$-0.1$, $1.7$]}
\expandafter\def\csname pn@wiki/novelquery/EM/ext/full\endcsname{$+0.8$ [$-0.1$, $1.7$]}
\expandafter\def\csname pn@wiki/novelquery/EM/nsig\endcsname{2}
\expandafter\def\csname pn@wiki/novelquery/EM/pool\endcsname{$+1.2$}
\expandafter\def\csname pn@wiki/novelquery/EM/pool/abs\endcsname{1.2}
\expandafter\def\csname pn@wiki/novelquery/EM/pool/ci\endcsname{[$0.4$, $2.0$]}
\expandafter\def\csname pn@wiki/novelquery/EM/pool/full\endcsname{$+1.2$ [$0.4$, $2.0$]}
\expandafter\def\csname pn@wiki/novelquery/EM/range\endcsname{$-0.7$ to $+2.8$}
\expandafter\def\csname pn@wiki/novelquery/ans/Decomp\endcsname{$+8.0$}
\expandafter\def\csname pn@wiki/novelquery/ans/Decomp/abs\endcsname{8.0}
\expandafter\def\csname pn@wiki/novelquery/ans/Decomp/ci\endcsname{[$6.4$, $9.5$]}
\expandafter\def\csname pn@wiki/novelquery/ans/Decomp/full\endcsname{$+8.0$ [$6.4$, $9.5$]}
\expandafter\def\csname pn@wiki/novelquery/ans/IRCoT\endcsname{$+0.8$}
\expandafter\def\csname pn@wiki/novelquery/ans/IRCoT/abs\endcsname{0.8}
\expandafter\def\csname pn@wiki/novelquery/ans/IRCoT/ci\endcsname{[$-0.5$, $2.0$]}
\expandafter\def\csname pn@wiki/novelquery/ans/IRCoT/full\endcsname{$+0.8$ [$-0.5$, $2.0$]}
\expandafter\def\csname pn@wiki/novelquery/ans/ReAct\endcsname{$+7.1$}
\expandafter\def\csname pn@wiki/novelquery/ans/ReAct/abs\endcsname{7.1}
\expandafter\def\csname pn@wiki/novelquery/ans/ReAct/ci\endcsname{[$5.3$, $8.8$]}
\expandafter\def\csname pn@wiki/novelquery/ans/ReAct/full\endcsname{$+7.1$ [$5.3$, $8.8$]}
\expandafter\def\csname pn@wiki/novelquery/ans/SR1\endcsname{$+1.2$}
\expandafter\def\csname pn@wiki/novelquery/ans/SR1/abs\endcsname{1.2}
\expandafter\def\csname pn@wiki/novelquery/ans/SR1/ci\endcsname{[$0.4$, $2.1$]}
\expandafter\def\csname pn@wiki/novelquery/ans/SR1/full\endcsname{$+1.2$ [$0.4$, $2.1$]}
\expandafter\def\csname pn@wiki/novelquery/ans/ext\endcsname{$+3.0$}
\expandafter\def\csname pn@wiki/novelquery/ans/ext/abs\endcsname{3.0}
\expandafter\def\csname pn@wiki/novelquery/ans/ext/ci\endcsname{[$2.1$, $3.9$]}
\expandafter\def\csname pn@wiki/novelquery/ans/ext/full\endcsname{$+3.0$ [$2.1$, $3.9$]}
\expandafter\def\csname pn@wiki/novelquery/ans/nsig\endcsname{3}
\expandafter\def\csname pn@wiki/novelquery/ans/pool\endcsname{$+4.2$}
\expandafter\def\csname pn@wiki/novelquery/ans/pool/abs\endcsname{4.2}
\expandafter\def\csname pn@wiki/novelquery/ans/pool/ci\endcsname{[$3.4$, $5.1$]}
\expandafter\def\csname pn@wiki/novelquery/ans/pool/full\endcsname{$+4.2$ [$3.4$, $5.1$]}
\expandafter\def\csname pn@wiki/novelquery/ans/range\endcsname{$+0.8$ to $+8.0$}
\expandafter\def\csname pn@wiki/novelquery/complete/Decomp\endcsname{$+10.2$}
\expandafter\def\csname pn@wiki/novelquery/complete/Decomp/abs\endcsname{10.2}
\expandafter\def\csname pn@wiki/novelquery/complete/Decomp/ci\endcsname{[$8.6$, $11.9$]}
\expandafter\def\csname pn@wiki/novelquery/complete/Decomp/full\endcsname{$+10.2$ [$8.6$, $11.9$]}
\expandafter\def\csname pn@wiki/novelquery/complete/IRCoT\endcsname{$+3.3$}
\expandafter\def\csname pn@wiki/novelquery/complete/IRCoT/abs\endcsname{3.3}
\expandafter\def\csname pn@wiki/novelquery/complete/IRCoT/ci\endcsname{[$2.1$, $4.7$]}
\expandafter\def\csname pn@wiki/novelquery/complete/IRCoT/full\endcsname{$+3.3$ [$2.1$, $4.7$]}
\expandafter\def\csname pn@wiki/novelquery/complete/ReAct\endcsname{$+8.5$}
\expandafter\def\csname pn@wiki/novelquery/complete/ReAct/abs\endcsname{8.5}
\expandafter\def\csname pn@wiki/novelquery/complete/ReAct/ci\endcsname{[$6.7$, $10.3$]}
\expandafter\def\csname pn@wiki/novelquery/complete/ReAct/full\endcsname{$+8.5$ [$6.7$, $10.3$]}
\expandafter\def\csname pn@wiki/novelquery/complete/SR1\endcsname{$+0.3$}
\expandafter\def\csname pn@wiki/novelquery/complete/SR1/abs\endcsname{0.3}
\expandafter\def\csname pn@wiki/novelquery/complete/SR1/ci\endcsname{[$-0.7$, $1.2$]}
\expandafter\def\csname pn@wiki/novelquery/complete/SR1/full\endcsname{$+0.3$ [$-0.7$, $1.2$]}
\expandafter\def\csname pn@wiki/novelquery/complete/ext\endcsname{$+4.0$}
\expandafter\def\csname pn@wiki/novelquery/complete/ext/abs\endcsname{4.0}
\expandafter\def\csname pn@wiki/novelquery/complete/ext/ci\endcsname{[$3.1$, $5.0$]}
\expandafter\def\csname pn@wiki/novelquery/complete/ext/full\endcsname{$+4.0$ [$3.1$, $5.0$]}
\expandafter\def\csname pn@wiki/novelquery/complete/nsig\endcsname{3}
\expandafter\def\csname pn@wiki/novelquery/complete/pool\endcsname{$+5.6$}
\expandafter\def\csname pn@wiki/novelquery/complete/pool/abs\endcsname{5.6}
\expandafter\def\csname pn@wiki/novelquery/complete/pool/ci\endcsname{[$4.7$, $6.5$]}
\expandafter\def\csname pn@wiki/novelquery/complete/pool/full\endcsname{$+5.6$ [$4.7$, $6.5$]}
\expandafter\def\csname pn@wiki/novelquery/complete/range\endcsname{$+0.3$ to $+10.2$}
\expandafter\def\csname pn@wiki/order/EM/Decomp\endcsname{$+0.3$}
\expandafter\def\csname pn@wiki/order/EM/Decomp/abs\endcsname{0.3}
\expandafter\def\csname pn@wiki/order/EM/Decomp/ci\endcsname{[$-1.2$, $1.7$]}
\expandafter\def\csname pn@wiki/order/EM/Decomp/full\endcsname{$+0.3$ [$-1.2$, $1.7$]}
\expandafter\def\csname pn@wiki/order/EM/IRCoT\endcsname{$+1.5$}
\expandafter\def\csname pn@wiki/order/EM/IRCoT/abs\endcsname{1.5}
\expandafter\def\csname pn@wiki/order/EM/IRCoT/ci\endcsname{[$-0.1$, $3.1$]}
\expandafter\def\csname pn@wiki/order/EM/IRCoT/full\endcsname{$+1.5$ [$-0.1$, $3.1$]}
\expandafter\def\csname pn@wiki/order/EM/ReAct\endcsname{$+0.9$}
\expandafter\def\csname pn@wiki/order/EM/ReAct/abs\endcsname{0.9}
\expandafter\def\csname pn@wiki/order/EM/ReAct/ci\endcsname{[$-0.7$, $2.4$]}
\expandafter\def\csname pn@wiki/order/EM/ReAct/full\endcsname{$+0.9$ [$-0.7$, $2.4$]}
\expandafter\def\csname pn@wiki/order/EM/SR1\endcsname{$+1.4$}
\expandafter\def\csname pn@wiki/order/EM/SR1/abs\endcsname{1.4}
\expandafter\def\csname pn@wiki/order/EM/SR1/ci\endcsname{[$-0.1$, $3.0$]}
\expandafter\def\csname pn@wiki/order/EM/SR1/full\endcsname{$+1.4$ [$-0.1$, $3.0$]}
\expandafter\def\csname pn@wiki/order/EM/ext\endcsname{$+1.3$}
\expandafter\def\csname pn@wiki/order/EM/ext/abs\endcsname{1.3}
\expandafter\def\csname pn@wiki/order/EM/ext/ci\endcsname{[$0.2$, $2.3$]}
\expandafter\def\csname pn@wiki/order/EM/ext/full\endcsname{$+1.3$ [$0.2$, $2.3$]}
\expandafter\def\csname pn@wiki/order/EM/nsig\endcsname{0}
\expandafter\def\csname pn@wiki/order/EM/pool\endcsname{$+1.0$}
\expandafter\def\csname pn@wiki/order/EM/pool/abs\endcsname{1.0}
\expandafter\def\csname pn@wiki/order/EM/pool/ci\endcsname{[$0.1$, $2.0$]}
\expandafter\def\csname pn@wiki/order/EM/pool/full\endcsname{$+1.0$ [$0.1$, $2.0$]}
\expandafter\def\csname pn@wiki/order/EM/range\endcsname{$+0.3$ to $+1.5$}
\expandafter\def\csname pn@wiki/order/ans/Decomp\endcsname{$0.0$}
\expandafter\def\csname pn@wiki/order/ans/Decomp/abs\endcsname{0.0}
\expandafter\def\csname pn@wiki/order/ans/Decomp/ci\endcsname{[$0.0$, $0.0$]}
\expandafter\def\csname pn@wiki/order/ans/Decomp/full\endcsname{$0.0$ [$0.0$, $0.0$]}
\expandafter\def\csname pn@wiki/order/ans/IRCoT\endcsname{$0.0$}
\expandafter\def\csname pn@wiki/order/ans/IRCoT/abs\endcsname{0.0}
\expandafter\def\csname pn@wiki/order/ans/IRCoT/ci\endcsname{[$0.0$, $0.0$]}
\expandafter\def\csname pn@wiki/order/ans/IRCoT/full\endcsname{$0.0$ [$0.0$, $0.0$]}
\expandafter\def\csname pn@wiki/order/ans/ReAct\endcsname{$0.0$}
\expandafter\def\csname pn@wiki/order/ans/ReAct/abs\endcsname{0.0}
\expandafter\def\csname pn@wiki/order/ans/ReAct/ci\endcsname{[$0.0$, $0.0$]}
\expandafter\def\csname pn@wiki/order/ans/ReAct/full\endcsname{$0.0$ [$0.0$, $0.0$]}
\expandafter\def\csname pn@wiki/order/ans/SR1\endcsname{$0.0$}
\expandafter\def\csname pn@wiki/order/ans/SR1/abs\endcsname{0.0}
\expandafter\def\csname pn@wiki/order/ans/SR1/ci\endcsname{[$0.0$, $0.0$]}
\expandafter\def\csname pn@wiki/order/ans/SR1/full\endcsname{$0.0$ [$0.0$, $0.0$]}
\expandafter\def\csname pn@wiki/order/ans/ext\endcsname{$0.0$}
\expandafter\def\csname pn@wiki/order/ans/ext/abs\endcsname{0.0}
\expandafter\def\csname pn@wiki/order/ans/ext/ci\endcsname{[$0.0$, $0.0$]}
\expandafter\def\csname pn@wiki/order/ans/ext/full\endcsname{$0.0$ [$0.0$, $0.0$]}
\expandafter\def\csname pn@wiki/order/ans/nsig\endcsname{0}
\expandafter\def\csname pn@wiki/order/ans/pool\endcsname{$0.0$}
\expandafter\def\csname pn@wiki/order/ans/pool/abs\endcsname{0.0}
\expandafter\def\csname pn@wiki/order/ans/pool/ci\endcsname{[$0.0$, $0.0$]}
\expandafter\def\csname pn@wiki/order/ans/pool/full\endcsname{$0.0$ [$0.0$, $0.0$]}
\expandafter\def\csname pn@wiki/order/ans/range\endcsname{$0.0$ to $0.0$}
\expandafter\def\csname pn@wiki/order/complete/Decomp\endcsname{$0.0$}
\expandafter\def\csname pn@wiki/order/complete/Decomp/abs\endcsname{0.0}
\expandafter\def\csname pn@wiki/order/complete/Decomp/ci\endcsname{[$0.0$, $0.0$]}
\expandafter\def\csname pn@wiki/order/complete/Decomp/full\endcsname{$0.0$ [$0.0$, $0.0$]}
\expandafter\def\csname pn@wiki/order/complete/IRCoT\endcsname{$0.0$}
\expandafter\def\csname pn@wiki/order/complete/IRCoT/abs\endcsname{0.0}
\expandafter\def\csname pn@wiki/order/complete/IRCoT/ci\endcsname{[$0.0$, $0.0$]}
\expandafter\def\csname pn@wiki/order/complete/IRCoT/full\endcsname{$0.0$ [$0.0$, $0.0$]}
\expandafter\def\csname pn@wiki/order/complete/ReAct\endcsname{$0.0$}
\expandafter\def\csname pn@wiki/order/complete/ReAct/abs\endcsname{0.0}
\expandafter\def\csname pn@wiki/order/complete/ReAct/ci\endcsname{[$0.0$, $0.0$]}
\expandafter\def\csname pn@wiki/order/complete/ReAct/full\endcsname{$0.0$ [$0.0$, $0.0$]}
\expandafter\def\csname pn@wiki/order/complete/SR1\endcsname{$0.0$}
\expandafter\def\csname pn@wiki/order/complete/SR1/abs\endcsname{0.0}
\expandafter\def\csname pn@wiki/order/complete/SR1/ci\endcsname{[$0.0$, $0.0$]}
\expandafter\def\csname pn@wiki/order/complete/SR1/full\endcsname{$0.0$ [$0.0$, $0.0$]}
\expandafter\def\csname pn@wiki/order/complete/ext\endcsname{$0.0$}
\expandafter\def\csname pn@wiki/order/complete/ext/abs\endcsname{0.0}
\expandafter\def\csname pn@wiki/order/complete/ext/ci\endcsname{[$0.0$, $0.0$]}
\expandafter\def\csname pn@wiki/order/complete/ext/full\endcsname{$0.0$ [$0.0$, $0.0$]}
\expandafter\def\csname pn@wiki/order/complete/nsig\endcsname{0}
\expandafter\def\csname pn@wiki/order/complete/pool\endcsname{$0.0$}
\expandafter\def\csname pn@wiki/order/complete/pool/abs\endcsname{0.0}
\expandafter\def\csname pn@wiki/order/complete/pool/ci\endcsname{[$0.0$, $0.0$]}
\expandafter\def\csname pn@wiki/order/complete/pool/full\endcsname{$0.0$ [$0.0$, $0.0$]}
\expandafter\def\csname pn@wiki/order/complete/range\endcsname{$0.0$ to $0.0$}
\expandafter\def\csname pn@wiki/qt/bridgecomparison/detour/EM\endcsname{$+1.9$}
\expandafter\def\csname pn@wiki/qt/bridgecomparison/detour/EM/abs\endcsname{1.9}
\expandafter\def\csname pn@wiki/qt/bridgecomparison/detour/EM/ci\endcsname{[$-1.2$, $5.0$]}
\expandafter\def\csname pn@wiki/qt/bridgecomparison/detour/EM/full\endcsname{$+1.9$ [$-1.2$, $5.0$]}
\expandafter\def\csname pn@wiki/qt/bridgecomparison/detour/complete\endcsname{$+10.4$}
\expandafter\def\csname pn@wiki/qt/bridgecomparison/detour/complete/abs\endcsname{10.4}
\expandafter\def\csname pn@wiki/qt/bridgecomparison/detour/complete/ci\endcsname{[$8.2$, $12.7$]}
\expandafter\def\csname pn@wiki/qt/bridgecomparison/detour/complete/full\endcsname{$+10.4$ [$8.2$, $12.7$]}
\expandafter\def\csname pn@wiki/qt/bridgecomparison/n\endcsname{303}
\expandafter\def\csname pn@wiki/qt/bridgecomparison/route/EM\endcsname{$+0.3$}
\expandafter\def\csname pn@wiki/qt/bridgecomparison/route/EM/abs\endcsname{0.3}
\expandafter\def\csname pn@wiki/qt/bridgecomparison/route/EM/ci\endcsname{[$-1.3$, $2.0$]}
\expandafter\def\csname pn@wiki/qt/bridgecomparison/route/EM/full\endcsname{$+0.3$ [$-1.3$, $2.0$]}
\expandafter\def\csname pn@wiki/qt/bridgecomparison/route/complete\endcsname{$+4.9$}
\expandafter\def\csname pn@wiki/qt/bridgecomparison/route/complete/abs\endcsname{4.9}
\expandafter\def\csname pn@wiki/qt/bridgecomparison/route/complete/ci\endcsname{[$3.5$, $6.4$]}
\expandafter\def\csname pn@wiki/qt/bridgecomparison/route/complete/full\endcsname{$+4.9$ [$3.5$, $6.4$]}
\expandafter\def\csname pn@wiki/qt/bridgecomparison/window/EM\endcsname{$+1.6$}
\expandafter\def\csname pn@wiki/qt/bridgecomparison/window/EM/abs\endcsname{1.6}
\expandafter\def\csname pn@wiki/qt/bridgecomparison/window/EM/ci\endcsname{[$-1.3$, $4.5$]}
\expandafter\def\csname pn@wiki/qt/bridgecomparison/window/EM/full\endcsname{$+1.6$ [$-1.3$, $4.5$]}
\expandafter\def\csname pn@wiki/qt/bridgecomparison/window/complete\endcsname{$+5.5$}
\expandafter\def\csname pn@wiki/qt/bridgecomparison/window/complete/abs\endcsname{5.5}
\expandafter\def\csname pn@wiki/qt/bridgecomparison/window/complete/ci\endcsname{[$3.9$, $7.2$]}
\expandafter\def\csname pn@wiki/qt/bridgecomparison/window/complete/full\endcsname{$+5.5$ [$3.9$, $7.2$]}
\expandafter\def\csname pn@wiki/qt/comparison/detour/EM\endcsname{$+3.1$}
\expandafter\def\csname pn@wiki/qt/comparison/detour/EM/abs\endcsname{3.1}
\expandafter\def\csname pn@wiki/qt/comparison/detour/EM/ci\endcsname{[$0.5$, $5.7$]}
\expandafter\def\csname pn@wiki/qt/comparison/detour/EM/full\endcsname{$+3.1$ [$0.5$, $5.7$]}
\expandafter\def\csname pn@wiki/qt/comparison/detour/complete\endcsname{$+6.5$}
\expandafter\def\csname pn@wiki/qt/comparison/detour/complete/abs\endcsname{6.5}
\expandafter\def\csname pn@wiki/qt/comparison/detour/complete/ci\endcsname{[$4.9$, $8.3$]}
\expandafter\def\csname pn@wiki/qt/comparison/detour/complete/full\endcsname{$+6.5$ [$4.9$, $8.3$]}
\expandafter\def\csname pn@wiki/qt/comparison/n\endcsname{382}
\expandafter\def\csname pn@wiki/qt/comparison/route/EM\endcsname{$-0.3$}
\expandafter\def\csname pn@wiki/qt/comparison/route/EM/abs\endcsname{0.3}
\expandafter\def\csname pn@wiki/qt/comparison/route/EM/ci\endcsname{[$-1.8$, $1.4$]}
\expandafter\def\csname pn@wiki/qt/comparison/route/EM/full\endcsname{$-0.3$ [$-1.8$, $1.4$]}
\expandafter\def\csname pn@wiki/qt/comparison/route/complete\endcsname{$0.0$}
\expandafter\def\csname pn@wiki/qt/comparison/route/complete/abs\endcsname{0.0}
\expandafter\def\csname pn@wiki/qt/comparison/route/complete/ci\endcsname{[$-0.7$, $0.6$]}
\expandafter\def\csname pn@wiki/qt/comparison/route/complete/full\endcsname{$0.0$ [$-0.7$, $0.6$]}
\expandafter\def\csname pn@wiki/qt/comparison/window/EM\endcsname{$+3.4$}
\expandafter\def\csname pn@wiki/qt/comparison/window/EM/abs\endcsname{3.4}
\expandafter\def\csname pn@wiki/qt/comparison/window/EM/ci\endcsname{[$0.8$, $6.0$]}
\expandafter\def\csname pn@wiki/qt/comparison/window/EM/full\endcsname{$+3.4$ [$0.8$, $6.0$]}
\expandafter\def\csname pn@wiki/qt/comparison/window/complete\endcsname{$+6.5$}
\expandafter\def\csname pn@wiki/qt/comparison/window/complete/abs\endcsname{6.5}
\expandafter\def\csname pn@wiki/qt/comparison/window/complete/ci\endcsname{[$4.8$, $8.4$]}
\expandafter\def\csname pn@wiki/qt/comparison/window/complete/full\endcsname{$+6.5$ [$4.8$, $8.4$]}
\expandafter\def\csname pn@wiki/qt/compositional/detour/EM\endcsname{$+4.5$}
\expandafter\def\csname pn@wiki/qt/compositional/detour/EM/abs\endcsname{4.5}
\expandafter\def\csname pn@wiki/qt/compositional/detour/EM/ci\endcsname{[$3.1$, $5.9$]}
\expandafter\def\csname pn@wiki/qt/compositional/detour/EM/full\endcsname{$+4.5$ [$3.1$, $5.9$]}
\expandafter\def\csname pn@wiki/qt/compositional/detour/complete\endcsname{$+14.5$}
\expandafter\def\csname pn@wiki/qt/compositional/detour/complete/abs\endcsname{14.5}
\expandafter\def\csname pn@wiki/qt/compositional/detour/complete/ci\endcsname{[$12.8$, $16.3$]}
\expandafter\def\csname pn@wiki/qt/compositional/detour/complete/full\endcsname{$+14.5$ [$12.8$, $16.3$]}
\expandafter\def\csname pn@wiki/qt/compositional/n\endcsname{623}
\expandafter\def\csname pn@wiki/qt/compositional/route/EM\endcsname{$+2.5$}
\expandafter\def\csname pn@wiki/qt/compositional/route/EM/abs\endcsname{2.5}
\expandafter\def\csname pn@wiki/qt/compositional/route/EM/ci\endcsname{[$1.5$, $3.5$]}
\expandafter\def\csname pn@wiki/qt/compositional/route/EM/full\endcsname{$+2.5$ [$1.5$, $3.5$]}
\expandafter\def\csname pn@wiki/qt/compositional/route/complete\endcsname{$+10.6$}
\expandafter\def\csname pn@wiki/qt/compositional/route/complete/abs\endcsname{10.6}
\expandafter\def\csname pn@wiki/qt/compositional/route/complete/ci\endcsname{[$9.1$, $12.1$]}
\expandafter\def\csname pn@wiki/qt/compositional/route/complete/full\endcsname{$+10.6$ [$9.1$, $12.1$]}
\expandafter\def\csname pn@wiki/qt/compositional/window/EM\endcsname{$+1.9$}
\expandafter\def\csname pn@wiki/qt/compositional/window/EM/abs\endcsname{1.9}
\expandafter\def\csname pn@wiki/qt/compositional/window/EM/ci\endcsname{[$0.8$, $3.1$]}
\expandafter\def\csname pn@wiki/qt/compositional/window/EM/full\endcsname{$+1.9$ [$0.8$, $3.1$]}
\expandafter\def\csname pn@wiki/qt/compositional/window/complete\endcsname{$+3.9$}
\expandafter\def\csname pn@wiki/qt/compositional/window/complete/abs\endcsname{3.9}
\expandafter\def\csname pn@wiki/qt/compositional/window/complete/ci\endcsname{[$3.0$, $4.9$]}
\expandafter\def\csname pn@wiki/qt/compositional/window/complete/full\endcsname{$+3.9$ [$3.0$, $4.9$]}
\expandafter\def\csname pn@wiki/qt/inference/detour/EM\endcsname{$+3.9$}
\expandafter\def\csname pn@wiki/qt/inference/detour/EM/abs\endcsname{3.9}
\expandafter\def\csname pn@wiki/qt/inference/detour/EM/ci\endcsname{[$1.4$, $6.5$]}
\expandafter\def\csname pn@wiki/qt/inference/detour/EM/full\endcsname{$+3.9$ [$1.4$, $6.5$]}
\expandafter\def\csname pn@wiki/qt/inference/detour/complete\endcsname{$+20.3$}
\expandafter\def\csname pn@wiki/qt/inference/detour/complete/abs\endcsname{20.3}
\expandafter\def\csname pn@wiki/qt/inference/detour/complete/ci\endcsname{[$16.7$, $24.0$]}
\expandafter\def\csname pn@wiki/qt/inference/detour/complete/full\endcsname{$+20.3$ [$16.7$, $24.0$]}
\expandafter\def\csname pn@wiki/qt/inference/n\endcsname{192}
\expandafter\def\csname pn@wiki/qt/inference/route/EM\endcsname{$+3.8$}
\expandafter\def\csname pn@wiki/qt/inference/route/EM/abs\endcsname{3.8}
\expandafter\def\csname pn@wiki/qt/inference/route/EM/ci\endcsname{[$1.8$, $5.9$]}
\expandafter\def\csname pn@wiki/qt/inference/route/EM/full\endcsname{$+3.8$ [$1.8$, $5.9$]}
\expandafter\def\csname pn@wiki/qt/inference/route/complete\endcsname{$+12.9$}
\expandafter\def\csname pn@wiki/qt/inference/route/complete/abs\endcsname{12.9}
\expandafter\def\csname pn@wiki/qt/inference/route/complete/ci\endcsname{[$9.9$, $16.0$]}
\expandafter\def\csname pn@wiki/qt/inference/route/complete/full\endcsname{$+12.9$ [$9.9$, $16.0$]}
\expandafter\def\csname pn@wiki/qt/inference/window/EM\endcsname{$+0.1$}
\expandafter\def\csname pn@wiki/qt/inference/window/EM/abs\endcsname{0.1}
\expandafter\def\csname pn@wiki/qt/inference/window/EM/ci\endcsname{[$-1.7$, $2.1$]}
\expandafter\def\csname pn@wiki/qt/inference/window/EM/full\endcsname{$+0.1$ [$-1.7$, $2.1$]}
\expandafter\def\csname pn@wiki/qt/inference/window/complete\endcsname{$+7.4$}
\expandafter\def\csname pn@wiki/qt/inference/window/complete/abs\endcsname{7.4}
\expandafter\def\csname pn@wiki/qt/inference/window/complete/ci\endcsname{[$5.2$, $9.9$]}
\expandafter\def\csname pn@wiki/qt/inference/window/complete/full\endcsname{$+7.4$ [$5.2$, $9.9$]}
\expandafter\def\csname pn@wiki/randomroute/EM/Decomp\endcsname{$+1.8$}
\expandafter\def\csname pn@wiki/randomroute/EM/Decomp/abs\endcsname{1.8}
\expandafter\def\csname pn@wiki/randomroute/EM/Decomp/ci\endcsname{[$0.7$, $2.8$]}
\expandafter\def\csname pn@wiki/randomroute/EM/Decomp/full\endcsname{$+1.8$ [$0.7$, $2.8$]}
\expandafter\def\csname pn@wiki/randomroute/EM/IRCoT\endcsname{$+1.0$}
\expandafter\def\csname pn@wiki/randomroute/EM/IRCoT/abs\endcsname{1.0}
\expandafter\def\csname pn@wiki/randomroute/EM/IRCoT/ci\endcsname{[$-0.2$, $2.3$]}
\expandafter\def\csname pn@wiki/randomroute/EM/IRCoT/full\endcsname{$+1.0$ [$-0.2$, $2.3$]}
\expandafter\def\csname pn@wiki/randomroute/EM/ReAct\endcsname{$+2.0$}
\expandafter\def\csname pn@wiki/randomroute/EM/ReAct/abs\endcsname{2.0}
\expandafter\def\csname pn@wiki/randomroute/EM/ReAct/ci\endcsname{[$0.9$, $3.2$]}
\expandafter\def\csname pn@wiki/randomroute/EM/ReAct/full\endcsname{$+2.0$ [$0.9$, $3.2$]}
\expandafter\def\csname pn@wiki/randomroute/EM/SR1\endcsname{$+0.1$}
\expandafter\def\csname pn@wiki/randomroute/EM/SR1/abs\endcsname{0.1}
\expandafter\def\csname pn@wiki/randomroute/EM/SR1/ci\endcsname{[$-0.5$, $0.7$]}
\expandafter\def\csname pn@wiki/randomroute/EM/SR1/full\endcsname{$+0.1$ [$-0.5$, $0.7$]}
\expandafter\def\csname pn@wiki/randomroute/EM/ext\endcsname{$+1.0$}
\expandafter\def\csname pn@wiki/randomroute/EM/ext/abs\endcsname{1.0}
\expandafter\def\csname pn@wiki/randomroute/EM/ext/ci\endcsname{[$0.4$, $1.7$]}
\expandafter\def\csname pn@wiki/randomroute/EM/ext/full\endcsname{$+1.0$ [$0.4$, $1.7$]}
\expandafter\def\csname pn@wiki/randomroute/EM/nsig\endcsname{2}
\expandafter\def\csname pn@wiki/randomroute/EM/pool\endcsname{$+1.2$}
\expandafter\def\csname pn@wiki/randomroute/EM/pool/abs\endcsname{1.0}
\expandafter\def\csname pn@wiki/randomroute/EM/pool/ci\endcsname{[$0.6$, $1.8$]}
\expandafter\def\csname pn@wiki/randomroute/EM/pool/full\endcsname{$+1.2$ [$0.6$, $1.8$]}
\expandafter\def\csname pn@wiki/randomroute/EM/range\endcsname{$+0.1$ to $+2.0$}
\expandafter\def\csname pn@wiki/randomroute/ans/Decomp\endcsname{$+6.2$}
\expandafter\def\csname pn@wiki/randomroute/ans/Decomp/abs\endcsname{6.2}
\expandafter\def\csname pn@wiki/randomroute/ans/Decomp/ci\endcsname{[$5.0$, $7.5$]}
\expandafter\def\csname pn@wiki/randomroute/ans/Decomp/full\endcsname{$+6.2$ [$5.0$, $7.5$]}
\expandafter\def\csname pn@wiki/randomroute/ans/IRCoT\endcsname{$+0.9$}
\expandafter\def\csname pn@wiki/randomroute/ans/IRCoT/abs\endcsname{0.9}
\expandafter\def\csname pn@wiki/randomroute/ans/IRCoT/ci\endcsname{[$0.1$, $1.8$]}
\expandafter\def\csname pn@wiki/randomroute/ans/IRCoT/full\endcsname{$+0.9$ [$0.1$, $1.8$]}
\expandafter\def\csname pn@wiki/randomroute/ans/ReAct\endcsname{$+4.3$}
\expandafter\def\csname pn@wiki/randomroute/ans/ReAct/abs\endcsname{4.3}
\expandafter\def\csname pn@wiki/randomroute/ans/ReAct/ci\endcsname{[$3.1$, $5.5$]}
\expandafter\def\csname pn@wiki/randomroute/ans/ReAct/full\endcsname{$+4.3$ [$3.1$, $5.5$]}
\expandafter\def\csname pn@wiki/randomroute/ans/SR1\endcsname{$+1.1$}
\expandafter\def\csname pn@wiki/randomroute/ans/SR1/abs\endcsname{1.1}
\expandafter\def\csname pn@wiki/randomroute/ans/SR1/ci\endcsname{[$0.5$, $1.7$]}
\expandafter\def\csname pn@wiki/randomroute/ans/SR1/full\endcsname{$+1.1$ [$0.5$, $1.7$]}
\expandafter\def\csname pn@wiki/randomroute/ans/ext\endcsname{$+2.1$}
\expandafter\def\csname pn@wiki/randomroute/ans/ext/abs\endcsname{2.1}
\expandafter\def\csname pn@wiki/randomroute/ans/ext/ci\endcsname{[$1.5$, $2.7$]}
\expandafter\def\csname pn@wiki/randomroute/ans/ext/full\endcsname{$+2.1$ [$1.5$, $2.7$]}
\expandafter\def\csname pn@wiki/randomroute/ans/nsig\endcsname{4}
\expandafter\def\csname pn@wiki/randomroute/ans/pool\endcsname{$+3.1$}
\expandafter\def\csname pn@wiki/randomroute/ans/pool/abs\endcsname{3.1}
\expandafter\def\csname pn@wiki/randomroute/ans/pool/ci\endcsname{[$2.5$, $3.8$]}
\expandafter\def\csname pn@wiki/randomroute/ans/pool/full\endcsname{$+3.1$ [$2.5$, $3.8$]}
\expandafter\def\csname pn@wiki/randomroute/ans/range\endcsname{$+0.9$ to $+6.2$}
\expandafter\def\csname pn@wiki/randomroute/complete/Decomp\endcsname{$+8.8$}
\expandafter\def\csname pn@wiki/randomroute/complete/Decomp/abs\endcsname{8.8}
\expandafter\def\csname pn@wiki/randomroute/complete/Decomp/ci\endcsname{[$7.4$, $10.2$]}
\expandafter\def\csname pn@wiki/randomroute/complete/Decomp/full\endcsname{$+8.8$ [$7.4$, $10.2$]}
\expandafter\def\csname pn@wiki/randomroute/complete/IRCoT\endcsname{$+2.6$}
\expandafter\def\csname pn@wiki/randomroute/complete/IRCoT/abs\endcsname{2.6}
\expandafter\def\csname pn@wiki/randomroute/complete/IRCoT/ci\endcsname{[$1.7$, $3.5$]}
\expandafter\def\csname pn@wiki/randomroute/complete/IRCoT/full\endcsname{$+2.6$ [$1.7$, $3.5$]}
\expandafter\def\csname pn@wiki/randomroute/complete/ReAct\endcsname{$+6.8$}
\expandafter\def\csname pn@wiki/randomroute/complete/ReAct/abs\endcsname{6.8}
\expandafter\def\csname pn@wiki/randomroute/complete/ReAct/ci\endcsname{[$5.6$, $8.1$]}
\expandafter\def\csname pn@wiki/randomroute/complete/ReAct/full\endcsname{$+6.8$ [$5.6$, $8.1$]}
\expandafter\def\csname pn@wiki/randomroute/complete/SR1\endcsname{$+1.2$}
\expandafter\def\csname pn@wiki/randomroute/complete/SR1/abs\endcsname{1.2}
\expandafter\def\csname pn@wiki/randomroute/complete/SR1/ci\endcsname{[$0.6$, $1.8$]}
\expandafter\def\csname pn@wiki/randomroute/complete/SR1/full\endcsname{$+1.2$ [$0.6$, $1.8$]}
\expandafter\def\csname pn@wiki/randomroute/complete/ext\endcsname{$+3.5$}
\expandafter\def\csname pn@wiki/randomroute/complete/ext/abs\endcsname{3.5}
\expandafter\def\csname pn@wiki/randomroute/complete/ext/ci\endcsname{[$2.9$, $4.2$]}
\expandafter\def\csname pn@wiki/randomroute/complete/ext/full\endcsname{$+3.5$ [$2.9$, $4.2$]}
\expandafter\def\csname pn@wiki/randomroute/complete/nsig\endcsname{4}
\expandafter\def\csname pn@wiki/randomroute/complete/pool\endcsname{$+4.8$}
\expandafter\def\csname pn@wiki/randomroute/complete/pool/abs\endcsname{4.8}
\expandafter\def\csname pn@wiki/randomroute/complete/pool/ci\endcsname{[$4.2$, $5.5$]}
\expandafter\def\csname pn@wiki/randomroute/complete/pool/full\endcsname{$+4.8$ [$4.2$, $5.5$]}
\expandafter\def\csname pn@wiki/randomroute/complete/range\endcsname{$+1.2$ to $+8.8$}
\expandafter\def\csname pn@wiki/rerankVSrrf/EM/Decomp\endcsname{$+2.9$}
\expandafter\def\csname pn@wiki/rerankVSrrf/EM/Decomp/abs\endcsname{2.9}
\expandafter\def\csname pn@wiki/rerankVSrrf/EM/Decomp/ci\endcsname{[$1.0$, $4.8$]}
\expandafter\def\csname pn@wiki/rerankVSrrf/EM/Decomp/full\endcsname{$+2.9$ [$1.0$, $4.8$]}
\expandafter\def\csname pn@wiki/rerankVSrrf/EM/IRCoT\endcsname{$+2.2$}
\expandafter\def\csname pn@wiki/rerankVSrrf/EM/IRCoT/abs\endcsname{2.2}
\expandafter\def\csname pn@wiki/rerankVSrrf/EM/IRCoT/ci\endcsname{[$-0.1$, $4.5$]}
\expandafter\def\csname pn@wiki/rerankVSrrf/EM/IRCoT/full\endcsname{$+2.2$ [$-0.1$, $4.5$]}
\expandafter\def\csname pn@wiki/rerankVSrrf/EM/ReAct\endcsname{$+5.2$}
\expandafter\def\csname pn@wiki/rerankVSrrf/EM/ReAct/abs\endcsname{5.2}
\expandafter\def\csname pn@wiki/rerankVSrrf/EM/ReAct/ci\endcsname{[$3.1$, $7.3$]}
\expandafter\def\csname pn@wiki/rerankVSrrf/EM/ReAct/full\endcsname{$+5.2$ [$3.1$, $7.3$]}
\expandafter\def\csname pn@wiki/rerankVSrrf/EM/SR1\endcsname{$+0.2$}
\expandafter\def\csname pn@wiki/rerankVSrrf/EM/SR1/abs\endcsname{0.2}
\expandafter\def\csname pn@wiki/rerankVSrrf/EM/SR1/ci\endcsname{[$-1.9$, $2.5$]}
\expandafter\def\csname pn@wiki/rerankVSrrf/EM/SR1/full\endcsname{$+0.2$ [$-1.9$, $2.5$]}
\expandafter\def\csname pn@wiki/rerankVSrrf/EM/ext\endcsname{$+2.5$}
\expandafter\def\csname pn@wiki/rerankVSrrf/EM/ext/abs\endcsname{2.5}
\expandafter\def\csname pn@wiki/rerankVSrrf/EM/ext/ci\endcsname{[$1.0$, $4.1$]}
\expandafter\def\csname pn@wiki/rerankVSrrf/EM/ext/full\endcsname{$+2.5$ [$1.0$, $4.1$]}
\expandafter\def\csname pn@wiki/rerankVSrrf/EM/nsig\endcsname{2}
\expandafter\def\csname pn@wiki/rerankVSrrf/EM/pool\endcsname{$+2.6$}
\expandafter\def\csname pn@wiki/rerankVSrrf/EM/pool/abs\endcsname{2.6}
\expandafter\def\csname pn@wiki/rerankVSrrf/EM/pool/ci\endcsname{[$1.2$, $4.0$]}
\expandafter\def\csname pn@wiki/rerankVSrrf/EM/pool/full\endcsname{$+2.6$ [$1.2$, $4.0$]}
\expandafter\def\csname pn@wiki/rerankVSrrf/EM/range\endcsname{$+0.2$ to $+5.2$}
\expandafter\def\csname pn@wiki/rerankVSrrf/ans/Decomp\endcsname{$+1.7$}
\expandafter\def\csname pn@wiki/rerankVSrrf/ans/Decomp/abs\endcsname{1.7}
\expandafter\def\csname pn@wiki/rerankVSrrf/ans/Decomp/ci\endcsname{[$0.3$, $3.2$]}
\expandafter\def\csname pn@wiki/rerankVSrrf/ans/Decomp/full\endcsname{$+1.7$ [$0.3$, $3.2$]}
\expandafter\def\csname pn@wiki/rerankVSrrf/ans/IRCoT\endcsname{$+5.3$}
\expandafter\def\csname pn@wiki/rerankVSrrf/ans/IRCoT/abs\endcsname{5.3}
\expandafter\def\csname pn@wiki/rerankVSrrf/ans/IRCoT/ci\endcsname{[$3.3$, $7.4$]}
\expandafter\def\csname pn@wiki/rerankVSrrf/ans/IRCoT/full\endcsname{$+5.3$ [$3.3$, $7.4$]}
\expandafter\def\csname pn@wiki/rerankVSrrf/ans/ReAct\endcsname{$+8.9$}
\expandafter\def\csname pn@wiki/rerankVSrrf/ans/ReAct/abs\endcsname{8.9}
\expandafter\def\csname pn@wiki/rerankVSrrf/ans/ReAct/ci\endcsname{[$6.8$, $10.9$]}
\expandafter\def\csname pn@wiki/rerankVSrrf/ans/ReAct/full\endcsname{$+8.9$ [$6.8$, $10.9$]}
\expandafter\def\csname pn@wiki/rerankVSrrf/ans/SR1\endcsname{$+1.4$}
\expandafter\def\csname pn@wiki/rerankVSrrf/ans/SR1/abs\endcsname{1.4}
\expandafter\def\csname pn@wiki/rerankVSrrf/ans/SR1/ci\endcsname{[$-0.4$, $3.1$]}
\expandafter\def\csname pn@wiki/rerankVSrrf/ans/SR1/full\endcsname{$+1.4$ [$-0.4$, $3.1$]}
\expandafter\def\csname pn@wiki/rerankVSrrf/ans/ext\endcsname{$+5.2$}
\expandafter\def\csname pn@wiki/rerankVSrrf/ans/ext/abs\endcsname{5.2}
\expandafter\def\csname pn@wiki/rerankVSrrf/ans/ext/ci\endcsname{[$3.9$, $6.5$]}
\expandafter\def\csname pn@wiki/rerankVSrrf/ans/ext/full\endcsname{$+5.2$ [$3.9$, $6.5$]}
\expandafter\def\csname pn@wiki/rerankVSrrf/ans/nsig\endcsname{3}
\expandafter\def\csname pn@wiki/rerankVSrrf/ans/pool\endcsname{$+4.3$}
\expandafter\def\csname pn@wiki/rerankVSrrf/ans/pool/abs\endcsname{4.3}
\expandafter\def\csname pn@wiki/rerankVSrrf/ans/pool/ci\endcsname{[$3.2$, $5.4$]}
\expandafter\def\csname pn@wiki/rerankVSrrf/ans/pool/full\endcsname{$+4.3$ [$3.2$, $5.4$]}
\expandafter\def\csname pn@wiki/rerankVSrrf/ans/range\endcsname{$+1.4$ to $+8.9$}
\expandafter\def\csname pn@wiki/rerankVSrrf/complete/Decomp\endcsname{$+0.4$}
\expandafter\def\csname pn@wiki/rerankVSrrf/complete/Decomp/abs\endcsname{0.4}
\expandafter\def\csname pn@wiki/rerankVSrrf/complete/Decomp/ci\endcsname{[$-0.9$, $1.7$]}
\expandafter\def\csname pn@wiki/rerankVSrrf/complete/Decomp/full\endcsname{$+0.4$ [$-0.9$, $1.7$]}
\expandafter\def\csname pn@wiki/rerankVSrrf/complete/IRCoT\endcsname{$+5.3$}
\expandafter\def\csname pn@wiki/rerankVSrrf/complete/IRCoT/abs\endcsname{5.3}
\expandafter\def\csname pn@wiki/rerankVSrrf/complete/IRCoT/ci\endcsname{[$2.9$, $7.7$]}
\expandafter\def\csname pn@wiki/rerankVSrrf/complete/IRCoT/full\endcsname{$+5.3$ [$2.9$, $7.7$]}
\expandafter\def\csname pn@wiki/rerankVSrrf/complete/ReAct\endcsname{$+8.3$}
\expandafter\def\csname pn@wiki/rerankVSrrf/complete/ReAct/abs\endcsname{8.3}
\expandafter\def\csname pn@wiki/rerankVSrrf/complete/ReAct/ci\endcsname{[$6.3$, $10.3$]}
\expandafter\def\csname pn@wiki/rerankVSrrf/complete/ReAct/full\endcsname{$+8.3$ [$6.3$, $10.3$]}
\expandafter\def\csname pn@wiki/rerankVSrrf/complete/SR1\endcsname{$+1.0$}
\expandafter\def\csname pn@wiki/rerankVSrrf/complete/SR1/abs\endcsname{1.0}
\expandafter\def\csname pn@wiki/rerankVSrrf/complete/SR1/ci\endcsname{[$-1.2$, $3.1$]}
\expandafter\def\csname pn@wiki/rerankVSrrf/complete/SR1/full\endcsname{$+1.0$ [$-1.2$, $3.1$]}
\expandafter\def\csname pn@wiki/rerankVSrrf/complete/ext\endcsname{$+4.9$}
\expandafter\def\csname pn@wiki/rerankVSrrf/complete/ext/abs\endcsname{4.9}
\expandafter\def\csname pn@wiki/rerankVSrrf/complete/ext/ci\endcsname{[$3.3$, $6.4$]}
\expandafter\def\csname pn@wiki/rerankVSrrf/complete/ext/full\endcsname{$+4.9$ [$3.3$, $6.4$]}
\expandafter\def\csname pn@wiki/rerankVSrrf/complete/nsig\endcsname{2}
\expandafter\def\csname pn@wiki/rerankVSrrf/complete/pool\endcsname{$+3.8$}
\expandafter\def\csname pn@wiki/rerankVSrrf/complete/pool/abs\endcsname{3.8}
\expandafter\def\csname pn@wiki/rerankVSrrf/complete/pool/ci\endcsname{[$2.5$, $5.0$]}
\expandafter\def\csname pn@wiki/rerankVSrrf/complete/pool/full\endcsname{$+3.8$ [$2.5$, $5.0$]}
\expandafter\def\csname pn@wiki/rerankVSrrf/complete/range\endcsname{$+0.4$ to $+8.3$}
\expandafter\def\csname pn@wiki/rerankcalls/mean\endcsname{27.5}
\expandafter\def\csname pn@wiki/rerankcalls/range\endcsname{25--32}
\expandafter\def\csname pn@wiki/rerankwin/EM/Decomp\endcsname{$+0.2$}
\expandafter\def\csname pn@wiki/rerankwin/EM/Decomp/abs\endcsname{0.2}
\expandafter\def\csname pn@wiki/rerankwin/EM/Decomp/ci\endcsname{[$-1.4$, $1.8$]}
\expandafter\def\csname pn@wiki/rerankwin/EM/Decomp/full\endcsname{$+0.2$ [$-1.4$, $1.8$]}
\expandafter\def\csname pn@wiki/rerankwin/EM/IRCoT\endcsname{$+1.0$}
\expandafter\def\csname pn@wiki/rerankwin/EM/IRCoT/abs\endcsname{1.0}
\expandafter\def\csname pn@wiki/rerankwin/EM/IRCoT/ci\endcsname{[$-0.8$, $2.8$]}
\expandafter\def\csname pn@wiki/rerankwin/EM/IRCoT/full\endcsname{$+1.0$ [$-0.8$, $2.8$]}
\expandafter\def\csname pn@wiki/rerankwin/EM/ReAct\endcsname{$+1.1$}
\expandafter\def\csname pn@wiki/rerankwin/EM/ReAct/abs\endcsname{1.1}
\expandafter\def\csname pn@wiki/rerankwin/EM/ReAct/ci\endcsname{[$-0.6$, $2.9$]}
\expandafter\def\csname pn@wiki/rerankwin/EM/ReAct/full\endcsname{$+1.1$ [$-0.6$, $2.9$]}
\expandafter\def\csname pn@wiki/rerankwin/EM/SR1\endcsname{$+1.4$}
\expandafter\def\csname pn@wiki/rerankwin/EM/SR1/abs\endcsname{1.4}
\expandafter\def\csname pn@wiki/rerankwin/EM/SR1/ci\endcsname{[$-0.5$, $3.4$]}
\expandafter\def\csname pn@wiki/rerankwin/EM/SR1/full\endcsname{$+1.4$ [$-0.5$, $3.4$]}
\expandafter\def\csname pn@wiki/rerankwin/EM/ext\endcsname{$+1.2$}
\expandafter\def\csname pn@wiki/rerankwin/EM/ext/abs\endcsname{1.2}
\expandafter\def\csname pn@wiki/rerankwin/EM/ext/ci\endcsname{[$-0.2$, $2.5$]}
\expandafter\def\csname pn@wiki/rerankwin/EM/ext/full\endcsname{$+1.2$ [$-0.2$, $2.5$]}
\expandafter\def\csname pn@wiki/rerankwin/EM/nsig\endcsname{0}
\expandafter\def\csname pn@wiki/rerankwin/EM/pool\endcsname{$+0.9$}
\expandafter\def\csname pn@wiki/rerankwin/EM/pool/abs\endcsname{0.9}
\expandafter\def\csname pn@wiki/rerankwin/EM/pool/ci\endcsname{[$-0.2$, $2.1$]}
\expandafter\def\csname pn@wiki/rerankwin/EM/pool/full\endcsname{$+0.9$ [$-0.2$, $2.1$]}
\expandafter\def\csname pn@wiki/rerankwin/EM/range\endcsname{$+0.2$ to $+1.4$}
\expandafter\def\csname pn@wiki/rerankwin/ans/Decomp\endcsname{$+1.7$}
\expandafter\def\csname pn@wiki/rerankwin/ans/Decomp/abs\endcsname{1.7}
\expandafter\def\csname pn@wiki/rerankwin/ans/Decomp/ci\endcsname{[$0.9$, $2.6$]}
\expandafter\def\csname pn@wiki/rerankwin/ans/Decomp/full\endcsname{$+1.7$ [$0.9$, $2.6$]}
\expandafter\def\csname pn@wiki/rerankwin/ans/IRCoT\endcsname{$+2.1$}
\expandafter\def\csname pn@wiki/rerankwin/ans/IRCoT/abs\endcsname{2.1}
\expandafter\def\csname pn@wiki/rerankwin/ans/IRCoT/ci\endcsname{[$1.2$, $3.1$]}
\expandafter\def\csname pn@wiki/rerankwin/ans/IRCoT/full\endcsname{$+2.1$ [$1.2$, $3.1$]}
\expandafter\def\csname pn@wiki/rerankwin/ans/ReAct\endcsname{$+2.6$}
\expandafter\def\csname pn@wiki/rerankwin/ans/ReAct/abs\endcsname{2.6}
\expandafter\def\csname pn@wiki/rerankwin/ans/ReAct/ci\endcsname{[$1.6$, $3.7$]}
\expandafter\def\csname pn@wiki/rerankwin/ans/ReAct/full\endcsname{$+2.6$ [$1.6$, $3.7$]}
\expandafter\def\csname pn@wiki/rerankwin/ans/SR1\endcsname{$+2.2$}
\expandafter\def\csname pn@wiki/rerankwin/ans/SR1/abs\endcsname{2.2}
\expandafter\def\csname pn@wiki/rerankwin/ans/SR1/ci\endcsname{[$1.1$, $3.3$]}
\expandafter\def\csname pn@wiki/rerankwin/ans/SR1/full\endcsname{$+2.2$ [$1.1$, $3.3$]}
\expandafter\def\csname pn@wiki/rerankwin/ans/ext\endcsname{$+2.3$}
\expandafter\def\csname pn@wiki/rerankwin/ans/ext/abs\endcsname{2.3}
\expandafter\def\csname pn@wiki/rerankwin/ans/ext/ci\endcsname{[$1.6$, $3.0$]}
\expandafter\def\csname pn@wiki/rerankwin/ans/ext/full\endcsname{$+2.3$ [$1.6$, $3.0$]}
\expandafter\def\csname pn@wiki/rerankwin/ans/nsig\endcsname{4}
\expandafter\def\csname pn@wiki/rerankwin/ans/pool\endcsname{$+2.2$}
\expandafter\def\csname pn@wiki/rerankwin/ans/pool/abs\endcsname{2.2}
\expandafter\def\csname pn@wiki/rerankwin/ans/pool/ci\endcsname{[$1.6$, $2.8$]}
\expandafter\def\csname pn@wiki/rerankwin/ans/pool/full\endcsname{$+2.2$ [$1.6$, $2.8$]}
\expandafter\def\csname pn@wiki/rerankwin/ans/range\endcsname{$+1.7$ to $+2.6$}
\expandafter\def\csname pn@wiki/rerankwin/complete/Decomp\endcsname{$+2.8$}
\expandafter\def\csname pn@wiki/rerankwin/complete/Decomp/abs\endcsname{2.8}
\expandafter\def\csname pn@wiki/rerankwin/complete/Decomp/ci\endcsname{[$2.0$, $3.7$]}
\expandafter\def\csname pn@wiki/rerankwin/complete/Decomp/full\endcsname{$+2.8$ [$2.0$, $3.7$]}
\expandafter\def\csname pn@wiki/rerankwin/complete/IRCoT\endcsname{$+5.7$}
\expandafter\def\csname pn@wiki/rerankwin/complete/IRCoT/abs\endcsname{5.7}
\expandafter\def\csname pn@wiki/rerankwin/complete/IRCoT/ci\endcsname{[$4.5$, $7.0$]}
\expandafter\def\csname pn@wiki/rerankwin/complete/IRCoT/full\endcsname{$+5.7$ [$4.5$, $7.0$]}
\expandafter\def\csname pn@wiki/rerankwin/complete/ReAct\endcsname{$+5.1$}
\expandafter\def\csname pn@wiki/rerankwin/complete/ReAct/abs\endcsname{5.1}
\expandafter\def\csname pn@wiki/rerankwin/complete/ReAct/ci\endcsname{[$3.9$, $6.3$]}
\expandafter\def\csname pn@wiki/rerankwin/complete/ReAct/full\endcsname{$+5.1$ [$3.9$, $6.3$]}
\expandafter\def\csname pn@wiki/rerankwin/complete/SR1\endcsname{$+6.5$}
\expandafter\def\csname pn@wiki/rerankwin/complete/SR1/abs\endcsname{6.5}
\expandafter\def\csname pn@wiki/rerankwin/complete/SR1/ci\endcsname{[$5.1$, $7.9$]}
\expandafter\def\csname pn@wiki/rerankwin/complete/SR1/full\endcsname{$+6.5$ [$5.1$, $7.9$]}
\expandafter\def\csname pn@wiki/rerankwin/complete/ext\endcsname{$+5.8$}
\expandafter\def\csname pn@wiki/rerankwin/complete/ext/abs\endcsname{5.8}
\expandafter\def\csname pn@wiki/rerankwin/complete/ext/ci\endcsname{[$4.9$, $6.7$]}
\expandafter\def\csname pn@wiki/rerankwin/complete/ext/full\endcsname{$+5.8$ [$4.9$, $6.7$]}
\expandafter\def\csname pn@wiki/rerankwin/complete/nsig\endcsname{4}
\expandafter\def\csname pn@wiki/rerankwin/complete/pool\endcsname{$+5.0$}
\expandafter\def\csname pn@wiki/rerankwin/complete/pool/abs\endcsname{5.0}
\expandafter\def\csname pn@wiki/rerankwin/complete/pool/ci\endcsname{[$4.3$, $5.8$]}
\expandafter\def\csname pn@wiki/rerankwin/complete/pool/full\endcsname{$+5.0$ [$4.3$, $5.8$]}
\expandafter\def\csname pn@wiki/rerankwin/complete/range\endcsname{$+2.8$ to $+6.5$}
\expandafter\def\csname pn@wiki/rewired/EM/Decomp\endcsname{$+0.8$}
\expandafter\def\csname pn@wiki/rewired/EM/Decomp/abs\endcsname{0.8}
\expandafter\def\csname pn@wiki/rewired/EM/Decomp/ci\endcsname{[$-0.1$, $1.8$]}
\expandafter\def\csname pn@wiki/rewired/EM/Decomp/full\endcsname{$+0.8$ [$-0.1$, $1.8$]}
\expandafter\def\csname pn@wiki/rewired/EM/IRCoT\endcsname{$+0.8$}
\expandafter\def\csname pn@wiki/rewired/EM/IRCoT/abs\endcsname{0.8}
\expandafter\def\csname pn@wiki/rewired/EM/IRCoT/ci\endcsname{[$-0.5$, $2.1$]}
\expandafter\def\csname pn@wiki/rewired/EM/IRCoT/full\endcsname{$+0.8$ [$-0.5$, $2.1$]}
\expandafter\def\csname pn@wiki/rewired/EM/ReAct\endcsname{$0.0$}
\expandafter\def\csname pn@wiki/rewired/EM/ReAct/abs\endcsname{0.0}
\expandafter\def\csname pn@wiki/rewired/EM/ReAct/ci\endcsname{[$-1.2$, $1.2$]}
\expandafter\def\csname pn@wiki/rewired/EM/ReAct/full\endcsname{$0.0$ [$-1.2$, $1.2$]}
\expandafter\def\csname pn@wiki/rewired/EM/SR1\endcsname{$+1.5$}
\expandafter\def\csname pn@wiki/rewired/EM/SR1/abs\endcsname{1.5}
\expandafter\def\csname pn@wiki/rewired/EM/SR1/ci\endcsname{[$0.2$, $2.7$]}
\expandafter\def\csname pn@wiki/rewired/EM/SR1/full\endcsname{$+1.5$ [$0.2$, $2.7$]}
\expandafter\def\csname pn@wiki/rewired/EM/ext\endcsname{$+0.8$}
\expandafter\def\csname pn@wiki/rewired/EM/ext/abs\endcsname{0.8}
\expandafter\def\csname pn@wiki/rewired/EM/ext/ci\endcsname{[$0.0$, $1.6$]}
\expandafter\def\csname pn@wiki/rewired/EM/ext/full\endcsname{$+0.8$ [$0.0$, $1.6$]}
\expandafter\def\csname pn@wiki/rewired/EM/nsig\endcsname{0}
\expandafter\def\csname pn@wiki/rewired/EM/pool\endcsname{$+0.8$}
\expandafter\def\csname pn@wiki/rewired/EM/pool/abs\endcsname{0.8}
\expandafter\def\csname pn@wiki/rewired/EM/pool/ci\endcsname{[$0.1$, $1.4$]}
\expandafter\def\csname pn@wiki/rewired/EM/pool/full\endcsname{$+0.8$ [$0.1$, $1.4$]}
\expandafter\def\csname pn@wiki/rewired/EM/range\endcsname{$0.0$ to $+1.5$}
\expandafter\def\csname pn@wiki/rewired/ans/Decomp\endcsname{$+0.5$}
\expandafter\def\csname pn@wiki/rewired/ans/Decomp/abs\endcsname{0.5}
\expandafter\def\csname pn@wiki/rewired/ans/Decomp/ci\endcsname{[$0.0$, $1.0$]}
\expandafter\def\csname pn@wiki/rewired/ans/Decomp/full\endcsname{$+0.5$ [$0.0$, $1.0$]}
\expandafter\def\csname pn@wiki/rewired/ans/IRCoT\endcsname{$+0.5$}
\expandafter\def\csname pn@wiki/rewired/ans/IRCoT/abs\endcsname{0.5}
\expandafter\def\csname pn@wiki/rewired/ans/IRCoT/ci\endcsname{[$0.1$, $0.9$]}
\expandafter\def\csname pn@wiki/rewired/ans/IRCoT/full\endcsname{$+0.5$ [$0.1$, $0.9$]}
\expandafter\def\csname pn@wiki/rewired/ans/ReAct\endcsname{$+0.4$}
\expandafter\def\csname pn@wiki/rewired/ans/ReAct/abs\endcsname{0.4}
\expandafter\def\csname pn@wiki/rewired/ans/ReAct/ci\endcsname{[$-0.1$, $0.9$]}
\expandafter\def\csname pn@wiki/rewired/ans/ReAct/full\endcsname{$+0.4$ [$-0.1$, $0.9$]}
\expandafter\def\csname pn@wiki/rewired/ans/SR1\endcsname{$+0.6$}
\expandafter\def\csname pn@wiki/rewired/ans/SR1/abs\endcsname{0.6}
\expandafter\def\csname pn@wiki/rewired/ans/SR1/ci\endcsname{[$0.1$, $1.1$]}
\expandafter\def\csname pn@wiki/rewired/ans/SR1/full\endcsname{$+0.6$ [$0.1$, $1.1$]}
\expandafter\def\csname pn@wiki/rewired/ans/ext\endcsname{$+0.5$}
\expandafter\def\csname pn@wiki/rewired/ans/ext/abs\endcsname{0.5}
\expandafter\def\csname pn@wiki/rewired/ans/ext/ci\endcsname{[$0.2$, $0.8$]}
\expandafter\def\csname pn@wiki/rewired/ans/ext/full\endcsname{$+0.5$ [$0.2$, $0.8$]}
\expandafter\def\csname pn@wiki/rewired/ans/nsig\endcsname{0}
\expandafter\def\csname pn@wiki/rewired/ans/pool\endcsname{$+0.5$}
\expandafter\def\csname pn@wiki/rewired/ans/pool/abs\endcsname{0.5}
\expandafter\def\csname pn@wiki/rewired/ans/pool/ci\endcsname{[$0.2$, $0.7$]}
\expandafter\def\csname pn@wiki/rewired/ans/pool/full\endcsname{$+0.5$ [$0.2$, $0.7$]}
\expandafter\def\csname pn@wiki/rewired/ans/range\endcsname{$+0.4$ to $+0.6$}
\expandafter\def\csname pn@wiki/rewired/complete/Decomp\endcsname{$+0.9$}
\expandafter\def\csname pn@wiki/rewired/complete/Decomp/abs\endcsname{0.9}
\expandafter\def\csname pn@wiki/rewired/complete/Decomp/ci\endcsname{[$0.5$, $1.5$]}
\expandafter\def\csname pn@wiki/rewired/complete/Decomp/full\endcsname{$+0.9$ [$0.5$, $1.5$]}
\expandafter\def\csname pn@wiki/rewired/complete/IRCoT\endcsname{$+0.7$}
\expandafter\def\csname pn@wiki/rewired/complete/IRCoT/abs\endcsname{0.7}
\expandafter\def\csname pn@wiki/rewired/complete/IRCoT/ci\endcsname{[$0.2$, $1.3$]}
\expandafter\def\csname pn@wiki/rewired/complete/IRCoT/full\endcsname{$+0.7$ [$0.2$, $1.3$]}
\expandafter\def\csname pn@wiki/rewired/complete/ReAct\endcsname{$+1.1$}
\expandafter\def\csname pn@wiki/rewired/complete/ReAct/abs\endcsname{1.1}
\expandafter\def\csname pn@wiki/rewired/complete/ReAct/ci\endcsname{[$0.5$, $1.8$]}
\expandafter\def\csname pn@wiki/rewired/complete/ReAct/full\endcsname{$+1.1$ [$0.5$, $1.8$]}
\expandafter\def\csname pn@wiki/rewired/complete/SR1\endcsname{$+1.5$}
\expandafter\def\csname pn@wiki/rewired/complete/SR1/abs\endcsname{1.5}
\expandafter\def\csname pn@wiki/rewired/complete/SR1/ci\endcsname{[$0.9$, $2.1$]}
\expandafter\def\csname pn@wiki/rewired/complete/SR1/full\endcsname{$+1.5$ [$0.9$, $2.1$]}
\expandafter\def\csname pn@wiki/rewired/complete/ext\endcsname{$+1.1$}
\expandafter\def\csname pn@wiki/rewired/complete/ext/abs\endcsname{1.1}
\expandafter\def\csname pn@wiki/rewired/complete/ext/ci\endcsname{[$0.7$, $1.5$]}
\expandafter\def\csname pn@wiki/rewired/complete/ext/full\endcsname{$+1.1$ [$0.7$, $1.5$]}
\expandafter\def\csname pn@wiki/rewired/complete/nsig\endcsname{4}
\expandafter\def\csname pn@wiki/rewired/complete/pool\endcsname{$+1.1$}
\expandafter\def\csname pn@wiki/rewired/complete/pool/abs\endcsname{1.1}
\expandafter\def\csname pn@wiki/rewired/complete/pool/ci\endcsname{[$0.8$, $1.4$]}
\expandafter\def\csname pn@wiki/rewired/complete/pool/full\endcsname{$+1.1$ [$0.8$, $1.4$]}
\expandafter\def\csname pn@wiki/rewired/complete/range\endcsname{$+0.7$ to $+1.5$}
\expandafter\def\csname pn@wiki/rewired10/ans/Decomp\endcsname{$+0.6$}
\expandafter\def\csname pn@wiki/rewired10/ans/Decomp/abs\endcsname{0.6}
\expandafter\def\csname pn@wiki/rewired10/ans/Decomp/ci\endcsname{[$0.2$, $1.0$]}
\expandafter\def\csname pn@wiki/rewired10/ans/Decomp/full\endcsname{$+0.6$ [$0.2$, $1.0$]}
\expandafter\def\csname pn@wiki/rewired10/ans/IRCoT\endcsname{$+0.3$}
\expandafter\def\csname pn@wiki/rewired10/ans/IRCoT/abs\endcsname{0.3}
\expandafter\def\csname pn@wiki/rewired10/ans/IRCoT/ci\endcsname{[$0.0$, $0.6$]}
\expandafter\def\csname pn@wiki/rewired10/ans/IRCoT/full\endcsname{$+0.3$ [$0.0$, $0.6$]}
\expandafter\def\csname pn@wiki/rewired10/ans/ReAct\endcsname{$+0.5$}
\expandafter\def\csname pn@wiki/rewired10/ans/ReAct/abs\endcsname{0.5}
\expandafter\def\csname pn@wiki/rewired10/ans/ReAct/ci\endcsname{[$0.1$, $0.9$]}
\expandafter\def\csname pn@wiki/rewired10/ans/ReAct/full\endcsname{$+0.5$ [$0.1$, $0.9$]}
\expandafter\def\csname pn@wiki/rewired10/ans/SR1\endcsname{$+0.5$}
\expandafter\def\csname pn@wiki/rewired10/ans/SR1/abs\endcsname{0.5}
\expandafter\def\csname pn@wiki/rewired10/ans/SR1/ci\endcsname{[$0.1$, $0.9$]}
\expandafter\def\csname pn@wiki/rewired10/ans/SR1/full\endcsname{$+0.5$ [$0.1$, $0.9$]}
\expandafter\def\csname pn@wiki/rewired10/ans/ext\endcsname{$+0.4$}
\expandafter\def\csname pn@wiki/rewired10/ans/ext/abs\endcsname{0.4}
\expandafter\def\csname pn@wiki/rewired10/ans/ext/ci\endcsname{[$0.2$, $0.7$]}
\expandafter\def\csname pn@wiki/rewired10/ans/ext/full\endcsname{$+0.4$ [$0.2$, $0.7$]}
\expandafter\def\csname pn@wiki/rewired10/ans/nsig\endcsname{3}
\expandafter\def\csname pn@wiki/rewired10/ans/pool\endcsname{$+0.5$}
\expandafter\def\csname pn@wiki/rewired10/ans/pool/abs\endcsname{0.5}
\expandafter\def\csname pn@wiki/rewired10/ans/pool/ci\endcsname{[$0.3$, $0.7$]}
\expandafter\def\csname pn@wiki/rewired10/ans/pool/full\endcsname{$+0.5$ [$0.3$, $0.7$]}
\expandafter\def\csname pn@wiki/rewired10/ans/range\endcsname{$+0.3$ to $+0.6$}
\expandafter\def\csname pn@wiki/rewired10/complete/Decomp\endcsname{$+1.0$}
\expandafter\def\csname pn@wiki/rewired10/complete/Decomp/abs\endcsname{1.0}
\expandafter\def\csname pn@wiki/rewired10/complete/Decomp/ci\endcsname{[$0.6$, $1.4$]}
\expandafter\def\csname pn@wiki/rewired10/complete/Decomp/full\endcsname{$+1.0$ [$0.6$, $1.4$]}
\expandafter\def\csname pn@wiki/rewired10/complete/IRCoT\endcsname{$+0.6$}
\expandafter\def\csname pn@wiki/rewired10/complete/IRCoT/abs\endcsname{0.6}
\expandafter\def\csname pn@wiki/rewired10/complete/IRCoT/ci\endcsname{[$0.2$, $1.0$]}
\expandafter\def\csname pn@wiki/rewired10/complete/IRCoT/full\endcsname{$+0.6$ [$0.2$, $1.0$]}
\expandafter\def\csname pn@wiki/rewired10/complete/ReAct\endcsname{$+1.4$}
\expandafter\def\csname pn@wiki/rewired10/complete/ReAct/abs\endcsname{1.4}
\expandafter\def\csname pn@wiki/rewired10/complete/ReAct/ci\endcsname{[$0.9$, $1.9$]}
\expandafter\def\csname pn@wiki/rewired10/complete/ReAct/full\endcsname{$+1.4$ [$0.9$, $1.9$]}
\expandafter\def\csname pn@wiki/rewired10/complete/SR1\endcsname{$+1.3$}
\expandafter\def\csname pn@wiki/rewired10/complete/SR1/abs\endcsname{1.3}
\expandafter\def\csname pn@wiki/rewired10/complete/SR1/ci\endcsname{[$0.9$, $1.8$]}
\expandafter\def\csname pn@wiki/rewired10/complete/SR1/full\endcsname{$+1.3$ [$0.9$, $1.8$]}
\expandafter\def\csname pn@wiki/rewired10/complete/ext\endcsname{$+1.1$}
\expandafter\def\csname pn@wiki/rewired10/complete/ext/abs\endcsname{1.1}
\expandafter\def\csname pn@wiki/rewired10/complete/ext/ci\endcsname{[$0.8$, $1.4$]}
\expandafter\def\csname pn@wiki/rewired10/complete/ext/full\endcsname{$+1.1$ [$0.8$, $1.4$]}
\expandafter\def\csname pn@wiki/rewired10/complete/nsig\endcsname{4}
\expandafter\def\csname pn@wiki/rewired10/complete/pool\endcsname{$+1.1$}
\expandafter\def\csname pn@wiki/rewired10/complete/pool/abs\endcsname{1.1}
\expandafter\def\csname pn@wiki/rewired10/complete/pool/ci\endcsname{[$0.8$, $1.3$]}
\expandafter\def\csname pn@wiki/rewired10/complete/pool/full\endcsname{$+1.1$ [$0.8$, $1.3$]}
\expandafter\def\csname pn@wiki/rewired10/complete/range\endcsname{$+0.6$ to $+1.4$}
\expandafter\def\csname pn@wiki/route/EM/Decomp\endcsname{$+2.1$}
\expandafter\def\csname pn@wiki/route/EM/Decomp/abs\endcsname{2.1}
\expandafter\def\csname pn@wiki/route/EM/Decomp/ci\endcsname{[$0.9$, $3.3$]}
\expandafter\def\csname pn@wiki/route/EM/Decomp/full\endcsname{$+2.1$ [$0.9$, $3.3$]}
\expandafter\def\csname pn@wiki/route/EM/IRCoT\endcsname{$-0.1$}
\expandafter\def\csname pn@wiki/route/EM/IRCoT/abs\endcsname{0.1}
\expandafter\def\csname pn@wiki/route/EM/IRCoT/ci\endcsname{[$-1.6$, $1.5$]}
\expandafter\def\csname pn@wiki/route/EM/IRCoT/full\endcsname{$-0.1$ [$-1.6$, $1.5$]}
\expandafter\def\csname pn@wiki/route/EM/IRCoTSR1\endcsname{$+0.1$}
\expandafter\def\csname pn@wiki/route/EM/IRCoTSR1/abs\endcsname{0.1}
\expandafter\def\csname pn@wiki/route/EM/IRCoTSR1/ci\endcsname{[$-0.9$, $1.0$]}
\expandafter\def\csname pn@wiki/route/EM/IRCoTSR1/full\endcsname{$+0.1$ [$-0.9$, $1.0$]}
\expandafter\def\csname pn@wiki/route/EM/ReAct\endcsname{$+0.4$}
\expandafter\def\csname pn@wiki/route/EM/ReAct/abs\endcsname{3.9}
\expandafter\def\csname pn@wiki/route/EM/ReAct/ci\endcsname{[$2.5$, $5.3$]}
\expandafter\def\csname pn@wiki/route/EM/ReAct/full\endcsname{$+3.9$ [$2.5$, $5.3$]}
\expandafter\def\csname pn@wiki/route/EM/SR1\endcsname{$+0.2$}
\expandafter\def\csname pn@wiki/route/EM/SR1/abs\endcsname{0.2}
\expandafter\def\csname pn@wiki/route/EM/SR1/ci\endcsname{[$-0.7$, $1.0$]}
\expandafter\def\csname pn@wiki/route/EM/SR1/full\endcsname{$+0.2$ [$-0.7$, $1.0$]}
\expandafter\def\csname pn@wiki/route/EM/ext\endcsname{$+1.3$}
\expandafter\def\csname pn@wiki/route/EM/ext/abs\endcsname{1.3}
\expandafter\def\csname pn@wiki/route/EM/ext/ci\endcsname{[$0.5$, $2.2$]}
\expandafter\def\csname pn@wiki/route/EM/ext/full\endcsname{$+1.3$ [$0.5$, $2.2$]}
\expandafter\def\csname pn@wiki/route/EM/nsig\endcsname{2}
\expandafter\def\csname pn@wiki/route/EM/pool\endcsname{$+1.5$}
\expandafter\def\csname pn@wiki/route/EM/pool/abs\endcsname{1.5}
\expandafter\def\csname pn@wiki/route/EM/pool/ci\endcsname{[$0.8$, $2.3$]}
\expandafter\def\csname pn@wiki/route/EM/pool/full\endcsname{$+1.5$ [$0.8$, $2.3$]}
\expandafter\def\csname pn@wiki/route/EM/range\endcsname{$-0.1$ to $+3.9$}
\expandafter\def\csname pn@wiki/route/ans/Decomp\endcsname{$+8.7$}
\expandafter\def\csname pn@wiki/route/ans/Decomp/abs\endcsname{8.7}
\expandafter\def\csname pn@wiki/route/ans/Decomp/ci\endcsname{[$7.2$, $10.3$]}
\expandafter\def\csname pn@wiki/route/ans/Decomp/full\endcsname{$+8.7$ [$7.2$, $10.3$]}
\expandafter\def\csname pn@wiki/route/ans/IRCoT\endcsname{$+2.1$}
\expandafter\def\csname pn@wiki/route/ans/IRCoT/abs\endcsname{2.1}
\expandafter\def\csname pn@wiki/route/ans/IRCoT/ci\endcsname{[$1.1$, $3.2$]}
\expandafter\def\csname pn@wiki/route/ans/IRCoT/full\endcsname{$+2.1$ [$1.1$, $3.2$]}
\expandafter\def\csname pn@wiki/route/ans/ReAct\endcsname{$+7.7$}
\expandafter\def\csname pn@wiki/route/ans/ReAct/abs\endcsname{7.7}
\expandafter\def\csname pn@wiki/route/ans/ReAct/ci\endcsname{[$6.1$, $9.4$]}
\expandafter\def\csname pn@wiki/route/ans/ReAct/full\endcsname{$+7.7$ [$6.1$, $9.4$]}
\expandafter\def\csname pn@wiki/route/ans/SR1\endcsname{$+1.4$}
\expandafter\def\csname pn@wiki/route/ans/SR1/abs\endcsname{1.4}
\expandafter\def\csname pn@wiki/route/ans/SR1/ci\endcsname{[$0.7$, $2.2$]}
\expandafter\def\csname pn@wiki/route/ans/SR1/full\endcsname{$+1.4$ [$0.7$, $2.2$]}
\expandafter\def\csname pn@wiki/route/ans/ext\endcsname{$+3.8$}
\expandafter\def\csname pn@wiki/route/ans/ext/abs\endcsname{3.8}
\expandafter\def\csname pn@wiki/route/ans/ext/ci\endcsname{[$3.0$, $4.6$]}
\expandafter\def\csname pn@wiki/route/ans/ext/full\endcsname{$+3.8$ [$3.0$, $4.6$]}
\expandafter\def\csname pn@wiki/route/ans/nsig\endcsname{4}
\expandafter\def\csname pn@wiki/route/ans/pool\endcsname{$+5.0$}
\expandafter\def\csname pn@wiki/route/ans/pool/abs\endcsname{5.0}
\expandafter\def\csname pn@wiki/route/ans/pool/ci\endcsname{[$4.2$, $5.8$]}
\expandafter\def\csname pn@wiki/route/ans/pool/full\endcsname{$+5.0$ [$4.2$, $5.8$]}
\expandafter\def\csname pn@wiki/route/ans/range\endcsname{$+1.4$ to $+8.7$}
\expandafter\def\csname pn@wiki/route/complete/Decomp\endcsname{$+11.3$}
\expandafter\def\csname pn@wiki/route/complete/Decomp/abs\endcsname{11.3}
\expandafter\def\csname pn@wiki/route/complete/Decomp/ci\endcsname{[$9.7$, $12.9$]}
\expandafter\def\csname pn@wiki/route/complete/Decomp/full\endcsname{$+11.3$ [$9.7$, $12.9$]}
\expandafter\def\csname pn@wiki/route/complete/IRCoT\endcsname{$+4.5$}
\expandafter\def\csname pn@wiki/route/complete/IRCoT/abs\endcsname{4.5}
\expandafter\def\csname pn@wiki/route/complete/IRCoT/ci\endcsname{[$3.4$, $5.6$]}
\expandafter\def\csname pn@wiki/route/complete/IRCoT/full\endcsname{$+4.5$ [$3.4$, $5.6$]}
\expandafter\def\csname pn@wiki/route/complete/ReAct\endcsname{$+10.9$}
\expandafter\def\csname pn@wiki/route/complete/ReAct/abs\endcsname{10.9}
\expandafter\def\csname pn@wiki/route/complete/ReAct/ci\endcsname{[$9.2$, $12.6$]}
\expandafter\def\csname pn@wiki/route/complete/ReAct/full\endcsname{$+10.9$ [$9.2$, $12.6$]}
\expandafter\def\csname pn@wiki/route/complete/SR1\endcsname{$+1.5$}
\expandafter\def\csname pn@wiki/route/complete/SR1/abs\endcsname{1.5}
\expandafter\def\csname pn@wiki/route/complete/SR1/ci\endcsname{[$0.7$, $2.3$]}
\expandafter\def\csname pn@wiki/route/complete/SR1/full\endcsname{$+1.5$ [$0.7$, $2.3$]}
\expandafter\def\csname pn@wiki/route/complete/ext\endcsname{$+5.6$}
\expandafter\def\csname pn@wiki/route/complete/ext/abs\endcsname{5.6}
\expandafter\def\csname pn@wiki/route/complete/ext/ci\endcsname{[$4.8$, $6.5$]}
\expandafter\def\csname pn@wiki/route/complete/ext/full\endcsname{$+5.6$ [$4.8$, $6.5$]}
\expandafter\def\csname pn@wiki/route/complete/nsig\endcsname{4}
\expandafter\def\csname pn@wiki/route/complete/pool\endcsname{$+7.0$}
\expandafter\def\csname pn@wiki/route/complete/pool/abs\endcsname{7.0}
\expandafter\def\csname pn@wiki/route/complete/pool/ci\endcsname{[$6.2$, $7.9$]}
\expandafter\def\csname pn@wiki/route/complete/pool/full\endcsname{$+7.0$ [$6.2$, $7.9$]}
\expandafter\def\csname pn@wiki/route/complete/range\endcsname{$+1.5$ to $+11.3$}
\expandafter\def\csname pn@wiki/routed/Decomp\endcsname{0.65}
\expandafter\def\csname pn@wiki/routed/IRCoT\endcsname{1.87}
\expandafter\def\csname pn@wiki/routed/ReAct\endcsname{1.59}
\expandafter\def\csname pn@wiki/routed/SR1\endcsname{0.42}
\expandafter\def\csname pn@wiki/routedshare/Decomp\endcsname{12.9}
\expandafter\def\csname pn@wiki/routedshare/IRCoT\endcsname{37.5}
\expandafter\def\csname pn@wiki/routedshare/ReAct\endcsname{34.3}
\expandafter\def\csname pn@wiki/routedshare/SR1\endcsname{15.1}
\expandafter\def\csname pn@wiki/routeonly/EM/Decomp\endcsname{$+1.3$}
\expandafter\def\csname pn@wiki/routeonly/EM/Decomp/abs\endcsname{1.3}
\expandafter\def\csname pn@wiki/routeonly/EM/Decomp/ci\endcsname{[$0.3$, $2.3$]}
\expandafter\def\csname pn@wiki/routeonly/EM/Decomp/full\endcsname{$+1.3$ [$0.3$, $2.3$]}
\expandafter\def\csname pn@wiki/routeonly/EM/IRCoT\endcsname{$+1.5$}
\expandafter\def\csname pn@wiki/routeonly/EM/IRCoT/abs\endcsname{1.5}
\expandafter\def\csname pn@wiki/routeonly/EM/IRCoT/ci\endcsname{[$0.1$, $2.9$]}
\expandafter\def\csname pn@wiki/routeonly/EM/IRCoT/full\endcsname{$+1.5$ [$0.1$, $2.9$]}
\expandafter\def\csname pn@wiki/routeonly/EM/ReAct\endcsname{$+2.8$}
\expandafter\def\csname pn@wiki/routeonly/EM/ReAct/abs\endcsname{2.8}
\expandafter\def\csname pn@wiki/routeonly/EM/ReAct/ci\endcsname{[$1.5$, $4.1$]}
\expandafter\def\csname pn@wiki/routeonly/EM/ReAct/full\endcsname{$+2.8$ [$1.5$, $4.1$]}
\expandafter\def\csname pn@wiki/routeonly/EM/SR1\endcsname{$+0.1$}
\expandafter\def\csname pn@wiki/routeonly/EM/SR1/abs\endcsname{0.1}
\expandafter\def\csname pn@wiki/routeonly/EM/SR1/ci\endcsname{[$-0.6$, $0.9$]}
\expandafter\def\csname pn@wiki/routeonly/EM/SR1/full\endcsname{$+0.1$ [$-0.6$, $0.9$]}
\expandafter\def\csname pn@wiki/routeonly/EM/ext\endcsname{$+1.5$}
\expandafter\def\csname pn@wiki/routeonly/EM/ext/abs\endcsname{1.5}
\expandafter\def\csname pn@wiki/routeonly/EM/ext/ci\endcsname{[$0.8$, $2.2$]}
\expandafter\def\csname pn@wiki/routeonly/EM/ext/full\endcsname{$+1.5$ [$0.8$, $2.2$]}
\expandafter\def\csname pn@wiki/routeonly/EM/nsig\endcsname{2}
\expandafter\def\csname pn@wiki/routeonly/EM/pool\endcsname{$+1.4$}
\expandafter\def\csname pn@wiki/routeonly/EM/pool/abs\endcsname{1.4}
\expandafter\def\csname pn@wiki/routeonly/EM/pool/ci\endcsname{[$0.8$, $2.0$]}
\expandafter\def\csname pn@wiki/routeonly/EM/pool/full\endcsname{$+1.4$ [$0.8$, $2.0$]}
\expandafter\def\csname pn@wiki/routeonly/EM/range\endcsname{$+0.1$ to $+2.8$}
\expandafter\def\csname pn@wiki/routeonly/ans/Decomp\endcsname{$+6.7$}
\expandafter\def\csname pn@wiki/routeonly/ans/Decomp/abs\endcsname{6.7}
\expandafter\def\csname pn@wiki/routeonly/ans/Decomp/ci\endcsname{[$5.3$, $8.2$]}
\expandafter\def\csname pn@wiki/routeonly/ans/Decomp/full\endcsname{$+6.7$ [$5.3$, $8.2$]}
\expandafter\def\csname pn@wiki/routeonly/ans/IRCoT\endcsname{$+1.1$}
\expandafter\def\csname pn@wiki/routeonly/ans/IRCoT/abs\endcsname{1.1}
\expandafter\def\csname pn@wiki/routeonly/ans/IRCoT/ci\endcsname{[$0.0$, $2.0$]}
\expandafter\def\csname pn@wiki/routeonly/ans/IRCoT/full\endcsname{$+1.1$ [$0.0$, $2.0$]}
\expandafter\def\csname pn@wiki/routeonly/ans/ReAct\endcsname{$+5.5$}
\expandafter\def\csname pn@wiki/routeonly/ans/ReAct/abs\endcsname{5.5}
\expandafter\def\csname pn@wiki/routeonly/ans/ReAct/ci\endcsname{[$3.9$, $7.1$]}
\expandafter\def\csname pn@wiki/routeonly/ans/ReAct/full\endcsname{$+5.5$ [$3.9$, $7.1$]}
\expandafter\def\csname pn@wiki/routeonly/ans/SR1\endcsname{$+0.9$}
\expandafter\def\csname pn@wiki/routeonly/ans/SR1/abs\endcsname{0.9}
\expandafter\def\csname pn@wiki/routeonly/ans/SR1/ci\endcsname{[$0.1$, $1.7$]}
\expandafter\def\csname pn@wiki/routeonly/ans/SR1/full\endcsname{$+0.9$ [$0.1$, $1.7$]}
\expandafter\def\csname pn@wiki/routeonly/ans/ext\endcsname{$+2.5$}
\expandafter\def\csname pn@wiki/routeonly/ans/ext/abs\endcsname{2.5}
\expandafter\def\csname pn@wiki/routeonly/ans/ext/ci\endcsname{[$1.7$, $3.3$]}
\expandafter\def\csname pn@wiki/routeonly/ans/ext/full\endcsname{$+2.5$ [$1.7$, $3.3$]}
\expandafter\def\csname pn@wiki/routeonly/ans/nsig\endcsname{2}
\expandafter\def\csname pn@wiki/routeonly/ans/pool\endcsname{$+3.5$}
\expandafter\def\csname pn@wiki/routeonly/ans/pool/abs\endcsname{3.5}
\expandafter\def\csname pn@wiki/routeonly/ans/pool/ci\endcsname{[$2.8$, $4.3$]}
\expandafter\def\csname pn@wiki/routeonly/ans/pool/full\endcsname{$+3.5$ [$2.8$, $4.3$]}
\expandafter\def\csname pn@wiki/routeonly/ans/range\endcsname{$+0.9$ to $+6.7$}
\expandafter\def\csname pn@wiki/routeonly/complete/Decomp\endcsname{$+7.8$}
\expandafter\def\csname pn@wiki/routeonly/complete/Decomp/abs\endcsname{7.8}
\expandafter\def\csname pn@wiki/routeonly/complete/Decomp/ci\endcsname{[$6.3$, $9.3$]}
\expandafter\def\csname pn@wiki/routeonly/complete/Decomp/full\endcsname{$+7.8$ [$6.3$, $9.3$]}
\expandafter\def\csname pn@wiki/routeonly/complete/IRCoT\endcsname{$+2.3$}
\expandafter\def\csname pn@wiki/routeonly/complete/IRCoT/abs\endcsname{2.3}
\expandafter\def\csname pn@wiki/routeonly/complete/IRCoT/ci\endcsname{[$1.1$, $3.5$]}
\expandafter\def\csname pn@wiki/routeonly/complete/IRCoT/full\endcsname{$+2.3$ [$1.1$, $3.5$]}
\expandafter\def\csname pn@wiki/routeonly/complete/ReAct\endcsname{$+7.1$}
\expandafter\def\csname pn@wiki/routeonly/complete/ReAct/abs\endcsname{7.1}
\expandafter\def\csname pn@wiki/routeonly/complete/ReAct/ci\endcsname{[$5.5$, $8.7$]}
\expandafter\def\csname pn@wiki/routeonly/complete/ReAct/full\endcsname{$+7.1$ [$5.5$, $8.7$]}
\expandafter\def\csname pn@wiki/routeonly/complete/SR1\endcsname{$+0.7$}
\expandafter\def\csname pn@wiki/routeonly/complete/SR1/abs\endcsname{0.7}
\expandafter\def\csname pn@wiki/routeonly/complete/SR1/ci\endcsname{[$-0.2$, $1.5$]}
\expandafter\def\csname pn@wiki/routeonly/complete/SR1/full\endcsname{$+0.7$ [$-0.2$, $1.5$]}
\expandafter\def\csname pn@wiki/routeonly/complete/ext\endcsname{$+3.4$}
\expandafter\def\csname pn@wiki/routeonly/complete/ext/abs\endcsname{3.4}
\expandafter\def\csname pn@wiki/routeonly/complete/ext/ci\endcsname{[$2.5$, $4.2$]}
\expandafter\def\csname pn@wiki/routeonly/complete/ext/full\endcsname{$+3.4$ [$2.5$, $4.2$]}
\expandafter\def\csname pn@wiki/routeonly/complete/nsig\endcsname{3}
\expandafter\def\csname pn@wiki/routeonly/complete/pool\endcsname{$+4.5$}
\expandafter\def\csname pn@wiki/routeonly/complete/pool/abs\endcsname{4.5}
\expandafter\def\csname pn@wiki/routeonly/complete/pool/ci\endcsname{[$3.7$, $5.3$]}
\expandafter\def\csname pn@wiki/routeonly/complete/pool/full\endcsname{$+4.5$ [$3.7$, $5.3$]}
\expandafter\def\csname pn@wiki/routeonly/complete/range\endcsname{$+0.7$ to $+7.8$}
\expandafter\def\csname pn@wiki/rule/concatdd\endcsname{38.2}
\expandafter\def\csname pn@wiki/rule/concatnodd\endcsname{30.1}
\expandafter\def\csname pn@wiki/rule/coverage\endcsname{62.5}
\expandafter\def\csname pn@wiki/rule/g2dd\endcsname{55.9}
\expandafter\def\csname pn@wiki/rule/g2nodd\endcsname{54.7}
\expandafter\def\csname pn@wiki/rule/rrfdd\endcsname{52.8}
\expandafter\def\csname pn@wiki/rule/rrfnodd\endcsname{48.1}
\expandafter\def\csname pn@wiki/rule/scoremaxdd\endcsname{48.6}
\expandafter\def\csname pn@wiki/rule/scoremaxnodd\endcsname{45.9}
\expandafter\def\csname pn@wiki/rule/top2dd\endcsname{44.1}
\expandafter\def\csname pn@wiki/rule/top2nodd\endcsname{44.1}
\expandafter\def\csname pn@wiki/rule/top3dd\endcsname{50.6}
\expandafter\def\csname pn@wiki/rule/top3nodd\endcsname{49.7}
\expandafter\def\csname pn@wiki/simgraph/EM/Decomp\endcsname{$+0.6$}
\expandafter\def\csname pn@wiki/simgraph/EM/Decomp/abs\endcsname{0.6}
\expandafter\def\csname pn@wiki/simgraph/EM/Decomp/ci\endcsname{[$-0.7$, $1.9$]}
\expandafter\def\csname pn@wiki/simgraph/EM/Decomp/full\endcsname{$+0.6$ [$-0.7$, $1.9$]}
\expandafter\def\csname pn@wiki/simgraph/EM/IRCoT\endcsname{$+0.5$}
\expandafter\def\csname pn@wiki/simgraph/EM/IRCoT/abs\endcsname{0.5}
\expandafter\def\csname pn@wiki/simgraph/EM/IRCoT/ci\endcsname{[$-0.9$, $2.1$]}
\expandafter\def\csname pn@wiki/simgraph/EM/IRCoT/full\endcsname{$+0.5$ [$-0.9$, $2.1$]}
\expandafter\def\csname pn@wiki/simgraph/EM/ReAct\endcsname{$+0.1$}
\expandafter\def\csname pn@wiki/simgraph/EM/ReAct/abs\endcsname{0.1}
\expandafter\def\csname pn@wiki/simgraph/EM/ReAct/ci\endcsname{[$-1.2$, $1.5$]}
\expandafter\def\csname pn@wiki/simgraph/EM/ReAct/full\endcsname{$+0.1$ [$-1.2$, $1.5$]}
\expandafter\def\csname pn@wiki/simgraph/EM/SR1\endcsname{$+1.7$}
\expandafter\def\csname pn@wiki/simgraph/EM/SR1/abs\endcsname{1.7}
\expandafter\def\csname pn@wiki/simgraph/EM/SR1/ci\endcsname{[$0.1$, $3.3$]}
\expandafter\def\csname pn@wiki/simgraph/EM/SR1/full\endcsname{$+1.7$ [$0.1$, $3.3$]}
\expandafter\def\csname pn@wiki/simgraph/EM/ext\endcsname{$+0.8$}
\expandafter\def\csname pn@wiki/simgraph/EM/ext/abs\endcsname{0.8}
\expandafter\def\csname pn@wiki/simgraph/EM/ext/ci\endcsname{[$-0.2$, $1.8$]}
\expandafter\def\csname pn@wiki/simgraph/EM/ext/full\endcsname{$+0.8$ [$-0.2$, $1.8$]}
\expandafter\def\csname pn@wiki/simgraph/EM/nsig\endcsname{0}
\expandafter\def\csname pn@wiki/simgraph/EM/pool\endcsname{$+0.8$}
\expandafter\def\csname pn@wiki/simgraph/EM/pool/abs\endcsname{0.8}
\expandafter\def\csname pn@wiki/simgraph/EM/pool/ci\endcsname{[$-0.1$, $1.6$]}
\expandafter\def\csname pn@wiki/simgraph/EM/pool/full\endcsname{$+0.8$ [$-0.1$, $1.6$]}
\expandafter\def\csname pn@wiki/simgraph/EM/range\endcsname{$+0.1$ to $+1.7$}
\expandafter\def\csname pn@wiki/simgraph/ans/Decomp\endcsname{$+2.2$}
\expandafter\def\csname pn@wiki/simgraph/ans/Decomp/abs\endcsname{2.2}
\expandafter\def\csname pn@wiki/simgraph/ans/Decomp/ci\endcsname{[$1.3$, $3.1$]}
\expandafter\def\csname pn@wiki/simgraph/ans/Decomp/full\endcsname{$+2.2$ [$1.3$, $3.1$]}
\expandafter\def\csname pn@wiki/simgraph/ans/IRCoT\endcsname{$+0.7$}
\expandafter\def\csname pn@wiki/simgraph/ans/IRCoT/abs\endcsname{0.7}
\expandafter\def\csname pn@wiki/simgraph/ans/IRCoT/ci\endcsname{[$0.3$, $1.1$]}
\expandafter\def\csname pn@wiki/simgraph/ans/IRCoT/full\endcsname{$+0.7$ [$0.3$, $1.1$]}
\expandafter\def\csname pn@wiki/simgraph/ans/ReAct\endcsname{$+1.4$}
\expandafter\def\csname pn@wiki/simgraph/ans/ReAct/abs\endcsname{1.4}
\expandafter\def\csname pn@wiki/simgraph/ans/ReAct/ci\endcsname{[$0.6$, $2.1$]}
\expandafter\def\csname pn@wiki/simgraph/ans/ReAct/full\endcsname{$+1.4$ [$0.6$, $2.1$]}
\expandafter\def\csname pn@wiki/simgraph/ans/SR1\endcsname{$+1.4$}
\expandafter\def\csname pn@wiki/simgraph/ans/SR1/abs\endcsname{1.4}
\expandafter\def\csname pn@wiki/simgraph/ans/SR1/ci\endcsname{[$0.5$, $2.2$]}
\expandafter\def\csname pn@wiki/simgraph/ans/SR1/full\endcsname{$+1.4$ [$0.5$, $2.2$]}
\expandafter\def\csname pn@wiki/simgraph/ans/ext\endcsname{$+1.1$}
\expandafter\def\csname pn@wiki/simgraph/ans/ext/abs\endcsname{1.1}
\expandafter\def\csname pn@wiki/simgraph/ans/ext/ci\endcsname{[$0.7$, $1.6$]}
\expandafter\def\csname pn@wiki/simgraph/ans/ext/full\endcsname{$+1.1$ [$0.7$, $1.6$]}
\expandafter\def\csname pn@wiki/simgraph/ans/nsig\endcsname{4}
\expandafter\def\csname pn@wiki/simgraph/ans/pool\endcsname{$+1.4$}
\expandafter\def\csname pn@wiki/simgraph/ans/pool/abs\endcsname{1.4}
\expandafter\def\csname pn@wiki/simgraph/ans/pool/ci\endcsname{[$1.0$, $1.8$]}
\expandafter\def\csname pn@wiki/simgraph/ans/pool/full\endcsname{$+1.4$ [$1.0$, $1.8$]}
\expandafter\def\csname pn@wiki/simgraph/ans/range\endcsname{$+0.7$ to $+2.2$}
\expandafter\def\csname pn@wiki/simgraph/complete/Decomp\endcsname{$+2.6$}
\expandafter\def\csname pn@wiki/simgraph/complete/Decomp/abs\endcsname{2.6}
\expandafter\def\csname pn@wiki/simgraph/complete/Decomp/ci\endcsname{[$1.8$, $3.5$]}
\expandafter\def\csname pn@wiki/simgraph/complete/Decomp/full\endcsname{$+2.6$ [$1.8$, $3.5$]}
\expandafter\def\csname pn@wiki/simgraph/complete/IRCoT\endcsname{$+1.8$}
\expandafter\def\csname pn@wiki/simgraph/complete/IRCoT/abs\endcsname{1.8}
\expandafter\def\csname pn@wiki/simgraph/complete/IRCoT/ci\endcsname{[$1.1$, $2.6$]}
\expandafter\def\csname pn@wiki/simgraph/complete/IRCoT/full\endcsname{$+1.8$ [$1.1$, $2.6$]}
\expandafter\def\csname pn@wiki/simgraph/complete/ReAct\endcsname{$+3.9$}
\expandafter\def\csname pn@wiki/simgraph/complete/ReAct/abs\endcsname{3.9}
\expandafter\def\csname pn@wiki/simgraph/complete/ReAct/ci\endcsname{[$2.8$, $4.9$]}
\expandafter\def\csname pn@wiki/simgraph/complete/ReAct/full\endcsname{$+3.9$ [$2.8$, $4.9$]}
\expandafter\def\csname pn@wiki/simgraph/complete/SR1\endcsname{$+5.6$}
\expandafter\def\csname pn@wiki/simgraph/complete/SR1/abs\endcsname{5.6}
\expandafter\def\csname pn@wiki/simgraph/complete/SR1/ci\endcsname{[$4.3$, $6.9$]}
\expandafter\def\csname pn@wiki/simgraph/complete/SR1/full\endcsname{$+5.6$ [$4.3$, $6.9$]}
\expandafter\def\csname pn@wiki/simgraph/complete/ext\endcsname{$+3.8$}
\expandafter\def\csname pn@wiki/simgraph/complete/ext/abs\endcsname{3.8}
\expandafter\def\csname pn@wiki/simgraph/complete/ext/ci\endcsname{[$3.1$, $4.5$]}
\expandafter\def\csname pn@wiki/simgraph/complete/ext/full\endcsname{$+3.8$ [$3.1$, $4.5$]}
\expandafter\def\csname pn@wiki/simgraph/complete/nsig\endcsname{4}
\expandafter\def\csname pn@wiki/simgraph/complete/pool\endcsname{$+3.5$}
\expandafter\def\csname pn@wiki/simgraph/complete/pool/abs\endcsname{3.5}
\expandafter\def\csname pn@wiki/simgraph/complete/pool/ci\endcsname{[$2.9$, $4.1$]}
\expandafter\def\csname pn@wiki/simgraph/complete/pool/full\endcsname{$+3.5$ [$2.9$, $4.1$]}
\expandafter\def\csname pn@wiki/simgraph/complete/range\endcsname{$+1.8$ to $+5.6$}
\expandafter\def\csname pn@wiki/sr1native\endcsname{44.6}
\expandafter\def\csname pn@wiki/sr1own/detour\endcsname{49.1}
\expandafter\def\csname pn@wiki/sr1own/detourVSinterleave\endcsname{$+1.9$}
\expandafter\def\csname pn@wiki/sr1own/detourVSinterleave/abs\endcsname{1.9}
\expandafter\def\csname pn@wiki/sr1own/detourVSinterleave/ci\endcsname{[$-0.2$, $3.9$]}
\expandafter\def\csname pn@wiki/sr1own/detourVSinterleave/full\endcsname{$+1.9$ [$-0.2$, $3.9$]}
\expandafter\def\csname pn@wiki/sr1own/detourVSnative\endcsname{$+4.5$}
\expandafter\def\csname pn@wiki/sr1own/detourVSnative/abs\endcsname{4.5}
\expandafter\def\csname pn@wiki/sr1own/detourVSnative/ci\endcsname{[$2.0$, $7.1$]}
\expandafter\def\csname pn@wiki/sr1own/detourVSnative/full\endcsname{$+4.5$ [$2.0$, $7.1$]}
\expandafter\def\csname pn@wiki/sr1own/interleave\endcsname{47.3}
\expandafter\def\csname pn@wiki/sr1own/interleaveVSnative\endcsname{$+2.7$}
\expandafter\def\csname pn@wiki/sr1own/interleaveVSnative/abs\endcsname{2.7}
\expandafter\def\csname pn@wiki/sr1own/interleaveVSnative/ci\endcsname{[$0.2$, $5.1$]}
\expandafter\def\csname pn@wiki/sr1own/interleaveVSnative/full\endcsname{$+2.7$ [$0.2$, $5.1$]}
\expandafter\def\csname pn@wiki/sr1own/native\endcsname{44.6}
\expandafter\def\csname pn@wiki/sr1own/rrf\endcsname{45.1}
\expandafter\def\csname pn@wiki/sr1own/rrfVSnative\endcsname{$+0.5$}
\expandafter\def\csname pn@wiki/sr1own/rrfVSnative/abs\endcsname{0.5}
\expandafter\def\csname pn@wiki/sr1own/rrfVSnative/ci\endcsname{[$-2.1$, $3.1$]}
\expandafter\def\csname pn@wiki/sr1own/rrfVSnative/full\endcsname{$+0.5$ [$-2.1$, $3.1$]}
\expandafter\def\csname pn@wiki/sr1own/window\endcsname{49.6}
\expandafter\def\csname pn@wiki/sr1own/windowVSinterleave\endcsname{$+2.3$}
\expandafter\def\csname pn@wiki/sr1own/windowVSinterleave/abs\endcsname{2.3}
\expandafter\def\csname pn@wiki/sr1own/windowVSinterleave/ci\endcsname{[$0.3$, $4.3$]}
\expandafter\def\csname pn@wiki/sr1own/windowVSinterleave/full\endcsname{$+2.3$ [$0.3$, $4.3$]}
\expandafter\def\csname pn@wiki/sr1own/windowVSnative\endcsname{$+5.0$}
\expandafter\def\csname pn@wiki/sr1own/windowVSnative/abs\endcsname{5.0}
\expandafter\def\csname pn@wiki/sr1own/windowVSnative/ci\endcsname{[$2.5$, $7.5$]}
\expandafter\def\csname pn@wiki/sr1own/windowVSnative/full\endcsname{$+5.0$ [$2.5$, $7.5$]}
\expandafter\def\csname pn@wiki/titleroute/EM/Decomp\endcsname{$+1.1$}
\expandafter\def\csname pn@wiki/titleroute/EM/Decomp/abs\endcsname{1.1}
\expandafter\def\csname pn@wiki/titleroute/EM/Decomp/ci\endcsname{[$0.1$, $2.1$]}
\expandafter\def\csname pn@wiki/titleroute/EM/Decomp/full\endcsname{$+1.1$ [$0.1$, $2.1$]}
\expandafter\def\csname pn@wiki/titleroute/EM/IRCoT\endcsname{$-0.2$}
\expandafter\def\csname pn@wiki/titleroute/EM/IRCoT/abs\endcsname{0.2}
\expandafter\def\csname pn@wiki/titleroute/EM/IRCoT/ci\endcsname{[$-1.5$, $1.1$]}
\expandafter\def\csname pn@wiki/titleroute/EM/IRCoT/full\endcsname{$-0.2$ [$-1.5$, $1.1$]}
\expandafter\def\csname pn@wiki/titleroute/EM/ReAct\endcsname{$+1.4$}
\expandafter\def\csname pn@wiki/titleroute/EM/ReAct/abs\endcsname{1.4}
\expandafter\def\csname pn@wiki/titleroute/EM/ReAct/ci\endcsname{[$0.1$, $2.7$]}
\expandafter\def\csname pn@wiki/titleroute/EM/ReAct/full\endcsname{$+1.4$ [$0.1$, $2.7$]}
\expandafter\def\csname pn@wiki/titleroute/EM/SR1\endcsname{$-0.3$}
\expandafter\def\csname pn@wiki/titleroute/EM/SR1/abs\endcsname{0.3}
\expandafter\def\csname pn@wiki/titleroute/EM/SR1/ci\endcsname{[$-1.0$, $0.4$]}
\expandafter\def\csname pn@wiki/titleroute/EM/SR1/full\endcsname{$-0.3$ [$-1.0$, $0.4$]}
\expandafter\def\csname pn@wiki/titleroute/EM/ext\endcsname{$+0.3$}
\expandafter\def\csname pn@wiki/titleroute/EM/ext/abs\endcsname{0.3}
\expandafter\def\csname pn@wiki/titleroute/EM/ext/ci\endcsname{[$-0.4$, $1.0$]}
\expandafter\def\csname pn@wiki/titleroute/EM/ext/full\endcsname{$+0.3$ [$-0.4$, $1.0$]}
\expandafter\def\csname pn@wiki/titleroute/EM/nsig\endcsname{0}
\expandafter\def\csname pn@wiki/titleroute/EM/pool\endcsname{$+0.5$}
\expandafter\def\csname pn@wiki/titleroute/EM/pool/abs\endcsname{0.5}
\expandafter\def\csname pn@wiki/titleroute/EM/pool/ci\endcsname{[$-0.1$, $1.1$]}
\expandafter\def\csname pn@wiki/titleroute/EM/pool/full\endcsname{$+0.5$ [$-0.1$, $1.1$]}
\expandafter\def\csname pn@wiki/titleroute/EM/range\endcsname{$-0.3$ to $+1.4$}
\expandafter\def\csname pn@wiki/titleroute/ans/Decomp\endcsname{$+3.5$}
\expandafter\def\csname pn@wiki/titleroute/ans/Decomp/abs\endcsname{3.5}
\expandafter\def\csname pn@wiki/titleroute/ans/Decomp/ci\endcsname{[$2.3$, $4.6$]}
\expandafter\def\csname pn@wiki/titleroute/ans/Decomp/full\endcsname{$+3.5$ [$2.3$, $4.6$]}
\expandafter\def\csname pn@wiki/titleroute/ans/IRCoT\endcsname{$-0.1$}
\expandafter\def\csname pn@wiki/titleroute/ans/IRCoT/abs\endcsname{0.1}
\expandafter\def\csname pn@wiki/titleroute/ans/IRCoT/ci\endcsname{[$-0.9$, $0.7$]}
\expandafter\def\csname pn@wiki/titleroute/ans/IRCoT/full\endcsname{$-0.1$ [$-0.9$, $0.7$]}
\expandafter\def\csname pn@wiki/titleroute/ans/ReAct\endcsname{$+1.4$}
\expandafter\def\csname pn@wiki/titleroute/ans/ReAct/abs\endcsname{1.4}
\expandafter\def\csname pn@wiki/titleroute/ans/ReAct/ci\endcsname{[$0.3$, $2.4$]}
\expandafter\def\csname pn@wiki/titleroute/ans/ReAct/full\endcsname{$+1.4$ [$0.3$, $2.4$]}
\expandafter\def\csname pn@wiki/titleroute/ans/SR1\endcsname{$+0.4$}
\expandafter\def\csname pn@wiki/titleroute/ans/SR1/abs\endcsname{0.4}
\expandafter\def\csname pn@wiki/titleroute/ans/SR1/ci\endcsname{[$-0.2$, $0.9$]}
\expandafter\def\csname pn@wiki/titleroute/ans/SR1/full\endcsname{$+0.4$ [$-0.2$, $0.9$]}
\expandafter\def\csname pn@wiki/titleroute/ans/ext\endcsname{$+0.6$}
\expandafter\def\csname pn@wiki/titleroute/ans/ext/abs\endcsname{0.6}
\expandafter\def\csname pn@wiki/titleroute/ans/ext/ci\endcsname{[$0.0$, $1.1$]}
\expandafter\def\csname pn@wiki/titleroute/ans/ext/full\endcsname{$+0.6$ [$0.0$, $1.1$]}
\expandafter\def\csname pn@wiki/titleroute/ans/nsig\endcsname{2}
\expandafter\def\csname pn@wiki/titleroute/ans/pool\endcsname{$+1.3$}
\expandafter\def\csname pn@wiki/titleroute/ans/pool/abs\endcsname{1.3}
\expandafter\def\csname pn@wiki/titleroute/ans/pool/ci\endcsname{[$0.8$, $1.8$]}
\expandafter\def\csname pn@wiki/titleroute/ans/pool/full\endcsname{$+1.3$ [$0.8$, $1.8$]}
\expandafter\def\csname pn@wiki/titleroute/ans/range\endcsname{$-0.1$ to $+3.5$}
\expandafter\def\csname pn@wiki/titleroute/complete/Decomp\endcsname{$+5.7$}
\expandafter\def\csname pn@wiki/titleroute/complete/Decomp/abs\endcsname{5.7}
\expandafter\def\csname pn@wiki/titleroute/complete/Decomp/ci\endcsname{[$4.4$, $7.0$]}
\expandafter\def\csname pn@wiki/titleroute/complete/Decomp/full\endcsname{$+5.7$ [$4.4$, $7.0$]}
\expandafter\def\csname pn@wiki/titleroute/complete/IRCoT\endcsname{$+0.6$}
\expandafter\def\csname pn@wiki/titleroute/complete/IRCoT/abs\endcsname{0.6}
\expandafter\def\csname pn@wiki/titleroute/complete/IRCoT/ci\endcsname{[$-0.3$, $1.5$]}
\expandafter\def\csname pn@wiki/titleroute/complete/IRCoT/full\endcsname{$+0.6$ [$-0.3$, $1.5$]}
\expandafter\def\csname pn@wiki/titleroute/complete/ReAct\endcsname{$+3.1$}
\expandafter\def\csname pn@wiki/titleroute/complete/ReAct/abs\endcsname{3.1}
\expandafter\def\csname pn@wiki/titleroute/complete/ReAct/ci\endcsname{[$1.9$, $4.3$]}
\expandafter\def\csname pn@wiki/titleroute/complete/ReAct/full\endcsname{$+3.1$ [$1.9$, $4.3$]}
\expandafter\def\csname pn@wiki/titleroute/complete/SR1\endcsname{$+0.5$}
\expandafter\def\csname pn@wiki/titleroute/complete/SR1/abs\endcsname{0.5}
\expandafter\def\csname pn@wiki/titleroute/complete/SR1/ci\endcsname{[$-0.1$, $1.1$]}
\expandafter\def\csname pn@wiki/titleroute/complete/SR1/full\endcsname{$+0.5$ [$-0.1$, $1.1$]}
\expandafter\def\csname pn@wiki/titleroute/complete/ext\endcsname{$+1.4$}
\expandafter\def\csname pn@wiki/titleroute/complete/ext/abs\endcsname{1.4}
\expandafter\def\csname pn@wiki/titleroute/complete/ext/ci\endcsname{[$0.8$, $2.0$]}
\expandafter\def\csname pn@wiki/titleroute/complete/ext/full\endcsname{$+1.4$ [$0.8$, $2.0$]}
\expandafter\def\csname pn@wiki/titleroute/complete/nsig\endcsname{2}
\expandafter\def\csname pn@wiki/titleroute/complete/pool\endcsname{$+2.5$}
\expandafter\def\csname pn@wiki/titleroute/complete/pool/abs\endcsname{2.5}
\expandafter\def\csname pn@wiki/titleroute/complete/pool/ci\endcsname{[$1.9$, $3.0$]}
\expandafter\def\csname pn@wiki/titleroute/complete/pool/full\endcsname{$+2.5$ [$1.9$, $3.0$]}
\expandafter\def\csname pn@wiki/titleroute/complete/range\endcsname{$+0.5$ to $+5.7$}
\expandafter\def\csname pn@wiki/vsrrf/EM/Decomp\endcsname{$+6.0$}
\expandafter\def\csname pn@wiki/vsrrf/EM/Decomp/abs\endcsname{6.0}
\expandafter\def\csname pn@wiki/vsrrf/EM/Decomp/ci\endcsname{[$4.0$, $8.1$]}
\expandafter\def\csname pn@wiki/vsrrf/EM/Decomp/full\endcsname{$+6.0$ [$4.0$, $8.1$]}
\expandafter\def\csname pn@wiki/vsrrf/EM/IRCoT\endcsname{$+3.9$}
\expandafter\def\csname pn@wiki/vsrrf/EM/IRCoT/abs\endcsname{3.9}
\expandafter\def\csname pn@wiki/vsrrf/EM/IRCoT/ci\endcsname{[$1.5$, $6.3$]}
\expandafter\def\csname pn@wiki/vsrrf/EM/IRCoT/full\endcsname{$+3.9$ [$1.5$, $6.3$]}
\expandafter\def\csname pn@wiki/vsrrf/EM/ReAct\endcsname{$+8.0$}
\expandafter\def\csname pn@wiki/vsrrf/EM/ReAct/abs\endcsname{8.0}
\expandafter\def\csname pn@wiki/vsrrf/EM/ReAct/ci\endcsname{[$5.9$, $10.2$]}
\expandafter\def\csname pn@wiki/vsrrf/EM/ReAct/full\endcsname{$+8.0$ [$5.9$, $10.2$]}
\expandafter\def\csname pn@wiki/vsrrf/EM/SR1\endcsname{$+2.6$}
\expandafter\def\csname pn@wiki/vsrrf/EM/SR1/abs\endcsname{2.6}
\expandafter\def\csname pn@wiki/vsrrf/EM/SR1/ci\endcsname{[$0.5$, $4.7$]}
\expandafter\def\csname pn@wiki/vsrrf/EM/SR1/full\endcsname{$+2.6$ [$0.5$, $4.7$]}
\expandafter\def\csname pn@wiki/vsrrf/EM/ext\endcsname{$+4.8$}
\expandafter\def\csname pn@wiki/vsrrf/EM/ext/abs\endcsname{4.8}
\expandafter\def\csname pn@wiki/vsrrf/EM/ext/ci\endcsname{[$3.4$, $6.3$]}
\expandafter\def\csname pn@wiki/vsrrf/EM/ext/full\endcsname{$+4.8$ [$3.4$, $6.3$]}
\expandafter\def\csname pn@wiki/vsrrf/EM/nsig\endcsname{4}
\expandafter\def\csname pn@wiki/vsrrf/EM/pool\endcsname{$+5.1$}
\expandafter\def\csname pn@wiki/vsrrf/EM/pool/abs\endcsname{5.1}
\expandafter\def\csname pn@wiki/vsrrf/EM/pool/ci\endcsname{[$3.8$, $6.4$]}
\expandafter\def\csname pn@wiki/vsrrf/EM/pool/full\endcsname{$+5.1$ [$3.8$, $6.4$]}
\expandafter\def\csname pn@wiki/vsrrf/EM/range\endcsname{$+2.6$ to $+8.0$}
\expandafter\def\csname pn@wiki/vsrrf/ans/Decomp\endcsname{$+11.6$}
\expandafter\def\csname pn@wiki/vsrrf/ans/Decomp/abs\endcsname{11.6}
\expandafter\def\csname pn@wiki/vsrrf/ans/Decomp/ci\endcsname{[$9.7$, $13.6$]}
\expandafter\def\csname pn@wiki/vsrrf/ans/Decomp/full\endcsname{$+11.6$ [$9.7$, $13.6$]}
\expandafter\def\csname pn@wiki/vsrrf/ans/IRCoT\endcsname{$+12.2$}
\expandafter\def\csname pn@wiki/vsrrf/ans/IRCoT/abs\endcsname{12.2}
\expandafter\def\csname pn@wiki/vsrrf/ans/IRCoT/ci\endcsname{[$10.2$, $14.1$]}
\expandafter\def\csname pn@wiki/vsrrf/ans/IRCoT/full\endcsname{$+12.2$ [$10.2$, $14.1$]}
\expandafter\def\csname pn@wiki/vsrrf/ans/ReAct\endcsname{$+19.5$}
\expandafter\def\csname pn@wiki/vsrrf/ans/ReAct/abs\endcsname{19.5}
\expandafter\def\csname pn@wiki/vsrrf/ans/ReAct/ci\endcsname{[$17.2$, $21.8$]}
\expandafter\def\csname pn@wiki/vsrrf/ans/ReAct/full\endcsname{$+19.5$ [$17.2$, $21.8$]}
\expandafter\def\csname pn@wiki/vsrrf/ans/SR1\endcsname{$+7.4$}
\expandafter\def\csname pn@wiki/vsrrf/ans/SR1/abs\endcsname{7.4}
\expandafter\def\csname pn@wiki/vsrrf/ans/SR1/ci\endcsname{[$5.9$, $9.0$]}
\expandafter\def\csname pn@wiki/vsrrf/ans/SR1/full\endcsname{$+7.4$ [$5.9$, $9.0$]}
\expandafter\def\csname pn@wiki/vsrrf/ans/ext\endcsname{$+13.0$}
\expandafter\def\csname pn@wiki/vsrrf/ans/ext/abs\endcsname{13.0}
\expandafter\def\csname pn@wiki/vsrrf/ans/ext/ci\endcsname{[$11.8$, $14.3$]}
\expandafter\def\csname pn@wiki/vsrrf/ans/ext/full\endcsname{$+13.0$ [$11.8$, $14.3$]}
\expandafter\def\csname pn@wiki/vsrrf/ans/nsig\endcsname{4}
\expandafter\def\csname pn@wiki/vsrrf/ans/pool\endcsname{$+12.7$}
\expandafter\def\csname pn@wiki/vsrrf/ans/pool/abs\endcsname{12.7}
\expandafter\def\csname pn@wiki/vsrrf/ans/pool/ci\endcsname{[$11.5$, $13.8$]}
\expandafter\def\csname pn@wiki/vsrrf/ans/pool/full\endcsname{$+12.7$ [$11.5$, $13.8$]}
\expandafter\def\csname pn@wiki/vsrrf/ans/range\endcsname{$+7.4$ to $+19.5$}
\expandafter\def\csname pn@wiki/vsrrf/complete/Decomp\endcsname{$+15.3$}
\expandafter\def\csname pn@wiki/vsrrf/complete/Decomp/abs\endcsname{15.3}
\expandafter\def\csname pn@wiki/vsrrf/complete/Decomp/ci\endcsname{[$13.5$, $17.2$]}
\expandafter\def\csname pn@wiki/vsrrf/complete/Decomp/full\endcsname{$+15.3$ [$13.5$, $17.2$]}
\expandafter\def\csname pn@wiki/vsrrf/complete/IRCoT\endcsname{$+22.7$}
\expandafter\def\csname pn@wiki/vsrrf/complete/IRCoT/abs\endcsname{22.7}
\expandafter\def\csname pn@wiki/vsrrf/complete/IRCoT/ci\endcsname{[$20.5$, $24.9$]}
\expandafter\def\csname pn@wiki/vsrrf/complete/IRCoT/full\endcsname{$+22.7$ [$20.5$, $24.9$]}
\expandafter\def\csname pn@wiki/vsrrf/complete/ReAct\endcsname{$+27.1$}
\expandafter\def\csname pn@wiki/vsrrf/complete/ReAct/abs\endcsname{27.1}
\expandafter\def\csname pn@wiki/vsrrf/complete/ReAct/ci\endcsname{[$24.8$, $29.4$]}
\expandafter\def\csname pn@wiki/vsrrf/complete/ReAct/full\endcsname{$+27.1$ [$24.8$, $29.4$]}
\expandafter\def\csname pn@wiki/vsrrf/complete/SR1\endcsname{$+15.5$}
\expandafter\def\csname pn@wiki/vsrrf/complete/SR1/abs\endcsname{15.5}
\expandafter\def\csname pn@wiki/vsrrf/complete/SR1/ci\endcsname{[$13.6$, $17.5$]}
\expandafter\def\csname pn@wiki/vsrrf/complete/SR1/full\endcsname{$+15.5$ [$13.6$, $17.5$]}
\expandafter\def\csname pn@wiki/vsrrf/complete/ext\endcsname{$+21.8$}
\expandafter\def\csname pn@wiki/vsrrf/complete/ext/abs\endcsname{21.8}
\expandafter\def\csname pn@wiki/vsrrf/complete/ext/ci\endcsname{[$20.4$, $23.2$]}
\expandafter\def\csname pn@wiki/vsrrf/complete/ext/full\endcsname{$+21.8$ [$20.4$, $23.2$]}
\expandafter\def\csname pn@wiki/vsrrf/complete/nsig\endcsname{4}
\expandafter\def\csname pn@wiki/vsrrf/complete/pool\endcsname{$+20.2$}
\expandafter\def\csname pn@wiki/vsrrf/complete/pool/abs\endcsname{20.2}
\expandafter\def\csname pn@wiki/vsrrf/complete/pool/ci\endcsname{[$19.0$, $21.4$]}
\expandafter\def\csname pn@wiki/vsrrf/complete/pool/full\endcsname{$+20.2$ [$19.0$, $21.4$]}
\expandafter\def\csname pn@wiki/vsrrf/complete/range\endcsname{$+15.3$ to $+27.1$}
\expandafter\def\csname pn@wiki/winVSrerank/EM/Decomp\endcsname{$+1.0$}
\expandafter\def\csname pn@wiki/winVSrerank/EM/Decomp/abs\endcsname{1.0}
\expandafter\def\csname pn@wiki/winVSrerank/EM/Decomp/ci\endcsname{[$-0.9$, $2.9$]}
\expandafter\def\csname pn@wiki/winVSrerank/EM/Decomp/full\endcsname{$+1.0$ [$-0.9$, $2.9$]}
\expandafter\def\csname pn@wiki/winVSrerank/EM/IRCoT\endcsname{$+1.7$}
\expandafter\def\csname pn@wiki/winVSrerank/EM/IRCoT/abs\endcsname{1.7}
\expandafter\def\csname pn@wiki/winVSrerank/EM/IRCoT/ci\endcsname{[$-0.3$, $3.7$]}
\expandafter\def\csname pn@wiki/winVSrerank/EM/IRCoT/full\endcsname{$+1.7$ [$-0.3$, $3.7$]}
\expandafter\def\csname pn@wiki/winVSrerank/EM/ReAct\endcsname{$-1.1$}
\expandafter\def\csname pn@wiki/winVSrerank/EM/ReAct/abs\endcsname{1.1}
\expandafter\def\csname pn@wiki/winVSrerank/EM/ReAct/ci\endcsname{[$-3.1$, $0.9$]}
\expandafter\def\csname pn@wiki/winVSrerank/EM/ReAct/full\endcsname{$-1.1$ [$-3.1$, $0.9$]}
\expandafter\def\csname pn@wiki/winVSrerank/EM/SR1\endcsname{$+2.2$}
\expandafter\def\csname pn@wiki/winVSrerank/EM/SR1/abs\endcsname{2.2}
\expandafter\def\csname pn@wiki/winVSrerank/EM/SR1/ci\endcsname{[$0.1$, $4.3$]}
\expandafter\def\csname pn@wiki/winVSrerank/EM/SR1/full\endcsname{$+2.2$ [$0.1$, $4.3$]}
\expandafter\def\csname pn@wiki/winVSrerank/EM/ext\endcsname{$+1.0$}
\expandafter\def\csname pn@wiki/winVSrerank/EM/ext/abs\endcsname{1.0}
\expandafter\def\csname pn@wiki/winVSrerank/EM/ext/ci\endcsname{[$-0.6$, $2.5$]}
\expandafter\def\csname pn@wiki/winVSrerank/EM/ext/full\endcsname{$+1.0$ [$-0.6$, $2.5$]}
\expandafter\def\csname pn@wiki/winVSrerank/EM/nsig\endcsname{0}
\expandafter\def\csname pn@wiki/winVSrerank/EM/pool\endcsname{$+1.0$}
\expandafter\def\csname pn@wiki/winVSrerank/EM/pool/abs\endcsname{1.0}
\expandafter\def\csname pn@wiki/winVSrerank/EM/pool/ci\endcsname{[$-0.4$, $2.4$]}
\expandafter\def\csname pn@wiki/winVSrerank/EM/pool/full\endcsname{$+1.0$ [$-0.4$, $2.4$]}
\expandafter\def\csname pn@wiki/winVSrerank/EM/range\endcsname{$-1.1$ to $+2.2$}
\expandafter\def\csname pn@wiki/winVSrerank/ans/Decomp\endcsname{$+1.2$}
\expandafter\def\csname pn@wiki/winVSrerank/ans/Decomp/abs\endcsname{1.2}
\expandafter\def\csname pn@wiki/winVSrerank/ans/Decomp/ci\endcsname{[$-0.1$, $2.5$]}
\expandafter\def\csname pn@wiki/winVSrerank/ans/Decomp/full\endcsname{$+1.2$ [$-0.1$, $2.5$]}
\expandafter\def\csname pn@wiki/winVSrerank/ans/IRCoT\endcsname{$+4.7$}
\expandafter\def\csname pn@wiki/winVSrerank/ans/IRCoT/abs\endcsname{4.7}
\expandafter\def\csname pn@wiki/winVSrerank/ans/IRCoT/ci\endcsname{[$3.4$, $6.1$]}
\expandafter\def\csname pn@wiki/winVSrerank/ans/IRCoT/full\endcsname{$+4.7$ [$3.4$, $6.1$]}
\expandafter\def\csname pn@wiki/winVSrerank/ans/ReAct\endcsname{$+2.9$}
\expandafter\def\csname pn@wiki/winVSrerank/ans/ReAct/abs\endcsname{2.9}
\expandafter\def\csname pn@wiki/winVSrerank/ans/ReAct/ci\endcsname{[$1.5$, $4.3$]}
\expandafter\def\csname pn@wiki/winVSrerank/ans/ReAct/full\endcsname{$+2.9$ [$1.5$, $4.3$]}
\expandafter\def\csname pn@wiki/winVSrerank/ans/SR1\endcsname{$+4.6$}
\expandafter\def\csname pn@wiki/winVSrerank/ans/SR1/abs\endcsname{4.6}
\expandafter\def\csname pn@wiki/winVSrerank/ans/SR1/ci\endcsname{[$3.1$, $6.2$]}
\expandafter\def\csname pn@wiki/winVSrerank/ans/SR1/full\endcsname{$+4.6$ [$3.1$, $6.2$]}
\expandafter\def\csname pn@wiki/winVSrerank/ans/ext\endcsname{$+4.1$}
\expandafter\def\csname pn@wiki/winVSrerank/ans/ext/abs\endcsname{4.1}
\expandafter\def\csname pn@wiki/winVSrerank/ans/ext/ci\endcsname{[$3.0$, $5.2$]}
\expandafter\def\csname pn@wiki/winVSrerank/ans/ext/full\endcsname{$+4.1$ [$3.0$, $5.2$]}
\expandafter\def\csname pn@wiki/winVSrerank/ans/nsig\endcsname{3}
\expandafter\def\csname pn@wiki/winVSrerank/ans/pool\endcsname{$+3.4$}
\expandafter\def\csname pn@wiki/winVSrerank/ans/pool/abs\endcsname{3.4}
\expandafter\def\csname pn@wiki/winVSrerank/ans/pool/ci\endcsname{[$2.4$, $4.3$]}
\expandafter\def\csname pn@wiki/winVSrerank/ans/pool/full\endcsname{$+3.4$ [$2.4$, $4.3$]}
\expandafter\def\csname pn@wiki/winVSrerank/ans/range\endcsname{$+1.2$ to $+4.7$}
\expandafter\def\csname pn@wiki/winVSrerank/complete/Decomp\endcsname{$+3.6$}
\expandafter\def\csname pn@wiki/winVSrerank/complete/Decomp/abs\endcsname{3.6}
\expandafter\def\csname pn@wiki/winVSrerank/complete/Decomp/ci\endcsname{[$2.5$, $4.7$]}
\expandafter\def\csname pn@wiki/winVSrerank/complete/Decomp/full\endcsname{$+3.6$ [$2.5$, $4.7$]}
\expandafter\def\csname pn@wiki/winVSrerank/complete/IRCoT\endcsname{$+12.9$}
\expandafter\def\csname pn@wiki/winVSrerank/complete/IRCoT/abs\endcsname{12.9}
\expandafter\def\csname pn@wiki/winVSrerank/complete/IRCoT/ci\endcsname{[$11.2$, $14.7$]}
\expandafter\def\csname pn@wiki/winVSrerank/complete/IRCoT/full\endcsname{$+12.9$ [$11.2$, $14.7$]}
\expandafter\def\csname pn@wiki/winVSrerank/complete/ReAct\endcsname{$+7.9$}
\expandafter\def\csname pn@wiki/winVSrerank/complete/ReAct/abs\endcsname{7.9}
\expandafter\def\csname pn@wiki/winVSrerank/complete/ReAct/ci\endcsname{[$6.5$, $9.4$]}
\expandafter\def\csname pn@wiki/winVSrerank/complete/ReAct/full\endcsname{$+7.9$ [$6.5$, $9.4$]}
\expandafter\def\csname pn@wiki/winVSrerank/complete/SR1\endcsname{$+13.1$}
\expandafter\def\csname pn@wiki/winVSrerank/complete/SR1/abs\endcsname{13.1}
\expandafter\def\csname pn@wiki/winVSrerank/complete/SR1/ci\endcsname{[$11.3$, $14.9$]}
\expandafter\def\csname pn@wiki/winVSrerank/complete/SR1/full\endcsname{$+13.1$ [$11.3$, $14.9$]}
\expandafter\def\csname pn@wiki/winVSrerank/complete/ext\endcsname{$+11.3$}
\expandafter\def\csname pn@wiki/winVSrerank/complete/ext/abs\endcsname{11.3}
\expandafter\def\csname pn@wiki/winVSrerank/complete/ext/ci\endcsname{[$10.1$, $12.6$]}
\expandafter\def\csname pn@wiki/winVSrerank/complete/ext/full\endcsname{$+11.3$ [$10.1$, $12.6$]}
\expandafter\def\csname pn@wiki/winVSrerank/complete/nsig\endcsname{4}
\expandafter\def\csname pn@wiki/winVSrerank/complete/pool\endcsname{$+9.4$}
\expandafter\def\csname pn@wiki/winVSrerank/complete/pool/abs\endcsname{9.4}
\expandafter\def\csname pn@wiki/winVSrerank/complete/pool/ci\endcsname{[$8.4$, $10.5$]}
\expandafter\def\csname pn@wiki/winVSrerank/complete/pool/full\endcsname{$+9.4$ [$8.4$, $10.5$]}
\expandafter\def\csname pn@wiki/winVSrerank/complete/range\endcsname{$+3.6$ to $+13.1$}
\expandafter\def\csname pn@wiki/wincpu/range\endcsname{0.3--0.4}
\expandafter\def\csname pn@wiki/window/EM/Decomp\endcsname{$+0.8$}
\expandafter\def\csname pn@wiki/window/EM/Decomp/abs\endcsname{0.8}
\expandafter\def\csname pn@wiki/window/EM/Decomp/ci\endcsname{[$-0.7$, $2.3$]}
\expandafter\def\csname pn@wiki/window/EM/Decomp/full\endcsname{$+0.8$ [$-0.7$, $2.3$]}
\expandafter\def\csname pn@wiki/window/EM/IRCoT\endcsname{$+2.0$}
\expandafter\def\csname pn@wiki/window/EM/IRCoT/abs\endcsname{2.0}
\expandafter\def\csname pn@wiki/window/EM/IRCoT/ci\endcsname{[$0.3$, $3.7$]}
\expandafter\def\csname pn@wiki/window/EM/IRCoT/full\endcsname{$+2.0$ [$0.3$, $3.7$]}
\expandafter\def\csname pn@wiki/window/EM/ReAct\endcsname{$+3.1$}
\expandafter\def\csname pn@wiki/window/EM/ReAct/abs\endcsname{3.1}
\expandafter\def\csname pn@wiki/window/EM/ReAct/ci\endcsname{[$1.4$, $4.7$]}
\expandafter\def\csname pn@wiki/window/EM/ReAct/full\endcsname{$+3.1$ [$1.4$, $4.7$]}
\expandafter\def\csname pn@wiki/window/EM/SR1\endcsname{$+2.1$}
\expandafter\def\csname pn@wiki/window/EM/SR1/abs\endcsname{2.1}
\expandafter\def\csname pn@wiki/window/EM/SR1/ci\endcsname{[$0.5$, $3.8$]}
\expandafter\def\csname pn@wiki/window/EM/SR1/full\endcsname{$+2.1$ [$0.5$, $3.8$]}
\expandafter\def\csname pn@wiki/window/EM/ext\endcsname{$+2.4$}
\expandafter\def\csname pn@wiki/window/EM/ext/abs\endcsname{2.4}
\expandafter\def\csname pn@wiki/window/EM/ext/ci\endcsname{[$1.3$, $3.6$]}
\expandafter\def\csname pn@wiki/window/EM/ext/full\endcsname{$+2.4$ [$1.3$, $3.6$]}
\expandafter\def\csname pn@wiki/window/EM/nsig\endcsname{3}
\expandafter\def\csname pn@wiki/window/EM/pool\endcsname{$+2.0$}
\expandafter\def\csname pn@wiki/window/EM/pool/abs\endcsname{2.0}
\expandafter\def\csname pn@wiki/window/EM/pool/ci\endcsname{[$1.0$, $3.1$]}
\expandafter\def\csname pn@wiki/window/EM/pool/full\endcsname{$+2.0$ [$1.0$, $3.1$]}
\expandafter\def\csname pn@wiki/window/EM/range\endcsname{$+0.8$ to $+3.1$}
\expandafter\def\csname pn@wiki/window/ans/Decomp\endcsname{$+3.6$}
\expandafter\def\csname pn@wiki/window/ans/Decomp/abs\endcsname{3.6}
\expandafter\def\csname pn@wiki/window/ans/Decomp/ci\endcsname{[$2.6$, $4.7$]}
\expandafter\def\csname pn@wiki/window/ans/Decomp/full\endcsname{$+3.6$ [$2.6$, $4.7$]}
\expandafter\def\csname pn@wiki/window/ans/IRCoT\endcsname{$+1.8$}
\expandafter\def\csname pn@wiki/window/ans/IRCoT/abs\endcsname{1.8}
\expandafter\def\csname pn@wiki/window/ans/IRCoT/ci\endcsname{[$1.1$, $2.6$]}
\expandafter\def\csname pn@wiki/window/ans/IRCoT/full\endcsname{$+1.8$ [$1.1$, $2.6$]}
\expandafter\def\csname pn@wiki/window/ans/ReAct\endcsname{$+3.5$}
\expandafter\def\csname pn@wiki/window/ans/ReAct/abs\endcsname{3.5}
\expandafter\def\csname pn@wiki/window/ans/ReAct/ci\endcsname{[$2.5$, $4.7$]}
\expandafter\def\csname pn@wiki/window/ans/ReAct/full\endcsname{$+3.5$ [$2.5$, $4.7$]}
\expandafter\def\csname pn@wiki/window/ans/SR1\endcsname{$+1.7$}
\expandafter\def\csname pn@wiki/window/ans/SR1/abs\endcsname{1.7}
\expandafter\def\csname pn@wiki/window/ans/SR1/ci\endcsname{[$0.8$, $2.7$]}
\expandafter\def\csname pn@wiki/window/ans/SR1/full\endcsname{$+1.7$ [$0.8$, $2.7$]}
\expandafter\def\csname pn@wiki/window/ans/ext\endcsname{$+2.4$}
\expandafter\def\csname pn@wiki/window/ans/ext/abs\endcsname{2.4}
\expandafter\def\csname pn@wiki/window/ans/ext/ci\endcsname{[$1.7$, $3.0$]}
\expandafter\def\csname pn@wiki/window/ans/ext/full\endcsname{$+2.4$ [$1.7$, $3.0$]}
\expandafter\def\csname pn@wiki/window/ans/nsig\endcsname{4}
\expandafter\def\csname pn@wiki/window/ans/pool\endcsname{$+2.7$}
\expandafter\def\csname pn@wiki/window/ans/pool/abs\endcsname{2.7}
\expandafter\def\csname pn@wiki/window/ans/pool/ci\endcsname{[$2.1$, $3.3$]}
\expandafter\def\csname pn@wiki/window/ans/pool/full\endcsname{$+2.7$ [$2.1$, $3.3$]}
\expandafter\def\csname pn@wiki/window/ans/range\endcsname{$+1.7$ to $+3.6$}
\expandafter\def\csname pn@wiki/window/complete/Decomp\endcsname{$+5.1$}
\expandafter\def\csname pn@wiki/window/complete/Decomp/abs\endcsname{5.1}
\expandafter\def\csname pn@wiki/window/complete/Decomp/ci\endcsname{[$4.0$, $6.3$]}
\expandafter\def\csname pn@wiki/window/complete/Decomp/full\endcsname{$+5.1$ [$4.0$, $6.3$]}
\expandafter\def\csname pn@wiki/window/complete/IRCoT\endcsname{$+3.8$}
\expandafter\def\csname pn@wiki/window/complete/IRCoT/abs\endcsname{3.8}
\expandafter\def\csname pn@wiki/window/complete/IRCoT/ci\endcsname{[$2.8$, $4.9$]}
\expandafter\def\csname pn@wiki/window/complete/IRCoT/full\endcsname{$+3.8$ [$2.8$, $4.9$]}
\expandafter\def\csname pn@wiki/window/complete/ReAct\endcsname{$+6.3$}
\expandafter\def\csname pn@wiki/window/complete/ReAct/abs\endcsname{6.3}
\expandafter\def\csname pn@wiki/window/complete/ReAct/ci\endcsname{[$5.1$, $7.5$]}
\expandafter\def\csname pn@wiki/window/complete/ReAct/full\endcsname{$+6.3$ [$5.1$, $7.5$]}
\expandafter\def\csname pn@wiki/window/complete/SR1\endcsname{$+6.3$}
\expandafter\def\csname pn@wiki/window/complete/SR1/abs\endcsname{6.3}
\expandafter\def\csname pn@wiki/window/complete/SR1/ci\endcsname{[$4.9$, $7.7$]}
\expandafter\def\csname pn@wiki/window/complete/SR1/full\endcsname{$+6.3$ [$4.9$, $7.7$]}
\expandafter\def\csname pn@wiki/window/complete/ext\endcsname{$+5.4$}
\expandafter\def\csname pn@wiki/window/complete/ext/abs\endcsname{5.4}
\expandafter\def\csname pn@wiki/window/complete/ext/ci\endcsname{[$4.6$, $6.3$]}
\expandafter\def\csname pn@wiki/window/complete/ext/full\endcsname{$+5.4$ [$4.6$, $6.3$]}
\expandafter\def\csname pn@wiki/window/complete/nsig\endcsname{4}
\expandafter\def\csname pn@wiki/window/complete/pool\endcsname{$+5.4$}
\expandafter\def\csname pn@wiki/window/complete/pool/abs\endcsname{5.4}
\expandafter\def\csname pn@wiki/window/complete/pool/ci\endcsname{[$4.6$, $6.1$]}
\expandafter\def\csname pn@wiki/window/complete/pool/full\endcsname{$+5.4$ [$4.6$, $6.1$]}
\expandafter\def\csname pn@wiki/window/complete/range\endcsname{$+3.8$ to $+6.3$}

\setcopyright{none}
\renewcommand\footnotetextcopyrightpermission[1]{}
\acmConference{}{}{}
\acmBooktitle{}
\acmDOI{}
\acmISBN{}

\begin{document}
\title[DetouR: Plugging Association Routes into Search Agents]{DetouR: Plugging Association Routes into Search Agents for Multi-hop Question Answering}

\author{Tingwei Chen}
\authornote{Equal contribution.}
\affiliation{\institution{}\country{}}
\author{Yunxiao Shi}
\authornotemark[1]
\authornote{Corresponding author.}
\affiliation{\institution{}\country{}}
\author{Zhendong Chu}
\affiliation{\institution{}\country{}}
\author{Zhiying Deng}
\affiliation{\institution{}\country{}}
\author{Min Xu}
\affiliation{\institution{}\country{}}
\author{Qingsong Wen}
\affiliation{\institution{}\country{}}
\renewcommand{\shortauthors}{Chen and Shi, et al.}

\begin{abstract}

Search agents answer multi-hop questions by interleaving reasoning with retrieval,
deciding at each step what to search next. Existing agents follow \emph{semantic routes}:
they infer what is missing and phrase it as a query, reaching only what the model can
name. Yet the decisive page is often one the question never names, frequently just one
hyperlink away from evidence already retrieved. Following such \emph{association routes}
reaches more gold pages than semantic search under the same budget. Exploiting them
raises two challenges: obtaining the association structure without costly offline graph
construction, and keeping linked evidence in the reader's limited context, where
relevance- and fusion-based selection discards it. We propose \method, a plug-in that
reuses the links arriving with retrieved pages and adds no search or LLM call. \reroute
replaces near-repeat queries with routes to linked pages, chosen by a lightweight learned
scorer; \chainctx runs personalized PageRank over these links so that evidence chains
reach the reader intact. Across four agents, \method improves EM by
\pnum{wiki/detour/EM/pool/abs} points on 2Wiki and \pnum{hot/detour/EM/pool/abs} on
HotpotQA on average. We also find that heavier LLM models struggle with this non-semantic evidence, adding little over
these simple selectors or even falling behind. In summary, our work reveals association routes as a cheap complement to semantic
routes that current search agents overlook: added as a plug-in,
they improve every agent we test, in replay and in live runs, at no
extra search or LLM cost.

\end{abstract}


\maketitle
\pagestyle{plain}   %

\section{Introduction}
\label{sec:intro}

\begin{figure*}[t]
  \centering
  \includegraphics[width=0.95\linewidth]{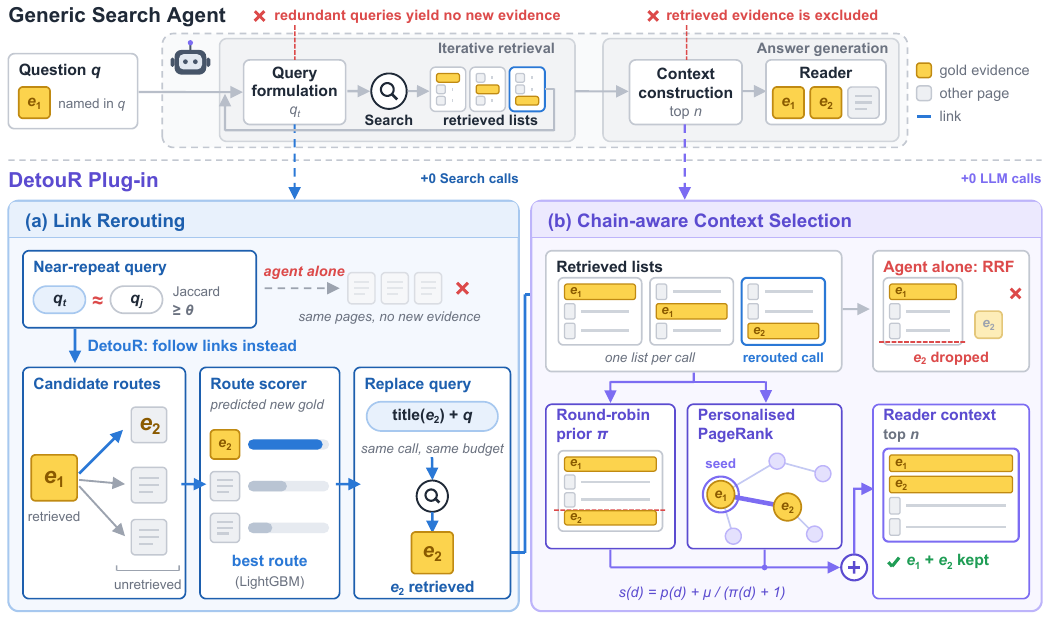}
  \vspace{-3mm}
\caption{Overview of \method. \emph{Top:} a generic search agent; red notes mark where it loses evidence. \emph{Bottom:} \method plugs into query formulation and context construction without extra calls. (a)~A near-repeat query is replaced by a query for $e_2$, linked from $e_1$. (b)~RRF drops $e_2$; personalized PageRank from $e_1$ keeps both.}
  \vspace{-3mm}
  \label{fig:needroute}
\end{figure*}

Retrieval-augmented generation (RAG) strengthens the correctness of a large language model's
answers by retrieving passages from external collections, allowing the model to respond with
knowledge beyond what it has memorized in its parameters~\cite{lewis2020rag}. Search agents
carry this idea further: they interleave reasoning with search and
decide at each step what to look up next~\cite{nakano2021webgpt,li2025searcho1}. Multi-hop question
answering has become the canonical testbed for this capability~\cite{yang2018hotpotqa,ho2020constructing},
because its evidence forms a chain: each hop reveals an intermediate entity, unstated in the
question, that leads to the next. How completely an agent assembles this chain largely
determines how well it answers.

A rich line of work has studied how such agents should decide what to retrieve.
IRCoT~\cite{trivedi2023ircot} interleaves retrieval with chain-of-thought reasoning, issuing
each newly generated reasoning sentence as the next query. ReAct~\cite{yao2023react}
alternates free-form thoughts with explicit search actions, so that retrieval follows an
evolving plan. Self-Ask~\cite{press2023selfask} makes decomposition explicit, posing and
answering follow-up sub-questions before composing the final
answer~\cite{zhou2023leasttomost}. Query rewriting and expansion methods
reformulate the request into queries better matched to the
retriever~\cite{ma2023queryrewriting,wang2023query2doc,gao2023hyde}. Most recently,
Search-R1~\cite{jin2025searchr1} and its successors~\cite{song2025r1searcher,song2025r1searcherpp}
learn when and what to search end to end with reinforcement learning.

Different as they are, these methods share one premise: the agent infers the implicit intent
behind the request, articulates what it still lacks, and expresses that need in words a
retriever can match. \textbf{We call the path by which an agent arrives at a piece of evidence a
\emph{retrieval route}, and this dominant kind a \emph{semantic route}, since its next target
is derived from meaning.} Its blind spot is that it reaches only what the model can name, while
the decisive page in a multi-hop question is often the one the question does not name.
Yet evidence can also be reached by a route that is not semantic at all. For a 2Wiki question
asking where Nancy Lincoln's husband was buried (\figref{fig:needroute}), the first search
already returns Nancy Lincoln's page, which links to her husband, Thomas Lincoln; the agent
never inferred that he was the next person to look up and pursued Joseph Hanks instead. \textbf{We call
such a step an \emph{association route}: its next target is read off an explicit association
from an item already retrieved, not inferred from what the request means.} The case is not
isolated. On HotpotQA, \pnum{assoc/onehop/hotpotqa/range}\% of the gold pages that four agents
never retrieve are linked from a page in their own top-three results, and under the same search
budget association routes reach \pnum{assoc/reach/hotpotqa/assoc}\% of the gold pages against
\pnum{assoc/reach/hotpotqa/sem}\% for semantic routes (\secref{sec:assoc}).

Such associations are woven into how information is organized. The Web already wires entities
together through hyperlinks, knowledge graphs and related-item lists~\cite{west2012wayfinding,eberhard2025llmndr,shi2025ocg}. Graph-based retrieval,
which walks hyperlinks~\cite{asai2020pathretriever,xiong2021mdr} or knowledge
graphs~\cite{sun2024tog,gutierrez2024hipporag,edge2024graphrag}, likewise recovers multi-hop
evidence that text similarity misses. This leads to our research question: \textbf{\emph{can a search
agent follow non-semantic association routes, grounded in the evidence it has already
retrieved, to gather more complete evidence and answer multi-hop questions more accurately?}}

Answering this question raises two challenges. The first is how to obtain the association
structure. Graph-based RAG extracts it offline from the entire corpus with an
LLM~\cite{edge2024graphrag,gutierrez2024hipporag}, and link-aware retrievers are trained on one
collection's links~\cite{asai2020pathretriever}; both are costly to build and maintain at Web
scale. 
Fortunately, the Web already provides this structure: pages embed hyperlinks and
recommender-generated related-item lists to aid navigation and
discovery~\cite{craswell2007clickgraph}, and the 5.2M Wikipedia abstracts behind HotpotQA alone
contain 19.5M hyperlinks.
These links arrive with the very pages an agent retrieves, so the graph can be
exploited during search without building anything in advance.

The second challenge is that retrieved evidence is not necessarily read. The reader sees only a
fixed budget of passages, selected by relevance to the
question~\cite{nogueira2020monot5,sun2023rankgpt,zhang2025qwen3embedding} or by rank fusion
such as RRF~\cite{cormack2009rrf,rackauckas2024ragfusion}, which favors passages that rank
high and recur across result lists~\cite{vogt1999fusion}. Evidence reached through an
association loses on both counts: it shares little with the question and is usually found by a
single call, so even after Thomas Lincoln is retrieved, RRF leaves him out
(\figref{fig:needroute}). Worse, as near-repeat queries accumulate, RRF contexts become
\emph{less} complete with further search (\secref{sec:lost}). Yet what makes Thomas Lincoln
relevant is his link to Nancy Lincoln, whom the question names. Personalized
PageRank~\cite{page1999pagerank,haveliwala2002topic,jeh2003scaling} captures exactly this:
seeded at the question's pages and paired with a Round-robin
merge~\cite{voorhees1995fusion,callan1995distributed} that ignores repetition, it scores a chain
as a whole rather than each passage in isolation.

Driven by these insights, we propose \textbf{\method, which lets a search agent take a short detour
through a link when the direct semantic route stalls.} \method is not a new search agent but a
plug-in for the two decisions that matter most for an agent's evidence: query formulation, which determines what is retrieved, and context construction,
which determines what is read. \reroute detects a follow-up query that nearly
repeats an earlier one and replaces it, within the same call, with a route to a page linked
from the retrieved evidence, selected by a lightweight learned estimate of the new evidence
it will bring. \chainctx runs personalized PageRank over the links
among retrieved passages from the pages the question names and combines it with a Round-robin
prior, so that the passages of a chain enter the reader's context together. Neither component
adds a search or LLM call, and the agent and reader stay unchanged.

We attach \method to four agents that search in different ways (IRCoT, ReAct, Search-R1 and
a plan-then-search agent) on HotpotQA and 2Wiki. Without any extra call, \method improves all
four agents on both collections, by \pnum{wiki/detour/EM/pool/abs} EM points on 2Wiki and
\pnum{hot/detour/EM/pool/abs} on HotpotQA on average. Further analysis reveals three key findings. \textbf{\ding{172} Reaching evidence pays off only when it is kept}: rerouting adds
only 0.8 EM points on 2Wiki and 0.3 on HotpotQA, and Chain-aware Context Selection adds 2.5 and
0.8. But the full \method gains 4.0 and 1.4, more than the 3.3 and 1.1 their separate sum.
\textbf{\ding{173} Lightweight selectors rival heavy LLM models}: in Link Rerouting, a
LightGBM is enough to choose which link to follow, since re-ranking its top routes
with cross-encoders of up to 0.6B parameters changes EM by at most 0.2 points. In Chain-aware
Context Selection, PageRank needs no model call, yet it beats an 8B reranker and an LLM judge in
complete contexts on 2Wiki (64.5\% vs.\ 57.9\% and 53.8\%) and matches the reranker on HotpotQA
(67.9\% $\approx$ 68.1\%; the judge reaches 61.8\%).
\textbf{\ding{174} \method nearly eliminates evidence lost at the merge}: a retrieved gold page
misses the reader's context in only 0.2\% (2Wiki) and 1.0\% (HotpotQA) of its wrong answers.

In summary, our contributions are:
\begin{itemize}[leftmargin=*,itemsep=1pt,topsep=2pt]
\item We identify an overlooked limitation of search agents: relying on
semantic routes alone, they miss evidence one link away and lose
retrieved evidence before it reaches the reader.
\item We introduce association routes as a non-semantic complement,
realized with a LightGBM route scorer that improves coverage and
personalized PageRank that improves retention. Both plug into existing
agents without extra search or LLM calls.
\item Extensive experiments on four agents and two collections trace
the gains to the association structure itself.
\end{itemize}

\section{Related Work}
\label{sec:related}

\textit{Search agents and redundant search.}
Search agents alternate reasoning with retrieval~\cite{trivedi2023ircot,yao2023react}, decompose the question~\cite{press2023selfask}, or learn when and what to search with reinforcement learning~\cite{jin2025searchr1,song2025r1searcher}. To reduce repeated search, ReSP adds an LLM summarizer~\cite{jiang2025resp}, and Search Wisely~\cite{wu2025searchwisely}, HiPRAG~\cite{wu2025hiprag}, StepSearch~\cite{zheng2025stepsearch} and OTC~\cite{wang2025otc} penalize redundant calls in training. Other methods add planners~\cite{chen2026stride,li2026par2rag}, evidence controllers~\cite{zhou2026dynakrag}, parallel queries~\cite{liu2026multisearch} or budget control~\cite{java2025frugalrag,fang2026budgetcontrol}. All of these change the agent's policy, so each agent has to be retrained or redesigned. \method only rewrites a flagged query and reorders the final context, so it needs no retraining. Its route scorer is a form of query performance prediction~\cite{carmel2010qpp}.

\textit{Routes from what has been retrieved.}
Pseudo-relevance feedback expands the next query with terms from retrieved text~\cite{rocchio1971relevance,lavrenko2001relevance,abduljaleel2004rm3}, Wikipedia~\cite{xu2009wikiprf}, anchor text~\cite{dang2010anchor} or linked entities~\cite{dalton2014eqfe}, and HyDE~\cite{gao2023hyde}, GenRead~\cite{yu2023genread} and FLARE~\cite{jiang2023flare} retrieve with model-generated text. All of them derive the next query from the question's wording, whereas an association route searches for a page that a retrieved page links to. Multi-hop retrievers follow hyperlinks and bridge entities~\cite{xiong2019bridge,asai2020pathretriever,xiong2021mdr,zhang2024beam}, but they are trained on one collection and replace the agent's retriever. Browsing agents click links~\cite{nakano2021webgpt,zhou2024webarena} and Think-on-Graph~\cite{sun2024tog} walks a knowledge graph, both with a model call at every step. \method needs neither. It rewrites only calls that would repeat an earlier query and picks the link with a lightweight scorer.

\textit{Graph propagation, fusion and context selection.}
Personalized PageRank biases a random walk toward seed nodes~\cite{page1999pagerank,haveliwala2002topic,jeh2003scaling}. Kurland and Lee rerank retrieved documents by centrality~\cite{kurland2005pagerank}, HippoRAG seeds personalized PageRank with the query's entities over a graph that an LLM extracts from the whole corpus in advance~\cite{gutierrez2024hipporag}, and other systems build or encode graph indexes~\cite{edge2024graphrag,wang2026sprig,bacellar2026phasegraph,qi2026cage} or walk a user--item graph~\cite{qiu2025gcrs}. Chain-aware Context Selection runs personalized PageRank over the corpus links among the passages an agent retrieved, after any agent and without model calls. List fusion rewards documents that several lists retrieve (the chorus effect~\cite{vogt1999fusion,fox1994combination}). RRF~\cite{cormack2009rrf} merges query variants in RAG-Fusion~\cite{rackauckas2024ragfusion}, and Round-robin merging comes from distributed retrieval~\cite{voorhees1995fusion,callan1995distributed}. A page reached through a single route appears in only one list, so fusion of this kind tends to drop it. Rerankers~\cite{nogueira2020monot5,sun2023rankgpt,zhang2025qwen3embedding} and compressors~\cite{xu2024recomp} score each passage separately, with a model call per passage.

Prior methods change the agent or its retriever, or build a graph offline. \method works with the links that retrieved pages already carry, without extra search or LLM calls.

\section{Preliminaries and Motivation}
\label{sec:needroute}

\subsection{Definitions}
\label{sec:task}

A search agent answers a request $q$ by interleaving reasoning with
retrieval over a collection $\mathcal{D}$.
Each retrieval call returns a ranked list of the top-$k$ passages.
The process starts with $q_0=q$ and $L_0=R_k(q_0)$.
The agent then reasons over the retrieved evidence and decides
whether further retrieval is needed.
After $T\le B$ follow-up queries, the retrieved pool is
$\mathcal{D}_T=\bigcup_{t=0}^{T}L_t$,
where $B$ is the retrieval budget.

Whether an agent can use a piece of evidence depends on two steps:
\emph{query formulation}, which determines what is retrieved, and
\emph{context construction}, which determines which retrieved passages
reach the reader. Evidence reaches the reader only if it is first
retrieved and then selected, so these two steps together govern the
agent's access to evidence (\secref{sec:overview}). We write both in a
common notation.
\textbf{\emph{(1) Query formulation.}}
At each follow-up step $t$, the agent discovers a retrieval target
$z_t=\gamma(q,s_{t-1})$ and realizes it as a query
$q_t=\phi(q,s_{t-1},z_t)$, which retrieves $L_t=R_k(q_t)$. Here
$\gamma$ is target discovery, $\phi$ is query realization, and
$s_{t-1}$ summarizes the request and the information gathered by
earlier searches. The target may be an unresolved information need or
a specific item to look up. We separate target discovery from query
realization because the two kinds of routes defined below differ only
in the former.
\textbf{\emph{(2) Context construction.}}
After the last search, a selection step $\beta$ chooses at most $n$
passages from the retrieved pool,
$W=\beta(q;L_0,\ldots,L_T)\subseteq\mathcal{D}_T$ with $|W|\le n$,
and the reader answers $\hat a=\rho(q,W)$. Agents may also manage
their context during intermediate reasoning; we treat this as part of
the agent state, and $\beta$ denotes the final selection.

\textit{Retrieval routes.}
We distinguish two ways of discovering subsequent retrieval
targets, semantic routes and association routes.
Let $\mathcal{V}$ denote the set of items (passages) in the collection.
An item is \emph{grounded} if it appears in a retrieved list available
to the agent, regardless of whether the LLM explicitly identifies it.
Let $G_{t-1}\subseteq\mathcal{V}$ denote the set of grounded items
before retrieval step $t$. Let
$\mathcal{E}\subseteq\mathcal{V}\times\mathcal{V}$ be an association
graph representing relations between items.

\begin{definition}[Semantic Route]
\label{def:semantic-route}
A retrieval step $t$ follows a semantic route if its target
$z_t=\gamma_{\mathrm{sem}}(q,s_{t-1})$ is discovered through the
LLM's interpretation of the request and available evidence,
without following an explicit, precomputed item-association link.
The query $q_t=\phi(q,s_{t-1},z_t)$ is then formulated to
retrieve information about the target.
Semantic routes include request rewriting, sub-question
decomposition, and identifying entities for further investigation.
\end{definition}

\begin{definition}[Association Route]
\label{def:association-route}
A retrieval step $t$ follows an association route if the agent
discovers a target item $m\notin G_{t-1}$ by following an available
association link $(a,m)\in\mathcal{E}$ from a grounded item
$a\in G_{t-1}$. The agent then sets $z_t=m$ and formulates
$q_t=\phi(q,s_{t-1},m)$ using the linked item's name together
with relevant terms from the original request.
The target $m$ is not yet grounded when the association link is
followed.
\end{definition}

The distinction between the two routes lies in how the next
retrieval target is discovered. A semantic route derives its
target through LLM reasoning, whereas an association route
obtains its target by following an explicit item
association. For example, for the question of \figref{fig:method}, a
semantic route must infer that Nancy Lincoln's husband is the next
entity to search for, whereas an association route obtains that target
directly from the link.

\subsection{Information Coverage and Retention}
\label{sec:overview}

We measure evidence availability using the annotated gold passages
for each request. Let $\mathcal{G}_q$ denote this set. A request is
covered if all gold passages are retrieved,
$\mathcal{G}_q \subseteq \mathcal{D}_T$, and its final context is
complete if all gold passages are passed to the reader,
$\mathcal{G}_q \subseteq W$. Since $W$ is selected from the retrieved
passages, the probability of obtaining a complete context can be
decomposed as
\begin{equation}
\Pr[\mathcal{G}_q \subseteq W]
=
\Pr[\mathcal{G}_q \subseteq \mathcal{D}_T]
\cdot
\Pr[\mathcal{G}_q \subseteq W
\mid
\mathcal{G}_q \subseteq \mathcal{D}_T],
\label{eq:complete_context}
\end{equation}
where $\Pr[\cdot]$ denotes the fraction of requests in the dataset
satisfying the corresponding event. The first term is the coverage
rate, i.e., the fraction of requests for which all annotated gold
passages are retrieved. The second term is the retention rate, i.e.,
among covered requests, the fraction for which all annotated gold
passages are retained in the final context. Coverage depends on which
queries are issued, while retention depends on which retrieved
passages are kept for the reader. A retrieved gold passage is lost
during context construction when it belongs to $\mathcal{D}_T$ but
not to $W$.

\begin{table}[t]
  \caption{Reach of semantic and association routes (\%), with the best five follow-up calls of each type chosen by an oracle.}
  \label{tab:reach}
  \centering
  \setlength{\aboverulesep}{0pt}\setlength{\belowrulesep}{0pt}\renewcommand{\arraystretch}{1.1}
  \setlength{\tabcolsep}{2.4pt}
  \begin{tabular}{@{}l l r rrr r@{}}
\toprule
 & & & \multicolumn{3}{c}{Best 5 calls reach (\%)} & Missed, \\
\cmidrule(lr){4-6}
Collection & Edges & Req. & Sem. & Assoc. & Both & 1 hop (\%) \\
\midrule
HotpotQA & Hyperlinks & 66.6 & 76.7 & \textbf{85.8} & 91.1 & 53--73 \\
2Wiki & Titles & 61.2 & 69.4 & \textbf{83.9} & 88.1 & 39--78 \\
\bottomrule
\end{tabular}

\end{table}

\subsection{Missing Evidence Is Often One Hop Away}
\label{sec:assoc}

\textit{Experimental Setting.}
We measure the reach of the two route types on HotpotQA and 2Wiki. On HotpotQA, the association
graph is given by Wikipedia hyperlinks; 2Wiki has none, so we add an edge
from paragraph $a$ to paragraph $b$ when the title of $b$ appears in $a$.
Both
route types use the same BM25 retriever~\cite{robertson2009bm25} and five follow-up searches. For
each route type, an oracle selects from its candidate queries the five
that retrieve the most gold items; route reach is the fraction of gold
items these five searches retrieve. In \tabref{tab:reach}, Req.\ is the request alone and Both combines the two route types. The last column is the share of the gold pages an agent never retrieves that are linked from its top three results, as a range over four agents.

\textit{Results.}
Association routes reach more gold evidence than semantic routes on
HotpotQA and 2Wiki, and combining both reaches the most
(\tabref{tab:reach}). Among the gold pages that the agents fail to
retrieve, \pnum{assoc/onehop/hotpotqa/range}\% on HotpotQA and
\pnum{assoc/onehop/2wiki/range}\% on 2Wiki are linked from a page in the
top three of an earlier result. 

\textit{Findings.}
Missing evidence is often reachable from pages the agent has already
retrieved. Since both route types use the same retriever and budget,
the gap lies in how the next target is chosen: agents reason over text
and do not reliably follow the associations that retrieved evidence
exposes. The two route types reach different evidence, so association
routes complement semantic search rather than replace it. Exploiting
associations under a fixed budget therefore requires selecting which
links to follow, since the reach above assumes an oracle choice, and
spending search calls on them without displacing useful semantic queries. This motivate the \secref{sec:route}.

\subsection{Retrieved Evidence Can Be Lost During Context Construction}
\label{sec:lost}

\begin{figure}[t]
  \centering
  \includegraphics[width=\columnwidth]{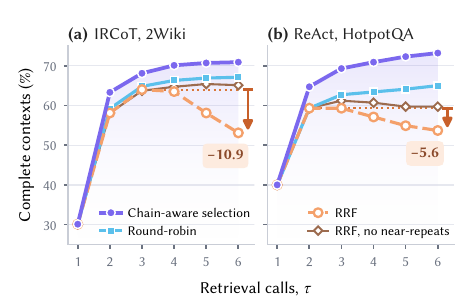}%
  \vspace{-3mm}
\caption{Context completeness as the agent makes more retrieval calls. Chain-aware selection is the context selector of \method (\secref{sec:context}).}
  \label{fig:prefix}
\vspace{-4mm}
\end{figure}

\textit{Experimental Setting.}
We test whether more searches always give the reader a more complete
context. For each request, we take the agent's first $\tau$ retrieval
calls $L_0,\dots,L_{\tau-1}$, where $L_0$ is retrieved by the request
itself, and build a context of $n{=}10$ passages from these lists of
$k{=}10$ passages each. The retrieved pool only grows with $\tau$, so
coverage cannot decrease, and by \eqnref{eq:complete_context} any drop
in the share of complete contexts comes from lower retention. We compare
two merges. Reciprocal rank fusion (RRF)~\cite{cormack2009rrf}, common
in multi-query RAG~\cite{rackauckas2024ragfusion}, scores a passage by
summing $1/(c+r)$ over the lists in which it appears at rank $r$
($c{=}59$), so passages retrieved by several lists rank higher.
Round-robin merging interleaves the lists and removes
duplicates~\cite{voorhees1995fusion,callan1995distributed}; it ignores
repetition and serves as a control. A follow-up query is a near-repeat
if it equals an earlier query or has a content-word Jaccard similarity
of at least 0.8 with one. To test whether near-repeats cause the drop,
we also apply RRF after removing the lists returned by near-repeat calls
(``RRF, no near-repeats''). \figref{fig:prefix} shows
IRCoT~\cite{trivedi2023ircot} on 2Wiki and ReAct~\cite{yao2023react} on
HotpotQA; \tabref{tab:repeat-retention} averages over all four agents.

\textit{Results.}
Under RRF, the share of complete contexts peaks after two or three calls
and then falls, by \pnum{pre/wiki/IRCoT/drop} points for IRCoT on 2Wiki
and \pnum{pre/hot/ReAct/drop} points for ReAct on HotpotQA, while
Round-robin merging keeps rising (\figref{fig:prefix}). At the end of
search, RRF excludes a retrieved gold passage that Round-robin keeps in
18.6\% of 2Wiki and 13.9\% of HotpotQA questions.

\begin{table}[t]
  \caption{Near-repeat follow-up calls compared with all other calls, mean of four agents.}
  \label{tab:repeat-retention}
  \centering \setlength{\tabcolsep}{4pt}
  \setlength{\aboverulesep}{0pt}\setlength{\belowrulesep}{0pt}\renewcommand{\arraystretch}{1.1}
  \begin{tabular}{@{}lcccc@{}}
\toprule
& \multicolumn{2}{c}{2Wiki} & \multicolumn{2}{c}{HotpotQA} \\
\cmidrule(lr){2-3}\cmidrule(l){4-5}
& Repeat & Other & Repeat & Other \\
\midrule
Share of calls (\%) & 28 & 72 & 22 & 78 \\
Seen passages (\%)  & 87 & 43 & 83 & 41 \\
New gold per call   & 0.01 & 0.20 & 0.01 & 0.11 \\
\bottomrule
\end{tabular}

  \vspace{-5mm}
\end{table}

\tabref{tab:repeat-retention} shows that near-repeats make up about a
quarter of follow-up calls. Most passages they return were already
retrieved, and they add almost no new gold passages. Removing their
lists almost removes the decline: from \pnum{pre/wiki/IRCoT/drop} to 0.3
points for IRCoT on 2Wiki, and from \pnum{pre/hot/ReAct/drop} to 1.5
points for ReAct on HotpotQA.

\textit{Findings.}
More retrieval does not guarantee a more complete context: beyond the
first few calls, losses in retention outweigh gains in coverage. The
loss arises when repeated search meets a merge that rewards repetition,
so passages the agent already has push out gold passages that only one
list returns. Round-robin merging keeps improving under the same context
budget, so the budget alone does not explain the loss. Near-repeat calls
thus add little evidence while harming retention. A page reached through
an association is new when retrieved and appears in a single list, so a
repetition-based merge is likely to drop it; keeping such evidence
requires selecting context by how passages relate to the rest of the
evidence rather than by how often they are retrieved. This motivate the \secref{sec:context}.

\begin{figure*}[t]
  \centering
\includegraphics[width=\textwidth]{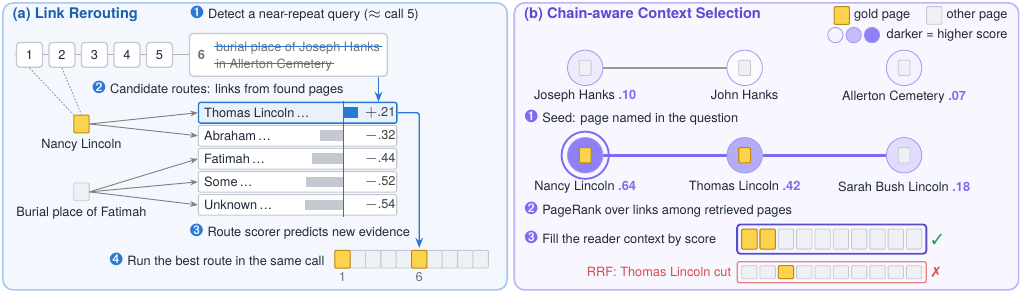}%
  \caption{\method on a 2Wiki question, ``Where was Nancy Lincoln's husband buried?'' (ReAct; squares: gold pages). (a) Link Rerouting replaces call~6, a near-repeat of call~5, with a route to \emph{Thomas Lincoln}. (b) Chain-aware Context Selection keeps him, while RRF cuts him; shading shows $s(d)$.}
  \vspace{-2mm}
  \label{fig:method}
\end{figure*}

\section{\method}
\label{sec:method}
\method is a plug-in that raises
the probability of a complete reader context, $\Pr[\mathcal{G}_q\subseteq W]$,
without extra retrieval budget. Its two components target the two factors
of \eqnref{eq:complete_context}. \emph{Link Rerouting} improves
coverage. It redirects follow-up calls flagged as near-repeats onto
association routes, and because a rerouted call replaces the original
call, the number of calls stays the same. \emph{Chain-aware Context
Selection} improves retention by keeping related evidence together when
it builds the reader context.

\subsection{Link Rerouting}
\label{sec:route}

Link Rerouting identifies near-repeat queries and redirects them
through existing associations in the retrieved evidence. A follow-up
query $q_t$ is flagged if it matches an earlier query after
normalization, or if
\[
\max_{0\le j<t}
\frac{|w(q_t)\cap w(q_j)|}{|w(q_t)\cup w(q_j)|}
\ge \theta,
\]
where $w(x)$ denotes the set of lower-cased content words after
stopword removal, $q_0$ is the original question, and $\theta=0.8$.
For a flagged query, we construct candidate association routes
$\mathcal{C}_t$ from links on top-ranked retrieved passages, called
anchors. Each linked page not yet retrieved provides a candidate route
formed by its title and the first six content words of $q$.
Previously executed routes are excluded.

We rank candidate routes by their expected contribution to the
retrieved evidence. For training, the gain of a route $r$ in search
state $s$ is defined as the number of newly retrieved gold passages,
$g(r\mid s)=|(R_k(r)\cap\mathcal{G}_q)\setminus\mathcal{D}(s)|$.
We use a lightweight LightGBM ranker~\cite{ke2017lightgbm},
trained with LambdaRank on route gains recorded from training
search trajectories, to estimate this gain for candidate routes.
The ranker uses features of the route, original question, and
search history.

When a query is flagged and candidate routes are available, we
replace it with the highest-scoring route:
\[
q'_t =
\begin{cases}
\arg\max_{r\in\mathcal{C}_t}\hat g(r\mid s_{t-1}),
& \text{if flagged and }\mathcal{C}_t\neq\emptyset,\\
q_t, & \text{otherwise.}
\end{cases}
\]
The selected route is executed through the agent's existing retrieval
interface, and the resulting evidence is returned to the agent for
its subsequent decisions.

\subsection{Chain-aware Context Selection}
\label{sec:context}

Chain-aware Context Selection combines Round-robin merging, which ignores repetition, with personalized PageRank over associations. After the agent's final search, we form a
prior order $\pi$ by interleaving the distinct retrieval lists in
Round-robin order, keeping each passage once. Equivalently, passages
are ordered by their best rank across the lists, with ties broken by
list order.

We then identify seeds from retrieved passages corresponding to items
mentioned in the original request. If no such passages are found,
we use the highest-ranked passages in $\pi$. Using existing corpus
associations, we build a graph over the retrieved passages and run
personalized PageRank~\cite{haveliwala2002topic,jeh2003scaling}
from the seeds to obtain an association score $p(d)$ for each passage.
We combine this score with the prior ranking as
$s(d)=p(d)+\mu/(\pi(d)+1)$, where $\pi(d)$ is the 0-based position
of passage $d$ and $\mu$ controls the contribution of the prior.
The final context contains the top $n$ passages ranked by $s(d)$,
with near-duplicates removed.

\providecommand{\skel}[1]{\textcolor{blue}{[#1]}}
\providecommand{\tbd}[1]{\textcolor{red}{#1}}

\section{Experimental Setup}
\label{sec:exp}\label{sec:setup}

\subsection{Datasets}
HotpotQA~\cite{yang2018hotpotqa} comes with 5.23M Wikipedia abstracts and
19.5M hyperlinks, which serve as its associations. For 2WikiMultiHopQA
(2Wiki)~\cite{ho2020constructing} we use the 415K-paragraph collection of
IRCoT~\cite{trivedi2023ircot}; it has no hyperlinks, so paragraph $a$
links to paragraph $b$ when $b$'s title occurs in $a$
(\secref{sec:assoc}, \appref{app:impl}). As the official test answers are not public, we follow prior work~\cite{trivedi2023ircot,jin2025searchr1} and evaluate on development questions, which have gold answers and supporting passages. We sample 1{,}500 development questions per
collection, drawn at random from the development split before the
experiments; the route scorer is trained only on questions from the
training split.

\subsection{Budget-matched Comparison}
Every agent first searches with the question itself and may then make
at most five follow-up searches, each returning 10 passages. 
An agent
alone and the same agent with \method share the retriever, this limit
and the reader budget. A rerouted call replaces the
agent's call instead of adding one, and the plug-in makes no LLM call.

\subsection{Baselines}
\label{sec:baselines}
We compare each agent (\secref{sec:hosts}) alone with the same agent plus \method. Alone, an agent's
retrieved lists are merged into the reader's 10 passages by
Round-robin~\cite{voorhees1995fusion,callan1995distributed}, which
interleaves the lists by rank and removes duplicates. We use it as the
default context construction because it keeps more gold passages than
RRF~\cite{cormack2009rrf}, the other common merge (\secref{sec:lost}). To test
whether a costlier context selector would do as well, we also compare two
off-the-shelf selectors that score passages with a model. The LLM judge
asks Qwen2.5-7B-Instruct for each passage whether it helps answer the
question, and Qwen3-Reranker-8B~\cite{zhang2025qwen3embedding} uses its
released weights. Both score the distinct passages among the top
40 of the RRF merge, and the ten highest-scoring passages form the
context.
For the rerouted call, we compare the route scorer with no rerouting, a
random link, title match and cross-encoder rerankers
(\secref{sec:rerank}).

\subsection{Evaluation}

\subsubsection{Plug-in replay}
We run each agent alone once and record its queries, results and
stopping point; every method is evaluated on this recorded run. A
rerouted call adds its results to the retrieved pool, but the agent's
later queries stay as recorded, so all methods make the same number of
searches. Replay is exact for Plan, whose queries are fixed in advance,
and for every comparison without rerouting (\tabref{tab:selectors}).
Results are from replay unless marked as live.

\subsubsection{Live agent}
\label{sec:full-observation}
Live, the agent sees what a rerouted call found and may change course;
we run IRCoT, ReAct and Search-R1 live with \method on the same
questions, against the same \emph{Alone} runs. A live run may make at
most as many searches as the agent alone made on that question, and it
reuses the recorded responses up to the first rerouted call.

\subsubsection{Metric}
We report exact match (EM) of the reader's answer and the share of
\emph{complete} contexts, which hold every annotated gold passage or a
near-duplicate of it (\secref{sec:overview}). For each comparison,
we estimate the mean paired difference and its 95\% confidence interval
with a paired bootstrap over questions (10{,}000 resamples); four-agent averages resample questions jointly. In the tables, $^{\dagger}$ marks a difference whose 95\% confidence interval excludes zero, against the agent alone unless the caption names another reference.

\subsection{Implementation}
\label{sec:impl}
\label{sec:config}
All settings are shared by all agents and collections (details in
\appref{app:impl}, variations in \appref{app:robust}). Retrieval uses BM25. Link Rerouting flags a follow-up
call at $\theta=0.8$; one LightGBM LambdaRank
scorer~\cite{ke2017lightgbm} per collection ranks the candidate routes,
trained on training questions disjoint from evaluation. Chain-aware
Context Selection uses at most four seeds, excludes hubs with corpus
in-degree above 1{,}000, and runs personalized PageRank with $\alpha=0.3$
and prior weight $\mu=0.2$.

\subsubsection{Computational overhead}
Neither component adds an LLM or retrieval call, and both run on a CPU.
The route scorer occupies 46\,kB (2Wiki) and 122\,kB (HotpotQA); the
association graphs, stored as memory-mapped compressed-sparse-row arrays,
occupy 8.4\,MB and 119.8\,MB beyond the corpus and index the agent
already uses. The model-based selectors of \secref{sec:baselines} run a 7B or 8B model on about \pnum{rev/x7/passages} passages
per question.

\subsubsection{Agents}
\label{sec:hosts}
We attach \method to four agents (\tabref{tab:agents} in \appref{app:agents}). IRCoT~\cite{trivedi2023ircot} issues each reasoning sentence as a query and makes five follow-up searches; ReAct~\cite{yao2023react} chooses between SEARCH and FINISH; Search-R1~\cite{jin2025searchr1} uses the released 7B PPO checkpoint, trained with E5~\cite{wang2022e5} on NQ~\cite{kwiatkowski2019nq} and HotpotQA, with BM25 in place of E5; and Plan, our own baseline, writes three sub-questions and two keyword queries in one LLM call. IRCoT, ReAct and Plan run on Qwen2.5-7B-Instruct~\cite{qwen25}.

\subsubsection{Reader}
The reader, Qwen2.5-7B-Instruct, answers from 10 passages of 600
characters~\cite{trivedi2023ircot} in every QA comparison.

\section{Results}
\label{sec:results}

The results answer four research questions.
\begin{itemize}[leftmargin=*,itemsep=1pt,topsep=2pt]
\item[] \textbf{RQ1 (Effectiveness).} With the same search calls and reader, does \method answer more questions correctly for every agent, in replay and live? (\secref{sec:main})
\item[] \textbf{RQ2 (Contribution).} How much does each component
contribute, and do the two depend on each other? (\secref{sec:ablation})
\item[] \textbf{RQ3 (Mechanism).} Do the links cause the gain, or would reordering, model scoring or a stronger route selector do as well? (\secref{sec:mechanism})
\item[] \textbf{RQ4 (Limits).} What limits the gain, and on which
questions and agents does \method help most? (\secref{sec:reachgain})
\end{itemize}

\begin{table}[t]
  \caption{EM (\%) of each agent alone and with \method, 1{,}500 questions per collection.}
  \label{tab:main}
  \centering
  \providecommand{\gain}[1]{\,{\textcolor{gainC}{$+$#1}}}
\providecommand{\sig}{$^{\dagger}$}
\ifdefined\tabMainFirst\else\newlength\tabMainFirst\newlength\tabMainRest\fi
\begingroup
\setlength{\aboverulesep}{0pt}\setlength{\belowrulesep}{0pt}\renewcommand{\arraystretch}{1.1}
\renewcommand{\sig}{\rlap{$^{\dagger}$}}
\setlength{\tabcolsep}{4pt}
\settowidth{\tabMainFirst}{Search-R1}\settowidth{\tabMainRest}{\textit{HotpotQA}}
\ifdim\tabMainRest>\tabMainFirst\setlength{\tabMainFirst}{\tabMainRest}\fi
\setlength{\tabMainRest}{\dimexpr0.85\columnwidth-\tabMainFirst-8\tabcolsep\relax}
\newcommand{\planboth}[1]{\multicolumn{2}{>{\columncolor{detourC!8}[\tabcolsep][0pt]\centering\arraybackslash}p{\dimexpr0.58\tabMainRest+2\tabcolsep\relax}@{}}{#1}}
\newcommand{\dsrow}[1]{\multicolumn{5}{@{}>{\columncolor{black!7}[0pt][0pt]}c@{}}{\textbf{#1}}\\}
\begin{tabular}{@{}p{\tabMainFirst}
  >{\centering\arraybackslash}p{0.25\tabMainRest}
  >{\centering\arraybackslash}p{0.17\tabMainRest}
  >{\columncolor{detourC!8}\centering\arraybackslash}p{0.29\tabMainRest}
  >{\columncolor{detourC!8}[\tabcolsep][0pt]\centering\arraybackslash}p{0.29\tabMainRest}@{}}
\toprule
Agent & Calls & Alone & Replay & Live \\
\midrule
\dsrow{2Wiki}
IRCoT      & 6.0       & 33.5 & 37.8\gain{4.3}\sig          & \textbf{38.0}\gain{4.5}\sig \\
ReAct      & 5.10/5.05 & 32.8 & \textbf{36.7}\gain{3.9}\sig & \textbf{36.7}\gain{3.9}\sig \\
Search-R1  & 3.81/3.69 & 34.1 & \textbf{36.5}\gain{2.3}\sig & 36.1\gain{2.0}\sig          \\
Plan$^{*}$ & 6.0       & 26.1 & \planboth{\textbf{31.5}\gain{5.3}\sig}                    \\
Mean       &           & 31.6 & \textbf{35.6}\gain{4.0}\sig & \textbf{35.6}\gain{3.9}\sig \\
\midrule
\dsrow{HotpotQA}
IRCoT      & 6.0       & 43.4 & 44.7\gain{1.3}              & \textbf{45.1}\gain{1.7}\sig \\
ReAct      & 5.10/5.06 & 42.3 & \textbf{44.1}\gain{1.9}\sig & 44.0\gain{1.7}\sig          \\
Search-R1  & 3.39/3.33 & 41.9 & \textbf{42.5}\gain{0.6}     & 42.1\gain{0.2}              \\
Plan$^{*}$ & 6.0       & 35.0 & \planboth{\textbf{36.8}\gain{1.8}\sig}                    \\
Mean       &           & 40.6 & \textbf{42.0}\gain{1.4}\sig & \textbf{42.0}\gain{1.4}\sig          \\
\bottomrule
\end{tabular}
\endgroup

\end{table}

\subsection{Answer Accuracy under Matched Budgets}
\label{sec:main}

\method raises EM for every agent on both collections without any extra call, in replay and in live runs (significantly in 13 of 16 cells).
Under replay (\tabref{tab:main}), averaged over the four agents, \method adds
\pnum{wiki/detour/EM/pool/abs} points to the agents alone on 2Wiki and
\pnum{hot/detour/EM/pool/abs} on HotpotQA. On 2Wiki the gain ranges from
$+2.3$ for Search-R1 to $+5.3$ for Plan; on HotpotQA it is smallest for
Search-R1 ($+0.6$), the agent that searches least. Against RRF merging (\tabref{tab:selectors}), the gains would be $+5.5$ and $+2.4$ points.

The gain holds when the agents run live. On
the same questions, \method improves EM over the agents alone by
3.9 points on 2Wiki and 1.4 on HotpotQA,
averaged over the four agents (Plan's queries are fixed, so its live run equals its replay). Per agent, the live gains differ from the
replayed ones by at most 0.4 points and on average by none, so replay and live runs give the same estimate. ReAct and Search-R1 make slightly fewer searches with \method (Calls in \tabref{tab:main}). Live, rerouting adds 1.5 and
0.6 points over Chain-aware Context Selection alone, both significant.

The gain also holds with 5, 10 or 20 passages, untruncated passages or a
Qwen2.5-32B-Instruct reader ($+4.3$ and $+2.0$), and changes by less than
one point with the rerouting threshold or the scorer's training data
(\appref{app:robust}). Under a dense retriever the EM gain shrinks (\secref{sec:reachgain}).

\subsection{Contribution of the Two Components}
\label{sec:ablation}

Both components are needed, and rerouting helps mainly once context selection keeps what it finds.
We remove each component of \method to measure what it contributes
(\tabref{tab:ablation}).
Without Link Rerouting, EM
drops by \pnum{wiki/route/EM/pool/abs} points on 2Wiki and
\pnum{hot/route/EM/pool/abs} on HotpotQA; without Chain-aware Context
Selection, it drops by \pnum{wiki/compose/EM/pool/abs} and
\pnum{hot/compose/EM/pool/abs}.

The two components depend on each other. On its own, rerouting adds almost nothing to EM ($+0.8$ and $+0.3$).
A page found by a rerouted call usually appears in only one list, which
merging tends to drop (\secref{sec:lost}). By \eqnref{eq:complete_context}, rerouting
raises coverage, but without better retention little of this gain
reaches the reader. Once context selection keeps such pages through
their links, rerouting adds $+1.5$ and $+0.6$ EM points, and its gain in
complete contexts grows from $+3.7$ to $+6.0$ points on 2Wiki and from
$+0.9$ to $+2.5$ on HotpotQA.

\begin{table}[t]
  \caption{EM and complete contexts (\%) with each component of \method switched off, mean of four agents in replay. Daggers mark differences from the full \method.}
  \label{tab:ablation}\label{tab:full-observation}
  \centering
  \ifdefined\tabCompR\else\newlength\tabCompR\newlength\tabCompS\newlength\tabCompW\newlength\tabCompH\newlength\tabCompX\fi
\providecommand{\tabdag}{\rlap{$^{\dagger}$}}
\begingroup
\setlength{\aboverulesep}{0pt}\setlength{\belowrulesep}{0pt}\renewcommand{\arraystretch}{1.1}
\setlength{\tabcolsep}{3pt}
\settowidth{\tabCompR}{Rerouting}\settowidth{\tabCompS}{selection}%
\settowidth{\tabCompW}{2Wiki}\settowidth{\tabCompH}{HotpotQA}%
\setlength{\tabCompX}{\dimexpr(0.85\columnwidth-\tabCompR-\tabCompS-2\tabCompW-2\tabCompH-10\tabcolsep)/6\relax}%
\addtolength{\tabCompR}{\tabCompX}\addtolength{\tabCompS}{\tabCompX}%
\addtolength{\tabCompW}{\tabCompX}\addtolength{\tabCompH}{\tabCompX}%
\newcommand{\detourrow}[6]{%
  \multicolumn{1}{@{}>{\columncolor{detourC!8}[0pt][\tabcolsep]\centering\arraybackslash}p{\tabCompR}}{#1} &
  \multicolumn{1}{>{\columncolor{detourC!8}\centering\arraybackslash}p{\tabCompS}}{#2} &
  \multicolumn{1}{>{\columncolor{detourC!8}\centering\arraybackslash}p{\tabCompW}}{#3} &
  \multicolumn{1}{>{\columncolor{detourC!8}\centering\arraybackslash}p{\tabCompH}}{#4} &
  \multicolumn{1}{>{\columncolor{detourC!8}\centering\arraybackslash}p{\tabCompW}}{#5} &
  \multicolumn{1}{>{\columncolor{detourC!8}[\tabcolsep][0pt]\centering\arraybackslash}p{\tabCompH}@{}}{#6} \\}
\begin{tabular}{@{}>{\centering\arraybackslash}p{\tabCompR}
  >{\centering\arraybackslash}p{\tabCompS}
  >{\centering\arraybackslash}p{\tabCompW}
  >{\centering\arraybackslash}p{\tabCompH}
  >{\centering\arraybackslash}p{\tabCompW}
  >{\centering\arraybackslash}p{\tabCompH}@{}}
\toprule
Link & Context & \multicolumn{2}{c}{EM} & \multicolumn{2}{c}{Complete (\%)} \\
\cmidrule(lr){3-4}\cmidrule(l){5-6}
Rerouting & Selection & 2Wiki & HotpotQA & 2Wiki & HotpotQA \\
\midrule
--          & --          & 31.6\tabdag & 40.6\tabdag & 59.3\tabdag & 60.2\tabdag \\
\ding{51}   & --          & 32.4\tabdag & 40.9\tabdag & 63.0\tabdag & 61.1\tabdag \\
--          & \ding{51}   & 34.1\tabdag & 41.4\tabdag & 64.5\tabdag & 67.9\tabdag \\
\detourrow{\ding{51}}{\ding{51}}{\textbf{35.6}}{\textbf{42.0}}{\textbf{70.4}}{\textbf{70.4}}
\bottomrule
\end{tabular}
\endgroup

\end{table}

\subsection{Why the Links Matter}
\label{sec:mechanism}

In both components, the gain comes from following the collection's links.
We start with context selection, on which rerouting depends (\secref{sec:ablation}).

\begin{table}[t]
\caption{Context selection on the agents' own lists without rerouting, mean of four agents. Daggers mark differences from Chain-aware.}
  \label{tab:selectors}\label{tab:design-controls}
  \centering
  \ifdefined\tabSelA\else\newlength\tabSelA\newlength\tabSelM\newlength\tabSelW\newlength\tabSelH\newlength\tabSelX\fi
\ifdefined\tabSelBox\else\newsavebox\tabSelBox\fi
\providecommand{\tabdag}{\rlap{$^{\dagger}$}}
\begingroup
\setlength{\aboverulesep}{0pt}\setlength{\belowrulesep}{0pt}\renewcommand{\arraystretch}{1.1}
\setlength{\tabcolsep}{1pt}
\def\tabSelWidth{0.85\columnwidth}%
\settowidth{\tabSelA}{\quad named pages, no links}\settowidth{\tabSelM}{27--28}%
\settowidth{\tabSelW}{2Wiki}\settowidth{\tabSelH}{HotpotQA}%
\setlength{\tabSelX}{\dimexpr(\tabSelWidth-\tabSelA-\tabSelM-2\tabSelW-2\tabSelH-10\tabcolsep)/6\relax}%
\ifdim\tabSelX<0pt\setlength{\tabSelX}{0pt}\fi
\addtolength{\tabSelA}{\tabSelX}\addtolength{\tabSelM}{\tabSelX}%
\addtolength{\tabSelW}{\tabSelX}\addtolength{\tabSelH}{\tabSelX}%
\newcommand{\detourrow}[6]{%
  \multicolumn{1}{@{}>{\columncolor{detourC!8}[0pt][\tabcolsep]\raggedright\arraybackslash}p{\tabSelA}}{#1} &
  \multicolumn{1}{>{\columncolor{detourC!8}\centering\arraybackslash}p{\tabSelM}}{#2} &
  \multicolumn{1}{>{\columncolor{detourC!8}\centering\arraybackslash}p{\tabSelW}}{#3} &
  \multicolumn{1}{>{\columncolor{detourC!8}\centering\arraybackslash}p{\tabSelH}}{#4} &
  \multicolumn{1}{>{\columncolor{detourC!8}\centering\arraybackslash}p{\tabSelW}}{#5} &
  \multicolumn{1}{>{\columncolor{detourC!8}[\tabcolsep][0pt]\centering\arraybackslash}p{\tabSelH}@{}}{#6} \\}
\sbox{\tabSelBox}{%
\begin{tabular}{@{}>{\raggedright\arraybackslash}p{\tabSelA}
  >{\centering\arraybackslash}p{\tabSelM}
  >{\centering\arraybackslash}p{\tabSelW}
  >{\centering\arraybackslash}p{\tabSelH}
  >{\centering\arraybackslash}p{\tabSelW}
  >{\centering\arraybackslash}p{\tabSelH}@{}}
\toprule
 & Model & \multicolumn{2}{c}{EM} & \multicolumn{2}{c}{Complete (\%)} \\
\cmidrule(lr){3-4}\cmidrule(l){5-6}
Context selection & calls & 2Wiki & HotpotQA & 2Wiki & HotpotQA \\
\midrule
RRF                         & 0 & 30.1\tabdag & 39.6\tabdag & 50.5\tabdag & 52.0\tabdag \\
Round-robin (alone)         & 0 & 31.6\tabdag & 40.6 & 59.3\tabdag & 60.2\tabdag \\
\detourrow{Chain-aware (\method)$^{\ddagger}$}{0}{\textbf{34.1}}{41.4}{\textbf{64.5}}{67.9}
\quad named pages, no links & 0 & 32.9\tabdag & 40.4\tabdag & 61.0\tabdag & 62.4\tabdag \\
\quad text-similarity graph & 0 & 32.7\tabdag & 40.2\tabdag & 61.1\tabdag & 63.4\tabdag \\
\quad rewired links         & 0 & 32.8\tabdag & 41.3 & 63.4\tabdag & 65.9\tabdag \\
\midrule
LLM judge (7B)              & 27--28 & 32.1\tabdag & 40.5 & 53.8\tabdag & 61.8\tabdag \\
\quad + link propagation    & 27--28 & \textbf{34.1} & 41.8 & 64.2 & \textbf{68.5}\tabdag \\
Qwen3-Reranker (8B)$^{\ddagger}$ & 27--28 & 32.6\tabdag & \textbf{42.0} & 57.9\tabdag & 68.1 \\
\bottomrule
\end{tabular}}%
\usebox{\tabSelBox}
\par\smallskip
\parbox{\wd\tabSelBox}{\footnotesize $^{\ddagger}$Retrieved gold pages that the question does not name, share kept in the
context: Chain-aware 97.8\% (2Wiki) and 96.2\% (HotpotQA); Qwen3-Reranker 81.2\% and 97.8\%.}
\endgroup

\end{table}

\subsubsection{Keeping the Evidence: Chain-aware Context Selection}
\label{sec:ctx-analysis}

In \tabref{tab:selectors}, model calls counts the passages a selector scores per question. Replacing the links costs EM. We keep everything else fixed and replace only the links
(\tabref{tab:selectors}, indented rows). We put the named pages first
without links, spread the scores over text similarity instead, or spread
them over randomly rewired links. Each change keeps fewer complete
contexts on both collections and costs about one EM point. The only case
where EM holds is rewired links on HotpotQA, and even there fewer
contexts are complete. The gain needs the collection's links, which connect the pages a question needs.

An LLM judge and Qwen3-Reranker-8B score each retrieved passage with a model. On 2Wiki, Chain-aware Context Selection keeps more complete contexts than either (64.5\% against at most 57.9\%) without any model call; on HotpotQA it ties with the reranker. With rerouting added, \method is 3.5 and 1.5 EM points above the judge, 3.0 points above the reranker on 2Wiki, and level with it on HotpotQA. Both models judge each passage against the question alone, so a second-hop page such as Thomas Lincoln (\figref{fig:method}), which the question never mentions, scores low; only its link to Nancy Lincoln shows that it belongs. Adding link propagation to the judge's ranking raises complete contexts from 53.8\% to 64.2\% on 2Wiki and from 61.8\% to 68.5\% on HotpotQA (\tabref{tab:selectors}). The gap lies in pages the question does not name, which Chain-aware Context Selection almost always keeps and the reranker drops in nearly a fifth of 2Wiki cases.

   \begin{figure}[t]
     \centering
     \begin{tikzpicture}[paperfig, x=1mm, y=1mm, font=\fontsize{7}{8.4}\selectfont,
  frm/.style={draw=black!45, line width=0.4pt}, grd/.style={draw=black!9, line width=0.3pt},
  tck/.style={inner sep=0pt, text=black!70, font=\fontsize{6.5}{7.5}\selectfont},
  val/.style={inner sep=0.35pt, fill=white, text=black!70, font=\fontsize{6.5}{7.5}\selectfont}]
  \draw[grd] (3.40,-28.00) -- (3.40,2.20);
  \draw[frm] (3.40,-28.00) -- ++(0,0.8);
  \node[tck, anchor=north] at (3.40,-28.90) {$-$1.5};
  \draw[grd] (7.65,-28.00) -- (7.65,2.20);
  \draw[frm] (7.65,-28.00) -- ++(0,0.8);
  \node[tck, anchor=north] at (7.65,-28.90) {$-$1};
  \draw[grd] (11.90,-28.00) -- (11.90,2.20);
  \draw[frm] (11.90,-28.00) -- ++(0,0.8);
  \node[tck, anchor=north] at (11.90,-28.90) {$-$0.5};
  \draw[grd] (16.15,-28.00) -- (16.15,2.20);
  \draw[frm] (16.15,-28.00) -- ++(0,0.8);
  \node[tck, anchor=north] at (16.15,-28.90) {0};
  \draw[grd] (20.40,-28.00) -- (20.40,2.20);
  \draw[frm] (20.40,-28.00) -- ++(0,0.8);
  \node[tck, anchor=north] at (20.40,-28.90) {+0.5};
  \draw[frm] (0.00,-28.00) -- (22.95,-28.00);
  \fill[black!47] (3.40,-1.20) rectangle (16.15,1.20);
  \node[val, anchor=west] at (3.85,0.00) {$-$1.5};   %
  \fill[black!47] (7.65,-5.20) rectangle (16.15,-2.80);
  \node[val, anchor=east] at (7.20,-4.00) {$-$1.0};
  \fill[black!47] (10.20,-9.20) rectangle (16.15,-6.80);
  \node[val, anchor=east] at (9.75,-8.00) {$-$0.7};
  \fill[detourC, draw=white, line width=0.25pt] (15.15,-13.80) -- (16.15,-12.80) -- (17.15,-13.80) -- (16.15,-14.80) -- cycle;
  \node[val, anchor=west, text=black] at (17.55,-13.80) {\textbf{35.6}};
  \fill[black!47] (16.15,-19.00) rectangle (17.85,-16.60);
  \node[val, anchor=west] at (18.30,-17.80) {+0.2};
  \fill[black!47] (16.15,-23.00) rectangle (17.85,-20.60);
  \node[val, anchor=west] at (18.30,-21.80) {+0.2};
  \fill[black!47] (14.45,-27.00) rectangle (16.15,-24.60);
  \node[val, anchor=east] at (14.00,-25.80) {$-$0.2};
  \draw[detourC, line width=0.7pt] (16.15,-28.00) -- (16.15,2.20);
  \node[inner sep=0pt, anchor=south] at (11.48,3.20) {\textbf{2Wiki}};
  \draw[grd] (29.85,-28.00) -- (29.85,2.20);
  \draw[frm] (29.85,-28.00) -- ++(0,0.8);
  \node[tck, anchor=north] at (29.85,-28.90) {$-$1.5};
  \draw[grd] (34.10,-28.00) -- (34.10,2.20);
  \draw[frm] (34.10,-28.00) -- ++(0,0.8);
  \node[tck, anchor=north] at (34.10,-28.90) {$-$1};
  \draw[grd] (38.35,-28.00) -- (38.35,2.20);
  \draw[frm] (38.35,-28.00) -- ++(0,0.8);
  \node[tck, anchor=north] at (38.35,-28.90) {$-$0.5};
  \draw[grd] (42.60,-28.00) -- (42.60,2.20);
  \draw[frm] (42.60,-28.00) -- ++(0,0.8);
  \node[tck, anchor=north] at (42.60,-28.90) {0};
  \draw[grd] (46.85,-28.00) -- (46.85,2.20);
  \draw[frm] (46.85,-28.00) -- ++(0,0.8);
  \node[tck, anchor=north] at (46.85,-28.90) {+0.5};
  \draw[frm] (26.45,-28.00) -- (49.40,-28.00);
  \fill[black!47] (37.50,-1.20) rectangle (42.60,1.20);
  \node[val, anchor=east] at (37.05,0.00) {$-$0.6};
  \fill[black!47] (39.20,-5.20) rectangle (42.60,-2.80);
  \node[val, anchor=east] at (38.75,-4.00) {$-$0.4};
  \fill[black!47] (41.75,-9.20) rectangle (42.60,-6.80);
  \node[val, anchor=east] at (41.30,-8.00) {$-$0.1};
  \fill[detourC, draw=white, line width=0.25pt] (41.60,-13.80) -- (42.60,-12.80) -- (43.60,-13.80) -- (42.60,-14.80) -- cycle;
  \node[val, anchor=west, text=black] at (44.00,-13.80) {\textbf{42.0}};
  \fill[black!47] (42.60,-19.00) rectangle (44.30,-16.60);
  \node[val, anchor=west] at (44.75,-17.80) {+0.2};
  \fill[black!47] (42.60,-23.00) rectangle (44.30,-20.60);
  \node[val, anchor=west] at (44.75,-21.80) {+0.2};
  \fill[black!47] (42.60,-27.00) rectangle (43.45,-24.60);
  \node[val, anchor=west] at (43.90,-25.80) {+0.1};
  \draw[detourC, line width=0.7pt] (42.60,-28.00) -- (42.60,2.20);
  \node[inner sep=0pt, anchor=south] at (37.93,3.20) {\textbf{HotpotQA}};
  \node[inner sep=0pt, anchor=east] at (-1.20,0.00) {No rerouting};
  \node[inner sep=0pt, anchor=east] at (-1.20,-4.00) {Random link};
  \node[inner sep=0pt, anchor=east] at (-1.20,-8.00) {Title match};
  \node[inner sep=0pt, anchor=east] at (-1.20,-13.80) {\textbf{Full \method{} (LightGBM)}};
  \node[inner sep=0pt, anchor=east] at (-1.20,-17.80) {+ MiniLM cross-encoder (22M)};
  \node[inner sep=0pt, anchor=east] at (-1.20,-21.80) {+ bge reranker (278M)};
  \node[inner sep=0pt, anchor=east] at (-1.20,-25.80) {+ Qwen3-Reranker (0.6B)};
  \node[inner sep=0pt, anchor=north] at (24.70,-32.60) {EM difference from the full \method{} (points)};
\end{tikzpicture}
\caption{EM change from the full \method{} under other ways of choosing the route, mean of four agents.}
\vspace{-5mm}
     \label{fig:rerank}
   \end{figure}
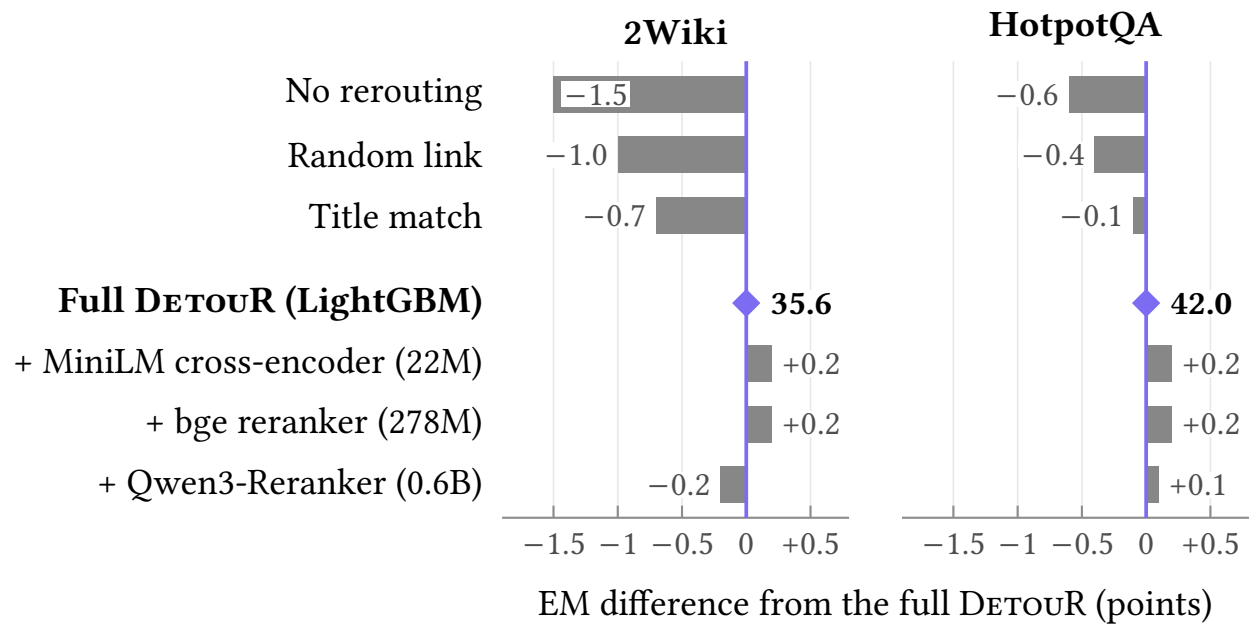

\subsubsection{Reaching the Evidence: Link Rerouting}
\label{sec:design}\label{sec:rerank}
When a query is flagged, the choice is which link to follow. We replace
the LightGBM route scorer with simpler choices (no rerouting, a random
link, title match) and re-rank its top five routes with three
cross-encoders: MiniLM~\cite{wang2020minilm} (22M),
bge~\cite{xiao2024cpack} (278M) and
Qwen3-Reranker~\cite{zhang2025qwen3embedding} (0.6B)
(\figref{fig:rerank}, per agent in \appref{app:rerank}).
Simpler route choices lose EM. Taking a random link from the same candidates,
or not rerouting at all, costs one to one and a half EM points on 2Wiki
and about half a point on HotpotQA (\figref{fig:rerank}). Title match,
which needs no training, comes within 0.7 points of our scorer on 2Wiki
and 0.1 on HotpotQA.

Re-ranking the scorer's top five routes with off-the-shelf cross-encoders of up to 0.6B parameters changes EM by at most 0.2 points. In
about 80\% of flagged calls, none of the candidate links leads to a new
gold page, so following a grounded link matters more than which one is
chosen, and a small scorer is enough.

\subsection{Limits and Where \method Helps Most}
\label{sec:reachgain}

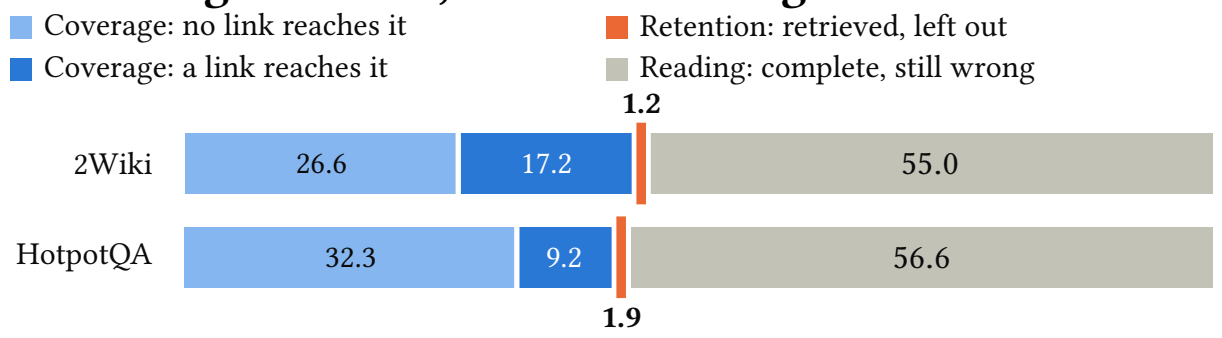
\begin{figure}[t]
  \centering
  \providecommand{\tbd}[1]{\textcolor{red}{#1}}
\begin{tikzpicture}[x=0.068cm, y=1cm, font=\scriptsize]
  \definecolor{efCovNo}{HTML}{86B6EF}
  \definecolor{efCovYes}{HTML}{2A78D6}
  \definecolor{efRet}{HTML}{EB6834}
  \definecolor{efRead}{HTML}{C3C2B7}
  \definecolor{efInk}{HTML}{0B0B0B}
  \definecolor{efMuted}{HTML}{52514E}
  \def\efH{0.40}

  \foreach \x/\y/\c/\t in {%
      -17/1.30/efCovNo/{Coverage: no link reaches it},
      -17/1.02/efCovYes/{Coverage: a link reaches it},
       41/1.30/efRet/{Retention: retrieved, left out},
       41/1.02/efRead/{Reading: complete, still wrong}} {
    \fill[\c] (\x,\y-0.075) rectangle +(2.1,0.15);
    \node[anchor=west, inner sep=0pt, text=efInk] at (\x+3.2,\y) {\t};
  }

  \newcommand{\efBar}[7]{%
    \pgfmathsetmacro{\efA}{#3}
    \pgfmathsetmacro{\efB}{#3+#4}
    \pgfmathsetmacro{\efC}{#3+#4+#5}
    \pgfmathsetmacro{\efM}{(\efB+\efC)/2}
    \node[anchor=east, text=efInk] at (-1.5,#1+\efH/2) {#2};
    \fill[efCovNo]  (0,#1)     rectangle (\efA,#1+\efH);
    \fill[efCovYes] (\efA,#1)  rectangle (\efB,#1+\efH);
    \fill[efRead]   (\efC,#1)  rectangle (100,#1+\efH);
    \node[text=efInk] at ({\efA/2},#1+\efH/2) {#3};
    \node[text=white] at ({(\efA+\efB)/2},#1+\efH/2) {#4};
    \pgfmathsetmacro{\efR}{100-\efC}
    \node[text=efInk] at ({(\efC+100)/2},#1+\efH/2) {\pgfmathprintnumber[fixed,precision=1,fixed zerofill]{\efR}};
    \draw[white, line width=1pt] (\efA,#1-0.01) -- (\efA,#1+\efH+0.01);
    \draw[white, line width=3.6pt] (\efM,#1-0.07) -- (\efM,#1+\efH+0.07);
    \draw[efRet, line width=1.8pt]  (\efM,#1-0.07) -- (\efM,#1+\efH+0.07);
    \ifx a#6
      \node[anchor=south, inner sep=1.5pt, text=efInk] at (\efM,#1+\efH+0.07) {\textbf{#7}};
    \else
      \node[anchor=north, inner sep=1.5pt, text=efInk] at (\efM,#1-0.07) {\textbf{#7}};
    \fi
  }

  \efBar{0.20}{2Wiki}{26.6}{17.2}{1.2}{a}{1.2}
  \efBar{-0.42}{HotpotQA}{32.3}{9.2}{1.9}{b}{1.9}
\end{tikzpicture}
  \Description{Two horizontal stacked bars, 2Wiki and HotpotQA, each splitting the wrong answers of DetouR into four parts, in percent. 2Wiki: 27.2 a gold page never retrieved and no link reaches it, 17.6 never retrieved but a link reaches it, 0.2 every gold page retrieved but one left out of the context, 55.0 complete context but a wrong answer. HotpotQA: 32.7, 9.3, 1.0 and 57.0.}
    \vspace{-2mm}
\caption{Breakdown of wrong answers with Chain-aware Context Selection (no rerouting) by where the evidence is lost.}
  \label{fig:funnel}
  \vspace{-4mm}
\end{figure}

Most remaining errors come from reading rather than from lost evidence
(\figref{fig:funnel}). Over the 6{,}000 agent--question pairs per
collection, we assign each wrong answer of the agents with Chain-aware Context Selection, before rerouting, to the step where its evidence is lost (\eqnref{eq:complete_context}): \emph{coverage misses},
where a gold page is not retrieved; \emph{retention misses}, where a
retrieved gold page is left out of the context; and \emph{reading
errors}, where the context is complete but the answer is wrong.
Retention misses are rare, at 1.2\% and 1.9\% of wrong answers,
because Chain-aware Context Selection keeps almost all retrieved
evidence (\secref{sec:ctx-analysis}). Few coverage misses can be fixed
by rerouting: only 39\% (2Wiki) and 22\% (HotpotQA) of them have the
missing page among the top five link targets of some flagged call. Even an oracle choice among these routes would raise EM by at most 3.1 and 2.8 points (\appref{app:impl}). Reading errors make up 55\% and 57\% of wrong
answers. Complete contexts do not guarantee correct answers: the reader
answers 43.8\% and 51.3\% of them correctly, and even given only the
gold pages it reaches 48.8 and 57.3 EM. Further gains therefore depend
more on the reader than on finding or keeping evidence.

The gain also depends on the retriever and on whether links exist. With
Search-R1's native E5 retriever, \method still adds complete contexts,
but its EM gain is significant only on HotpotQA (\tabref{tab:e5},
\appref{app:robust}); on live web search, where associations come only from noisy entity mentions, the average gain stays positive but small (\appref{app:web}). The gain is largest
for HotpotQA bridge questions and 2Wiki inference and compositional
questions, and rerouting helps most for IRCoT and Plan, which always
make five follow-up searches (\appref{app:gains}).

\section{Conclusion and Limitations}
\label{sec:conclusion}
\label{sec:limits}

Search agents lose evidence in two ways. They often leave available association routes unused, and rank fusion tends to push out retrieved pages that only one search found. \method plugs into an agent at these two points: it reroutes near-repeat queries along links of retrieved pages and keeps linked evidence in the reader's context, without extra retrieval or LLM calls. It raises EM for all four agents on both datasets, in replay and in live runs.

\method needs an association graph, so its gain is small on live web search, where results rarely keep their links, and our web test is small. The near-repeat test can also flag a query that shares most words with an earlier one but asks for a different fact. Better entity matching could carry association routes to open web search, where links must be inferred from page text, and since most remaining errors are reading errors, readers that use the links among passages are a natural next step.

\clearpage

\bibliographystyle{ACM-Reference-Format}
\bibliography{references}


\begin{thebibliography}{72}


\ifx \showCODEN    \undefined \def \showCODEN     #1{\unskip}     \fi
\ifx \showISBNx    \undefined \def \showISBNx     #1{\unskip}     \fi
\ifx \showISBNxiii \undefined \def \showISBNxiii  #1{\unskip}     \fi
\ifx \showISSN     \undefined \def \showISSN      #1{\unskip}     \fi
\ifx \showLCCN     \undefined \def \showLCCN      #1{\unskip}     \fi
\ifx \shownote     \undefined \def \shownote      #1{#1}          \fi
\ifx \showarticletitle \undefined \def \showarticletitle #1{#1}   \fi
\ifx \showURL      \undefined \def \showURL       {\relax}        \fi
\providecommand\bibfield[2]{#2}
\providecommand\bibinfo[2]{#2}
\providecommand\natexlab[1]{#1}
\providecommand\showeprint[2][]{arXiv:#2}
\makeatletter
\@ifundefined{NAT@parse@date}{}{\let\NAT@parse@date@orig\NAT@parse@date}
\@ifundefined{NAT@parse@date}{}{\def\NAT@parse@date#1#2#3#4#5#6@@{\NAT@parse@date@orig#1#2#3#4#5#6@@\def\NAT@tempyear{0000}\def\NAT@tempexlab{{?}}\ifx\NAT@year\NAT@tempyear\ifx\NAT@exlab\NAT@tempexlab\def\NAT@date{[n.\,d.]}\else\edef\NAT@date{[n.\,d.]\NAT@exlab}\fi\fi}}
\makeatother

\bibitem[Abdul-Jaleel et~al\mbox{.}(2004)]%
        {abduljaleel2004rm3}
\bibfield{author}{\bibinfo{person}{Nasreen Abdul-Jaleel}, \bibinfo{person}{James Allan}, \bibinfo{person}{W~Bruce Croft}, \bibinfo{person}{Fernando Diaz}, \bibinfo{person}{Leah Larkey}, \bibinfo{person}{Xiaoyan Li}, \bibinfo{person}{Mark~D Smucker}, {and} \bibinfo{person}{Courtney Wade}.} \bibinfo{year}{2004}\natexlab{}.
\newblock \showarticletitle{UMass at TREC 2004: Novelty and HARD}.
\newblock  (\bibinfo{year}{2004}).
\newblock


\bibitem[Asai et~al\mbox{.}(2020)]%
        {asai2020pathretriever}
\bibfield{author}{\bibinfo{person}{Akari Asai}, \bibinfo{person}{Kazuma Hashimoto}, \bibinfo{person}{Hannaneh Hajishirzi}, \bibinfo{person}{Richard Socher}, {and} \bibinfo{person}{Caiming Xiong}.} \bibinfo{year}{2020}\natexlab{}.
\newblock \showarticletitle{Learning to Retrieve Reasoning Paths over Wikipedia Graph for Question Answering}. In \bibinfo{booktitle}{\emph{International Conference on Learning Representations}}.
\newblock
\urldef\tempurl%
\url{https://openreview.net/forum?id=SJgVHkrYDH}
\showURL{%
\tempurl}


\bibitem[Bacellar(2026)]%
        {bacellar2026phasegraph}
\bibfield{author}{\bibinfo{person}{Andre Bacellar}.} \bibinfo{year}{2026}\natexlab{}.
\newblock \bibinfo{title}{Calibrated Fusion for Heterogeneous Graph-Vector Retrieval in Multi-Hop QA}.
\newblock
\showeprint[arxiv]{2603.28886}~[cs.IR]


\bibitem[Callan et~al\mbox{.}(1995)]%
        {callan1995distributed}
\bibfield{author}{\bibinfo{person}{James~P. Callan}, \bibinfo{person}{Zhihong Lu}, {and} \bibinfo{person}{W.~Bruce Croft}.} \bibinfo{year}{1995}\natexlab{}.
\newblock \showarticletitle{Searching distributed collections with inference networks}. In \bibinfo{booktitle}{\emph{Proceedings of the 18th annual international ACM SIGIR conference on Research and development in information retrieval - SIGIR ’95}}. \bibinfo{publisher}{ACM Press}, \bibinfo{pages}{21–28}.
\newblock
\href{https://doi.org/10.1145/215206.215328}{doi:\nolinkurl{10.1145/215206.215328}}


\bibitem[Carmel and Yom-Tov(2010)]%
        {carmel2010qpp}
\bibfield{author}{\bibinfo{person}{David Carmel} {and} \bibinfo{person}{Elad Yom-Tov}.} \bibinfo{year}{2010}\natexlab{}.
\newblock \bibinfo{booktitle}{\emph{Estimating the Query Difficulty for Information Retrieval}}.
\newblock \bibinfo{publisher}{Springer International Publishing}.
\newblock
\href{https://doi.org/10.1007/978-3-031-02272-2}{doi:\nolinkurl{10.1007/978-3-031-02272-2}}


\bibitem[Chen et~al\mbox{.}(2026)]%
        {chen2026stride}
\bibfield{author}{\bibinfo{person}{Wei Chen}, \bibinfo{person}{Lili Zhao}, \bibinfo{person}{Zhi Zheng}, \bibinfo{person}{HuiJun Hou}, {and} \bibinfo{person}{Tong Xu}.} \bibinfo{year}{2026}\natexlab{}.
\newblock \bibinfo{title}{STRIDE: Strategic Iterative Decision-Making for Retrieval-Augmented Multi-Hop Question Answering}.
\newblock
\showeprint[arxiv]{2604.17405}~[cs.AI]


\bibitem[Cormack et~al\mbox{.}(2009)]%
        {cormack2009rrf}
\bibfield{author}{\bibinfo{person}{Gordon~V. Cormack}, \bibinfo{person}{Charles L~A Clarke}, {and} \bibinfo{person}{Stefan Buettcher}.} \bibinfo{year}{2009}\natexlab{}.
\newblock \showarticletitle{Reciprocal rank fusion outperforms condorcet and individual rank learning methods}. In \bibinfo{booktitle}{\emph{Proceedings of the 32nd international ACM SIGIR conference on Research and development in information retrieval}}. \bibinfo{publisher}{ACM}, \bibinfo{pages}{758–759}.
\newblock
\href{https://doi.org/10.1145/1571941.1572114}{doi:\nolinkurl{10.1145/1571941.1572114}}


\bibitem[Craswell and Szummer(2007)]%
        {craswell2007clickgraph}
\bibfield{author}{\bibinfo{person}{Nick Craswell} {and} \bibinfo{person}{Martin Szummer}.} \bibinfo{year}{2007}\natexlab{}.
\newblock \showarticletitle{Random walks on the click graph}. In \bibinfo{booktitle}{\emph{Proceedings of the 30th annual international ACM SIGIR conference on Research and development in information retrieval}}. \bibinfo{publisher}{ACM}, \bibinfo{pages}{239–246}.
\newblock
\href{https://doi.org/10.1145/1277741.1277784}{doi:\nolinkurl{10.1145/1277741.1277784}}


\bibitem[Dalton et~al\mbox{.}(2014)]%
        {dalton2014eqfe}
\bibfield{author}{\bibinfo{person}{Jeffrey Dalton}, \bibinfo{person}{Laura Dietz}, {and} \bibinfo{person}{James Allan}.} \bibinfo{year}{2014}\natexlab{}.
\newblock \showarticletitle{Entity query feature expansion using knowledge base links}. In \bibinfo{booktitle}{\emph{Proceedings of the 37th international ACM SIGIR conference on Research \& development in information retrieval}}. \bibinfo{publisher}{ACM}, \bibinfo{pages}{365–374}.
\newblock
\href{https://doi.org/10.1145/2600428.2609628}{doi:\nolinkurl{10.1145/2600428.2609628}}


\bibitem[Dang and Croft(2010)]%
        {dang2010anchor}
\bibfield{author}{\bibinfo{person}{Van Dang} {and} \bibinfo{person}{Bruce~W. Croft}.} \bibinfo{year}{2010}\natexlab{}.
\newblock \showarticletitle{Query reformulation using anchor text}. In \bibinfo{booktitle}{\emph{Proceedings of the third ACM international conference on Web search and data mining}}. \bibinfo{publisher}{ACM}, \bibinfo{pages}{41–50}.
\newblock
\href{https://doi.org/10.1145/1718487.1718493}{doi:\nolinkurl{10.1145/1718487.1718493}}


\bibitem[Eberhard et~al\mbox{.}(2025)]%
        {eberhard2025llmndr}
\bibfield{author}{\bibinfo{person}{Lukas Eberhard}, \bibinfo{person}{Thorsten Ruprechter}, {and} \bibinfo{person}{Denis Helic}.} \bibinfo{year}{2025}\natexlab{}.
\newblock \showarticletitle{Large Language Models as Narrative-Driven Recommenders}. In \bibinfo{booktitle}{\emph{Proceedings of the ACM on Web Conference 2025}}. \bibinfo{publisher}{ACM}, \bibinfo{pages}{4543–4561}.
\newblock
\href{https://doi.org/10.1145/3696410.3714668}{doi:\nolinkurl{10.1145/3696410.3714668}}


\bibitem[Edge et~al\mbox{.}(2025)]%
        {edge2024graphrag}
\bibfield{author}{\bibinfo{person}{Darren Edge}, \bibinfo{person}{Ha Trinh}, \bibinfo{person}{Newman Cheng}, \bibinfo{person}{Joshua Bradley}, \bibinfo{person}{Alex Chao}, \bibinfo{person}{Apurva Mody}, \bibinfo{person}{Steven Truitt}, \bibinfo{person}{Dasha Metropolitansky}, \bibinfo{person}{Robert~Osazuwa Ness}, {and} \bibinfo{person}{Jonathan Larson}.} \bibinfo{year}{2025}\natexlab{}.
\newblock \bibinfo{title}{From Local to Global: A Graph RAG Approach to Query-Focused Summarization}.
\newblock
\showeprint[arxiv]{2404.16130}~[cs.CL]


\bibitem[Fang et~al\mbox{.}(2026)]%
        {fang2026budgetcontrol}
\bibfield{author}{\bibinfo{person}{Zhengru Fang}, \bibinfo{person}{Senkang~Forest Hu}, \bibinfo{person}{Zhonghao Chang}, \bibinfo{person}{Yu Guo}, \bibinfo{person}{Yihang Tao}, \bibinfo{person}{Hongyao Liu}, \bibinfo{person}{Mengzhe Ruan}, \bibinfo{person}{Jun Huang}, {and} \bibinfo{person}{Yuguang Fang}.} \bibinfo{year}{2026}\natexlab{}.
\newblock \bibinfo{title}{Inference-Time Budget Control for LLM Search Agents}.
\newblock
\showeprint[arxiv]{2605.05701}~[cs.AI]


\bibitem[Gao et~al\mbox{.}(2023)]%
        {gao2023hyde}
\bibfield{author}{\bibinfo{person}{Luyu Gao}, \bibinfo{person}{Xueguang Ma}, \bibinfo{person}{Jimmy Lin}, {and} \bibinfo{person}{Jamie Callan}.} \bibinfo{year}{2023}\natexlab{}.
\newblock \showarticletitle{Precise Zero-Shot Dense Retrieval without Relevance Labels}. In \bibinfo{booktitle}{\emph{Proceedings of the 61st Annual Meeting of the {A}ssociation for {C}omputational {L}inguistics (Volume 1: Long Papers)}}. \bibinfo{publisher}{Association for Computational Linguistics}, \bibinfo{address}{Toronto, Canada}, \bibinfo{pages}{1762--1777}.
\newblock
\href{https://doi.org/10.18653/v1/2023.acl-long.99}{doi:\nolinkurl{10.18653/v1/2023.acl-long.99}}


\bibitem[Gutiérrez et~al\mbox{.}(2024)]%
        {gutierrez2024hipporag}
\bibfield{author}{\bibinfo{person}{Bernal Gutiérrez}, \bibinfo{person}{Yiheng Shu}, \bibinfo{person}{Yu Gu}, \bibinfo{person}{Michihiro Yasunaga}, {and} \bibinfo{person}{Yu Su}.} \bibinfo{year}{2024}\natexlab{}.
\newblock \showarticletitle{HippoRAG: Neurobiologically Inspired Long-Term Memory for Large Language Models}. In \bibinfo{booktitle}{\emph{Advances in Neural Information Processing Systems 37}}. \bibinfo{publisher}{Neural Information Processing Systems Foundation, Inc. (NeurIPS)}, \bibinfo{pages}{59532–59569}.
\newblock
\href{https://doi.org/10.52202/079017-1902}{doi:\nolinkurl{10.52202/079017-1902}}


\bibitem[Haveliwala(2002)]%
        {haveliwala2002topic}
\bibfield{author}{\bibinfo{person}{Taher~H. Haveliwala}.} \bibinfo{year}{2002}\natexlab{}.
\newblock \showarticletitle{Topic-sensitive PageRank}. In \bibinfo{booktitle}{\emph{Proceedings of the 11th international conference on World Wide Web}}. \bibinfo{publisher}{ACM}, \bibinfo{pages}{517–526}.
\newblock
\href{https://doi.org/10.1145/511446.511513}{doi:\nolinkurl{10.1145/511446.511513}}


\bibitem[Ho et~al\mbox{.}(2020)]%
        {ho2020constructing}
\bibfield{author}{\bibinfo{person}{Xanh Ho}, \bibinfo{person}{Anh-Khoa Duong~Nguyen}, \bibinfo{person}{Saku Sugawara}, {and} \bibinfo{person}{Akiko Aizawa}.} \bibinfo{year}{2020}\natexlab{}.
\newblock \showarticletitle{Constructing A Multi-hop {QA} Dataset for Comprehensive Evaluation of Reasoning Steps}. In \bibinfo{booktitle}{\emph{Proceedings of the 28th International Conference on Computational Linguistics}}. \bibinfo{publisher}{International Committee on Computational Linguistics}, \bibinfo{address}{Barcelona, Spain (Online)}, \bibinfo{pages}{6609--6625}.
\newblock
\href{https://doi.org/10.18653/v1/2020.coling-main.580}{doi:\nolinkurl{10.18653/v1/2020.coling-main.580}}


\bibitem[Java et~al\mbox{.}(2026)]%
        {java2025frugalrag}
\bibfield{author}{\bibinfo{person}{Abhinav Java}, \bibinfo{person}{Srivathsan Koundinyan}, \bibinfo{person}{Nagarajan Natarajan}, {and} \bibinfo{person}{Amit Sharma}.} \bibinfo{year}{2026}\natexlab{}.
\newblock \bibinfo{title}{FrugalRAG: Less is More in RL Finetuning for Multi-Hop Question Answering}.
\newblock
\showeprint[arxiv]{2507.07634}~[cs.CL]


\bibitem[Jeh and Widom(2003)]%
        {jeh2003scaling}
\bibfield{author}{\bibinfo{person}{Glen Jeh} {and} \bibinfo{person}{Jennifer Widom}.} \bibinfo{year}{2003}\natexlab{}.
\newblock \showarticletitle{Scaling personalized web search}. In \bibinfo{booktitle}{\emph{Proceedings of the twelfth international conference on World Wide Web - WWW ’03}}. \bibinfo{publisher}{ACM Press}, \bibinfo{pages}{271}.
\newblock
\href{https://doi.org/10.1145/775152.775191}{doi:\nolinkurl{10.1145/775152.775191}}


\bibitem[Jiang et~al\mbox{.}(2025)]%
        {jiang2025resp}
\bibfield{author}{\bibinfo{person}{Zhouyu Jiang}, \bibinfo{person}{Mengshu Sun}, \bibinfo{person}{Lei Liang}, {and} \bibinfo{person}{Zhiqiang Zhang}.} \bibinfo{year}{2025}\natexlab{}.
\newblock \showarticletitle{Retrieve, Summarize, Plan: Advancing Multi-hop Question Answering with an Iterative Approach}. In \bibinfo{booktitle}{\emph{Companion Proceedings of the ACM on Web Conference 2025}}. \bibinfo{publisher}{ACM}, \bibinfo{pages}{1677–1686}.
\newblock
\href{https://doi.org/10.1145/3701716.3716889}{doi:\nolinkurl{10.1145/3701716.3716889}}


\bibitem[Jiang et~al\mbox{.}(2023)]%
        {jiang2023flare}
\bibfield{author}{\bibinfo{person}{Zhengbao Jiang}, \bibinfo{person}{Frank Xu}, \bibinfo{person}{Luyu Gao}, \bibinfo{person}{Zhiqing Sun}, \bibinfo{person}{Qian Liu}, \bibinfo{person}{Jane Dwivedi-Yu}, \bibinfo{person}{Yiming Yang}, \bibinfo{person}{Jamie Callan}, {and} \bibinfo{person}{Graham Neubig}.} \bibinfo{year}{2023}\natexlab{}.
\newblock \showarticletitle{Active Retrieval Augmented Generation}. In \bibinfo{booktitle}{\emph{Proceedings of the 2023 Conference on Empirical Methods in Natural Language Processing}}. \bibinfo{publisher}{Association for Computational Linguistics}, \bibinfo{address}{Singapore}, \bibinfo{pages}{7969--7992}.
\newblock
\href{https://doi.org/10.18653/v1/2023.emnlp-main.495}{doi:\nolinkurl{10.18653/v1/2023.emnlp-main.495}}


\bibitem[Jin et~al\mbox{.}(2025)]%
        {jin2025searchr1}
\bibfield{author}{\bibinfo{person}{Bowen Jin}, \bibinfo{person}{Hansi Zeng}, \bibinfo{person}{Zhenrui Yue}, \bibinfo{person}{Jinsung Yoon}, \bibinfo{person}{Sercan~O Arik}, \bibinfo{person}{Dong Wang}, \bibinfo{person}{Hamed Zamani}, {and} \bibinfo{person}{Jiawei Han}.} \bibinfo{year}{2025}\natexlab{}.
\newblock \showarticletitle{Search-R1: Training {LLM}s to Reason and Leverage Search Engines with Reinforcement Learning}. In \bibinfo{booktitle}{\emph{Second Conference on Language Modeling}}.
\newblock
\urldef\tempurl%
\url{https://openreview.net/forum?id=Rwhi91ideu}
\showURL{%
\tempurl}


\bibitem[Ke et~al\mbox{.}(2017)]%
        {ke2017lightgbm}
\bibfield{author}{\bibinfo{person}{Guolin Ke}, \bibinfo{person}{Qi Meng}, \bibinfo{person}{Thomas Finley}, \bibinfo{person}{Taifeng Wang}, \bibinfo{person}{Wei Chen}, \bibinfo{person}{Weidong Ma}, \bibinfo{person}{Qiwei Ye}, {and} \bibinfo{person}{Tie-Yan Liu}.} \bibinfo{year}{2017}\natexlab{}.
\newblock \showarticletitle{LightGBM: A Highly Efficient Gradient Boosting Decision Tree}. In \bibinfo{booktitle}{\emph{Advances in Neural Information Processing Systems}}, Vol.~\bibinfo{volume}{30}. \bibinfo{publisher}{Curran Associates, Inc.}
\newblock
\urldef\tempurl%
\url{https://proceedings.neurips.cc/paper_files/paper/2017/file/6449f44a102fde848669bdd9eb6b76fa-Paper.pdf}
\showURL{%
\tempurl}


\bibitem[Kurland and Lee(2005)]%
        {kurland2005pagerank}
\bibfield{author}{\bibinfo{person}{Oren Kurland} {and} \bibinfo{person}{Lillian Lee}.} \bibinfo{year}{2005}\natexlab{}.
\newblock \showarticletitle{PageRank without hyperlinks: structural re-ranking using links induced by language models}. In \bibinfo{booktitle}{\emph{Proceedings of the 28th annual international ACM SIGIR conference on Research and development in information retrieval}}. \bibinfo{publisher}{ACM}, \bibinfo{pages}{306–313}.
\newblock
\href{https://doi.org/10.1145/1076034.1076087}{doi:\nolinkurl{10.1145/1076034.1076087}}


\bibitem[Kwiatkowski et~al\mbox{.}(2019)]%
        {kwiatkowski2019nq}
\bibfield{author}{\bibinfo{person}{Tom Kwiatkowski}, \bibinfo{person}{Jennimaria Palomaki}, \bibinfo{person}{Olivia Redfield}, \bibinfo{person}{Michael Collins}, \bibinfo{person}{Ankur Parikh}, \bibinfo{person}{Chris Alberti}, \bibinfo{person}{Danielle Epstein}, \bibinfo{person}{Illia Polosukhin}, \bibinfo{person}{Jacob Devlin}, \bibinfo{person}{Kenton Lee}, \bibinfo{person}{Kristina Toutanova}, \bibinfo{person}{Llion Jones}, \bibinfo{person}{Matthew Kelcey}, \bibinfo{person}{Ming-Wei Chang}, \bibinfo{person}{Andrew~M. Dai}, \bibinfo{person}{Jakob Uszkoreit}, \bibinfo{person}{Quoc Le}, {and} \bibinfo{person}{Slav Petrov}.} \bibinfo{year}{2019}\natexlab{}.
\newblock \showarticletitle{Natural Questions: A Benchmark for Question Answering Research}.
\newblock \bibinfo{journal}{\emph{Transactions of the Association for Computational Linguistics}}  \bibinfo{volume}{7} (\bibinfo{year}{2019}), \bibinfo{pages}{452--466}.
\newblock
\href{https://doi.org/10.1162/tacl_a_00276}{doi:\nolinkurl{10.1162/tacl_a_00276}}


\bibitem[Lavrenko and Croft(2001)]%
        {lavrenko2001relevance}
\bibfield{author}{\bibinfo{person}{Victor Lavrenko} {and} \bibinfo{person}{W.~Bruce Croft}.} \bibinfo{year}{2001}\natexlab{}.
\newblock \showarticletitle{Relevance based language models}. In \bibinfo{booktitle}{\emph{Proceedings of the 24th annual international ACM SIGIR conference on Research and development in information retrieval}}. \bibinfo{publisher}{ACM}, \bibinfo{pages}{120–127}.
\newblock
\href{https://doi.org/10.1145/383952.383972}{doi:\nolinkurl{10.1145/383952.383972}}


\bibitem[Lewis et~al\mbox{.}(2020)]%
        {lewis2020rag}
\bibfield{author}{\bibinfo{person}{Patrick Lewis}, \bibinfo{person}{Ethan Perez}, \bibinfo{person}{Aleksandra Piktus}, \bibinfo{person}{Fabio Petroni}, \bibinfo{person}{Vladimir Karpukhin}, \bibinfo{person}{Naman Goyal}, \bibinfo{person}{Heinrich K\"{u}ttler}, \bibinfo{person}{Mike Lewis}, \bibinfo{person}{Wen-tau Yih}, \bibinfo{person}{Tim Rockt\"{a}schel}, \bibinfo{person}{Sebastian Riedel}, {and} \bibinfo{person}{Douwe Kiela}.} \bibinfo{year}{2020}\natexlab{}.
\newblock \showarticletitle{Retrieval-Augmented Generation for Knowledge-Intensive NLP Tasks}. In \bibinfo{booktitle}{\emph{Advances in Neural Information Processing Systems}}, Vol.~\bibinfo{volume}{33}. \bibinfo{publisher}{Curran Associates, Inc.}, \bibinfo{pages}{9459--9474}.
\newblock
\urldef\tempurl%
\url{https://proceedings.neurips.cc/paper_files/paper/2020/file/6b493230205f780e1bc26945df7481e5-Paper.pdf}
\showURL{%
\tempurl}


\bibitem[Li et~al\mbox{.}(2025)]%
        {li2025searcho1}
\bibfield{author}{\bibinfo{person}{Xiaoxi Li}, \bibinfo{person}{Guanting Dong}, \bibinfo{person}{Jiajie Jin}, \bibinfo{person}{Yuyao Zhang}, \bibinfo{person}{Yujia Zhou}, \bibinfo{person}{Yutao Zhu}, \bibinfo{person}{Peitian Zhang}, {and} \bibinfo{person}{Zhicheng Dou}.} \bibinfo{year}{2025}\natexlab{}.
\newblock \showarticletitle{{S}earch-o1: Agentic Search-Enhanced Large Reasoning Models}. In \bibinfo{booktitle}{\emph{Proceedings of the 2025 Conference on Empirical Methods in Natural Language Processing}}. \bibinfo{publisher}{Association for Computational Linguistics}, \bibinfo{address}{Suzhou, China}, \bibinfo{pages}{5420--5438}.
\newblock
\href{https://doi.org/10.18653/v1/2025.emnlp-main.276}{doi:\nolinkurl{10.18653/v1/2025.emnlp-main.276}}


\bibitem[Li et~al\mbox{.}(2026)]%
        {li2026par2rag}
\bibfield{author}{\bibinfo{person}{Xingyu Li}, \bibinfo{person}{Rongguang Wang}, \bibinfo{person}{Yuying Wang}, \bibinfo{person}{Mengqing Guo}, \bibinfo{person}{Chenyang Li}, \bibinfo{person}{Tao Sheng}, \bibinfo{person}{Sujith Ravi}, {and} \bibinfo{person}{Dan Roth}.} \bibinfo{year}{2026}\natexlab{}.
\newblock \showarticletitle{{PAR}$^2$-{RAG}: Planned Active Retrieval and Reasoning for Multi-Hop Question Answering}. In \bibinfo{booktitle}{\emph{Proceedings of the 64th Annual Meeting of the {A}ssociation for {C}omputational {L}inguistics (Volume 6: Industry Track)}}. \bibinfo{publisher}{Association for Computational Linguistics}, \bibinfo{address}{San Diego, California, USA}, \bibinfo{pages}{1727--1737}.
\newblock
\href{https://doi.org/10.18653/v1/2026.acl-industry.118}{doi:\nolinkurl{10.18653/v1/2026.acl-industry.118}}


\bibitem[Liu et~al\mbox{.}(2026)]%
        {liu2026multisearch}
\bibfield{author}{\bibinfo{person}{Jiabei Liu}, \bibinfo{person}{Wenyu Mao}, \bibinfo{person}{Junfei Tan}, \bibinfo{person}{Chunxu Shen}, \bibinfo{person}{Lingling Yi}, \bibinfo{person}{Jiancan Wu}, {and} \bibinfo{person}{Xiang Wang}.} \bibinfo{year}{2026}\natexlab{}.
\newblock \bibinfo{title}{Scaling Retrieval-Augmented Reasoning with Parallel Search and Explicit Merging}.
\newblock
\showeprint[arxiv]{2605.13534}~[cs.AI]


\bibitem[Ma et~al\mbox{.}(2023)]%
        {ma2023queryrewriting}
\bibfield{author}{\bibinfo{person}{Xinbei Ma}, \bibinfo{person}{Yeyun Gong}, \bibinfo{person}{Pengcheng He}, \bibinfo{person}{Hai Zhao}, {and} \bibinfo{person}{Nan Duan}.} \bibinfo{year}{2023}\natexlab{}.
\newblock \showarticletitle{Query Rewriting in Retrieval-Augmented Large Language Models}. In \bibinfo{booktitle}{\emph{Proceedings of the 2023 Conference on Empirical Methods in Natural Language Processing}}. \bibinfo{publisher}{Association for Computational Linguistics}, \bibinfo{address}{Singapore}, \bibinfo{pages}{5303--5315}.
\newblock
\href{https://doi.org/10.18653/v1/2023.emnlp-main.322}{doi:\nolinkurl{10.18653/v1/2023.emnlp-main.322}}


\bibitem[Nakano et~al\mbox{.}(2022)]%
        {nakano2021webgpt}
\bibfield{author}{\bibinfo{person}{Reiichiro Nakano}, \bibinfo{person}{Jacob Hilton}, \bibinfo{person}{Suchir Balaji}, \bibinfo{person}{Jeff Wu}, \bibinfo{person}{Long Ouyang}, \bibinfo{person}{Christina Kim}, \bibinfo{person}{Christopher Hesse}, \bibinfo{person}{Shantanu Jain}, \bibinfo{person}{Vineet Kosaraju}, \bibinfo{person}{William Saunders}, \bibinfo{person}{Xu Jiang}, \bibinfo{person}{Karl Cobbe}, \bibinfo{person}{Tyna Eloundou}, \bibinfo{person}{Gretchen Krueger}, \bibinfo{person}{Kevin Button}, \bibinfo{person}{Matthew Knight}, \bibinfo{person}{Benjamin Chess}, {and} \bibinfo{person}{John Schulman}.} \bibinfo{year}{2022}\natexlab{}.
\newblock \bibinfo{title}{WebGPT: Browser-assisted question-answering with human feedback}.
\newblock
\showeprint[arxiv]{2112.09332}~[cs.CL]


\bibitem[Nogueira et~al\mbox{.}(2020)]%
        {nogueira2020monot5}
\bibfield{author}{\bibinfo{person}{Rodrigo Nogueira}, \bibinfo{person}{Zhiying Jiang}, \bibinfo{person}{Ronak Pradeep}, {and} \bibinfo{person}{Jimmy Lin}.} \bibinfo{year}{2020}\natexlab{}.
\newblock \showarticletitle{Document Ranking with a Pretrained Sequence-to-Sequence Model}. In \bibinfo{booktitle}{\emph{Findings of the {A}ssociation for {C}omputational {L}inguistics: {EMNLP} 2020}}. \bibinfo{publisher}{Association for Computational Linguistics}, \bibinfo{address}{Online}, \bibinfo{pages}{708--718}.
\newblock
\href{https://doi.org/10.18653/v1/2020.findings-emnlp.63}{doi:\nolinkurl{10.18653/v1/2020.findings-emnlp.63}}


\bibitem[Page et~al\mbox{.}(1999)]%
        {page1999pagerank}
\bibfield{author}{\bibinfo{person}{Lawrence Page}, \bibinfo{person}{Sergey Brin}, \bibinfo{person}{Rajeev Motwani}, {and} \bibinfo{person}{Terry Winograd}.} \bibinfo{year}{1999}\natexlab{}.
\newblock \showarticletitle{The PageRank Citation Ranking : Bringing Order to the Web}. In \bibinfo{booktitle}{\emph{The Web Conference}}.
\newblock
\urldef\tempurl%
\url{https://api.semanticscholar.org/CorpusID:1508503}
\showURL{%
\tempurl}


\bibitem[Press et~al\mbox{.}(2023)]%
        {press2023selfask}
\bibfield{author}{\bibinfo{person}{Ofir Press}, \bibinfo{person}{Muru Zhang}, \bibinfo{person}{Sewon Min}, \bibinfo{person}{Ludwig Schmidt}, \bibinfo{person}{Noah Smith}, {and} \bibinfo{person}{Mike Lewis}.} \bibinfo{year}{2023}\natexlab{}.
\newblock \showarticletitle{Measuring and Narrowing the Compositionality Gap in Language Models}. In \bibinfo{booktitle}{\emph{Findings of the {A}ssociation for {C}omputational {L}inguistics: {EMNLP} 2023}}. \bibinfo{publisher}{Association for Computational Linguistics}, \bibinfo{address}{Singapore}, \bibinfo{pages}{5687--5711}.
\newblock
\href{https://doi.org/10.18653/v1/2023.findings-emnlp.378}{doi:\nolinkurl{10.18653/v1/2023.findings-emnlp.378}}


\bibitem[Qi et~al\mbox{.}(2026)]%
        {qi2026cage}
\bibfield{author}{\bibinfo{person}{Tong Qi}, \bibinfo{person}{Jingyu Wu}, \bibinfo{person}{Youbing Yin}, \bibinfo{person}{Spencer Hong}, \bibinfo{person}{Daben Liu}, {and} \bibinfo{person}{Erin Babinsky}.} \bibinfo{year}{2026}\natexlab{}.
\newblock \bibinfo{title}{CAGE: Coherence-Aware Graph Encoding for Retrieval-Augmented Generation}.
\newblock
\showeprint[arxiv]{2609.04647}~[cs.CL]


\bibitem[Qiu et~al\mbox{.}(2025)]%
        {qiu2025gcrs}
\bibfield{author}{\bibinfo{person}{Zhangchi Qiu}, \bibinfo{person}{Linhao Luo}, \bibinfo{person}{Zicheng Zhao}, \bibinfo{person}{Shirui Pan}, {and} \bibinfo{person}{Alan Wee-Chung Liew}.} \bibinfo{year}{2025}\natexlab{}.
\newblock \bibinfo{booktitle}{\emph{Graph Retrieval-Augmented LLM for Conversational Recommendation Systems}}.
\newblock \bibinfo{publisher}{Springer Nature Singapore}, \bibinfo{pages}{344–355}.
\newblock
\showISBNx{9789819681808}
\showISSN{1611-3349}
\href{https://doi.org/10.1007/978-981-96-8180-8_27}{doi:\nolinkurl{10.1007/978-981-96-8180-8_27}}


\bibitem[Qwen et~al\mbox{.}(2025)]%
        {qwen25}
\bibfield{author}{\bibinfo{person}{Qwen}, \bibinfo{person}{:}, \bibinfo{person}{An Yang}, \bibinfo{person}{Baosong Yang}, \bibinfo{person}{Beichen Zhang}, \bibinfo{person}{Binyuan Hui}, \bibinfo{person}{Bo Zheng}, \bibinfo{person}{Bowen Yu}, \bibinfo{person}{Chengyuan Li}, \bibinfo{person}{Dayiheng Liu}, \bibinfo{person}{Fei Huang}, \bibinfo{person}{Haoran Wei}, \bibinfo{person}{Huan Lin}, \bibinfo{person}{Jian Yang}, \bibinfo{person}{Jianhong Tu}, \bibinfo{person}{Jianwei Zhang}, \bibinfo{person}{Jianxin Yang}, \bibinfo{person}{Jiaxi Yang}, \bibinfo{person}{Jingren Zhou}, \bibinfo{person}{Junyang Lin}, \bibinfo{person}{Kai Dang}, \bibinfo{person}{Keming Lu}, \bibinfo{person}{Keqin Bao}, \bibinfo{person}{Kexin Yang}, \bibinfo{person}{Le Yu}, \bibinfo{person}{Mei Li}, \bibinfo{person}{Mingfeng Xue}, \bibinfo{person}{Pei Zhang}, \bibinfo{person}{Qin Zhu}, \bibinfo{person}{Rui Men}, \bibinfo{person}{Runji Lin}, \bibinfo{person}{Tianhao Li}, \bibinfo{person}{Tianyi Tang}, \bibinfo{person}{Tingyu Xia}, \bibinfo{person}{Xingzhang Ren}, \bibinfo{person}{Xuancheng Ren}, \bibinfo{person}{Yang Fan}, \bibinfo{person}{Yang Su}, \bibinfo{person}{Yichang Zhang}, \bibinfo{person}{Yu Wan}, \bibinfo{person}{Yuqiong Liu}, \bibinfo{person}{Zeyu Cui}, \bibinfo{person}{Zhenru Zhang}, {and} \bibinfo{person}{Zihan Qiu}.} \bibinfo{year}{2025}\natexlab{}.
\newblock \bibinfo{title}{Qwen2.5 Technical Report}.
\newblock
\showeprint[arxiv]{2412.15115}~[cs.CL]


\bibitem[Rackauckas(2024)]%
        {rackauckas2024ragfusion}
\bibfield{author}{\bibinfo{person}{Zackary Rackauckas}.} \bibinfo{year}{2024}\natexlab{}.
\newblock \showarticletitle{Rag-Fusion: A New Take on Retrieval Augmented Generation}.
\newblock \bibinfo{journal}{\emph{International Journal on Natural Language Computing}} \bibinfo{volume}{13}, \bibinfo{number}{1} (\bibinfo{year}{2024}), \bibinfo{pages}{37–47}.
\newblock
\href{https://doi.org/10.5121/ijnlc.2024.13103}{doi:\nolinkurl{10.5121/ijnlc.2024.13103}}


\bibitem[Robertson and Zaragoza(2009)]%
        {robertson2009bm25}
\bibfield{author}{\bibinfo{person}{Stephen Robertson} {and} \bibinfo{person}{Hugo Zaragoza}.} \bibinfo{year}{2009}\natexlab{}.
\newblock \showarticletitle{The Probabilistic Relevance Framework: BM25 and Beyond}.
\newblock \bibinfo{journal}{\emph{Foundations and Trends® in Information Retrieval}} \bibinfo{volume}{4}, \bibinfo{number}{1-2} (\bibinfo{year}{2009}), \bibinfo{pages}{1–174}.
\newblock
\href{https://doi.org/10.1561/1500000019}{doi:\nolinkurl{10.1561/1500000019}}


\bibitem[Rocchio~Jr(1971)]%
        {rocchio1971relevance}
\bibfield{author}{\bibinfo{person}{Joseph~John Rocchio~Jr}.} \bibinfo{year}{1971}\natexlab{}.
\newblock \showarticletitle{Relevance feedback in information retrieval}.
\newblock \bibinfo{journal}{\emph{The SMART retrieval system: experiments in automatic document processing}} (\bibinfo{year}{1971}).
\newblock


\bibitem[Shaw and Fox(1995)]%
        {fox1994combination}
\bibfield{author}{\bibinfo{person}{Joseph~A Shaw} {and} \bibinfo{person}{Edward~A Fox}.} \bibinfo{year}{1995}\natexlab{}.
\newblock \showarticletitle{Combination of multiple searches}.
\newblock \bibinfo{journal}{\emph{NIST special publication SP}} (\bibinfo{year}{1995}), \bibinfo{pages}{105--105}.
\newblock


\bibitem[Shi et~al\mbox{.}(2025)]%
        {shi2025ocg}
\bibfield{author}{\bibinfo{person}{Yunxiao Shi}, \bibinfo{person}{Haoning Shang}, \bibinfo{person}{Xing Zi}, \bibinfo{person}{Wujiang Xu}, \bibinfo{person}{Yue Feng}, {and} \bibinfo{person}{Min Xu}.} \bibinfo{year}{2025}\natexlab{}.
\newblock \showarticletitle{Answering Narrative-Driven Recommendation Queries via a Retrieve{--}Rank Paradigm and the {OCG}-Agent}. In \bibinfo{booktitle}{\emph{Proceedings of the 2025 Conference on Empirical Methods in Natural Language Processing}}. \bibinfo{publisher}{Association for Computational Linguistics}, \bibinfo{address}{Suzhou, China}, \bibinfo{pages}{13181--13202}.
\newblock
\href{https://doi.org/10.18653/v1/2025.emnlp-main.667}{doi:\nolinkurl{10.18653/v1/2025.emnlp-main.667}}


\bibitem[Song et~al\mbox{.}(2025a)]%
        {song2025r1searcher}
\bibfield{author}{\bibinfo{person}{Huatong Song}, \bibinfo{person}{Jinhao Jiang}, \bibinfo{person}{Yingqian Min}, \bibinfo{person}{Jie Chen}, \bibinfo{person}{Zhipeng Chen}, \bibinfo{person}{Wayne~Xin Zhao}, \bibinfo{person}{Lei Fang}, {and} \bibinfo{person}{Ji-Rong Wen}.} \bibinfo{year}{2025}\natexlab{a}.
\newblock \bibinfo{title}{R1-Searcher: Incentivizing the Search Capability in LLMs via Reinforcement Learning}.
\newblock
\showeprint[arxiv]{2503.05592}~[cs.AI]


\bibitem[Song et~al\mbox{.}(2025b)]%
        {song2025r1searcherpp}
\bibfield{author}{\bibinfo{person}{Huatong Song}, \bibinfo{person}{Jinhao Jiang}, \bibinfo{person}{Wenqing Tian}, \bibinfo{person}{Zhipeng Chen}, \bibinfo{person}{Yuhuan Wu}, \bibinfo{person}{Jiahao Zhao}, \bibinfo{person}{Yingqian Min}, \bibinfo{person}{Wayne~Xin Zhao}, \bibinfo{person}{Lei Fang}, {and} \bibinfo{person}{Ji-Rong Wen}.} \bibinfo{year}{2025}\natexlab{b}.
\newblock \bibinfo{title}{R1-Searcher++: Incentivizing the Dynamic Knowledge Acquisition of LLMs via Reinforcement Learning}.
\newblock
\showeprint[arxiv]{2505.17005}~[cs.CL]


\bibitem[Sun et~al\mbox{.}(2024)]%
        {sun2024tog}
\bibfield{author}{\bibinfo{person}{Jiashuo Sun}, \bibinfo{person}{Chengjin Xu}, \bibinfo{person}{Lumingyuan Tang}, \bibinfo{person}{Saizhuo Wang}, \bibinfo{person}{Chen Lin}, \bibinfo{person}{Yeyun Gong}, \bibinfo{person}{Lionel Ni}, \bibinfo{person}{Heung-Yeung Shum}, {and} \bibinfo{person}{Jian Guo}.} \bibinfo{year}{2024}\natexlab{}.
\newblock \showarticletitle{Think-on-Graph: Deep and Responsible Reasoning of Large Language Model on Knowledge Graph}. In \bibinfo{booktitle}{\emph{The Twelfth International Conference on Learning Representations}}.
\newblock
\urldef\tempurl%
\url{https://openreview.net/forum?id=nnVO1PvbTv}
\showURL{%
\tempurl}


\bibitem[Sun et~al\mbox{.}(2023)]%
        {sun2023rankgpt}
\bibfield{author}{\bibinfo{person}{Weiwei Sun}, \bibinfo{person}{Lingyong Yan}, \bibinfo{person}{Xinyu Ma}, \bibinfo{person}{Shuaiqiang Wang}, \bibinfo{person}{Pengjie Ren}, \bibinfo{person}{Zhumin Chen}, \bibinfo{person}{Dawei Yin}, {and} \bibinfo{person}{Zhaochun Ren}.} \bibinfo{year}{2023}\natexlab{}.
\newblock \showarticletitle{Is {C}hat{GPT} Good at Search? Investigating Large Language Models as Re-Ranking Agents}. In \bibinfo{booktitle}{\emph{Proceedings of the 2023 Conference on Empirical Methods in Natural Language Processing}}. \bibinfo{publisher}{Association for Computational Linguistics}, \bibinfo{address}{Singapore}, \bibinfo{pages}{14918--14937}.
\newblock
\href{https://doi.org/10.18653/v1/2023.emnlp-main.923}{doi:\nolinkurl{10.18653/v1/2023.emnlp-main.923}}


\bibitem[Trivedi et~al\mbox{.}(2023)]%
        {trivedi2023ircot}
\bibfield{author}{\bibinfo{person}{Harsh Trivedi}, \bibinfo{person}{Niranjan Balasubramanian}, \bibinfo{person}{Tushar Khot}, {and} \bibinfo{person}{Ashish Sabharwal}.} \bibinfo{year}{2023}\natexlab{}.
\newblock \showarticletitle{Interleaving Retrieval with Chain-of-Thought Reasoning for Knowledge-Intensive Multi-Step Questions}. In \bibinfo{booktitle}{\emph{Proceedings of the 61st Annual Meeting of the {A}ssociation for {C}omputational {L}inguistics (Volume 1: Long Papers)}}. \bibinfo{publisher}{Association for Computational Linguistics}, \bibinfo{address}{Toronto, Canada}, \bibinfo{pages}{10014--10037}.
\newblock
\href{https://doi.org/10.18653/v1/2023.acl-long.557}{doi:\nolinkurl{10.18653/v1/2023.acl-long.557}}


\bibitem[Vogt and Cottrell(1999)]%
        {vogt1999fusion}
\bibfield{author}{\bibinfo{person}{Christopher~C. Vogt} {and} \bibinfo{person}{Garrison~W. Cottrell}.} \bibinfo{year}{1999}\natexlab{}.
\newblock \showarticletitle{Fusion Via a Linear Combination of Scores}.
\newblock \bibinfo{journal}{\emph{Information Retrieval}} \bibinfo{volume}{1}, \bibinfo{number}{3} (\bibinfo{year}{1999}), \bibinfo{pages}{151–173}.
\newblock
\href{https://doi.org/10.1023/a:1009980820262}{doi:\nolinkurl{10.1023/a:1009980820262}}


\bibitem[Voorhees et~al\mbox{.}(1995)]%
        {voorhees1995fusion}
\bibfield{author}{\bibinfo{person}{Ellen~M. Voorhees}, \bibinfo{person}{Narendra~K. Gupta}, {and} \bibinfo{person}{Ben Johnson-Laird}.} \bibinfo{year}{1995}\natexlab{}.
\newblock \showarticletitle{Learning collection fusion strategies}. In \bibinfo{booktitle}{\emph{Proceedings of the 18th annual international ACM SIGIR conference on Research and development in information retrieval - SIGIR ’95}}. \bibinfo{publisher}{ACM Press}, \bibinfo{pages}{172–179}.
\newblock
\href{https://doi.org/10.1145/215206.215357}{doi:\nolinkurl{10.1145/215206.215357}}


\bibitem[Wang et~al\mbox{.}(2025)]%
        {wang2025otc}
\bibfield{author}{\bibinfo{person}{Hongru Wang}, \bibinfo{person}{Cheng Qian}, \bibinfo{person}{Wanjun Zhong}, \bibinfo{person}{Xiusi Chen}, \bibinfo{person}{Jiahao Qiu}, \bibinfo{person}{Shijue Huang}, \bibinfo{person}{Bowen Jin}, \bibinfo{person}{Mengdi Wang}, \bibinfo{person}{Kam-Fai Wong}, {and} \bibinfo{person}{Heng Ji}.} \bibinfo{year}{2025}\natexlab{}.
\newblock \bibinfo{title}{Acting Less is Reasoning More! Teaching Model to Act Efficiently}.
\newblock
\showeprint[arxiv]{2504.14870}~[cs.AI]


\bibitem[Wang et~al\mbox{.}(2024)]%
        {wang2022e5}
\bibfield{author}{\bibinfo{person}{Liang Wang}, \bibinfo{person}{Nan Yang}, \bibinfo{person}{Xiaolong Huang}, \bibinfo{person}{Binxing Jiao}, \bibinfo{person}{Linjun Yang}, \bibinfo{person}{Daxin Jiang}, \bibinfo{person}{Rangan Majumder}, {and} \bibinfo{person}{Furu Wei}.} \bibinfo{year}{2024}\natexlab{}.
\newblock \bibinfo{title}{Text Embeddings by Weakly-Supervised Contrastive Pre-training}.
\newblock
\showeprint[arxiv]{2212.03533}~[cs.CL]


\bibitem[Wang et~al\mbox{.}(2023)]%
        {wang2023query2doc}
\bibfield{author}{\bibinfo{person}{Liang Wang}, \bibinfo{person}{Nan Yang}, {and} \bibinfo{person}{Furu Wei}.} \bibinfo{year}{2023}\natexlab{}.
\newblock \showarticletitle{{Q}uery2doc: Query Expansion with Large Language Models}. In \bibinfo{booktitle}{\emph{Proceedings of the 2023 Conference on Empirical Methods in Natural Language Processing}}. \bibinfo{publisher}{Association for Computational Linguistics}, \bibinfo{address}{Singapore}, \bibinfo{pages}{9414--9423}.
\newblock
\href{https://doi.org/10.18653/v1/2023.emnlp-main.585}{doi:\nolinkurl{10.18653/v1/2023.emnlp-main.585}}


\bibitem[Wang(2025)]%
        {wang2026sprig}
\bibfield{author}{\bibinfo{person}{Qizhi Wang}.} \bibinfo{year}{2025}\natexlab{}.
\newblock \bibinfo{title}{Democratizing GraphRAG: Linear, CPU-Only Graph Retrieval for Multi-Hop QA}.
\newblock
\showeprint[arxiv]{2602.23372}~[cs.IR]


\bibitem[Wang et~al\mbox{.}(2020)]%
        {wang2020minilm}
\bibfield{author}{\bibinfo{person}{Wenhui Wang}, \bibinfo{person}{Furu Wei}, \bibinfo{person}{Li Dong}, \bibinfo{person}{Hangbo Bao}, \bibinfo{person}{Nan Yang}, {and} \bibinfo{person}{Ming Zhou}.} \bibinfo{year}{2020}\natexlab{}.
\newblock \showarticletitle{MiniLM: Deep Self-Attention Distillation for Task-Agnostic Compression of Pre-Trained Transformers}. In \bibinfo{booktitle}{\emph{Advances in Neural Information Processing Systems}}, Vol.~\bibinfo{volume}{33}. \bibinfo{publisher}{Curran Associates, Inc.}, \bibinfo{pages}{5776--5788}.
\newblock
\urldef\tempurl%
\url{https://proceedings.neurips.cc/paper_files/paper/2020/file/3f5ee243547dee91fbd053c1c4a845aa-Paper.pdf}
\showURL{%
\tempurl}


\bibitem[West and Leskovec(2012)]%
        {west2012wayfinding}
\bibfield{author}{\bibinfo{person}{Robert West} {and} \bibinfo{person}{Jure Leskovec}.} \bibinfo{year}{2012}\natexlab{}.
\newblock \showarticletitle{Human wayfinding in information networks}. In \bibinfo{booktitle}{\emph{Proceedings of the 21st international conference on World Wide Web}}. \bibinfo{publisher}{ACM}, \bibinfo{pages}{619–628}.
\newblock
\href{https://doi.org/10.1145/2187836.2187920}{doi:\nolinkurl{10.1145/2187836.2187920}}


\bibitem[Wu et~al\mbox{.}(2026)]%
        {wu2025hiprag}
\bibfield{author}{\bibinfo{person}{Peilin Wu}, \bibinfo{person}{Mian Zhang}, \bibinfo{person}{Kun Wan}, \bibinfo{person}{Wentian Zhao}, \bibinfo{person}{Kaiyu He}, \bibinfo{person}{Xinya Du}, {and} \bibinfo{person}{Zhiyu Chen}.} \bibinfo{year}{2026}\natexlab{}.
\newblock \bibinfo{title}{HiPRAG: Hierarchical Process Rewards for Efficient Agentic Retrieval Augmented Generation}.
\newblock
\showeprint[arxiv]{2510.07794}~[cs.CL]


\bibitem[Wu et~al\mbox{.}(2025)]%
        {wu2025searchwisely}
\bibfield{author}{\bibinfo{person}{Peilin Wu}, \bibinfo{person}{Mian Zhang}, \bibinfo{person}{Xinlu Zhang}, \bibinfo{person}{Xinya Du}, {and} \bibinfo{person}{Zhiyu Chen}.} \bibinfo{year}{2025}\natexlab{}.
\newblock \showarticletitle{Search Wisely: Mitigating Sub-optimal Agentic Searches By Reducing Uncertainty}. In \bibinfo{booktitle}{\emph{Proceedings of the 2025 Conference on Empirical Methods in Natural Language Processing}}. \bibinfo{publisher}{Association for Computational Linguistics}, \bibinfo{address}{Suzhou, China}, \bibinfo{pages}{19723--19734}.
\newblock
\href{https://doi.org/10.18653/v1/2025.emnlp-main.998}{doi:\nolinkurl{10.18653/v1/2025.emnlp-main.998}}


\bibitem[Xiao et~al\mbox{.}(2024)]%
        {xiao2024cpack}
\bibfield{author}{\bibinfo{person}{Shitao Xiao}, \bibinfo{person}{Zheng Liu}, \bibinfo{person}{Peitian Zhang}, \bibinfo{person}{Niklas Muennighoff}, \bibinfo{person}{Defu Lian}, {and} \bibinfo{person}{Jian-Yun Nie}.} \bibinfo{year}{2024}\natexlab{}.
\newblock \showarticletitle{C-Pack: Packed Resources For General Chinese Embeddings}. In \bibinfo{booktitle}{\emph{Proceedings of the 47th International ACM SIGIR Conference on Research and Development in Information Retrieval}}. \bibinfo{publisher}{ACM}, \bibinfo{pages}{641–649}.
\newblock
\href{https://doi.org/10.1145/3626772.3657878}{doi:\nolinkurl{10.1145/3626772.3657878}}


\bibitem[Xiong et~al\mbox{.}(2021)]%
        {xiong2021mdr}
\bibfield{author}{\bibinfo{person}{Wenhan Xiong}, \bibinfo{person}{Xiang Li}, \bibinfo{person}{Srini Iyer}, \bibinfo{person}{Jingfei Du}, \bibinfo{person}{Patrick Lewis}, \bibinfo{person}{William~Yang Wang}, \bibinfo{person}{Yashar Mehdad}, \bibinfo{person}{Scott Yih}, \bibinfo{person}{Sebastian Riedel}, \bibinfo{person}{Douwe Kiela}, {and} \bibinfo{person}{Barlas Oguz}.} \bibinfo{year}{2021}\natexlab{}.
\newblock \showarticletitle{Answering Complex Open-Domain Questions with Multi-Hop Dense Retrieval}. In \bibinfo{booktitle}{\emph{International Conference on Learning Representations}}.
\newblock
\urldef\tempurl%
\url{https://openreview.net/forum?id=EMHoBG0avc1}
\showURL{%
\tempurl}


\bibitem[Xiong et~al\mbox{.}(2019)]%
        {xiong2019bridge}
\bibfield{author}{\bibinfo{person}{Wenhan Xiong}, \bibinfo{person}{Mo Yu}, \bibinfo{person}{Xiaoxiao Guo}, \bibinfo{person}{Hong Wang}, \bibinfo{person}{Shiyu Chang}, \bibinfo{person}{Murray Campbell}, {and} \bibinfo{person}{William~Yang Wang}.} \bibinfo{year}{2019}\natexlab{}.
\newblock \showarticletitle{Simple yet Effective Bridge Reasoning for Open-Domain Multi-Hop Question Answering}. In \bibinfo{booktitle}{\emph{Proceedings of the 2nd Workshop on Machine Reading for Question Answering}}. \bibinfo{publisher}{Association for Computational Linguistics}, \bibinfo{address}{Hong Kong, China}, \bibinfo{pages}{48--52}.
\newblock
\href{https://doi.org/10.18653/v1/D19-5806}{doi:\nolinkurl{10.18653/v1/D19-5806}}


\bibitem[Xu et~al\mbox{.}(2024)]%
        {xu2024recomp}
\bibfield{author}{\bibinfo{person}{Fangyuan Xu}, \bibinfo{person}{Weijia Shi}, {and} \bibinfo{person}{Eunsol Choi}.} \bibinfo{year}{2024}\natexlab{}.
\newblock \showarticletitle{{RECOMP}: Improving Retrieval-Augmented {LM}s with Context Compression and Selective Augmentation}. In \bibinfo{booktitle}{\emph{The Twelfth International Conference on Learning Representations}}.
\newblock
\urldef\tempurl%
\url{https://openreview.net/forum?id=mlJLVigNHp}
\showURL{%
\tempurl}


\bibitem[Xu et~al\mbox{.}(2009)]%
        {xu2009wikiprf}
\bibfield{author}{\bibinfo{person}{Yang Xu}, \bibinfo{person}{Gareth~J.F. Jones}, {and} \bibinfo{person}{Bin Wang}.} \bibinfo{year}{2009}\natexlab{}.
\newblock \showarticletitle{Query dependent pseudo-relevance feedback based on wikipedia}. In \bibinfo{booktitle}{\emph{Proceedings of the 32nd international ACM SIGIR conference on Research and development in information retrieval}}. \bibinfo{publisher}{ACM}, \bibinfo{pages}{59–66}.
\newblock
\href{https://doi.org/10.1145/1571941.1571954}{doi:\nolinkurl{10.1145/1571941.1571954}}


\bibitem[Yang et~al\mbox{.}(2018)]%
        {yang2018hotpotqa}
\bibfield{author}{\bibinfo{person}{Zhilin Yang}, \bibinfo{person}{Peng Qi}, \bibinfo{person}{Saizheng Zhang}, \bibinfo{person}{Yoshua Bengio}, \bibinfo{person}{William Cohen}, \bibinfo{person}{Ruslan Salakhutdinov}, {and} \bibinfo{person}{Christopher~D. Manning}.} \bibinfo{year}{2018}\natexlab{}.
\newblock \showarticletitle{{H}otpot{QA}: A Dataset for Diverse, Explainable Multi-hop Question Answering}. In \bibinfo{booktitle}{\emph{Proceedings of the 2018 Conference on Empirical Methods in Natural Language Processing}}. \bibinfo{publisher}{Association for Computational Linguistics}, \bibinfo{address}{Brussels, Belgium}, \bibinfo{pages}{2369--2380}.
\newblock
\href{https://doi.org/10.18653/v1/D18-1259}{doi:\nolinkurl{10.18653/v1/D18-1259}}


\bibitem[Yao et~al\mbox{.}(2023)]%
        {yao2023react}
\bibfield{author}{\bibinfo{person}{Shunyu Yao}, \bibinfo{person}{Jeffrey Zhao}, \bibinfo{person}{Dian Yu}, \bibinfo{person}{Nan Du}, \bibinfo{person}{Izhak Shafran}, \bibinfo{person}{Karthik Narasimhan}, {and} \bibinfo{person}{Yuan Cao}.} \bibinfo{year}{2023}\natexlab{}.
\newblock \showarticletitle{{ReAct}: Synergizing Reasoning and Acting in Language Models}. In \bibinfo{booktitle}{\emph{International Conference on Learning Representations (ICLR)}}.
\newblock


\bibitem[Yu et~al\mbox{.}(2023)]%
        {yu2023genread}
\bibfield{author}{\bibinfo{person}{Wenhao Yu}, \bibinfo{person}{Dan Iter}, \bibinfo{person}{Shuohang Wang}, \bibinfo{person}{Yichong Xu}, \bibinfo{person}{Mingxuan Ju}, \bibinfo{person}{Soumya Sanyal}, \bibinfo{person}{Chenguang Zhu}, \bibinfo{person}{Michael Zeng}, {and} \bibinfo{person}{Meng Jiang}.} \bibinfo{year}{2023}\natexlab{}.
\newblock \showarticletitle{Generate rather than Retrieve: Large Language Models are Strong Context Generators}. In \bibinfo{booktitle}{\emph{The Eleventh International Conference on Learning Representations}}.
\newblock
\urldef\tempurl%
\url{https://openreview.net/forum?id=fB0hRu9GZUS}
\showURL{%
\tempurl}


\bibitem[Zhang et~al\mbox{.}(2024)]%
        {zhang2024beam}
\bibfield{author}{\bibinfo{person}{Jiahao Zhang}, \bibinfo{person}{Haiyang Zhang}, \bibinfo{person}{Dongmei Zhang}, \bibinfo{person}{Liu Yong}, {and} \bibinfo{person}{Shen Huang}.} \bibinfo{year}{2024}\natexlab{}.
\newblock \showarticletitle{End-to-End Beam Retrieval for Multi-Hop Question Answering}. In \bibinfo{booktitle}{\emph{Proceedings of the 2024 Conference of the {N}orth {A}merican Chapter of the {A}ssociation for {C}omputational {L}inguistics: Human Language Technologies (Volume 1: Long Papers)}}. \bibinfo{publisher}{Association for Computational Linguistics}, \bibinfo{address}{Mexico City, Mexico}, \bibinfo{pages}{1718--1731}.
\newblock
\href{https://doi.org/10.18653/v1/2024.naacl-long.96}{doi:\nolinkurl{10.18653/v1/2024.naacl-long.96}}


\bibitem[Zhang et~al\mbox{.}(2025)]%
        {zhang2025qwen3embedding}
\bibfield{author}{\bibinfo{person}{Yanzhao Zhang}, \bibinfo{person}{Mingxin Li}, \bibinfo{person}{Dingkun Long}, \bibinfo{person}{Xin Zhang}, \bibinfo{person}{Huan Lin}, \bibinfo{person}{Baosong Yang}, \bibinfo{person}{Pengjun Xie}, \bibinfo{person}{An Yang}, \bibinfo{person}{Dayiheng Liu}, \bibinfo{person}{Junyang Lin}, \bibinfo{person}{Fei Huang}, {and} \bibinfo{person}{Jingren Zhou}.} \bibinfo{year}{2025}\natexlab{}.
\newblock \bibinfo{title}{Qwen3 Embedding: Advancing Text Embedding and Reranking Through Foundation Models}.
\newblock
\showeprint[arxiv]{2506.05176}~[cs.CL]


\bibitem[Zheng et~al\mbox{.}(2025)]%
        {zheng2025stepsearch}
\bibfield{author}{\bibinfo{person}{Xuhui Zheng}, \bibinfo{person}{Kang An}, \bibinfo{person}{Ziliang Wang}, \bibinfo{person}{Yuhang Wang}, {and} \bibinfo{person}{Yichao Wu}.} \bibinfo{year}{2025}\natexlab{}.
\newblock \showarticletitle{{S}tep{S}earch: Igniting {LLM}s Search Ability via Step-Wise Proximal Policy Optimization}. In \bibinfo{booktitle}{\emph{Proceedings of the 2025 Conference on Empirical Methods in Natural Language Processing}}. \bibinfo{publisher}{Association for Computational Linguistics}, \bibinfo{address}{Suzhou, China}, \bibinfo{pages}{21805--21830}.
\newblock
\href{https://doi.org/10.18653/v1/2025.emnlp-main.1106}{doi:\nolinkurl{10.18653/v1/2025.emnlp-main.1106}}


\bibitem[Zhou et~al\mbox{.}(2026)]%
        {zhou2026dynakrag}
\bibfield{author}{\bibinfo{person}{Chenyu Zhou}, \bibinfo{person}{Yaqi Wu}, \bibinfo{person}{Xiaolei Guo}, \bibinfo{person}{Jiaqi Huang}, \bibinfo{person}{Xianfa Zhang}, \bibinfo{person}{Junxu Zhang}, \bibinfo{person}{Zhuo Yu}, \bibinfo{person}{Zhubo Shi}, \bibinfo{person}{Jianghao Lin}, {and} \bibinfo{person}{Dongdong Ge}.} \bibinfo{year}{2026}\natexlab{}.
\newblock \bibinfo{title}{DynaKRAG: A Unified Framework for Learnable Evidence Control in Multi-Hop Retrieval-Augmented Generation}.
\newblock
\showeprint[arxiv]{2607.06507}~[cs.CL]


\bibitem[Zhou et~al\mbox{.}(2023)]%
        {zhou2023leasttomost}
\bibfield{author}{\bibinfo{person}{Denny Zhou}, \bibinfo{person}{Nathanael Sch{\"a}rli}, \bibinfo{person}{Le Hou}, \bibinfo{person}{Jason Wei}, \bibinfo{person}{Nathan Scales}, \bibinfo{person}{Xuezhi Wang}, \bibinfo{person}{Dale Schuurmans}, \bibinfo{person}{Claire Cui}, \bibinfo{person}{Olivier Bousquet}, \bibinfo{person}{Quoc~V Le}, {and} \bibinfo{person}{Ed~H. Chi}.} \bibinfo{year}{2023}\natexlab{}.
\newblock \showarticletitle{Least-to-Most Prompting Enables Complex Reasoning in Large Language Models}. In \bibinfo{booktitle}{\emph{The Eleventh International Conference on Learning Representations}}.
\newblock
\urldef\tempurl%
\url{https://openreview.net/forum?id=WZH7099tgfM}
\showURL{%
\tempurl}


\bibitem[Zhou et~al\mbox{.}(2024)]%
        {zhou2024webarena}
\bibfield{author}{\bibinfo{person}{Shuyan Zhou}, \bibinfo{person}{Frank~F. Xu}, \bibinfo{person}{Hao Zhu}, \bibinfo{person}{Xuhui Zhou}, \bibinfo{person}{Robert Lo}, \bibinfo{person}{Abishek Sridhar}, \bibinfo{person}{Xianyi Cheng}, \bibinfo{person}{Tianyue Ou}, \bibinfo{person}{Yonatan Bisk}, \bibinfo{person}{Daniel Fried}, \bibinfo{person}{Uri Alon}, {and} \bibinfo{person}{Graham Neubig}.} \bibinfo{year}{2024}\natexlab{}.
\newblock \showarticletitle{WebArena: A Realistic Web Environment for Building Autonomous Agents}. In \bibinfo{booktitle}{\emph{The Twelfth International Conference on Learning Representations}}.
\newblock
\urldef\tempurl%
\url{https://openreview.net/forum?id=oKn9c6ytLx}
\showURL{%
\tempurl}


\end{thebibliography}
\appendix
\section{Implementation details}
\label{app:impl}
\textit{Routes and scorer.} Routes come from links on the top three passages of the question's list and the top two of each later list (at most 20 or 10 targets per passage, ranked by title-word overlap with the question; routes already run are removed). A route runs as a BM25 search for the target title plus the first six content words of the question. The scorer's 30 features describe the route (16, including IDF-based predictors~\cite{carmel2010qpp}), the question (2) and the search history (12).

\textit{Context construction.} Seeds are retrieved passages whose title, without parenthesized disambiguation, matches whole words of the question (at most four); otherwise the first three Round-robin passages. Pages with corpus in-degree above 1{,}000 are excluded, PageRank runs 30 iterations, and near-duplicates (same normalized title and first 300 characters) enter the context once.

\textit{Title-mention links.} On 2Wiki, $a$ links to $b$ when $b$'s title (at least four characters) occurs in $a$'s text as whole words. A Qwen2.5-72B judge on 150 random links finds that 41\% refer to the intended entity, but 99\% of links between two gold pages are correct; all sampled links into pages with over 1{,}000 in-links were wrong, and the hub cut removes them.

\textit{Upper bound for rerouting.} For every flagged call, an oracle keeps its query or takes one of the five top-ranked routes, a question counts as covered if some choice retrieves every gold page, and a newly covered question is credited with the answer of a reader given only the gold pages. The agent's later queries are kept as recorded.

\textit{Training data and leakage.} The scorer is trained on roll-outs over 1{,}500 HotpotQA and 3{,}000 2Wiki training questions under several call-selection policies, labeled by the gold pages each route would add (2.1M and 0.67M candidates). Training and evaluation questions are disjoint; 3.2\% (HotpotQA) and 8.0\% (2Wiki) of evaluation gold pages are also training gold pages, and no feature identifies a page.

\section{Agents in the study}
\label{app:agents}
\tabref{tab:agents} compares how each agent reads search results in its original paper and in our protocol. Answers come from one shared reader, so that an agent alone and the agent with \method differ only in their evidence. With IRCoT's original stopping rule, the 7B model answers right after the initial search on 80.5\% of 2Wiki and 94.3\% of HotpotQA questions, leaving almost no repeated queries for \method, and its EM drops by 5.9 and 8.5 points; we therefore ask it to name the next fact to look up instead of answering, so it always makes five follow-up searches.

\begin{table}[t]
  \caption{How each agent reads search results in its original paper and in our protocol. In our protocol every agent shares the same retriever and reader.}
  \label{tab:agents}
  \centering
  \footnotesize
  \setlength{\tabcolsep}{3pt}
  \begin{tabular}{@{}>{\raggedright}p{1.15cm}>{\raggedright}p{3.55cm}>{\raggedright\arraybackslash}p{3.3cm}@{}}
    \toprule
    Agent & Native evidence; who answers & In our protocol \\
    \midrule
    IRCoT & BM25 top $K{\le}8$ per step, ${\le}15$ in arrival order; separate reader & 5 follow-up calls, top 10 per call; sees up to 8 full paragraphs \\
    ReAct & first 5 sentences of one Wikipedia page per search; the agent & BM25 top 10, sees the top 5 as full paragraphs; ${\le}5$ follow-up searches \\
    Search-R1 & E5 top 3 passages per search, $B{=}4$; the RL-trained policy & BM25 top 10, policy sees top 3; ${\le}5$ follow-up searches \\
    Plan & Our planner: one LLM call writes three sub-questions and two keyword queries & the five queries in order after the initial search, results never read \\
    \bottomrule
  \end{tabular}
\end{table}

\section{Robustness checks}
\label{app:robust}
\providecommand{\tabdag}{\rlap{$^{\dagger}$}}
\begin{table}[t]
\caption{EM (\%) of the agents alone and with \method under other settings, mean of four agents. The default row is the main setting of \tabref{tab:main}, and other rows change one setting. The $\theta$ rows change the near-repeat threshold (default $0.8$).}
\label{tab:robust}\centering
\begingroup\setlength{\aboverulesep}{0pt}\setlength{\belowrulesep}{0pt}\renewcommand{\arraystretch}{1.1}\setlength{\tabcolsep}{4pt}
\begin{tabular}{@{}l cc cc@{}}
\toprule
 & \multicolumn{2}{c}{2Wiki} & \multicolumn{2}{c}{HotpotQA} \\
\cmidrule(lr){2-3}\cmidrule(l){4-5}
Setting & Alone & \method & Alone & \method \\
\midrule
Default ($n=10$, $\theta=0.8$) & 31.6 & 35.6\tabdag & 40.6 & 42.0\tabdag \\
\midrule
Context, $n=5$ passages  & 31.9 & 36.2\tabdag & 40.3 & 43.2\tabdag \\
Context, $n=20$ passages & 33.6 & 36.0\tabdag & 40.2 & 41.5\tabdag \\
\midrule
Reader, untruncated      & 32.9 & 36.7\tabdag & 40.9 & 42.6\tabdag \\
Reader, Qwen2.5-32B      & 46.5 & 50.8\tabdag & 47.2 & 49.2\tabdag \\
\midrule
$\theta=0.6$             & 31.6 & 36.4\tabdag & 40.6 & 42.4\tabdag \\
$\theta=0.9$             & 31.6 & 35.3\tabdag & 40.6 & 41.6\tabdag \\
\bottomrule
\end{tabular}
\endgroup
\end{table}

\providecommand{\tabdag}{\rlap{$^{\dagger}$}}
\begin{table}[t]
\caption{EM and complete contexts (\%) of Search-R1 with its native E5 retriever, with each component of \method switched on, replayed on 1{,}500 questions per collection. Seeds keep only the longest of nested title matches, and the route scorer is retrained on E5 runs.}
\label{tab:e5}\centering
\begingroup\setlength{\aboverulesep}{0pt}\setlength{\belowrulesep}{0pt}\renewcommand{\arraystretch}{1.1}\setlength{\tabcolsep}{4pt}
\begin{tabular}{@{}l cc cc@{}}
\toprule
 & \multicolumn{2}{c}{EM} & \multicolumn{2}{c}{Complete (\%)} \\
\cmidrule(lr){2-3}\cmidrule(l){4-5}
Setting & 2Wiki & HotpotQA & 2Wiki & HotpotQA \\
\midrule
Agent alone             & 36.9 & 42.9        & 83.2        & 70.0        \\
Context selection only  & 36.6 & 44.2        & 84.7 & 73.9 \\
Link Rerouting only     & 37.3 & 43.2        & 84.3 & 71.2 \\
\rowcolor{detourC!8} \method & 36.8 & 44.5\tabdag & 86.3 & 75.5 \\
\bottomrule
\end{tabular}
\endgroup
\end{table}

\begin{table*}[t]
  \caption{EM (\%) of each way of choosing the route, per agent. Everything else is the same as in \method. Cross-encoders re-rank the scorer's top five routes, and the Mean columns are those of \figref{fig:rerank}.}
  \label{tab:rerank-full}
  \centering
  \begin{tabular}{@{}l ccccc@{\hspace{12pt}}ccccc@{}}
\toprule
& \multicolumn{5}{c}{2Wiki} & \multicolumn{5}{c}{HotpotQA} \\
\cmidrule(lr){2-6}\cmidrule(l){7-11}
Route selection & IRCoT & ReAct & Search-R1 & Plan & Mean & IRCoT & ReAct & Search-R1 & Plan & Mean \\
\midrule
No rerouting & 35.6 & 36.3 & 36.3 & 28.3 & 34.1 & 43.9 & 43.5 & 42.4 & 35.9 & 41.4 \\
Random link & 36.5 & 35.8 & 36.4 & 29.7 & 34.6 & 44.4 & 43.8 & 42.2 & 36.0 & 41.6 \\
Title match & 36.2 & 37.1 & 36.7 & 29.7 & 34.9 & 44.3 & 44.2 & 42.3 & 36.9 & 41.9 \\
\rowcolor{detourC!8}
LightGBM (\method) & 37.8 & 36.7 & 36.5 & \textbf{31.5} & 35.6 & 44.7 & 44.1 & 42.5 & 36.8 & 42.0 \\
\quad + MiniLM cross-encoder (22M) & \textbf{38.1} & 37.0 & 36.6 & \textbf{31.5} & \textbf{35.8} & 44.5 & 43.9 & \textbf{43.1} & \textbf{37.5} & \textbf{42.2} \\
\quad + bge reranker (278M) & 37.9 & \textbf{37.6} & 36.6 & 31.2 & \textbf{35.8} & \textbf{44.8} & \textbf{44.4} & 42.6 & 37.2 & \textbf{42.2} \\
\quad + Qwen3-Reranker (0.6B) & 37.1 & 36.9 & \textbf{37.0} & 30.9 & 35.4 & 44.5 & 43.7 & 43.0 & 37.4 & 42.1 \\
\bottomrule
\end{tabular}

\end{table*}
\tabref{tab:robust} varies the default settings, which all agents and collections share.  Taking links from fewer or more passages per list, or training the route scorer on the other collection or on a tenth or a third of the 2Wiki training questions, changes EM by at most 0.3 points.

\textit{Dense retriever.} With Search-R1's native E5 retriever (\tabref{tab:e5}), only 12\% of calls are near-repeats and 83\% of the 2Wiki contexts are already complete, so \method adds complete contexts on both collections but EM only on HotpotQA. Under E5, seeds keep only the longest of nested title matches, because E5 also retrieves pages whose shorter titles sit inside the named entity (\emph{Iron Man (1951 film)} for \emph{The Iron Man}), and with the main-study seed rule these wrong-entity seeds lower 2Wiki EM by 1.1 points instead of 0.1. We chose this rule after inspecting evaluation questions. Retraining the scorer on E5, or also flagging calls whose results were mostly seen, changes EM by at most 0.4 points.

\textit{Why complete contexts rise without EM under E5.} With the main-study seed rule, context selection alone raises complete 2Wiki contexts by 1.5 points but lowers EM by 1.5. The loss comes from contexts that were already complete (71 answers turn wrong, 56 turn right), and rarely from removed evidence. Of the contexts that turn wrong, 38\% keep the same ten passages in a new order, and only 1\% of the passages pushed out contain the answer. Instead, these questions have 0.72 wrong-entity seeds on average (0.24 under BM25), and PageRank from them moves pages about the wrong entity to the front, which the 7B reader follows; with the same passages in Round-robin order, 53 of the 71 losses disappear, and Qwen2.5-72B as reader shows no net loss. A stronger retriever thus leaves context selection less to complete.

\section{Route selection per agent}
\label{app:rerank}
\tabref{tab:rerank-full} gives the per-agent results behind \figref{fig:rerank}.

\section{Gains by question type and agent}
\label{app:gains}
\textit{Question types.}
Across collections, association routes add more reach over semantic
routes on 2Wiki than on HotpotQA (14.5 against 9.1 points for the best
five calls; \tabref{tab:reach}), and \method's gain is also larger on
2Wiki (\tabref{tab:main}). The same holds across question types. On
HotpotQA, bridge questions, whose second page is named only in the
first, gain 1.8 EM points (significant), while
comparison questions, which name both pages, gain nothing
($-0.2$). On 2Wiki, the gain is largest
for inference and compositional questions ($+4.6$, $+4.4$), followed by
bridge-comparison ($+4.0$) and comparison questions ($+2.9$).

\textit{Agents.}
Rerouting matters most for IRCoT and for Plan, which writes all
its queries before reading any result. It matters least for ReAct and Search-R1, which decide after each search whether and what to search next ($+0.4$ and $+0.2$ on 2Wiki).
One reason is that routed pages partly substitute for later searches:
the agent finds 18\% and 14\% of the gold pages that rerouted calls
retrieve on its own, from 2--3\% for IRCoT to 68\% and 33\% for
Search-R1.

\section{Live web pilot}
\label{app:web}
To test \method beyond curated collections, we ran ReAct, IRCoT and Plan with live web search (DeepSeek Search, top 10 results with their page text; October 2026) on 200 questions per collection from the main study. Before the agent reads a result list, we drop pages that may contain a question with its answer, namely dataset mirrors (Hugging Face, GitHub and similar sites), question-answer and quiz sites, and pages that contain the question verbatim. Web results rarely link to one another and many sites block crawlers, so associations follow the 2Wiki rule on the returned text, and a page is associated with an entity when the entity's Wikipedia title occurs in it, dropping common words and titles that occur for many questions. Two rules are adapted to web pages. A result is a seed when its cleaned title contains a question entity, and routes are multi-word entity titles that co-occur with a question entity, ranked by co-occurrence and searched with the first six content words of the question. These rules were fixed on the first 50 HotpotQA questions and applied unchanged to 2Wiki and to all three agents. Search-R1 is not included because it needs its own released checkpoint. Queries longer than the search service's limit of 128 characters are truncated.

On the web, associations come only from entity mentions in the returned text and are much noisier than corpus links. Even so, \method keeps a positive average gain without extra calls. Averaged over the three agents, EM rises by 1.3 points on 2Wiki, mostly through context selection, and by 0.2 on HotpotQA (\tabref{tab:web}). With 200 questions per collection, neither gain is significant. The room for gains is smaller than on the curated collections, because web search already puts the answer in many contexts and result pages rarely keep their links, so rerouting adds little. Most remaining errors come from entities that share a name with the intended one and from comparison questions whose seeds cover only one of the two entities. These errors need better entity matching, and a better route choice does not fix them.
\providecommand{\tabdag}{\rlap{$^{\dagger}$}}
\begin{table}[t]
\caption{EM (\%) of three agents with live web search (DeepSeek Search), 200 questions per collection.}
\label{tab:web}\centering
\begingroup\setlength{\aboverulesep}{0pt}\setlength{\belowrulesep}{0pt}\renewcommand{\arraystretch}{1.1}\setlength{\tabcolsep}{4pt}
\begin{tabular}{@{}l cc@{}}
\toprule
Agent & 2Wiki & HotpotQA \\
\midrule
ReAct & 39.5 & 37.5 \\
\rowcolor{detourC!8} \quad + \method & 41.0 & 37.5 \\
IRCoT & 37.5 & 31.5 \\
\rowcolor{detourC!8} \quad + \method & 36.5 & 31.5 \\
Plan  & 30.0 & 31.5 \\
\rowcolor{detourC!8} \quad + \method & 33.5 & 32.0 \\
\midrule
Mean  & 35.7 & 33.5 \\
\rowcolor{detourC!8} \quad + \method & 37.0 & 33.7 \\
\bottomrule
\end{tabular}
\endgroup
\end{table}

\end{document}